\documentclass[preprint,12pt]{elsarticle}

\usepackage{lineno}
\usepackage{amsmath}
\usepackage{mathtools}
\usepackage{listings}
\usepackage{graphicx}
\usepackage[dvipsnames]{xcolor}
\usepackage{hyperref}
\usepackage{multirow}

\graphicspath{
  {images/}
}

\lstdefinestyle{fortranstyle}{
  language=Fortran,
  basicstyle=\ttfamily\fontsize{8pt}{9.6pt}\selectfont,
  keywordstyle=\color{blue},
  commentstyle=\color{cyan},
  stringstyle=\color{red},
}

\lstdefinestyle{cstyle}{
  language=C,
  basicstyle=\ttfamily\fontsize{8pt}{9.6pt}\selectfont,
  keywordstyle=\color{blue},
  commentstyle=\color{cyan},
  stringstyle=\color{red},
}

\lstdefinestyle{cppstyle}{
  language=C++,
  basicstyle=\ttfamily\fontsize{8pt}{9.6pt}\selectfont,
  keywordstyle=\color{blue},
  commentstyle=\color{cyan},
  stringstyle=\color{red},
}

\lstdefinestyle{cmakestyle}{
  language=bash,
  basicstyle=\ttfamily\fontsize{8pt}{9.6pt}\selectfont,
  keywordstyle=\color{blue},
  commentstyle=\color{cyan},
  stringstyle=\color{red},
}

\lstdefinestyle{bashstyle}{
  language=bash,
  basicstyle=\ttfamily\fontsize{8pt}{9.6pt}\selectfont,
  keywordstyle=\color{blue},
  commentstyle=\color{cyan},
  stringstyle=\color{red},
}

\newcommand{\tth}{\ttfamily\hyphenchar\font=45\relax}
\renewcommand{\_}{\textunderscore\discretionary{-}{}{}}

\newcommand{\laura}{{\sc Laura}$^{++}$}
\newcommand{\tfa}{{\sc TFA2}}
\newcommand{\tensorflow}{{\sc TensorFlow}}
\newcommand{\vecampfit}{{\sc VecAmpFit}}

\newcommand{\Dfun}[6]{D^{#1}_{#2\,#3}(#4, #5, #6)}
\newcommand{\dfun}[4]{d^{#1}_{#2\,#3}(#4)}

\newcommand{\vpho}{\gamma^*}
\newcommand{\elp}{e^+}
\newcommand{\elm}{e^-}
\newcommand{\mup}{\mu^+}
\newcommand{\mum}{\mu^-}
\newcommand{\lp}{\ell^+}
\newcommand{\lm}{\ell^-}
\newcommand{\pip}{\pi^+}
\newcommand{\pim}{\pi^-}
\newcommand{\piz}{\pi^0}
\newcommand{\piI}{\pi^{(1)}}
\newcommand{\piII}{\pi^{(2)}}
\newcommand{\rhoj}{\rho_J}

\newcommand{\ajI}{a_J^{(1)}}
\newcommand{\ajII}{a_J^{(2)}}
\newcommand{\fj}{f_J}
\newcommand{\phij}{\phi_J}
\newcommand{\kb}{\bar{K}}
\newcommand{\kp}{K^+}
\newcommand{\km}{K^-}

\newcommand{\ks}{K^0_S}

\newcommand{\kI}{K^{(1)}}
\newcommand{\kII}{\bar{K}^{(2)}}
\newcommand{\kst}{K^*}
\newcommand{\kstb}{\bar{K}^*}

\newcommand{\kstjIaI}{K_J^{*(11)}}
\newcommand{\kstjIaII}{K_J^{*(12)}}
\newcommand{\kstjIIaI}{\bar{K}_J^{*(21)}}
\newcommand{\kstjIIaII}{\bar{K}_J^{*(22)}}
\newcommand{\kj}{K_J}

\newcommand{\kjI}{K_J^{(1)}}
\newcommand{\kjII}{\bar{K}_J^{(2)}}
\newcommand{\Dz}{D^0}

\newcommand{\psiIIs}{\psi(2S)}
\newcommand{\Bp}{B^+}

\newcommand{\Bz}{B^0}

\newcommand{\kkpipi}{\elp \elm \to K \bar{K} \pi \pi}
\newcommand{\kpkmpippim}{\elp \elm \to \kp \km \pip \pim}

\newcommand{\kskspippim}{\elp \elm \to \ks \ks \pip \pim}

\newcommand{\kIkIIpiIpiII}{\elp \elm \to \kI \kII \piI \piII}
\newcommand{\kmpippiz}{\Dz \to \km \pip \piz}
\newcommand{\kpkpkm}{\Bp \to \kp \kp \km}

\newcommand{\mev}{\mathrm{MeV}}
\newcommand{\gevc}{\mathrm{GeV}/c}

\DeclareMathOperator{\Sig}{Sig}

\journal{Computer Physics Communications}

\begin{document}

\begin{frontmatter}

\title{\vecampfit{}: vectorized amplitude-analysis fitting library}

\author[a]{Kirill Chilikin\corref{author}}
\cortext[author] {Corresponding author.\\\textit{E-mail address:} K.A.Chilikin@inp.nsk.su}
\address[a]{Budker Institute of Nuclear Physics SB RAS, Novosibirsk, Russia}

\begin{abstract}
A new library \vecampfit{} for multidimensional amplitude analyses
in high-energy physics has been developed for an ongoing amplitude analysis
at Belle II experiment. It includes a fitter performing
likelihood calculation and explicitly-vectorized subprograms for amplitude
implementation. The fitter supports explicit gradient calculation and
simultaneous fitting of multiple data sets.
\end{abstract}


\end{frontmatter}


{\bf PROGRAM SUMMARY}

\begin{small}
\noindent
{\em Program Title: \vecampfit{}}                                          \\
{\em Developer's repository link:} {\url{https://gitlab.desy.de/chilikin/vecampfit}} \\
{\em Licensing provisions:} Apache License, Version 2.0       \\
{\em Programming language:} Fortran, C++, C                   \\
{\em Nature of problem:} Amplitude analyses require a large amount
of calculations, especially for evaluation of signal-density normalization
integrals. Thus, optimization of amplitude-analysis fitting programs is
important for efficient data analysis at high-energy-physics experiments. \\
{\em Solution method:} \vecampfit{} library is designed to achieve high
performance of the analysis code by vectorization of data storage and
amplitude calculations, providing a number of vectorized subroutines
for physical quantities and mathematical functions, and calculation
of likelihood gradient. Parallel calculations are possible
using {\sc OpenMP}. \\
\end{small}

\section{Introduction}

\subsection{Amplitude analyses and their computational requirements}

Particle decays, reactions between the beam and target, or beam-annihilation
processes are described by their complex amplitude.
The general form of the signal density for particle decays is
\begin{equation}
S(\Phi, p_S) = \sum\limits_f \sum\limits_{i j} \rho_{i j}
  \left[\sum\limits_k A_{i k f}(\Phi, p_S)\right]
  \left[\sum\limits_{k'} A_{j k' f}^*(\Phi, p_S)\right]
\label{eq:signal_density}
\end{equation}
where $\Phi$ is the decay phase space, $p_S$ are signal-density parameters,
$i$ and $j$ are the initial state, $k$ and $k'$ are the intermediate state,
$f$ is the final state, $\rho$ is the initial-state density matrix,
$A$ is the decay amplitude, and asterisk denotes complex conjugation.
Fitting experimental-data events to signal-density expressions similar
to Eq.~\eqref{eq:signal_density} is referred to as amplitude analysis.
Its simplest case is an analysis of a three-body decay of
a spin-zero particle such as $B$ or $D$
which is referred to as a Dalitz analysis and corresponds to
a two-dimensional phase space, which may be represented, for example,
as two squared invariant masses of the final-state particle
pairs $m_{12}^2$ and $m_{23}^2$.

For more complex cases, the decay phase space has more than two dimensions.
Such decays have four or more final-state particles e.g.
$D^0 \to \km \pi^+ \pi^+ \pi^-$, possibly including subsequent decays
of relatively narrow states
e.g. $\Bp \to J/\psi (\to \mu^+ \mu^-) \phi (\to \kp \km) \kp$,
or additional angular dependence due to non-zero spin of the initial
particle, e.g. $\elp \elm \to \kp \km \piz$.
Amplitude analyses of the decays with three or more phase-space dimensions are
referred to as multidimensional. Multidimensional amplitude analysis is
the most sensitive method for analyzing complex decay chains since it
allows to use all available information about the dependency of the signal
density on the phase-space variables. Such analyses provide many interesting
physics results. Some examples from the field of quarkonium spectroscopy,
which motivated this work, include determination of the $\chi_{c1}(3872)$
quantum numbers~\cite{LHCb:2013kgk,LHCb:2015jfc},
studies of the $T_{c\bar{c}1}(4430)^+$~\cite{Belle:2013shl,LHCb:2014zfx},
observation of multiple exotic states in
$\Bp \to J/\psi \phi \kp$~\cite{LHCb:2016axx,LHCb:2021uow},
determination of the $T_{b\bar{b}1}(10610)^+$ and $T_{b\bar{b}1}(10650)^+$
quantum numbers~\cite{Belle:2014vzn}, and others.

In a real experiment, the probability density function of the observed events
corresponding to the signal density given by Eq.~\eqref{eq:signal_density}
is modified by the detection efficiency, resolution, and presence
of background:
\begin{equation}
\begin{aligned}
\rho(\Phi, p_S, p_B) = &
  f_S \frac{\int R(\Phi, \Phi') \epsilon(\Phi') S(\Phi', p_S) d \Phi'}
           {\int \int R(\Phi, \Phi') \epsilon(\Phi') S(\Phi', p_S)
            d \Phi' d \Phi} \\
&
  + \sum\limits_k
      f_B^{(k)} \frac{\epsilon(\Phi) B^{(k)}(\Phi, p_B^{(k)})}
                     {\int \epsilon(\Phi) B^{(k)}(\Phi, p_B^{(k)}) d \Phi}, \\
\label{eq:probability_density_resolution}
\end{aligned}
\end{equation}
where $\Phi'$ is the phase space for true kinematic parameters,
$\Phi$ is the phase space for reconstructed kinematic parameters,
$f_S$ is the signal fraction, $k$ is the number of the background source,
$f_B^{(k)}$ is the background fraction,
$R$ is the resolution function, $\epsilon$ is the signal detection
efficiency, $B^{(k)}$ is the background density, and
$p_B^{(k)}$ are the background-density parameters.
Here the background density function is defined so that it needs
to be multiplied by the signal efficiency to get the actual
background distribution:
\begin{equation}
B^{(k)}_\text{(obs)}(\Phi, p_B^{(k)}) = B^{(k)}(\Phi, p_B^{(k)}) \epsilon(\Phi),
\label{eq:background_density_definition}
\end{equation}
where $B^{(k)}_\text{(obs)}$ is the observed background distribution.
It is not necessary to parameterize the efficiency if this background-density
definition is used.
Although usually a single background density function is used,
multiple background sources may be necessary, for example,
if a specific peaking-background source is described from Monte Carlo (MC)
simulation, while the remaining background is described by its distribution
in a control region.

The resolution term vanishes in the signal normalization integral
after performing integration over $\Phi$ first:
\begin{equation}
\begin{aligned}
\rho(\Phi, p_S, p_B) = &
  f_S \frac{\int R(\Phi, \Phi') \epsilon(\Phi') S(\Phi', p_S) d \Phi'}
    {\int \epsilon(\Phi') S(\Phi', p_S) d \Phi'} \\
&
  + \sum\limits_k
      f_B^{(k)} \frac{\epsilon(\Phi) B^{(k)}(\Phi, p_B^{(k)})}
                     {\int \epsilon(\Phi) B^{(k)}(\Phi, p_B^{(k)}) d \Phi}. \\
\end{aligned}
\label{eq:probability_density_integrated_resolution}
\end{equation}
Unbinned maximum-likelihood fits are performed using this general
density-function form as discussed in more detail in Sec.~\ref{sec:formalism}.
The background density is often determined from a control region before
the signal fit, although simultaneous fitting of the background and signal
regions is also possible. In this case the background normalization integral
needs to be calculated only one time before the fit. The signal normalization
integral needs to be calculated for each likelihood-function call.

In case of the isobar model (coherent sum of individual-resonance amplitudes)
the amplitude can be represented in the following form:
\begin{equation}
A = \sum\limits_k a_{i k f} R_{i k f}(\Phi, p_S),
\end{equation}
where $a_{i k f}$ are complex amplitudes and $R_{i k f}(\Phi, p_S)$ are resonant
shapes for the initial state $i$, intermediate state $k$, and final state $f$.
If the parameters of intermediate resonances are not varied, the signal
normalization integral reduces to
\begin{equation}
\sum\limits_f \sum\limits_{i j} \rho_{i j} \sum\limits_{k,k'}
  a_{i k f} a_{j k' f}^*
  \int \epsilon(\Phi) R_{i k f}(\Phi, p_S) R_{j k' f}^*(\Phi, p_S) d \Phi,
\end{equation}
where all integrals need to be calculated only once. Generally, however, such
reduction is impossible. If the lineshape of any individual contribution is
determined from the fit, then full integration needs to be performed 
at each minimization step.

This necessity of numeric calculation of the signal-density integral results in
fitting programs requiring a large processing time. Their optimization is
important for efficient experimental data analysis. The optimization is
a common task for multiple physics analyses, as integration methods and
various amplitude expressions are common for different analyses.
This motivates development of specialized amplitude-analysis frameworks.

\subsection{Optimization methods and implementation approaches}

There are several possible ways to reduce the fitting time.
The first one is optimization of the density-function calculation.
The density-function implementation needs to avoid any unnecessary calculations.
Values depending on the phase-space point but not fit parameters such as
angular distributions need to be calculated in advance and stored in buffers.
Similarly, values depending on fit parameters only need to be calculated
at the beginning of each likelihood-function call. Finally, the calculation
should avoid any conditional checks such as whether a resonance is included
into the model if it is possible to perform such checks at compilation time.
Another way of calculation optimization is usage of hardware parallelism.
One can use the fact that both data events and normalization
integrals require computation of the same density function for a large number
of phase-space points and utilize single-instruction multiple-data (SIMD) code.
SIMD instructions perform the same operation for each element of
a fixed-length vector register.

The second method is optimization of the minimization procedure.
Gradient-minimization algorithms such as the Fletcher's
algorithm~\cite{Fletcher:1970kfz} known as {\tth MIGRAD} in
{\sc MINUIT}~\cite{James:1975dr} perform calculations of gradient and
Hessian. Both gradient and Hessian can be calculated symbolically or obtained
from the signal-density functions by automatic-differentiation algorithms,
and then used by the minimization instead of the results of numerical
calculations. The function calls with gradient calculation take a longer time
but the necessary number of the calls becomes lower.

Finally, the execution time of a single fit can be reduced by performing
parallel calculations.
The parallel calculations can be performed using multiple threads on the
same machine, for example, via OpenMP~\cite{openmp}.
Another way is to perform parallel calculations with different processes
on different machines; the process communication can be implemented using
OpenMPI~\cite{gabriel04:_open_mpi}.
Finally, one can use offloading to GPUs, which also perform calculations
in parallel, for example, via OpenACC~\cite{openacc}.

There are several different approaches to amplitude-analysis
programming. Some frameworks, including
\laura{}~\cite{Back:2017zqt},
{\sc AmpTools}~\cite{matthew_shepherd_2024_10961168}, and
{\sc PAWIAN}~\cite{pawian} perform dynamic construction of the amplitude
at runtime from pre-programmed parts such as resonance lineshapes
using either configuration within the program itself (\laura{})
or via configuration files. The amplitude generator
{\sc AmpGen}~\cite{ampgen} generates C++ code for calculation of isobar-model
amplitude with contributions described in a configuration file
in covariant formalism or helicity formalism with partial-wave
amplitudes (canonical) formalism~\cite{JPAC:2017vtd}.
Finally, several frameworks
perform calculation of the amplitude specified as a mathematical function
by constructing computational graphs and performing their evaluation
using a computational backend such as
\tensorflow{}~\cite{tensorflow2015-whitepaper}.
{\sc TensorFlowAnalysis2}~\cite{tfa2} and {\sc TF-PWA}~\cite{Jiang:2024vbw}
are based on \tensorflow{} while
the ComPWA project~\cite{fritsch_2022_6908150} can use several different
computational backends including {\sc JAX}~\cite{jax2018github},
{\sc Numpy}~\cite{numpy}, and \tensorflow{}.

\subsection{\vecampfit{} library}

\vecampfit{} is a new fitting framework
developed for an ongoing multidimensional amplitude
analysis of $B \to \psi(2S) \pi K$ with a simultaneous
fit to all four isospin-related $B$ decay channels: $\Bz \to \psiIIs \pim \kp$,
$\Bp \to \psiIIs \pip \ks$, $\Bp \to \psiIIs \piz \kp$, and
$\Bz \to \psiIIs \piz \ks$, and reconstruction
of the $\psi(2S)$ in two different decay channels, $\lp \lm$ (combination
of $\elp \elm$ and $\mup \mum$) and $J/\psi \pip \pim$,
at Belle II experiment~\cite{Belle-II:2010dht}.
It is a continuation of amplitude-analysis work performed
at Belle~\cite{Belle:2000cnh,Belle:2012iwr} and Belle II since 2009
but no existing analysis is based on the library yet.
The original codebase has been created in 2017 from the code used for Belle
amplitude analyses~\cite{Belle:2013shl,Belle:2014nuw,Belle:2017egg}
and consisted of a C++ fitter, several non-vectorized physical and mathematical
functions written in then-current Fortran 2008, and the library Fortran module.
Most of the original code is no longer in active use:
the functions are vectorized and the fitter has been reimplemented
in Fortran.
Mixed-language programming is used: the amplitude calculation is implemented
in Fortran 2018, while the general fit control is performed by C++ code
(C++ 20 or later standard if it is used to build {\sc ROOT} is required).
Modern Fortran has been chosen for implementation because
it is more convenient for fixed-length vectorized calculations than C++.
Using modules and generic interfaces provides an easy way to change
the vector length without any code modifications.

Model configuration is performed at {\sc cmake} configuration and compilation
time using {\tth awk} templates and C preprocessor, resulting in specific-model
Fortran sources. Within a specific model, the library follows
imperative-programming approach: the signal and background density functions
need to be implemented by the user. Thus, \vecampfit{} is not limited to any
specific model; instead, it provides a fitter and functions necessary to build
the amplitude. It is possible to implement fits in any formalism extending
the set of \vecampfit{} physical functions as necessary with a limitation
that convolution with resolution functions is not supported yet;
it needs to be supported by the \vecampfit{} fitter itself and is planned
to be included in future library updates.
All \vecampfit{} subprograms intended to be used in amplitudes are
vectorized and operate on short data vectors of a fixed length.
This is the origin of the library name.
Values depending on the phase-space point are stored in vectorized buffers;
values depending on fit parameters only can also be bufferized.
The fitter supports minimization with explicit calculation of gradient,
which needs to be calculated and implemented by the user
(automatic differentiation is not supported), but not all library subprograms
have versions with gradient calculation yet.
Support for automatic differentiation has not been considered yet;
a possible way to do this is by using Enzyme automatic-differentiation
framework~\cite{enzyme_1,enzyme_2,enzyme_3} operating on LLVM internal
representation, however, {\sc flang} was not able to compile \vecampfit{}
until the very recent LLVM version 22.1.0 due to insufficient Fortran 2018
support. Thread-level parallel calculations are possible
using OpenMP, but parallel calculations on different machines
using OpenMPI are not supported. Offloading to GPUs is
supported but requires further optimization.
It requires a compiler that can perform offloading using OpenACC.

\vecampfit{} consists of several parts. Fortran processing tool
{\tth vecampfit-fortran} is used during building for automatic generation
of Fortran interfaces, C headers, and documentation.
The library {\tth vecampfit-real} contains subprograms for operations on
representation of real numbers such as extraction or setting of the mantissa,
transfer and type-conversion functions for integer and real numbers.
The library {\tth vecampfit-arbitrary-precision} implements integer arithmetic,
operations on arbitrary-precision integer, real, and complex numbers,
and functions calculated using arbitrary-precision arithmetics.
The arbitrary-precision operations are used to calculate mathematical
constants such as $e$ or $\pi$ and series coefficients used in implementation
of vectorized functions.

The main library {\tth vecampfit} contains the implementation of
the basic complex-vector type, mathematical constants and functions,
interpolation, linear algebra, numeric integration, statistical functions and
tests, physical functions useful for implementation of amplitudes related
to kinematics, high-energy physics, and quantum mechanics,
functions for generation of Monte Carlo events, particle data,
ROOT interface~\cite{Brun:1997pa}, and the fitter. In addition,
it implements specialized neural networks used for parameterization
of efficiency and background distributions, which are experimental and
are not described here.

The remainder of this paper is organized as follows.
The amplitude-analysis formalism is described in Sec.~\ref{sec:formalism}.
Example of $\kmpippiz$ Dalitz analysis is then presented
in Sec.~\ref{sec:kmpippiz}. This example is used to describe the details
of implementation of the amplitude-analysis formalism in \vecampfit{}
in Sec.~\ref{sec:formalism_implementation} and fitting procedure
in Sec.~\ref{sec:fitting_procedure}.
A complex example of analysis workflow including efficiency, resolution,
background, and signal fits
is presented in Sec.~\ref{sec:epem_kkpipi_workflow} using
a simultaneous fit of the processes $\kpkmpippim$ and $\kskspippim$
at multiple beam energies.
Performance studies are presented
in Sec.~\ref{sec:performance}.
\vecampfit{} auxiliary parts
and subprograms intended for implementation of density functions
are described in~\ref{sec:subprogram_overview}.
Derivation of $\kkpipi$ signal amplitude used in examples is presented
in~\ref{sec:kkpipi_amplitude}.

\section{Amplitude-analysis formalism}
\label{sec:formalism}

\subsection{Unbinned maximum-likelihood fit}

If resolution can be ignored, one can set
$R(\Phi, \Phi') = \delta(\Phi, \Phi')$ and the probability density given by
Eq.~\eqref{eq:probability_density_integrated_resolution} becomes
\begin{equation}
\rho(\Phi, p_S, p_B) =
  f_S \frac{\epsilon(\Phi) S(\Phi, p_S)}
           {\int \epsilon(\Phi) S(\Phi, p_S) d \Phi}
  + \sum\limits_k f_B^{(k)}
    \frac{\epsilon(\Phi) B^{(k)}(\Phi, p_B^{(k)})}
         {\int \epsilon(\Phi) B^{(k)}(\Phi, p_B^{(k)}) d \Phi}.
\label{eq:probability_density}
\end{equation}
If the resolution cannot be ignored, then it is necessary to perform
convolution of the signal density and resolution numerically.
Fitting with resolution is not supported in \vecampfit{} yet.

In case of a standard maximum-likelihood fit which does not determine
the normalization, the background fractions $f_B^{(k)}$ are fit parameters
and the signal fraction $f_S$ is calculated as
\begin{equation}
f_S = 1 - \sum\limits_k f_B^{(k)}.
\label{eq:signal_fraction}
\end{equation}
In case of an extended maximum-likelihood fit, the signal yield $N_S$ and
background yields $N_B^{(k)}$ are fit parameters. The fractions are given by
\begin{equation}
f_S = \frac{N_S}{N},\ f_B^{(k)} = \frac{N_B^{(k)}}{N},
\end{equation}
where $N$ is the total yield given by
\begin{equation}
N = N_S + \sum\limits_k N_B^{(k)}.
\end{equation}

The probability density for the yield is given by the Poisson distribution
\begin{equation}
\rho(N) = \frac{N^{N_\text{data}} \exp(-N)}{\Gamma(N_\text{data} + 1)}, \\
\label{eq:extended_probability_density}
\end{equation}
where $N_\text{data}$ is the number of observed data events and
factorial is replaced by the equivalent $\Gamma$ function to use $\ln \Gamma$
for logarithm calculation.

The number of background sources is set by the user; the default number is zero.
The background density has to be predetermined from a separate fit;
variation of the background parameters during signal fit
is not supported. The background fractions $f_B^{(k)}$ or yields $N_B^{(k)}$
should be fit parameters; their numbers needs to be set by the user
for each background source before loading of fit data.
The background-fraction parameter may either be set entirely free
or constrained from a separate fit, for example, to the invariant
mass of the parent particle or the difference between the reconstructed energy
and half of beam energy $\Delta E = E - \sqrt{s}/2$ in case of production of
a particle-antiparticle pair at a $e^+ e^-$ collider.

Three ways of calculation of the normalization integral
$\int \epsilon(\Phi) S(\Phi) d \Phi$ are implemented
in \vecampfit{}: Monte Carlo, grid, and formula integration.
In addition, calculation of the normalization integral can be disabled.
The integration method can be set by calling
{\tth vecampfit\_fitter\_set\_normalization\_integration\_type()};
the default method is MC integration.

\subsubsection{Monte Carlo integration}

In case of MC integration, the normalization integrals are calculated using
MC events. The MC events are generated using some density function
$d_\text{MC}(\Phi)$ and then passed through detector simulation, which changes
their distribution to $d_\text{MC}(\Phi) \epsilon(\Phi)$.
While common first choice is the uniform distribution $d_\text{MC}(\Phi) = 1$,
it is recommended to increase the number of events in the regions of higher
signal density such as narrow resonances or regions of rapid signal-density
change such as $\rho$-$\omega$ interference to improve integration
precision.

The signal normalization integral is given by
\begin{equation}
\int \epsilon(\Phi) S(\Phi, p_S) d \Phi =
  \frac{\bar{\epsilon}}{N_{\text{MC}}} \sum\limits_j w_j S(\Phi_j, p_S),
\end{equation}
where
\begin{equation}
\bar{\epsilon} = \int \epsilon(\Phi) d \Phi,
\end{equation}
$N_{\text{MC}}$ is the number of MC events,
the index $j$ runs over normalization MC events,
and $w_j$ are MC weights. The weights need to be assigned by the user and
their choice is not restricted by the library. It is necessary to take into
account the density function used for generation:
\begin{equation}
w_j \propto \frac{1}{d_\text{MC}(\Phi_j)},
\end{equation}
but the weights may also include other effects such as the difference
of the efficiency between the data and MC.
The weights may be scaled arbitrarily as their multiplication
by a common constant adds a constant to the likelihood function
and does not affect the fit results.

Example weight definition taking the difference
of the efficiency between the data and MC into account is
\begin{equation}
w_j = \frac{R(\Phi_j)}{d_\text{MC}(\Phi_j)},
\label{eq:weight}
\end{equation}
where $R(\Phi_j)$ is the ratio between the efficiencies in data and MC,
which can be determined as
\begin{equation}
R(\Phi) = \prod\limits_k \frac{\epsilon_\text{data}^{(k)}(\Phi)}
                              {\epsilon_\text{MC}^{(k)}(\Phi)},
\end{equation}
where $k$ is the final-state particle index, $\epsilon_\text{data}^{(k)}$
and $\epsilon_\text{MC}^{(k)}$ are the efficiencies in data and MC,
respectively. The efficiency ratios for individual final-state particles
need to be determined from control samples.

The function to be minimized for an unbinned maximum-likelihood fit is
\begin{equation}
\mathcal{F} =
\sum_i - 2 \ln \left[
  f_S \frac{S(\Phi_i, p_S)}{\sum\limits_j w_j S(\Phi_j, p_S)} +
  \sum\limits_k
    f_B^{(k)} \frac{B^{(k)}(\Phi_i, p_B^{(k)})}
                   {\sum\limits_j w_j B^{(k)}(\Phi_j, p_B^{(k)})}
\right] - 2 \ln \frac{\epsilon(\Phi_i) N_{\text{MC}}}{\bar{\epsilon}},
\label{eq:likelihood_0}
\end{equation}
where the index $i$ runs over data events. The second logarithm
in Eq.~\eqref{eq:likelihood_0} is constant and it can be omitted.
The \vecampfit{} implementation provides also a way to add data weights $w_i$
and parameter prior likelihoods:
\begin{equation}
\begin{aligned}
\mathcal{F} = &
\sum\limits_i - 2 w_i \ln \left[
  f_S \frac{S(\Phi_i, p_S)}{\sum\limits_j w_j S(\Phi_j, p_S)} +
  \sum\limits_k
    f_B^{(k)} \frac{B^{(k)}(\Phi_i, p_B^{(k)})}
                   {\sum\limits_j w_j B^{(k)}(\Phi_j, p_B^{(k)})}
\right] \\
& + \sum\limits_l C_l(p_S^{(n_l)}) + C(p_S),
\end{aligned}
\label{eq:likelihood_vecampfit}
\end{equation}
where $l$ is the prior-likelihood number, $n_l$ is the parameter number
for $l$-th prior-likelihood, $C_l$ is a library-defined
transformed prior-likelihood function ($-2 \ln L_\text{prior}$),
and $C$ is a user-defined transformed prior-likelihood function.
A common case of asymmetric-Gaussian prior likelihoods is implemented
in the library:
\begin{equation}
C_l(p_S^{(n_l)}) =
\begin{dcases*}
\left(\frac{p_S^{(n_l)} - (p_S^{(n_l)})_{(0)}}{\sigma_{n_l}^{(-)}}\right)^2 &
for $p_S^{(n_l)} < (p_S^{(n_l)})_{(0)}$, \\
\left(\frac{p_S^{(n_l)} - (p_S^{(n_l)})_{(0)}}{\sigma_{n_l}^{(+)}}\right)^2 &
for $p_S^{(n_l)} \ge (p_S^{(n_l)})_{(0)}$, \\
\end{dcases*}
\end{equation}
where $(p_S^{(n_l)})_{(0)}$ is the central value,
and $\sigma_{n_l}^{(-)}$ and $\sigma_{n_l}^{(+)}$ are the negative-side
and positive-side standard deviations, respectively.
The choice of phase-space variables is always user-defined;
it is described in more detail in Sec.~\ref{sec:event_parameter_buffers}.
The data weights are usually not necessary for actual amplitude analysis,
but they are needed, for example, for parameterization of a background
originating from a specific channel using its signal MC events
to take the difference between the reconstruction efficiency
in data and MC into account.
Errors returned by the fit are not correct if fitting with the data weights
and they need to be estimated separately from pseudoexperiments.

In case of an extended unbinned maximum likelihood fit the likelihood has an
additional term
\begin{equation}
\begin{aligned}
\Delta \mathcal{F} & =
- 2 \ln \frac{N^{N_\text{data}} \exp(-N)}{\Gamma(N_\text{data} + 1)}
& = 2 (N - N_\text{data} \ln(N)),
\end{aligned}
\end{equation}
where the constant term containing the logarithmic $\Gamma$ function is omitted.

The fit can also be performed simultaneously for several data sets.
The likelihood function is then
\begin{equation}
\begin{aligned}
\mathcal{F} = &
\sum_a \sum\limits_i - 2 w_i \ln \left[
  (f_S)_a \frac{S_a(\Phi_i, p_S)}{\sum\limits_j w_j S_a(\Phi_j, p_S)} +
  \sum\limits_k
    (f_B^{(k)})_a \frac{B^{(k)}_a(\Phi_i, p_B^{(k)})}
                       {\sum\limits_j w_j B^{(k)}_a(\Phi_j, p_B^{(k)})}
\right] \\
& + \sum\limits_l C_l(p_S^{(n_l)}) + C(p_S),
\end{aligned}
\label{eq:likelihood_vecampfit_simultaneous}
\end{equation}
where $a$ is the number of simultaneous-fit point, and the signal density
$S_a$, the background densities $B^{(k)}_a$, the signal fraction $(f_S)_a$,
and the background fractions $(f_B^{(k)})_a$
all depend on the simultaneous-fit point.

An alternative approach is to parameterize the efficiency $\epsilon(\Phi)$
by a separate procedure and use MC points without detector simulation
for fitting. This is equivalent to redefinition of the signal
and background density functions:
\begin{equation}
\begin{aligned}
S_\epsilon(\Phi, p_S) & = S(\Phi, p_S) \epsilon(\Phi), \\
B^{(k)}_\epsilon(\Phi, p_B^{(k)}) & =
  B^{(k)}(\Phi, p_B^{(k)}) \epsilon(\Phi) =
  B^{(k)}_\text{(obs)}(\Phi, p_B^{(k)}). \\
\end{aligned}
\label{eq:density_efficiency}
\end{equation}
The ratio between the efficiencies in data and MC $R(\Phi_j)$ is supposed to be
taken into account when parameterizing the efficiency in this case; however,
the weights related to non-uniform MC density function $d_\text{MC}(\Phi)$
still remain. The likelihood becomes
\begin{equation}
\begin{aligned}
\mathcal{F} = &
\sum\limits_i - 2 w_i \ln \left[
  f_S \frac{S_\epsilon(\Phi_i, p_S)}{\sum\limits_j w_j
  S_\epsilon(\Phi_j, p_S)} +
  \sum\limits_k
    f_B^{(k)} \frac{B^{(k)}_\epsilon(\Phi_i, p_B^{(k)})}
                   {\sum\limits_j w_j B^{(k)}_\epsilon(\Phi_j, p_B^{(k)})}
\right] \\
& + \sum\limits_l C_l(p_S^{(n_l)}) + C(p_S).
\end{aligned}
\label{eq:likelihood_parametric_efficiency}
\end{equation}
\vecampfit{} does not provide any special support for fitting
with parameterized efficiency, however, since the likelihood formula is
almost the same as Eq.~\eqref{eq:likelihood_vecampfit},
such fit can be performed by the user with the existing
fitter code after redefinition of the signal density function as given
by Eq.~\eqref{eq:density_efficiency} and appropriate redefinition of weights.
The background needs to be fitted using the same MC sample as utilized
for the signal fit; redefinition of the background-density parameterization
with respect to fitting without efficiency parameterization is not necessary.

\subsubsection{Grid integration}

The likelihood for grid integration is the same as the likelihood
for MC integration with parameterized efficiency given
by Eq.~\eqref{eq:likelihood_parametric_efficiency}, but the definitions
of weights and integration points are different. In case of
the grid integration, the index $j$ runs over grid points which need
to be defined by the user in the fitting program. The weights are given by
\begin{equation}
w_j = w_j^\text{(integration)} w_j^\text{(Jacobian)},
\end{equation}
where $w_j^\text{(integration)}$ are weights originating from
the integration method and $w_j^\text{(Jacobian)}$ are weights originating
from the Jacobian of the transformation between the phase-space and grid
coordinates.

For example, Gauss-Legendre quadrature is used for integration in the
$\kpkpkm$ example described in Sec.~\ref{sec:performance_laura};
the integration weights are the Gauss-Legendre
weights rescaled by the size of the integration region:
\begin{equation}
w_{j_1 j_2}^\text{(integration)} =
\frac{m_{\kp_1 \km}^\text{(max)} - m_{\kp_1 \km}^\text{(min)}}{2}
\frac{m_{\kp_2 \km}^\text{(max)} - m_{\kp_2 \km}^\text{(min)}}{2}
  w_{j_1}^\text{(GL)} w_{j_2}^\text{(GL)},
\end{equation}
where the indices $n = 1, 2$ correspond to the number of $\kp$,
$m_{\kp_n \km}^\text{(max)}$ and $m_{\kp_n \km}^\text{(min)}$
are the maximum and minimum invariant masses in the integration range,
respectively, and $w_{j_n}^\text{(GL)}$ are Gauss-Legendre weights
for integration over the interval $[-1,1]$.
The grid is defined in invariant-mass coordinates, thus
\begin{equation}
w_j^\text{(Jacobian)} =
  \frac{\partial m_{\kp_1 \km}^2}{\partial m_{\kp_1 \km}}
  \frac{\partial m_{\kp_2 \km}^2}{\partial m_{\kp_2 \km}}
  = 4 m_{\kp_1 \km} m_{\kp_2 \km}.
\end{equation}
The minimum and maximum invariant masses do not have to be the same as
the kinematic borders of the Dalitz plot because it may be split into
several integration regions.

\subsubsection{Formula integration}

In case of formula integration, the normalization integrals in
Eq.~\eqref{eq:probability_density} are calculated symbolically.
This method can only work for simple fits that do not require to take the
efficiency into account. Formula integration is only supported
for fits without background. Thus, the probability density is
\begin{equation}
\rho(\Phi, p_S) = \frac{S(\Phi, p_S)}{\int S(\Phi, p_S) d \Phi}
\label{eq:probability_density_symbolic}
\end{equation}
in case of formula integration. If this method is used, then the user
has to calculate the integral and provide the integration function by calling
{\tth vecampfit\_fitter\_set\_integration\_function()}.
The integration function $I(p_S)$ should only depend on fit parameters:
\begin{equation}
I(p_S) = \int S(\Phi, p_S) d \Phi.
\end{equation}

\subsubsection{Fit without normalization}

Finally, the calculation of the normalization integral can be disabled.
This is necessary in certain specific cases. For example, one can fit
the reconstruction efficiency by using generator-level MC events defining
the probability density function as
\begin{equation}
\rho(\Phi) =
\begin{dcases*}
\epsilon(\Phi)     & for reconstructed events, \\
1 - \epsilon(\Phi) & for non-reconstructed events. \\
\end{dcases*}
\end{equation}
The function defined in this way does not need any further normalization.

\subsection{Fit with explicit gradient calculation}

The likelihood function for Monte Carlo integration is given by
Eq.~\eqref{eq:likelihood_vecampfit}. Partial derivatives of the likelihood
are
\begin{equation}
\begin{aligned}
\frac{\partial \mathcal{F}}{\partial p_S^{(\alpha)}} = &
\sum\limits_i \frac{-2 f_S w_i}{
  f_S \frac{S(\Phi_i, p_S)}{\sum\limits_j w_j S(\Phi_j, p_S)} +
  \sum\limits_k f_B^{(k)} \frac{B^{(k)}(\Phi_i, p_B^{(k)})}
    {\sum\limits_j w_j B^{(k)}(\Phi_j, p_B^{(k)})}
  } \\
& \qquad\times
\left[ \frac{\frac{\partial S(\Phi_i, p_S)}{\partial p_S^{(\alpha)}}}
  {\sum\limits_j w_j S(\Phi_j, p_S)}
  - \frac{S(\Phi_i, p_S)}{\left(\sum\limits_j w_j S(\Phi_j, p_S)\right)^2}
  \left(\sum\limits_j w_j \frac{\partial S(\Phi_j, p_S)}
  {\partial p_S^{(\alpha)}}\right)
\right] \\
& + \delta_{n_k \alpha}
    \frac{\partial C_k(p_S^{(n_k)})}{\partial p_S^{(\alpha)}}
  + \frac{\partial C(p_S)}{\partial p_S^{(\alpha)}},
\end{aligned}
\label{eq:likelihood_vecampfit_derivatives}
\end{equation}
where $\alpha$ is the parameter number. Since the derivatives depend on
the signal-density derivatives, it is necessary to calculate
not only the signal density, but also the gradient
$\partial S(\Phi, p_S) / \partial p_S^{(\alpha)}$ for each data and
MC event. Two signal density functions need to be set, the first one
computing both signal density and its gradient simultaneously,
and the second one computing the signal density only.
Partial derivatives of the likelihood over the signal and background fractions
are
\begin{equation}
\begin{aligned}
\frac{\partial \mathcal{F}}{\partial f_S} & =
\sum\limits_i \frac{-2 \frac{S(\Phi_i, p_S)}
                            {\sum\limits_j w_j S(\Phi_j, p_S)} w_i}{
  f_S \frac{S(\Phi_i, p_S)}{\sum\limits_j w_j S(\Phi_j, p_S)} +
  \sum\limits_k
    f_B^{(k)} \frac{B^{(k)}(\Phi_i, p_B^{(k)})}
                   {\sum\limits_j w_j B^{(k)}(\Phi_j, p_B^{(k)})}
  }, \\
\frac{\partial \mathcal{F}}{\partial f_B^{(k)}} & =
\sum\limits_i \frac{-2 \frac{B^{(k)}(\Phi_i, p_B^{(k)})}
                            {\sum\limits_j w_j B^{(k)}(\Phi_j, p_B^{(k)})} w_i}{
  f_S \frac{S(\Phi_i, p_S)}{\sum\limits_j w_j S(\Phi_j, p_S)} +
  \sum\limits_k
    f_B^{(k)} \frac{B^{(k)}(\Phi_i, p_B^{(k)})}
                   {\sum\limits_j w_j B^{(k)}(\Phi_j, p_B^{(k)})}
  }. \\
\end{aligned}
\label{eq:likelihood_vecampfit_fraction_derivatives}
\end{equation}
In case of a standard maximum likelihood fit, the signal fraction is
not a fit parameter; it is calculated from background fractions
using Eq.~\eqref{eq:signal_fraction}. The partial derivatives over
the background fractions are given by
\begin{equation}
\left(\frac{\partial \mathcal{F}}{\partial f_B^{(k)}}\right)_\text{non-extended}
  = \frac{\partial \mathcal{F}}{\partial f_B^{(k)}}
    + \frac{\partial \mathcal{F}}{\partial f_S}
      \frac{\partial f_S}{\partial f_B^{(k)}}
  = \frac{\partial \mathcal{F}}{\partial f_B^{(k)}}
    - \frac{\partial \mathcal{F}}{\partial f_S}
\end{equation}

In case of an extended unbinned maximum likelihood fit, yields are fit
parameters.
Partial derivatives over the yields are expressed in terms of the
partial derivatives over the signal and background fraction by comparing
the results of differentiation of Eq.~\ref{eq:likelihood_vecampfit}.
After addition of the extended-likelihood term derivatives,
the partial derivatives over the yields are given by
\begin{equation}
\begin{aligned}
\frac{\partial \mathcal{F}}{\partial N_S} &=
  2 + \frac{1}{N} \frac{\partial \mathcal{F}}{\partial f_S}, \\
\frac{\partial \mathcal{F}}{\partial N_B^{(k)}} &=
  2 + \frac{1}{N} \frac{\partial \mathcal{F}}{\partial f_B^{(k)}}. \\
\end{aligned}
\end{equation}

\section{$\kmpippiz$ Dalitz analysis example}
\label{sec:kmpippiz}

The $\kmpippiz$ Dalitz analysis based on CLEO model~\cite{CLEO:2000fvk}.
As for any Dalitz analysis, the amplitude is defined in a two-dimensional
phase space $\Phi = (M_{\km \pip}^2, M_{\km \piz}^2)$. The decays
of intermediate resonances are described by the Breit-Wigner amplitude
\begin{align}
B_R^\text{(no $q$)}(m) & = \frac{\frac{F_R(q)}{F_R(q_0)}}{m_0^2 - m^2 - i m_0 \Gamma(m)}, \label{eq:breit_wigner_q_f_nophsp} \\
\Gamma(m) & = \Gamma_0 \left(\frac{q}{q_0}\right)^{2L + 1} \frac{m_0}{m}
\frac{F_R^2(q)}{F_R^2(q_0)}, \label{eq:breit_wigner_q_f_nophsp_gamma}
\end{align}
where $R$ is the decaying resonance, $m_0$ and $\Gamma_0$ are its mass
and width, respectively,
$m = \sqrt{s}$ is the invariant mass, $q$ and $q_0$ are the daughter
momenta for the invariant and resonance masses, respectively, $L$
is the angular momentum, and $F_R(q)$ and $F_R(q_0)$ are the
formfactors at the invariant and resonance masses, respectively.
The superscript ``(no $q$)'' refers to the absence of a phase-space factor
in the numerator. Blatt-Weisskopf
formfactors~\cite{blattweisskopf,VonHippel:1972fg} are used:
\begin{equation}
F_R^\text{(BW)}(q, r, L) = \frac{1}{|h_L^{(1)}(z)| z^{L + 1}},
\label{eq:blatt_weisskopf}
\end{equation}
where $r$ is the effective radius, $z = q r$ and $h_L^{(1)}(z)$
is the spherical Hankel function of the first kind. For particular values
of the angular momentum $L$, the Blatt-Weisskopf formfactors are
\begin{equation}
\begin{aligned}
F_R^\text{(BW)}(q, r, 0) & = 1, \\
F_R^\text{(BW)}(q, r, 1) & = \frac{1}{\sqrt{z^2 + 1}}, \\
F_R^\text{(BW)}(q, r, 2) & = \frac{1}{\sqrt{z^4 + 3 z^2 + 9}}, \\
F_R^\text{(BW)}(q, r, 3) & = \frac{1}{\sqrt{z^6 + 6 z^4 + 45 z^2 + 225}}, \\
F_R^\text{(BW)}(q, r, 4) & = \frac{1}{\sqrt{z^8 + 10 z^6 + 135 z^4 + 1575 z^2
                                      + 11025}}, \\
F_R^\text{(BW)}(q, r, 5) & = \frac{1}{\sqrt{z^{10} + 15 z^8 + 315 z^6 + 6300 z^4
                                      + 99225 z^2 + 893025}}. \\
\end{aligned}
\end{equation}

The amplitude is calculated using covariant formalism. The result is
\begin{equation}
\begin{aligned}
A(\Phi) = & a_{nr} e^{i\phi_{nr}}
+ \sum\limits_R a_{R} e^{i\phi_{R}}
  \frac{F_D^\text{(BW)}(q, r^{(D)}, J)}{F_D^\text{(BW)}(q_0, r^{(D)}, J)}
  B_R^\text{(no $q$)}(M_{AB}) {\cal A}_J(ABC|R) \\
\end{aligned}
\end{equation}
where $q$ and $q_0$ are the momenta of the decay $D \to R C$ ($R$ being
an intermediate resonance in the $AB$ system)
for the invariant mass of the combination $AB$ and the mass of
the intermediate resonance $R$, respectively, $r^{(D)}$ is the $D$
effective radius, $J$ is the $R$ spin, and the angular part is
\begin{equation}
\begin{aligned}
{\cal A}_0(ABC|R) & = 1; \\
{\cal A}_1(ABC|R) & = M^2_{AC} - M^2_{BC} + \frac{(M^2_D - M^2_C)(M^2_B -
M^2_A)}{M^2_{R}}. \\
\end{aligned}
\end{equation}
Depending on the intermediate combination, the following particle orders
are used: $(A, B, C) = (\pip, \piz, \km)$, $(\km, \pip, \piz)$,
$(\km, \piz, \pip)$. Since all initial and final-state particles have
zero spin, the signal density function is $S(\Phi) = |A(\Phi)|^2$.

\begin{figure}
{\centering
\includegraphics[width=6.7cm]{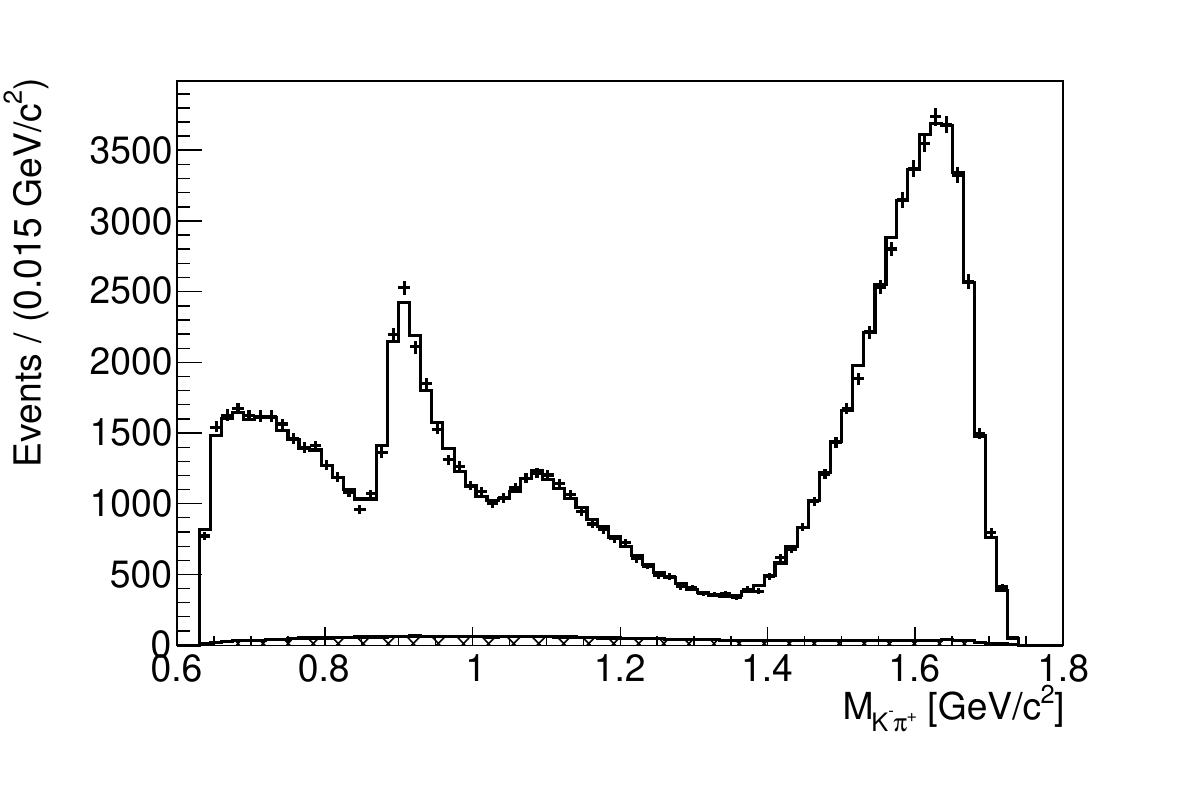}
\includegraphics[width=6.7cm]{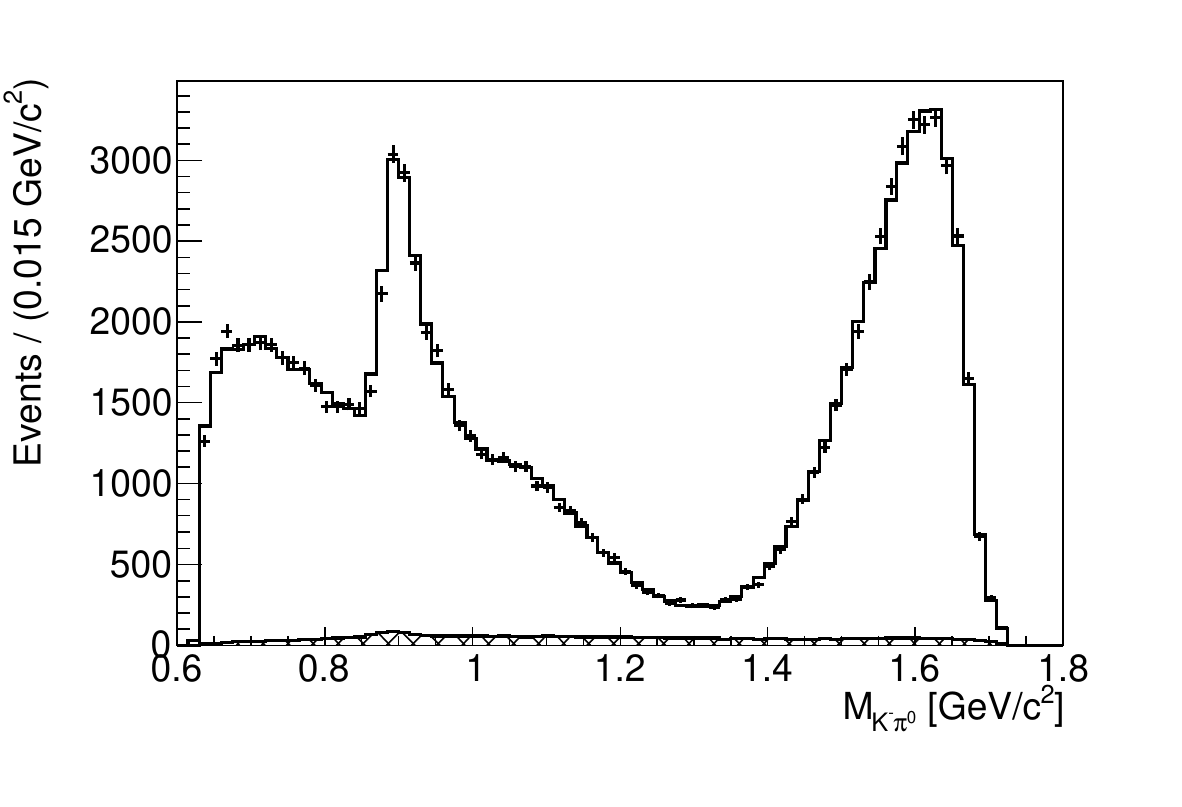} \\
\includegraphics[width=6.7cm]{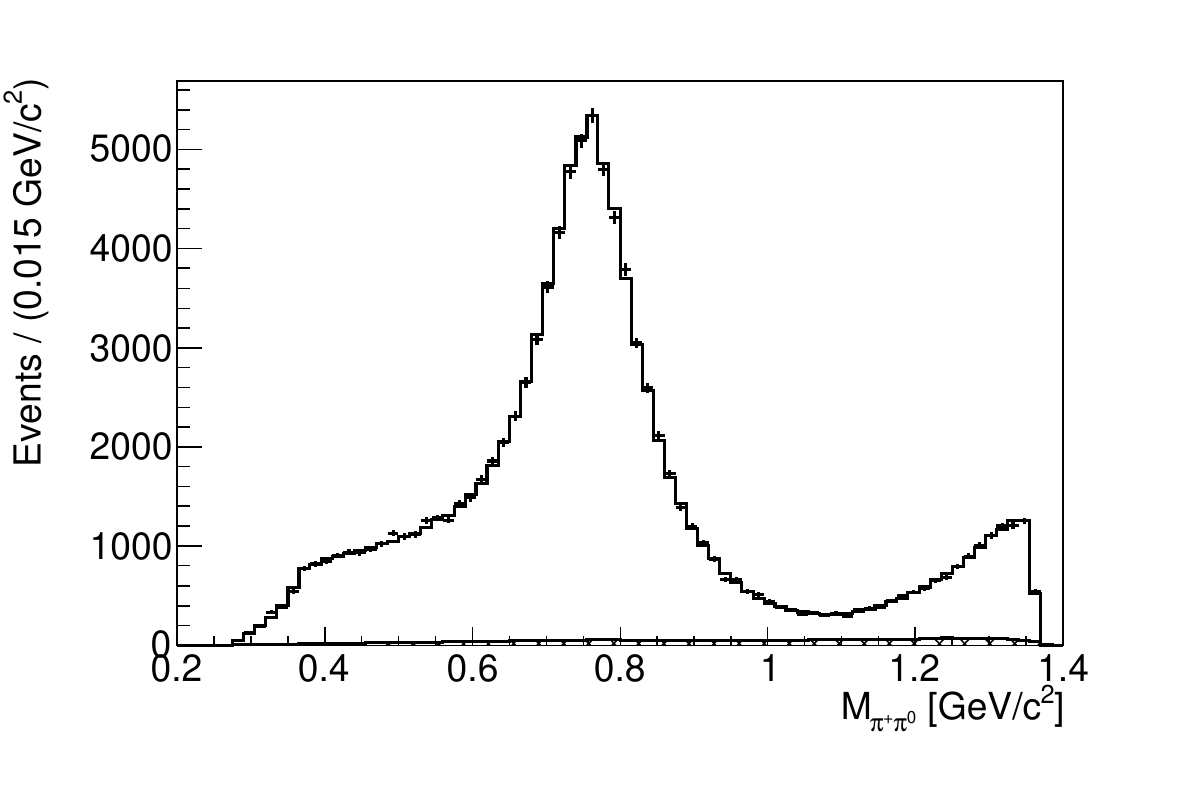} \\}
\caption{Results of fit to the $\kmpippiz$ signal distribution for
a pseudoexperiment. The points with error bars are pseudo-data,
the solid line is the fit result, and the hatched histograms are the background
contribution.}
\label{fig:dz_kmpippiz}
\end{figure}

The model includes 7 resonances: $\rho(770)^+$ and $\rho(1700)^+$ decaying
to $\pip \piz$; $\kst(892)^-$, $\kst_0(1430)^-$, and $\kst(1680)^-$ decaying
to $\km \piz$; $\kstb(892)^0$ and $\kstb_0(1430)^0$ decaying to $\km \pip$.
The $\Dz$ is below the threshold of the decay $\Dz \to \rho(1700)^+ \km$
for the $\rho(1700)^+$ nominal mass, thus, the formfactor normalization
to unity at the nominal mass is impossible. Since no other method is described
by CLEO, the exact normalization used for the $\rho(1700)^+$ cannot
be recovered. In the \vecampfit{} example, the normalization formfactor
$F_D^\text{(BW)}(q_0, r^{(D)}, J)$ is simply omitted for the $\rho(1700)^+$.
The results of CLEO fit are used as the default values for MC generation.
Results of the fit to a pseudoexperiment with 100000 signal events are shown
in Fig.~\ref{fig:dz_kmpippiz}. Note that only qualitative
comparison with CLEO is possible because the data and reconstruction efficiency
from CLEO analysis~\cite{CLEO:2000fvk} are not available, and the example
does not include reconstruction efficiency at all.

\begin{figure}
{\centering
\includegraphics[width=3.3cm]{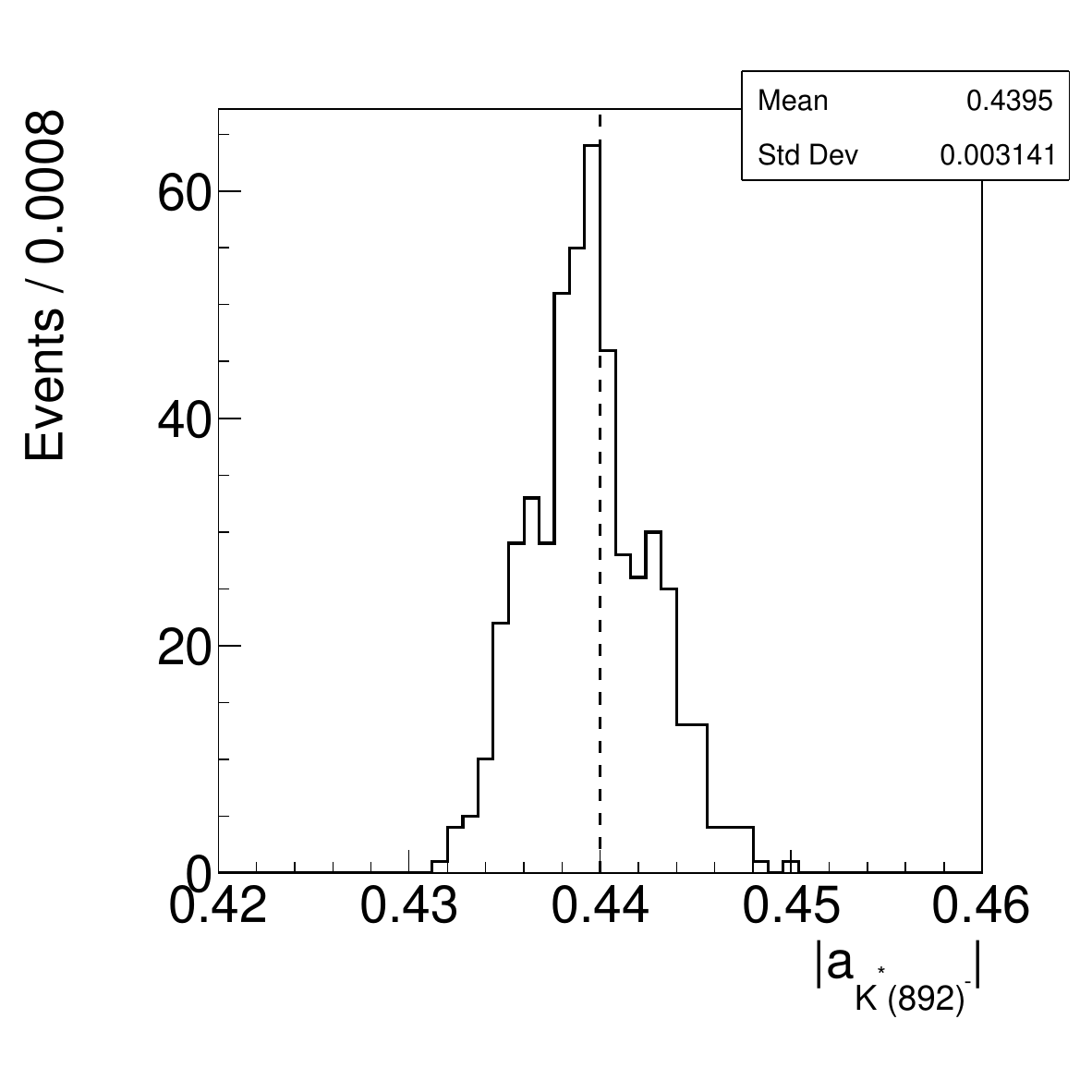}
\includegraphics[width=3.3cm]{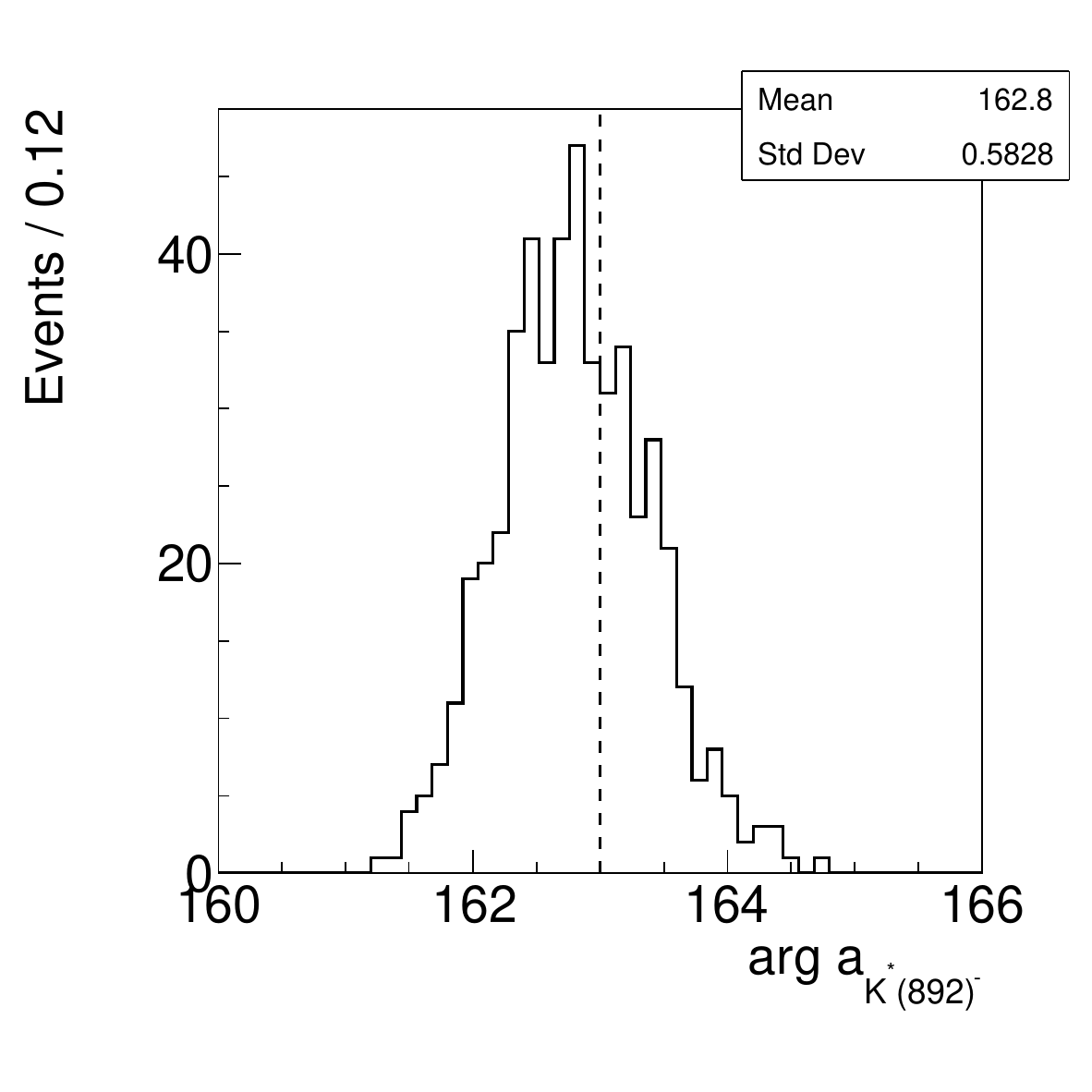}
\includegraphics[width=3.3cm]{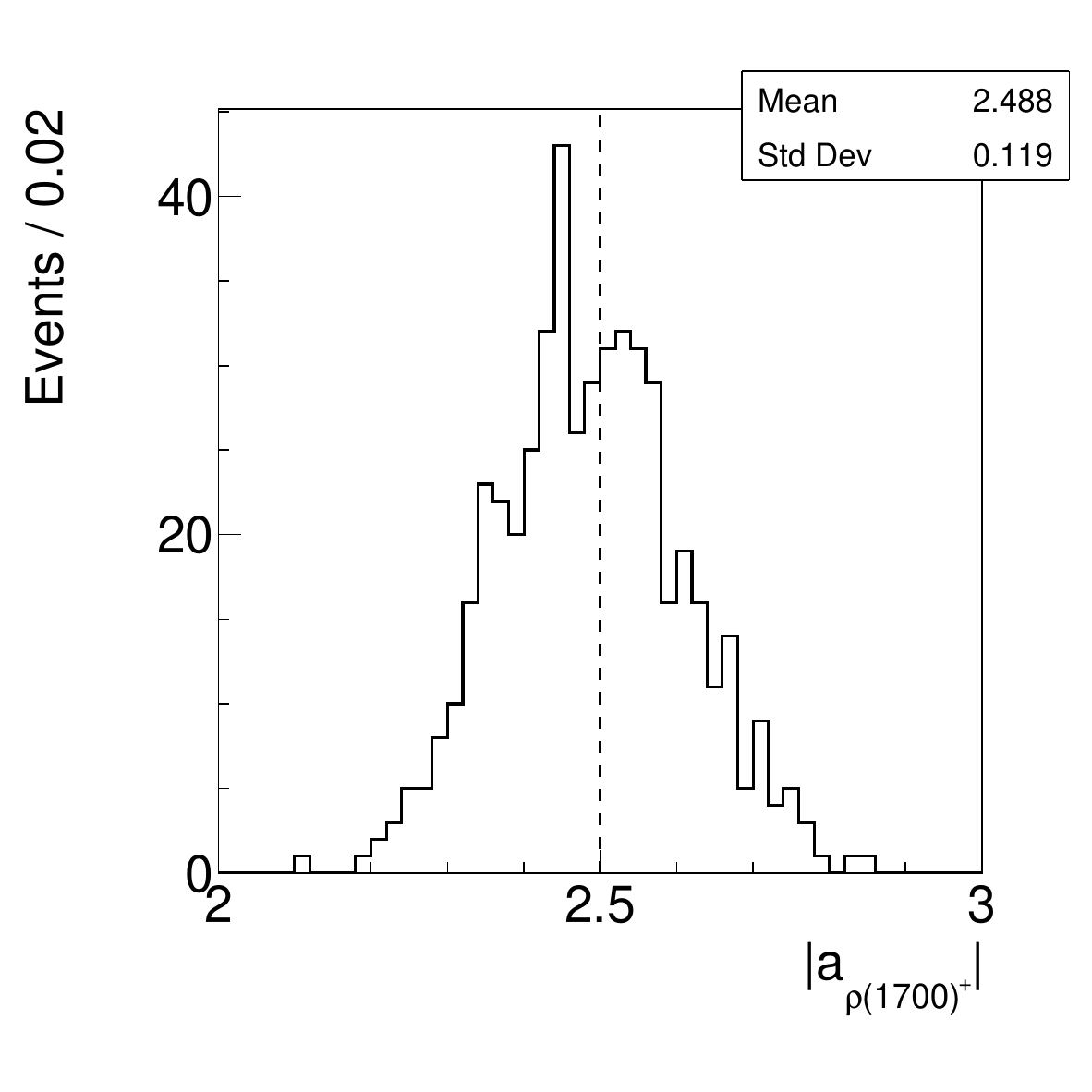}
\includegraphics[width=3.3cm]{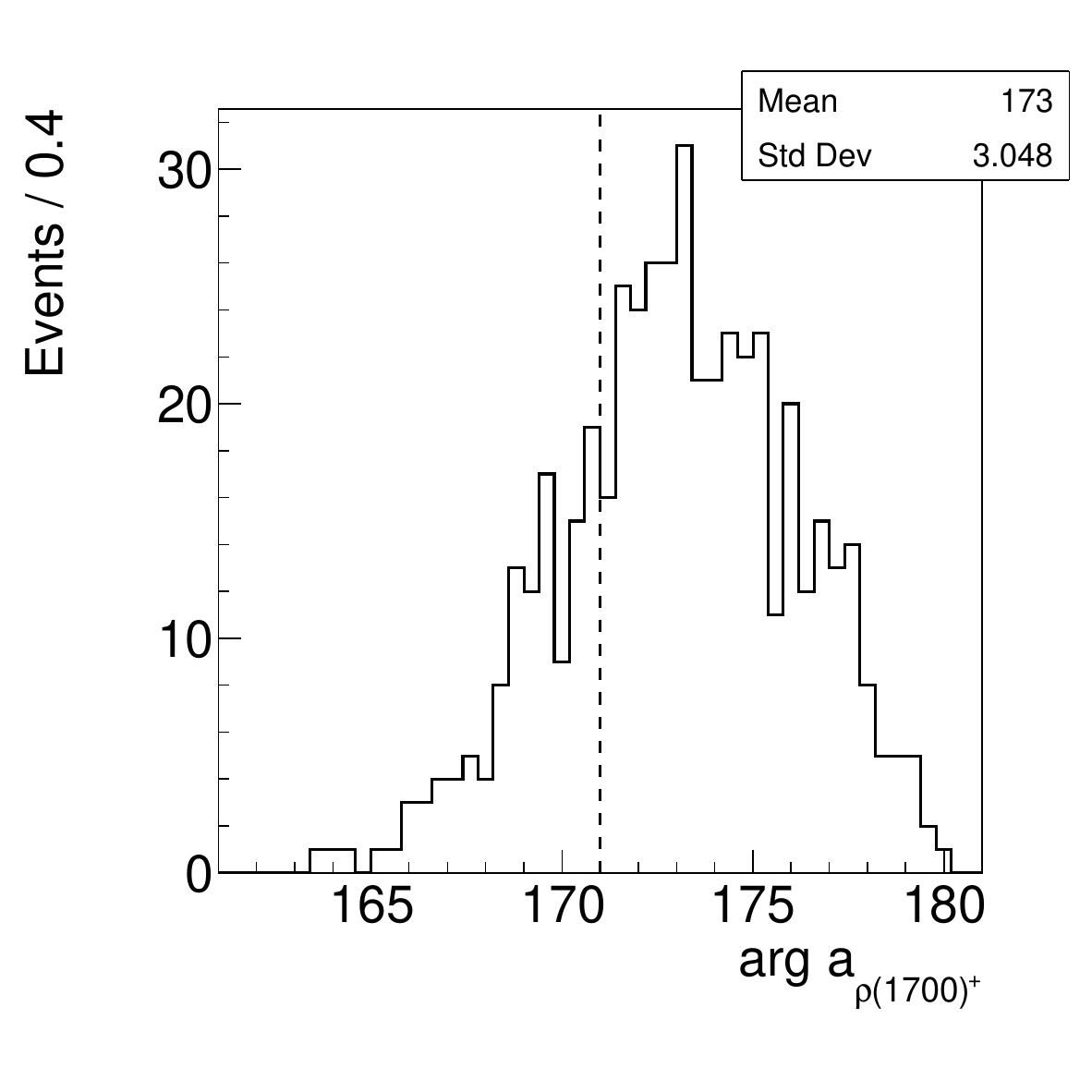} \\
\includegraphics[width=3.3cm]{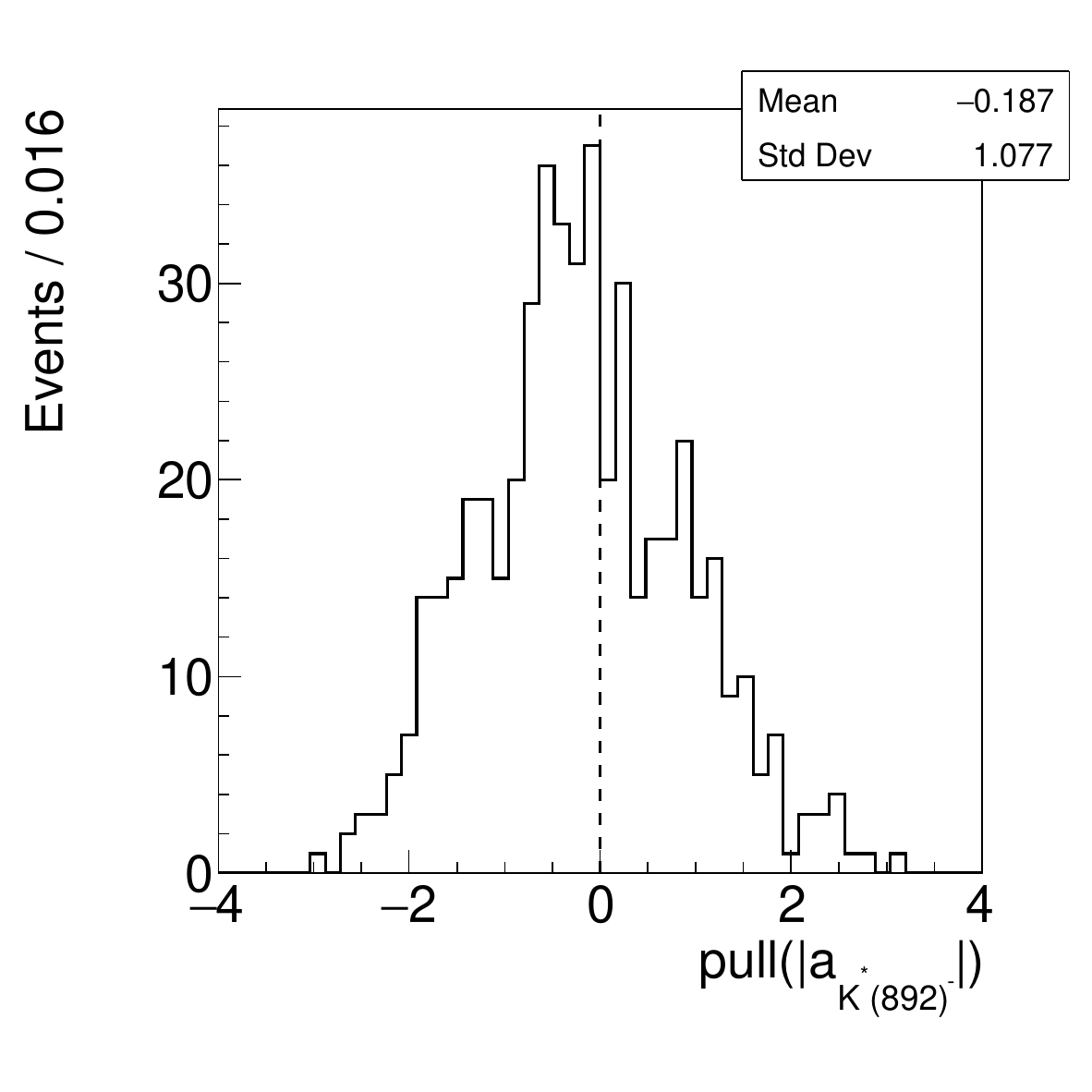}
\includegraphics[width=3.3cm]{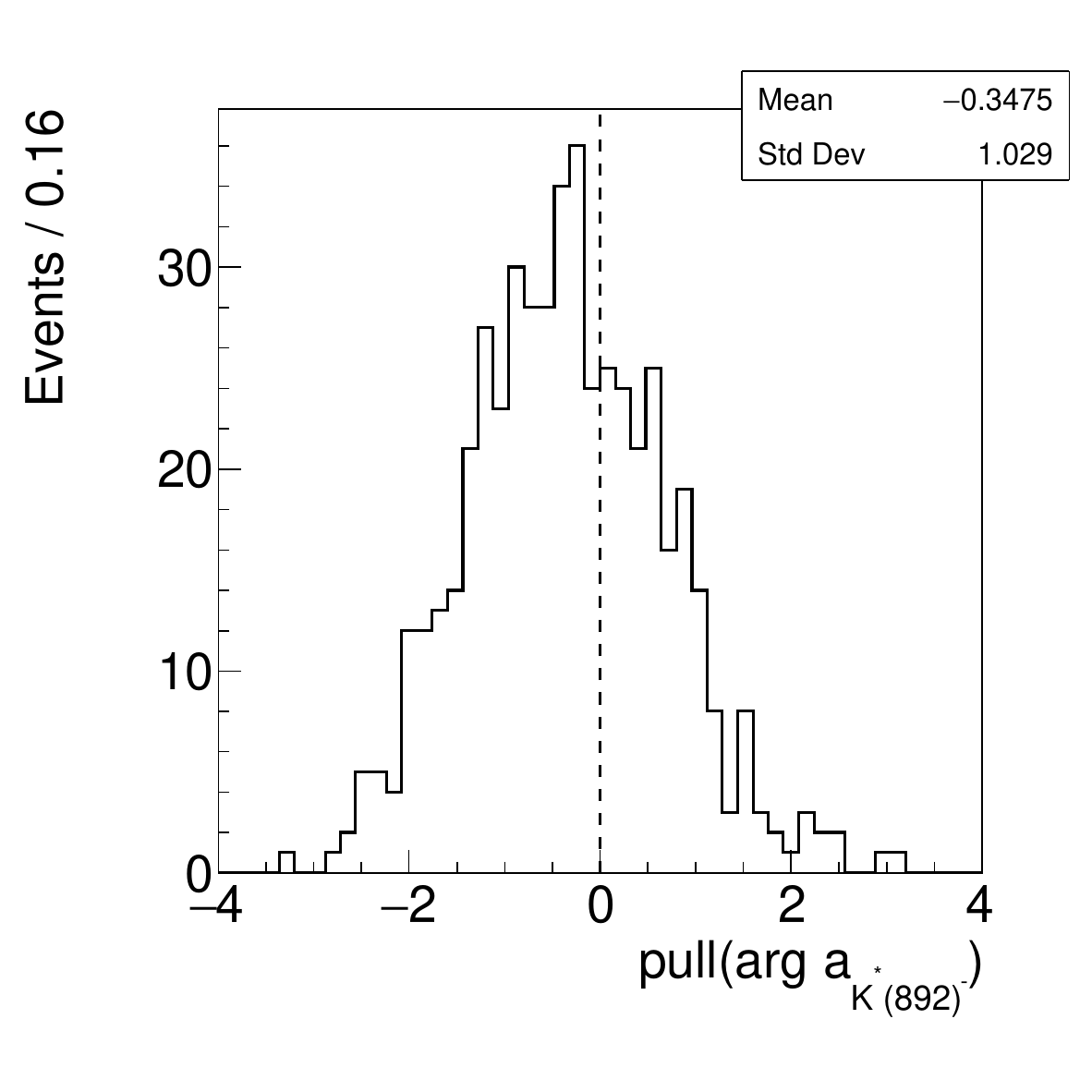}
\includegraphics[width=3.3cm]{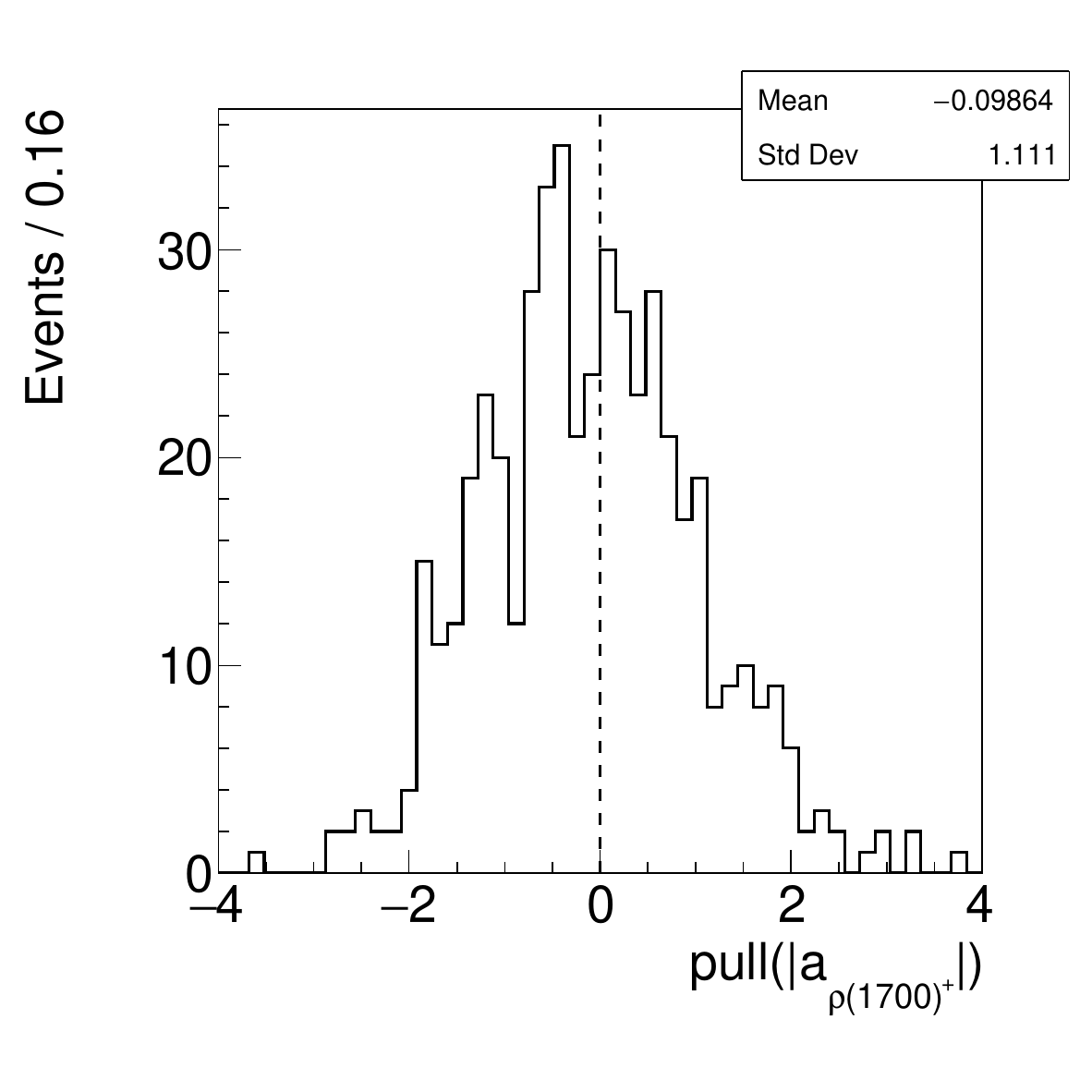}
\includegraphics[width=3.3cm]{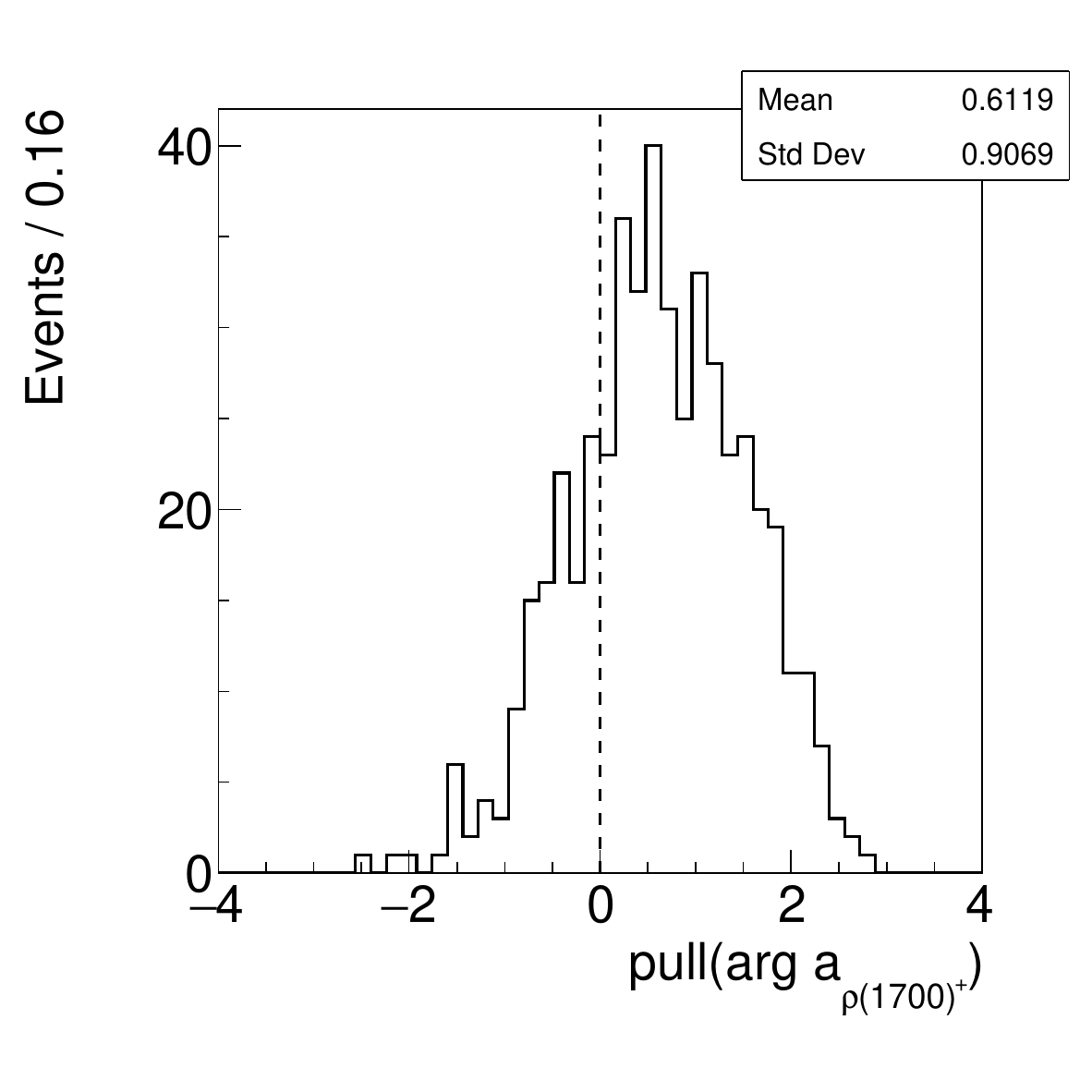} \\}
\caption{Resulting absolute value and phase of the amplitude
for the $\kst(892)^-$ and $\rho(1700)^+$ (top row) and the corresponding
pull distributions (bottom row) for a sample of 500 pseudoexperiments.
The true values are in the middle of the $x$-axis range
and they are marked by vertical dashed lines.}
\label{fig:dz_kmpippiz_input_output}
\end{figure}

To validate the generation and fitting procedure, 500 pseudoexperiments
are generated and the distributions of the resulting amplitude absolute values
and phases are checked. The distributions for the $\kst(892)^-$ and
the $\rho(1700)^+$ are shown in Fig.~\ref{fig:dz_kmpippiz_input_output}
together with the corresponding pull distributions. The model is found
to reproduce the input values with minor systematic shifts; for the presented
parameters, the largest shift of about 2 degrees is observed for the phase
of the $\rho(1700)^+$. It is much less than the statistical uncertainty from
the original CLEO analysis, which was equal to 6 degrees.

\section{Implementation of the amplitude-analysis formalism}
\label{sec:formalism_implementation}

\subsection{Vectorization of likelihood calculation}

The calculations are performed by the \vecampfit{} fitter in the following
way. The main likelihood formula given by Eq.~\eqref{eq:likelihood_vecampfit}
is modified to use event vectors:
\begin{equation}
\mathcal{F} =
\sum\limits_{i_1 = 0}^{\lfloor \frac{N_\text{data} - 1}{L} \rfloor}
\sum\limits_{i_2 = 1}^{L} \mathcal{L}_{i_1 i_2}
+ \sum\limits_k C_k(p_{n_k}) + C(p),
\end{equation}
where $\lfloor\ \rfloor$ is the floor function, $L$ is the vector length, and
$\mathcal{L}_{i_1 i_2}$ is the likelihood for the data event with the indices
$i_1$ and $i_2$ given by
\begin{equation}
\mathcal{L}_{i_1 i_2} =
\begin{dcases*}
  - 2 w_{i_1 i_2} \ln \left[
  f_S \frac{S(\Phi_{i_1 i_2}, p_S)}{I_S}
  + \sum\limits_k f_B^{(k)} \frac{B^{(k)}(\Phi_{i_1 i_2}, p_B^{(k)})}{I_{B^{(k)}}}
  \right]
  & if $i_1 L + i_2 \leq N_\text{data}$, \\
0 & if $i_1 L + i_2 > N_\text{data}$. \\
\end{dcases*}
\label{eq:likelihood_event}
\end{equation}
The normalization integrals $I_S$ and $I_B$ are given by
\begin{equation}
I_S = \sum\limits_{j_1 = 0}^{\lfloor \frac{N_\text{MC} - 1}{L} \rfloor}
    \sum\limits_{j_2 = 1}^{L}
    w_{j_1 j_2} S(\Phi_{j_1 j_2}, p_S),
\end{equation}
and
\begin{equation}
I_{B^{(k)}} = \sum\limits_{j_1 = 0}^{\lfloor \frac{N_\text{MC} - 1}{L} \rfloor}
    \sum\limits_{j_2 = 1}^{L}
    w_{j_1 j_2} B^{(k)}(\Phi_{j_1 j_2}, p_B^{(k)}),
\end{equation}
where the signal and background densities are set to 0 if
$j_1 L + j_2 > N_\text{MC}$.

In the case of scalar calculations, the likelihoods $\mathcal{L}_{i_1 i_2}$
in Eq.~\eqref{eq:likelihood_event} are calculated for each event and then
their values are summed. The \vecampfit{} library instead calculates
their values for the entire vector at once using vectorized signal and
background density functions. The likelihood vectors are given by
\begin{equation}
\vec{\mathcal{L}}_{i_1} =
  - 2 \vec{w}_{i_1} \ln \left[
  f_S \frac{\vec{S}(\vec{\Phi}_{i_1}, p_S)}{I_S}
  + \sum\limits_k f_B^{(k)}
    \frac{\vec{B}^{(k)}(\vec{\Phi}_{i_1}, p_B^{(k)})}{I_{B^{(k)}}}
  \right],
\end{equation}
Thus, the computations require user-defined vectorized signal and background
density functions $\vec{S}$ and $\vec{B}$ which take an event vector
$\vec{\Phi}$ as an argument. Both density functions and event vector have
a fixed vector length $L$. The calculation of MC normalization sums
$I_S$ and $I_{B^{(k)}}$ is vectorized in a similar way.

Vectorized \vecampfit{} subprograms and the fitter code are compiled
for several fixed vector lengths. The default list of vector lengths
includes $L = 1, 2, 4, 8, 16, 32, 64$; it can be modified when
configuring the library build if necessary.
Similarly, vectorized \vecampfit{} Fortran modules are provided
in several versions corresponding to a specific vector length, e.g.
{\tth VECAMPFIT\_HIGH\_ENERGY\_PHYSICS\_VECTOR\_\#LENGTH\#} with
{\tth \#LENGTH\#} expanding into one of the lengths from the list.
The modules provide generic interfaces for their subprograms so that
the names do not depend on the vector length.
For example, the subroutine for calculation of the simple constant-width
Breit-Wigner amplitude from the high-energy-physics
module can be called as {\tth BREIT\_WIGNER\_CONSTANT\_WIDTH\_M}
regardless of the vector length.

When compiling a fitting program, the vector length has to be chosen explicitly
by the user. In the examples, the length is chosen at configuration time and
it is controlled by a single {\sc cmake} variable {\tth VECTOR\_LENGTH}.
Global configuration files generated by {\sc cmake} provide length-independent
preprocessor definitions for Fortran module names so that the code does
not depend on the vector length. Examples of the preprocessor definitions
can be found in the file {\tth examples/dz\_kmpippiz/src/config.h.in}.

\subsection{Templates and model configuration}

Before discussion of \vecampfit{} implementation details and fitting procedure,
it is necessary to introduce creation of template sources which will be used
multiple times in code examples. The template sources are explained using
the $\kkpipi$ example, which is discussed in more detail
in Sec.~\ref{sec:epem_kkpipi_workflow}.

The \vecampfit{} library is designed so that the fitting programs
for each model in an analysis based on the library can be created by
a single build of the analysis code.
The density functions are similar but not necessarily the same for different
models. They may also depend on the process (i.e. $\kpkmpippim$ or
$\kskspippim$), experimental period, data sample, and decay channel.
There is no dependency on experimental period and no different decay channels
in the $\kkpipi$ example. A single source file is used
to compile the code for each necessary variant. The code
from all sources is then collected into a model library, which is used
to link fitting programs.

This requires modification of the source code depending on its variant.
Particularly, subroutine names should be different. Therefore, the source files
are implemented as templates.
Generation of the source files to be compiled
is performed by {\sc awk} script
{\tth awk/analysis\_template\_source.awk}. For example,
the background-density function is declared as
\begin{lstlisting}[style=fortranstyle]
  MODULE SUBROUTINE BACKGROUND_DENSITY_#MODEL#_#PROCESS#_#SAMPLE#(BD, EV, PAR, &
&     PAR_BUF) BIND(C)
\end{lstlisting}
and for the default model for process $\kpkmpippim$ at $\sqrt{s}$ = 1800 MeV
this declaration expands into
\begin{lstlisting}[style=fortranstyle]
  MODULE SUBROUTINE BACKGROUND_DENSITY_kpkmpippim_default_kpkmpippim_1800_mev( &
&     BD, EV, PAR, PAR_BUF) BIND(C)
\end{lstlisting}
The patterns enclosed by two number-sign symbols: {\tth \#MODEL\#},
{\tth \#PROCESS\#}, and {\tth \#SAMPLE\#} are replaced by {\sc awk}
by the actual values of model, process, and data-sample names, respectively.
Similar expressions are defined for running period, decay channel,
particle name, charge, spin, parity, parity letter (``P'' for positive parity
or ``M'' for negative parity), and various model lists.

Configuration of the model is performed by C preprocessor. For example,
the background configuration header
{\tth config\_background\_kpkmpippim\_default\_1800\_mev.h}
for the same energy point contains the polynomial order only:
\begin{lstlisting}[style=cstyle]
#ifndef VECAMPFIT_EXAMPLES_EPEM_KKPIPI_CONFIG_BACKGROUND_#MODEL#_#PROCESS#_#SAMPLE#_H
#define VECAMPFIT_EXAMPLES_EPEM_KKPIPI_CONFIG_BACKGROUND_#MODEL#_#PROCESS#_#SAMPLE#_H

// Configuration of background model for E = 1800 MeV.

// Polynomial order.
#define POLYNOMIAL_ORDER 2

#endif
\end{lstlisting}
First-level configuration headers are created manually for each model.
Higher-level configuration headers include the original headers and provide
other definitions based on the previous ones and also process,
experimental period, data sample, and decay channel.
Such headers are included from source files; similar to them,
the configuration headers are templates created for each necessary model
variant.
For background fit, the second-level header {\tth configuration\_background.h}
contains common process-based definitions only:
\begin{lstlisting}[style=cstyle]
#ifndef VECAMPFIT_EXAMPLES_EPEM_KKPIPI_CONFIGURATION_BACKGROUND_#MODEL#_#PROCESS#_#SAMPLE#_H
#define VECAMPFIT_EXAMPLES_EPEM_KKPIPI_CONFIGURATION_BACKGROUND_#MODEL#_#PROCESS#_#SAMPLE#_H

#include <vecampfit/examples/epem_kkpipi/models/#model#/#process#/#sample#/config_background.h>

// Configuration of background model.

// Process.
#include <vecampfit-example-models/epem_kkpipi/#process#/process.h>

#endif
\end{lstlisting}
Signal fit discussed in Sec.~\ref{sec:epem_kkpipi_signal_fit} provides
more complex example of model configuration.

Generation of the template sources is requested by {\sc cmake} functions
{\tth vecampfit\_template\_source\_*} from the file
{\tth cmake/analysis\_template\_source.cmake}.
For example, the generation of the background-density source files is
performed by
\begin{lstlisting}[style=cmakestyle]
vecampfit_template_source_model_process_sample(
  ${amplitude} ${fit_type} ${process_group} ${model_list} ${model_name}
  ${process} ${sample} ${fit_type}_density.F90 "" "vecampfit/examples"
  "${include_directories}" FALSE FALSE "" "" "" "" SOURCES generated_sources
)
\end{lstlisting}
The sources for all models are then collected by calling {\sc cmake} function
\begin{lstlisting}[style=cmakestyle]
vecampfit_collect_model_sources(
  ${amplitude} "efficiency;resolution;background" "${PROCESS}"
  "${MODEL_LIST}" "" get_fit_type_sources "" "" "" get_submodel_sources
) 
\end{lstlisting}
where the arguments {\tth get*} are {\sc cmake} functions creating
the lists of sources, header files, and files generated by {\sc awk}.

\subsection{Event and parameter buffers}
\label{sec:event_parameter_buffers}

The input data and MC events need to be stored in ROOT files
in trees~\cite{Brun:1997pa}; other input formats
are not supported.
By default, events are assumed to be stored as event class using one
tree branch. The name of the tree and event-class branch are passed
as arguments to the function {\tth vecampfit\_fitter\_load\_data\_file()};
its usage is described in Sec.~\ref{sec:fitting_procedure}.
The default storage format can be changed by specifying a user function
setting branch addresses for reading of the event from the data tree using
{\tth vecampfit\_fitter\_set\_branch\_addresses\_function()}.

The $\kmpippiz$ event class {\tth d0\_kmpippi0\_event}
contains four-momenta of the final-state particles:
\begin{lstlisting}[style=cppstyle]
class d0_kmpippi0_event : public TObject
{
	/* All member functions are omitted. */
        float km_momentum[4];
        float pip_momentum[4];
        float pi0_momentum[4];
}
\end{lstlisting}
The event-vector arguments $\vec{\Phi}$ are called event buffers
in the \vecampfit{} library. The signal event buffers in the
$\kmpippiz$ example contain all invariant masses,
momenta of the decays of the $D^0$ to an intermediate
resonance and a third particle, and momenta of the decays of
an intermediate resonance to a pair of final-state particles:
\begin{lstlisting}[style=fortranstyle]
  TYPE, BIND(C) :: SIGNAL_EVENT_BUFFER_#MODEL#
    REAL(C_REAL_KIND) M_KM_PIP(VECTOR_LENGTH)
    REAL(C_REAL_KIND) M_KM_PIZ(VECTOR_LENGTH)
    REAL(C_REAL_KIND) M_PIP_PIZ(VECTOR_LENGTH)
    REAL(C_REAL_KIND) Q_D0_RHOP_KM(VECTOR_LENGTH)
    REAL(C_REAL_KIND) Q_D0_KSTM_PIP(VECTOR_LENGTH)
    REAL(C_REAL_KIND) Q_D0_KSTZ_PIZ(VECTOR_LENGTH)
    REAL(C_REAL_KIND) Q_RHOP_PIP_PIZ(VECTOR_LENGTH)
    REAL(C_REAL_KIND) Q_KSTM_KM_PIZ(VECTOR_LENGTH)
    REAL(C_REAL_KIND) Q_KSTZ_KM_PIP(VECTOR_LENGTH)
  END TYPE
\end{lstlisting}
Here, {\tth C\_REAL\_KIND} is a C-interoperable real kind ({\tth C\_FLOAT}
or {\tth C\_DOUBLE}) and {\tth VECTOR\_LENGTH} is the vector length, which are
chosen during configuration. The suffix {\tth \#MODEL\#}
is the template part and it always has the value {\tth DZ\_KMPIPPIZ\_DEFAULT}
for the $\kmpippiz$ example.
To achieve optimal computational performance, event buffers need to store all
pre-calculated kinematic quantities required for signal or background
density evaluation that depend solely on event kinematics and not on fit
parameters.

The event buffers are filled once at the beginning of the fitting
upon data loading by a user-defined event-buffer-filling function:
\begin{lstlisting}[style=cppstyle]
void fill_signal_event_buffer_#model#(void *buf, const void *ev, int index)
{
        signal_event_buffer_#model# *buffer =
                (signal_event_buffer_#model#*)buf;
        const d0_kmpippi0_event *event = (const d0_kmpippi0_event*)ev;
        buffer->m_km_pip[index] = event->get_mass_km_pip();
        buffer->m_km_piz[index] = event->get_mass_km_pi0();
	/* The remaining code is omitted. */
}
\end{lstlisting}
The event-buffer type is used for signal-density implementation performed
in Fortran, however, the filling function needs to load data from a ROOT tree
and needs to be written in C++. Thus, conversion of the Fortran derived type
to a C struct is necessary. The conversion is performed by the tool
{\tth vecampfit-fortran} which is called via {\tth cmake} function
{\tth vecampfit\_generate\_c\_header()}. It is also able to create C function
declarations corresponding to C-interoperable Fortran functions or subroutines.
The C declaration of the event buffer is contained in the generated header
{\tth vecampfit/examples/dz\_kmpippiz/models/kmpippiz\_default/signal\_density\_generated.h} and looks like
\begin{lstlisting}[style=cppstyle]
struct signal_event_buffer_kmpippiz_default {
        double m_km_pip[8];
        double m_km_piz[8];
        double m_pip_piz[8];
        // Other member variables are omitted.
};
\end{lstlisting}

In addition to the event buffer, the density functions also take
a parameter buffer as an argument. The parameter buffer is a user-provided
derived type that is used to store values depending on the fit parameters only.
The signal parameter buffer
in the $\kmpippiz$ example contains buffers for individual resonances
and nonresonant amplitude:
\begin{lstlisting}[style=fortranstyle]
  TYPE, BIND(C) :: SIGNAL_PARAMETER_BUFFER_#MODEL#
    TYPE(SIGNAL_PARAMETER_BUFFER_AMPLITUDE_#MODEL#) A_D0_NR
    TYPE(SIGNAL_PARAMETER_BUFFER_RESONANCE_#MODEL#) RHO770P
    ! Other resonances are omitted.
  END TYPE
\end{lstlisting}
The buffers for individual resonances contain their complex amplitudes,
$m_0^{-2}$, $m_0^{-3}$, inverse momenta and formfactors at the resonance
mass and their derivatives:
\begin{lstlisting}[style=fortranstyle]
  TYPE, BIND(C) :: SIGNAL_PARAMETER_BUFFER_RESONANCE_#MODEL#
    TYPE(SIGNAL_PARAMETER_BUFFER_AMPLITUDE_#MODEL#) AMP_D_R
    REAL(C_REAL_KIND) INVERSE_M2
    REAL(C_REAL_KIND) INVERSE_M3
    REAL(C_REAL_KIND) INVERSE_Q0
    REAL(C_REAL_KIND) DINVERSE_Q0_DM0
    REAL(C_REAL_KIND) INVERSE_F0
    REAL(C_REAL_KIND) DINVERSE_F0_DM0
    REAL(C_REAL_KIND) INVERSE_F0_D
    REAL(C_REAL_KIND) DINVERSE_F0_D_DM0
  END TYPE
\end{lstlisting}

The parameter buffer is updated once at the beginning of each likelihood
evaluation in the sequential code region before density-function calls.
The signal parameter buffer is filled in the $\kmpippiz$ Dalitz analysis
by the subroutine
\begin{lstlisting}[style=fortranstyle]
  SUBROUTINE FILL_SIGNAL_PARAMETER_BUFFER_#MODEL#(PAR_BUF, PAR) BIND(C)
    USE, INTRINSIC :: ISO_C_BINDING
    IMPLICIT NONE
    TYPE(SIGNAL_PARAMETER_BUFFER_#MODEL#), INTENT(OUT) :: PAR_BUF
    REAL(C_REAL_KIND), INTENT(IN) :: PAR(DZ_KMPIPPIZ_SIGNAL_MODEL_PARAMETERS)
    CALL FILL_SIGNAL_PARAMETER_BUFFER_AMPLITUDE(PAR_BUF%A_D0_NR, PAR(D0_NR_A), &
&     PAR(D0_NR_P))
    CALL FILL_SIGNAL_PARAMETER_BUFFER_RHOP(PAR_BUF%RHO770P, &
&     PAR(RHO770P_PIP_PIZ_A), PAR(RHO770P_PIP_PIZ_P), PAR(RHO770P_MASS), &
&     PAR(RHO770P_R), PAR(D0_R), 1)
    ! Calls for other resonances are omitted.
  END SUBROUTINE
\end{lstlisting}
Actual filling is performed by individual-resonance subroutines, for example
\begin{lstlisting}[style=fortranstyle]
  SUBROUTINE FILL_SIGNAL_PARAMETER_BUFFER_AMPLITUDE(PAR_BUF, AMP_D_R_ABS, &
&     AMP_D_R_ARG)
    USE VECAMPFIT_MATHEMATICAL_CONSTANTS
    IMPLICIT NONE
    TYPE(SIGNAL_PARAMETER_BUFFER_AMPLITUDE_#MODEL#), INTENT(OUT) :: PAR_BUF
    REAL(REAL_KIND), INTENT(IN) :: AMP_D_R_ABS, AMP_D_R_ARG
    REAL(REAL_KIND), PARAMETER :: DEG_TO_RAD = &
&     REAL(DEGREE_IN_RADIANS_REAL64, REAL_KIND)
    PAR_BUF%DAMP_DABSA = EXP(CMPLX(0, AMP_D_R_ARG * DEG_TO_RAD, REAL_KIND))
    PAR_BUF%AMP = AMP_D_R_ABS * PAR_BUF%DAMP_DABSA 
    PAR_BUF%DAMP_DARGA = CMPLX(0, DEG_TO_RAD, REAL_KIND) * PAR_BUF%AMP
  END SUBROUTINE
\end{lstlisting}

The density-function subroutines have three input arguments: the event buffer,
fit parameters, and parameter buffer. They have one output argument,
the density vector. Examples of density-function implementation
can be found in the files {\tth background\_density.F90} and
{\tth signal\_density.F90} of the $\kmpippiz$ example. The signal density
looks like
\begin{lstlisting}[style=fortranstyle]
  !> Signal density.
  !>
  !> @param SD      Signal density.
  !> @param EV      Event buffer.
  !> @param PAR     Fit parameters.
  !> @param PAR_BUF Parameter buffer.
  SUBROUTINE SIGNAL_DENSITY_#MODEL#(SD, EV, PAR, PAR_BUF) BIND(C)
!$OMP DECLARE TARGET
    USE, INTRINSIC :: ISO_C_BINDING
    IMPLICIT NONE
!$ACC ROUTINE
    REAL(C_REAL_KIND), INTENT(OUT) :: SD(VECTOR_LENGTH)
    TYPE(SIGNAL_EVENT_BUFFER_#MODEL#), INTENT(IN) :: EV
    REAL(C_REAL_KIND), INTENT(IN) :: PAR(DZ_KMPIPPIZ_SIGNAL_MODEL_PARAMETERS)
    TYPE(SIGNAL_PARAMETER_BUFFER_#MODEL#), INTENT(IN) :: PAR_BUF
    ! Signal-density implementation is omitted.
  END SUBROUTINE
\end{lstlisting}
The density-function C declarations are created by the tool
{\tth vecampfit-fortran} similar to event and parameter buffers. The
declaration of the $\kmpippiz$ signal density looks like
\begin{lstlisting}[style=cppstyle]
void signal_density_kmpippiz_default(double *sd,
        const struct signal_event_buffer_kmpippiz_default *ev,
        const double *par,
        const struct signal_parameter_buffer_kmpippiz_default *par_buf);
\end{lstlisting}

Parallel calculation of the signal-density values is implemented
in Fortran derived types
{\tth VECAMPFIT\_DENSITY\_CALCULATOR\_\#TYPE\_KIND\#\_\#LENGTH\#}
using OpenMP parallel loops and
{\tth VECAMPFIT\_DENSITY\_CALCULATOR\_OFFLOADING\_\#TYPE\_KIND\#\_\#LENGTH\#}
using OpenACC parallel loops for calculations at CPU and GPU, respectively.
The output signal-density vector is a thread-local
variable defined in a {\tth BLOCK} construct within the loop,
while other arguments are constant ({\tth INTENT(IN)} in Fortran),
ensuring the thread safety.
All vectorized library subprograms used for the signal-density calculation
do not need global or static variables. They are explicitly forbidden
to modify the global state by declaring them {\tth PURE} Fortran subroutines.
There are several static variables in \vecampfit{} such as C++ instances
of the {\sc ROOT} random engine and the minimization class {\tth TMinuit},
or Fortran global fitter. None of these variables need to be used in
the parallel regions performing the signal-density evaluation.

\section{Fitting procedure}
\label{sec:fitting_procedure}

The $\kmpippiz$ Dalitz analysis provides examples
of a basic fitting procedure in the files {\tth fit\_signal.cc}
and {\tth fit\_background.cc} for the signal and background,
respectively. In addition, an example of fitting-procedure control from
Fortran is provided in the file {\tth fit\_background.F90}.
The C++-based signal fit is described in detail below.
The fit starts with the fitter initialization:
\begin{lstlisting}[style=cppstyle]
        /* Alignment. */
        const size_t alignment = 64;
        /* Initialize fitter. */
        SUBROUTINE_WITH_KIND_LENGTH(vecampfit_fitter_initialize)(
                1, DZ_KMPIPPIZ_SIGNAL_PARAMETERS, "dz_kmpippiz_event",
                sizeof(dz_kmpippiz_event), alignment, false,
                options->weighted_mc || options->grid_integration,
                options->offloading);
\end{lstlisting}
The macro {\tth SUBROUTINE\_WITH\_KIND\_LENGTH} expands into the subroutine
name followed by two suffixes denoting the variable type and kind,
and vector length.
The number of simultaneous-fit points is specified by the first argument.
The following arguments specify the number of fit parameters,
the name and size of the ROOT class which is used to store the event data,
alignment of event buffers allocated by the fitter,
whether the data and MC events are weighted,
and whether to use GPU offloading.

Fitter initialization is followed by setting the number of background sources
for each simultaneous-fit point:
\begin{lstlisting}[style=cppstyle]
        /* Set number of background sources. */
        res = vecampfit_fitter_set_n_background_sources(1, 1);
        if (res != 0)
                return -1;
\end{lstlisting}
Functions used for fitter setup return non-zero values in case
of incorrect input arguments or other errors like failed memory allocation.
An abnormal return code means that either the program is not correct
or an irrecoverable error happened. The fitting program should exit
in this case.

After setting the number of background sources, the sizes of the event and
parameter buffers, and functions for filling of the buffers can be set:
\begin{lstlisting}[style=cppstyle]
        /* Set function filling signal event buffer. */
        res = vecampfit_fitter_set_fill_signal_event_buffer(
                fill_signal_event_buffer_#model#,
                sizeof(signal_event_buffer_#model#), 1);
        if (res != 0)
                return -1;
        /* Set function filling background event buffer. */
        res = vecampfit_fitter_set_fill_background_event_buffer(
                fill_background_event_buffer_#model#,
                sizeof(background_event_buffer_kmpippiz_default), 1, 1);
        if (res != 0)
                return -1;
        /* Set function filling signal parameter buffer. */
        res = vecampfit_fitter_set_fill_signal_parameter_buffer(
                VECAMPFIT_FITTER_FILL_PARAMETER_BUFFER_FUNCTION(
                fill_signal_parameter_buffer_#model#),
                sizeof(signal_parameter_buffer_#model#));
        if (res != 0)
                return -1;
        /* Set function filling background parameter buffer. */
        res = vecampfit_fitter_set_fill_background_parameter_buffer(
                VECAMPFIT_FITTER_FILL_PARAMETER_BUFFER_FUNCTION(
                fill_background_parameter_buffer_kmpippiz_default),
                sizeof(background_parameter_buffer_kmpippiz_default), 1, 1);
        if (res != 0)
                return -1;
\end{lstlisting}
The parameter buffers are allocated by the corresponding
setting functions, while the event buffers are created by
{\tth vecampfit\_fitter\_load\_data\_file()} at the data-loading stage
described in the following. The user does not need to perform any explicit
management of the buffers other than providing the filling functions,
which are called by the fitter as necessary.

The next step is setting of the signal and background density functions:
\begin{lstlisting}[style=cppstyle]
        /* Set signal density. */
        res = vecampfit_fitter_set_signal_density(
                VECAMPFIT_FITTER_DENSITY_FUNCTION(signal_density_#model#), 1);
        if (res != 0)
                return -1;
        /* Set background density. */
        res = vecampfit_fitter_set_background_density(
                VECAMPFIT_FITTER_DENSITY_FUNCTION(
                background_density_kmpippiz_default), 1, 1);
        if (res != 0)
                return -1;
\end{lstlisting}

The background fractions $f_B^{(k)}$ from Eq.~\eqref{eq:likelihood_vecampfit}
should be fit parameters. Their numbers are set by calling
\begin{lstlisting}[style=cppstyle]
        /* Set background-fraction parameter. */
        res = vecampfit_fitter_set_background_fraction_parameter(
                BACKGROUND_FRACTION, 1, 1);
        if (res != 0)
                return -1;
\end{lstlisting}

If necessary, the integration type is then set. A function for filling
of event class from signal parameter buffer is set for grid integration:
\begin{lstlisting}[style=cppstyle]
        if (options->grid_integration) {
                /* Set integration type to grid. */
                res = vecampfit_fitter_set_normalization_integration_type(
                        INTEGRATION_TYPE_GRID);
                if (res != 0)
                        return -1;
                /*
                 * Set function filling event class from signal parameter
                 * buffer. It is used when writing the result ROOT file
                 * to fill signal events from the points of integration grid.
                 */
                vecampfit_fitter_set_fill_event_class(
                        dz_kmpippiz_fill_event_class, 1);
        }
\end{lstlisting}

The random engine is initialized for assignment of initial parameters:
\begin{lstlisting}[style=cppstyle]
        /* Set random seed for parameter initialization. */
        unsigned int random_seed = options->random_seed;
        if (!options->fixed_random_seed) {
                res = vecampfit_urandom(&random_seed, sizeof(int));
                if (res != 1)
                        return -1;
        }
        vecampfit_random_engine_set_seed(random_seed);
        printf("Random seed: %u\n", random_seed);
\end{lstlisting}
By default, the {\tth TRandom1} generator from ROOT with maximum
random-number quality parameter (4) is used. It is possible to replace
it by other {\tth TRandom} generators by calling {\tth
vecampfit\_set\_random\_generator()}. The random-number generator is static
and multi-thread generation is not supported; however, the generator is
typically necessary only for randomization of initial parameters which
does not need parallel execution.

The fitting then continues with loading of fit parameters:
\begin{lstlisting}[style=cppstyle]
        /* Read signal parameters. */
        std::string file_name;
        if (options->generate_mc) {
                file_name = std::string(VECAMPFIT_DATA_DIRECTORY) +
                        "/examples/dz_kmpippiz/parameters_signal_mc.dat";
        } else {
                file_name = std::string(VECAMPFIT_DATA_DIRECTORY) +
                        "/examples/dz_kmpippiz/parameters_signal.dat";
        }
        res = vecampfit_fitter_load_parameters_file(FUNCTION_TYPE_SIGNAL,
                DZ_KMPIPPIZ_SIGNAL_PARAMETERS, file_name.c_str(), 1);
        if (res != 0)
                return -1;
        /* Read background parameters. */
        file_name = std::string(VECAMPFIT_DATA_DIRECTORY) +
                "/examples/dz_kmpippiz/parameters_background.dat";
        res = vecampfit_fitter_load_parameters_file(FUNCTION_TYPE_BACKGROUND,
                DZ_KMPIPPIZ_BACKGROUND_MODEL_PARAMETERS,
                file_name.c_str(), 1);
        if (res != 0)
                return -1;
\end{lstlisting}
The parameters are stored in a text file with a single line corresponding
to each parameter and containing the parameter name, its initial value $I$,
initial numeric-minimization step, lower and upper bounds,
and random change $R$.
As in {\sc Minuit} which is used as the fitting engine, if both the lower and
upper bounds are equal to zero, then the parameter is not constrained.
When fitting, the initial value is chosen at random with a uniform
distribution in the range $(I - R, I + R)$.
The initial lines of the parameter file {\tth parameters\_signal.dat}
for the signal fit in the $\kmpippiz$ example are:
\begin{lstlisting}[style=bashstyle]
D_R          5.0 1E-3 0E0 0E0      0E0
RHOP1_M    0.770 1E-3 0E0 0E0      0E0
RHOP1_G   0.1507 1E-3 0E0 0E0      0E0
RHOP1_R      1.5 1E-3 0E0 0E0      0E0
\end{lstlisting}
The masses and widths of the resonances are taken from
the CLEO analysis in Ref~\cite{CLEO:2000fvk}.

The fit preparation finishes by reading the normalization Monte Carlo
and data events:
\begin{lstlisting}[style=cppstyle]
        /* Read normalization Monte Carlo. */
        if (!(options->generate_mc && options->normalization_mc &&
                        !options->weighted_mc_full_density) &&
                        !options->grid_integration) {
                if (options->weighted_mc) {
                        res = vecampfit_fitter_load_data_file(
                                "mc_signal_weighted.root", "tree", "event",
                                DATA_TYPE_MONTE_CARLO, 1, 0);
                } else {
                        res = vecampfit_fitter_load_data_file("mc_signal.root",
                                 "tree", "event", DATA_TYPE_MONTE_CARLO, 1, 0);
                }
                if (res != 0)
                        return -1;
        }
        /* Read data. */
        if (options->pseudoexperiment >= 0) {
                file_name = std::string("data_signal_") +
                        std::to_string(options->pseudoexperiment) + ".root";
        } else {
                file_name = "data_signal.root";
        }
        res = vecampfit_fitter_load_data_file(file_name.c_str(), "tree",
                "event", DATA_TYPE_DATA, 1, 0);
        if (res != 0)
                return -1;
\end{lstlisting}
The example provides a way to choose between Monte Carlo samples with uniform
distribution and weighted in accordance with the full density function.

The fitting is performed by the TMinuit class from ROOT
via \vecampfit{} interface. The fitting-command sequence is
\begin{lstlisting}[style=cppstyle]
        /* Fix all parameters. */
        vecampfit_minuit_interface_fix_range(1, SIGNAL_PARAMETERS);
        /* Release absolute values of the amplitudes. */
        vecampfit_minuit_interface_release(D0_NR_A);
        //vecampfit_minuit_interface_release(RHO770P_PIP_PIZ_A);
        vecampfit_minuit_interface_release(RHO1700P_PIP_PIZ_A);
        vecampfit_minuit_interface_release(KST892M_KM_PIZ_A);
        vecampfit_minuit_interface_release(KST01430M_KM_PIZ_A);
        vecampfit_minuit_interface_release(KST1680M_KM_PIZ_A);
        vecampfit_minuit_interface_release(KST892Z_KM_PIP_A);
        vecampfit_minuit_interface_release(KST01430Z_KM_PIP_A);
        vecampfit_minuit_interface_migrad(100000);
        /* Release phases of the amplitudes. */
        vecampfit_minuit_interface_release(D0_NR_P);
        //vecampfit_minuit_interface_release(RHO770P_PIP_PIZ_P);
        vecampfit_minuit_interface_release(RHO1700P_PIP_PIZ_P);
        vecampfit_minuit_interface_release(KST892M_KM_PIZ_P);
        vecampfit_minuit_interface_release(KST01430M_KM_PIZ_P);
        vecampfit_minuit_interface_release(KST1680M_KM_PIZ_P);
        vecampfit_minuit_interface_release(KST892Z_KM_PIP_P);
        vecampfit_minuit_interface_release(KST01430Z_KM_PIP_P);
        /* Release background fraction. */
        vecampfit_minuit_interface_release(BACKGROUND_FRACTION);
        vecampfit_minuit_interface_migrad(100000);
        /* Releasing of resonance masses and widths is omitted. */
\end{lstlisting}

Finally, the fit results are recorded to a ROOT file with density-function
values, contributions of each background source, and original events;
a parameter file suitable for another run of the fitting program;
a ROOT file with the resulting minimum value of the likelihood function
$\mathcal{F}$, parameters,
errors, and error matrix with its eigenvectors and eigenvalues;
and a text file with the same results except the error matrix.
\begin{lstlisting}[style=cppstyle]
        if (options->save_density) {
                vecampfit_fitter_write_density_root("signal_density.root", 1,
                        false, options->disable_normalization);
        }
        vecampfit_fitter_write_result_parameters(
                "parameters_signal_result.dat");
        if (options->pseudoexperiment >= 0) {
                file_name = std::string("signal_result_") +
                        std::to_string(options->pseudoexperiment) + ".root";
        } else {
                file_name = "signal_result.root";
        }
        vecampfit_write_fit_result_root(file_name.c_str());
        vecampfit_fitter_write_result_text("results.txt");
        if (options->offloading) {
                vecampfit_offloading_free(signal_density_device);
                vecampfit_offloading_free(signal_density_gradient_device);
        }
\end{lstlisting}

Various advanced aspects of the fitting procedure
are described together with the $\kkpipi$ analysis workflow
in Sec.~\ref{sec:epem_kkpipi_workflow}.

\section{Analysis workflow example}
\label{sec:epem_kkpipi_workflow}

The process $\kpkmpippim$ has been studied by {\sc BaBar}~\cite{BaBar:2011btv},
CMD-3~\cite{Shemyakin:2015cba} and BESIII~\cite{BESIII:2021ftf}.
The CMD-3 collaboration performed an amplitude fit, however, it was not
a full amplitude analysis because of additional restrictions on relative
phases of various amplitudes, which were only allowed to be equal to 0
or $\pi$. It was found that the dominant contributions to the amplitude are
$e^+ e^- \to K_1(1270)^+ \km$ and $e^+ e^- \to K_1(1400)^+ \km$. The fit has
been performed at 9 combined energy points with $\sqrt{s}$ from
1600 to 2000 MeV. One-dimensional studies by {\sc BaBar} also indicate
the presence of the $K_1(1270)^+$ and $K_1(1400)^+$ resonances.
The process $\kskspippim$ has been studied
by {\sc BaBar}~\cite{BaBar:2014uwz} and CMD-3~\cite{CMD-3:2019cvx}.

A simplified simultaneous amplitude analysis of the processes $\kpkmpippim$
and $\kskspippim$ is implemented as an example of analysis workflow.
It includes parameterizations of efficiency, resolution, background, and
fit to the signal events. The fit is performed at 5 different energy points
with $\sqrt{s} = $ 1800, 1850, 1900, 1950, and 2000 $\mev$ simultaneously.

\subsection{Formalism}

The decay chains are similar for both $\kpkmpippim$ and $\kskspippim$.
The final-state particles are denoted as $\kIkIIpiIpiII$.
There are four chains proceeding via three-body resonances:
$\vpho \to \kjI \kII$, $\kjI \to \kI \piI \piII$;
$\vpho \to \kjII \kI$, $\kjII \to \kII \piII \piI$;
$\vpho \to \ajI \piII$, $\ajI \to \piI \kI \kII$;
$\vpho \to \ajII \piI$, $\ajII \to \piII \kII \kI$.
The three-body state has three possible decay channels; for example,
for the first chain they are $\kjI \to \kstjIaI \piII$, $\kstjIaI \to \kI \piI$
(cannot proceed through conventional resonances for $\kpkmpippim$);
$\kjI \to \kstjIaII \piI$, $\kstjIaII \to \kI \piII$;
$\kjI \to \rhoj \kI$, $\rhoj \to \piI \piII$.
In addition, there are three decay chains with a pair of two-body resonances:
$\vpho \to \phij \fj$, $\phij \to \kI \kII$, $\fj \to \piI \piII$;
$\vpho \to \kstjIaI \kstjIIaII$, $\kstjIaI \to \kI \piI$,
$\kstjIIaII \to \kII \piII$; $\vpho \to \kstjIaII \kstjIIaI$,
$\kstjIaII \to \kI \piII$, $\kstjIIaI \to \kII \piI$.

The decay phase space is 7-dimensional. It is convenient to use the
$\vpho \to \phij \fj$ decay chain for resolution determination because
narrow resonances $\phi$ and $\omega$ are expected to be observed in
the $\kp \km$ and $\pip \pim$ invariant masses, respectively.
For this chain, the helicity-formalism variables are
\begin{equation}
\Phi = (\theta_{\vpho}^{(5)}, M_{\kI \kII}, \theta_{\phij}^{(5)},
  \varphi_{\kI}^{(5)}, M_{\piI \piII}, \theta_{\fj}^{(5)},
  \varphi_{\piI}^{(5)}),
\end{equation}
where $\theta$ and $\varphi$ are the helicity and azimuthal angles of
the particle specified by the subscript, respectively, and the index in
round brackets specifies the decay chain.
These variables are used for parameterization of efficiency, resolution,
and background.

Two-body resonances are described by the Breit-Wigner amplitude including
a phase-space factor:
\begin{equation}
\begin{aligned}
B_R(m) & =
 \frac{\left(\frac{q}{q_0}\right)^{L} \frac{F_L(q)}{F_L(q_0)}}
      {m_0^2 - m^2 - i m_0 \Gamma(m)}, \\
\end{aligned}
\label{eq:breit_wigner_q_f}
\end{equation}
where $\Gamma(m)$ is given by Eq.~\eqref{eq:breit_wigner_q_f_nophsp_gamma}.
Three-body resonances are described by the Breit-Wigner amplitude
with a constant width and the invariant mass in the imaginary part
of the denominator:
\begin{equation}
B_R^{(m)}(m) = \frac{1}{m_0^2 - m^2 - i m \Gamma_0}.
\end{equation}
This is a simplification suitable for an example but not for real data analysis.
In an actual analysis, it is necessary to take the dependence of the $\kj$
decay matrix element on $K \pi \pi$ invariant mass into account
and calculate the mass-dependent width similar to Ref.~\cite{LHCb:2017swu}.

The amplitude is calculated using helicity formalism (for actual analysis,
covariant formalism is usually preferred for annihilation processes
in this energy range). Detailed amplitude calculation
including exact definitions of all angles is presented
in~\ref{sec:kkpipi_amplitude}.
The result is
\begin{equation}
\begin{aligned}
A^{\vpho \to \kjI \kII}_{\lambda_{\vpho}}(\Phi) = &
\sum\limits_{\lambda_{\kjI}=-1,0,1}
  \dfun{1}{\lambda_{\vpho}}{\lambda_{\kjI}}{\theta_{\vpho}^{(1)}}
\sum\limits_{\kjI}
  H^{(\vpho \to \kjI \kII)}_{\lambda_{\kjI}\,0}
  B_{\kjI}^{(m)}(M_{\kI \piI \piII}) \\
& \times \left(
  A^{(\kjI \to \kstjIaI \piII)}_{\lambda_{\kjI}}(\Phi)
+ A^{(\kjI \to \kstjIaII \piI)}_{\lambda_{\kjI}}(\Phi)
+ A^{(\kjI \to \rhoj \kI)}_{\lambda_{\kjI}}(\Phi)\right), \\
\end{aligned}
\label{eq:amplitude_chain_1}
\end{equation}
where $\lambda$ is the helicity of the particle specified by the subscript,
$d$ are Wigner $d$ functions~\cite{wigner_1931}, $H$ are helicity amplitudes,
\begin{equation}
\begin{aligned}
A^{(\kjI \to \kstjIaI \piII)}_{\lambda_{\kjI}}(\Phi) = &
  e^{i \lambda_{\kjI} \varphi_{\kstjIaI}^{[1]}}
\sum\limits_{\lambda_{\kstjIaI} = -J(\kjI)}^{J(\kjI)}
  \dfun{J(\kjI)}{\lambda_{\kjI}}{\lambda_{\kstjIaI}}{\theta_{\kjI}^{[1]}} \\
& \times
\sum\limits_{J(\kstjIaI)}
  e^{i \lambda_{\kstjIaI} \varphi_{\kI}^{[1]}}
  \dfun{J(\kstjIaI)}{\lambda_{\kstjIaI}}{0}{\theta_{\kstjIaI}^{[1]}} \\
& \times
\sum\limits_{\kstjIaI}
  H^{(\kjI \to \kstjIaI \piII)}_{\lambda_{\kstjIaI}\,0}
  H^{(\kstjIaI \to \kI \piI)}_{0\,0}
  B_{\kstjIaI}(M_{\kI \piI}), \\
\end{aligned}
\label{eq:amplitude_chain1_1}
\end{equation}
\begin{equation}
\begin{aligned}
A^{(\kjI \to \kstjIaII \piI)}_{\lambda_{\kjI}}(\Phi) = &
  e^{i \lambda_{\kjI} \varphi_{\kstjIaII}^{[2]}}
\sum\limits_{\lambda_{\kstjIaII} = -J(\kjI)}^{J(\kjI)}
  \dfun{J(\kjI)}{\lambda_{\kjI}}{\lambda_{\kstjIaII}}{\theta_{\kjI}^{[2]}} \\
& \times
\sum\limits_{J(\kstjIaII)}
  e^{i \lambda_{\kstjIaII} \varphi_{\kI}^{[2]}}
  \dfun{J(\kstjIaII)}{\lambda_{\kstjIaII}}{0}{\theta_{\kstjIaII}^{[2]}} \\
& \times
\sum\limits_{\kstjIaII}
  H^{(\kjI \to \kstjIaII \piI)}_{\lambda_{\kstjIaII}\,0}
  H^{(\kstjIaII \to \kI \piII)}_{0\,0}
  B_{\kstjIaII}(M_{\kI \piII}), \\
\end{aligned}
\label{eq:amplitude_chain1_2}
\end{equation}
and
\begin{equation}
\begin{aligned}
A^{(\kjI \to \rhoj \kI)}_{\lambda_{\kjI}}(\Phi) = &
  e^{i \lambda_{\kjI} \varphi_{\rhoj}^{[3]}}
\sum\limits_{\lambda_{\rhoj} = -J(\kjI)}^{J(\kjI)}
  \dfun{J(\kjI)}{\lambda_{\kjI}}{\lambda_{\rhoj}}{\theta_{\kjI}^{[3]}} \\
& \times
\sum\limits_{J(\rhoj)}
  e^{i \lambda_{\rhoj} \varphi_{\piI}^{[3]}}
  \dfun{J(\rhoj)}{\lambda_{\rhoj}}{0}{\theta_{\rhoj}^{[3]}} \\
& \times
\sum\limits_{\rhoj}
  H^{(\kjI \to \rhoj \kI)}_{\lambda_{\rhoj}\,0}
  H^{(\rhoj \to \piI \piII)}_{0\,0}
  B_{\rhoj}(M_{\piI \piII}). \\
\end{aligned}
\label{eq:amplitude_chain1_3}
\end{equation}
Note that the summation over intermediate-resonance helicity in
Eqs.~\eqref{eq:amplitude_chain1_1}, \eqref{eq:amplitude_chain1_2},
and \eqref{eq:amplitude_chain1_3}
is performed after summation over intermediate resonances with arbitrary spin
and consequently includes all possible intermediate resonances and
their helicities. Single-resonance selection rules such as
$|\lambda_{\kstjIaI}| \leq J(\lambda_{\kstjIaI})$
in Eq.~\eqref{eq:amplitude_chain1_1} need to be applied during summation
over intermediate resonances.

The helicity amplitudes are calculated from partial-wave amplitudes,
which are used as actual fit parameters, using Eq.~(B5)
from Ref.~\cite{Jacob:1959at}. The partial-wave amplitudes
of isospin-related channels of the decays $\kst \to K \pi$
and $\rho \to \pi \pi$, as well as ratios between
the partial-wave amplitudes of isospin-related channels of the decays
$K_J \to K^* \pi$ and $K_J \to \rho K$ are fixed
at the corresponding Clebsch-Gordan coefficients.
For both $\kpkmpippim$ and $\kskspippim$, the second decay chain
$\vpho \to \kjII \kI$, $\kjII \to \kII \piII \piI$ is charge-conjugate to
the first one and its amplitude $A^{\vpho \to \kjII \kI}_{\lambda_{\vpho}}$
has the same partial-wave amplitudes. Only the dominant
contributions $e^+ e^- \to K_1(1270) \kb$ and $e^+ e^- \to K_1(1400) \kb$
are included into the example, thus, implementation of other decay chains
is not necessary.
Decays to $\rho(770) \kp$ and $\kst(892) \pip$ are included for both
$K_1(1270)$ and $K_1(1400)$.
The signal density function in case of a non-polarized beam is
\begin{equation}
S = \sum\limits_{\lambda_{\vpho} = -1, 1}
  |A^{\vpho \to \kjI \kII}_{\lambda_{\vpho}}(\Phi)
   + A^{\vpho \to \kjII \kI}_{\lambda_{\vpho}}(\Phi)|^2.
\end{equation}
The helicity of a virtual photon originating from
annihilation of ultrarelativistic electron and positron has two possible
values $\lambda_{\vpho} = -1, 1$; $\lambda_{\vpho} = 0$ is strongly suppressed.

\subsection{Efficiency parameterization and fits with explicit gradient calculation}
\label{sec:kkpipi_efficiency}

Parameterization of reconstruction efficiency is generally not necessary
for signal fitting because the likelihood function given by
Eq.~\eqref{eq:likelihood_vecampfit_simultaneous} takes the efficiency
into account via distribution of the normalization MC events passed through
detector simulation. Parameterization of the efficiency is used in the
$\kkpipi$ example for generation of resolution fit-result histograms.
It may also be necessary to parameterize the efficiency to generate
pseudoexperiments with the same distribution of signal events as in data.

The efficiency is evaluated by performing a fit to a generator-level
signal MC sample that includes information on whether the true signal event
was reconstructed. For real analysis,
such sample can be produced by passing generator-level events through full
Geant4 detector simulation~\cite{GEANT4:2002zbu,Allison:2006ve,Allison:2016lfl}.
A simple parameterization is used in the example instead.
The reconstruction efficiency is calculated as
a product of efficiencies for individual final-state particles:
\begin{equation}
\epsilon(\Phi^\text{(lab)}) = \prod\limits_{\kI, \kII, \piI, \piII}
  \epsilon_f(\Phi_f^\text{(lab)}),
\end{equation}
where $f$ is the final-state particle index, $\Phi_f^\text{(lab)}$
is the laboratory-frame phase space for individual final-state particle,
and $\Phi^\text{(lab)}$ is the laboratory-frame phase space for all
final-state particles. The laboratory-frame phase-space variables are
not independent due to kinematic constraints from the decay. For tracks
\begin{equation}
\begin{aligned}
\epsilon_f(\Phi_f^\text{(lab)}) = & 0.92 [1.0 - 0.2 \exp(- 5.0 p_t)] \\
& \times
  T(\cos \theta, -0.8, 200) T(- \cos \theta, 0.8, 200) T(p_t, 0.1, 200),
\end{aligned}
\end{equation}
where $p_t$ is the transverse momentum measured in $\gevc$,
$\theta$ is the polar angle in the laboratory frame and $T$
is the logistic function describing thresholds:
\begin{equation}
T(x, x_0, k) = \frac{1}{1 + \exp [- k (x - x_0)]}.
\end{equation}

The calibrated ratio between the efficiencies in data and MC is given by
\begin{equation}
R(\Phi^\text{(lab)}) = \prod\limits_{\kI, \kII, \piI, \piII}
  R_f(\Phi_f^\text{(lab)}),
\end{equation}
where for tracks
\begin{equation}
R_f = 0.99 - 0.02 \exp(-p_t).
\end{equation}
The efficiency is multiplied by this correction when generating data events
and it is used as weight for MC events.

The efficiency parameterization used for fitting is given by
\begin{equation}
E(\Phi^\text{(lab)}, \Phi) = \Sig\left(
  \Sig^{-1}(\min[\max(E_0(\Phi^\text{(lab)}), \epsilon_1), 1 - \epsilon_1]
  + P_1(\Phi) \right),
\label{eq:efficiency}
\end{equation}
where $\Phi$ is the decay phase space,
$\epsilon_1 = 1 \times 10^{-15}$, $P_1$ is a seven-dimensional
first-order polynomial, $\Sig$ is the sigmoid (standard logistic) function
\begin{equation}
\Sig(x) = \frac{1}{1 + e^{-x}},\ \Sig^{-1}(x) = - \ln(\frac{1}{x} - 1),
\end{equation}
and $E_0$ is the efficiency before correction given by
\begin{equation}
\begin{aligned}
E_0(\Phi^\text{(lab)}) = \prod\limits_{\kI, \kII, \piI, \piII}
  & P_n(x_{p_t}, \cos \theta)
  T(\cos \theta, (\cos \theta)_\text{min}, k_1) \\
& \times
  T(- \cos \theta, -(\cos \theta)_\text{max}, k_2)
  T(x_{p_t}, (x_{p_t})_\text{min}, k_3),
\end{aligned}
\end{equation}
where $P_n$ is a two-dimensional polynomial using Chebyshev polynomials as
a basis, its order $n$ being 4 for $\kpkmpippim$ and 8 for $\kskspippim$,
$x_{p_t}$ is linearly transformed transverse momentum:
\begin{equation}
x_{p_t} = \frac{p_t}{2 p_\text{max}} - 1,
\end{equation}
where $p_\text{max}$ is maximum momentum, which corresponds to
the minimum invariant mass of the combination of the other three particles,
$(\cos \theta)_\text{min}$, $(\cos \theta)_\text{max}$,
$(x_{p_t})_\text{min}$, and $k_i$ are threshold parameters.

An unbinned maximum likelihood fit is performed using generator-level events.
The likelihood is
\begin{equation}
{\mathcal L}(\Phi^\text{(lab)}_j, \Phi_j) =
\begin{dcases*}
-2 \ln g\left(\frac{E(\Phi^\text{(lab)}_j,\Phi_j)}{w_j}\right) & for reconstructed events, \\
-2 \ln \left[1 - g\left(\frac{E(\Phi^\text{(lab)}_j,\Phi_j)}{w_j}\right)\right] & for non-reconstructed events, \\
\end{dcases*}
\end{equation}
where $j$ is the MC event index, $w_j$ is the event weight, and
\begin{equation}
g(x) =
\begin{dcases*}
\epsilon_1 & for $x < \epsilon_1$, \\
x & for $\epsilon_1 \le x < 1 - \epsilon_2$, \\
1 - \epsilon_2 + \frac{\epsilon_2[x - (1 - \epsilon_2)]}
{\epsilon_2 + [x - (1 - \epsilon_2)]}  & for $1 - \epsilon_2 \le x$, \\
\end{dcases*}
\end{equation}
where $\epsilon_2 = 1 \times 10^{-5}$. The function $g(x)$ protects
against appearance of incorrect efficiency values during fitting.
If it is necessary to determine the efficiency in data, then
weights are defined as the ratio between the efficiencies in data and MC:
$w_j = R(\Phi_j) = E_\text{data}(\Phi_j)/E_\text{MC}(\Phi_j)$.
If it is necessary to determine the efficiency in MC, then $w_j = 1$.

\begin{figure}
{\centering
\includegraphics[width=3.3cm]{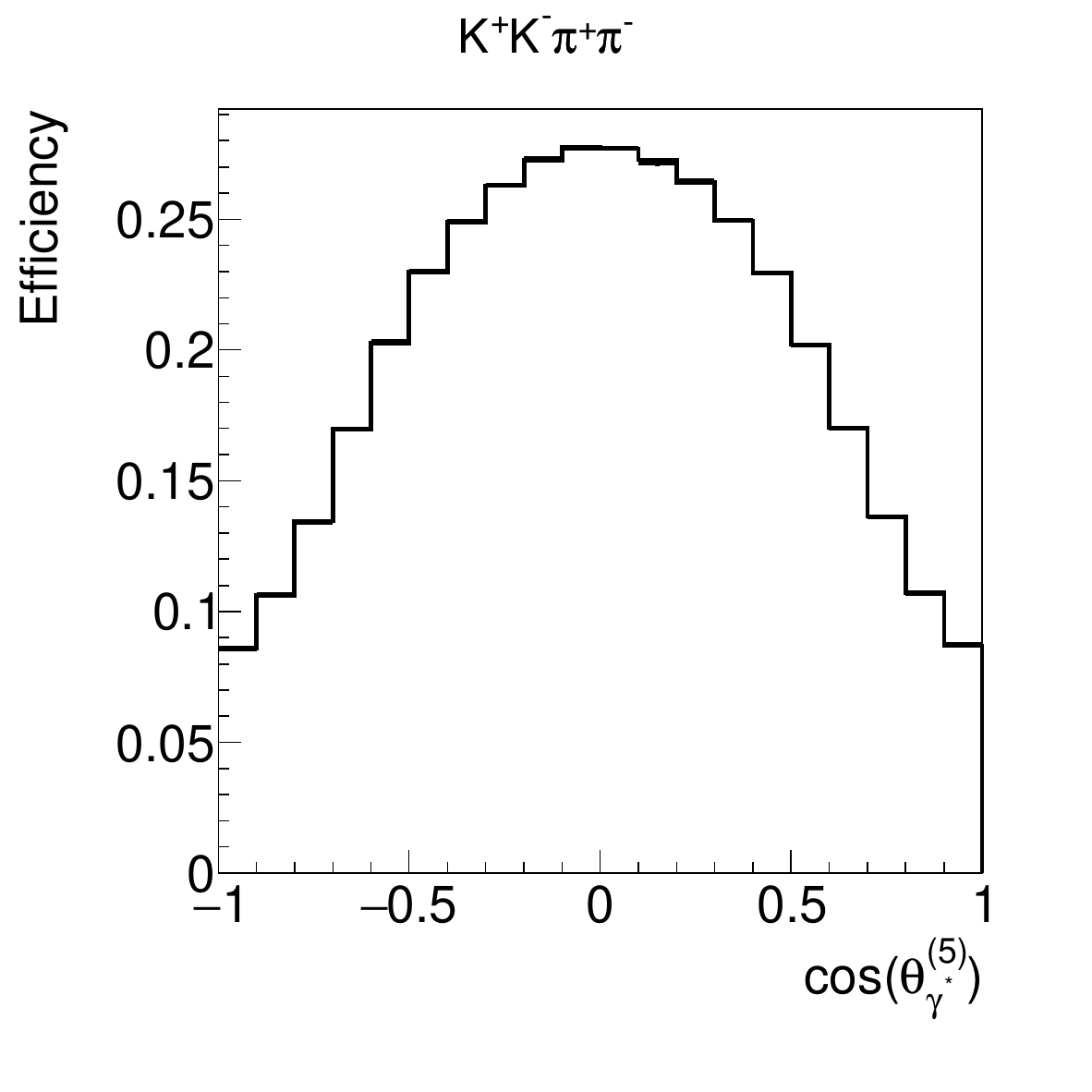}
\includegraphics[width=3.3cm]{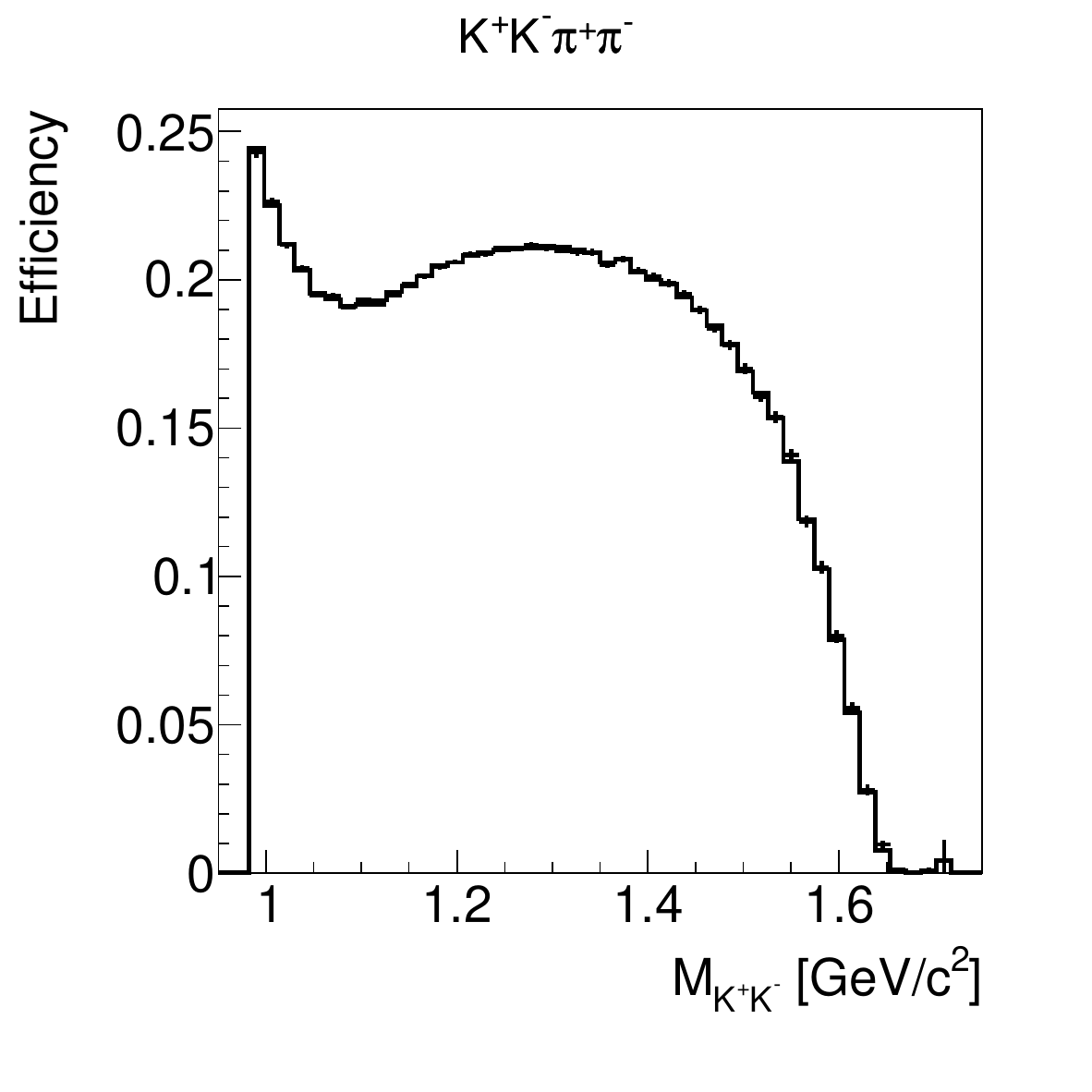}
\includegraphics[width=3.3cm]{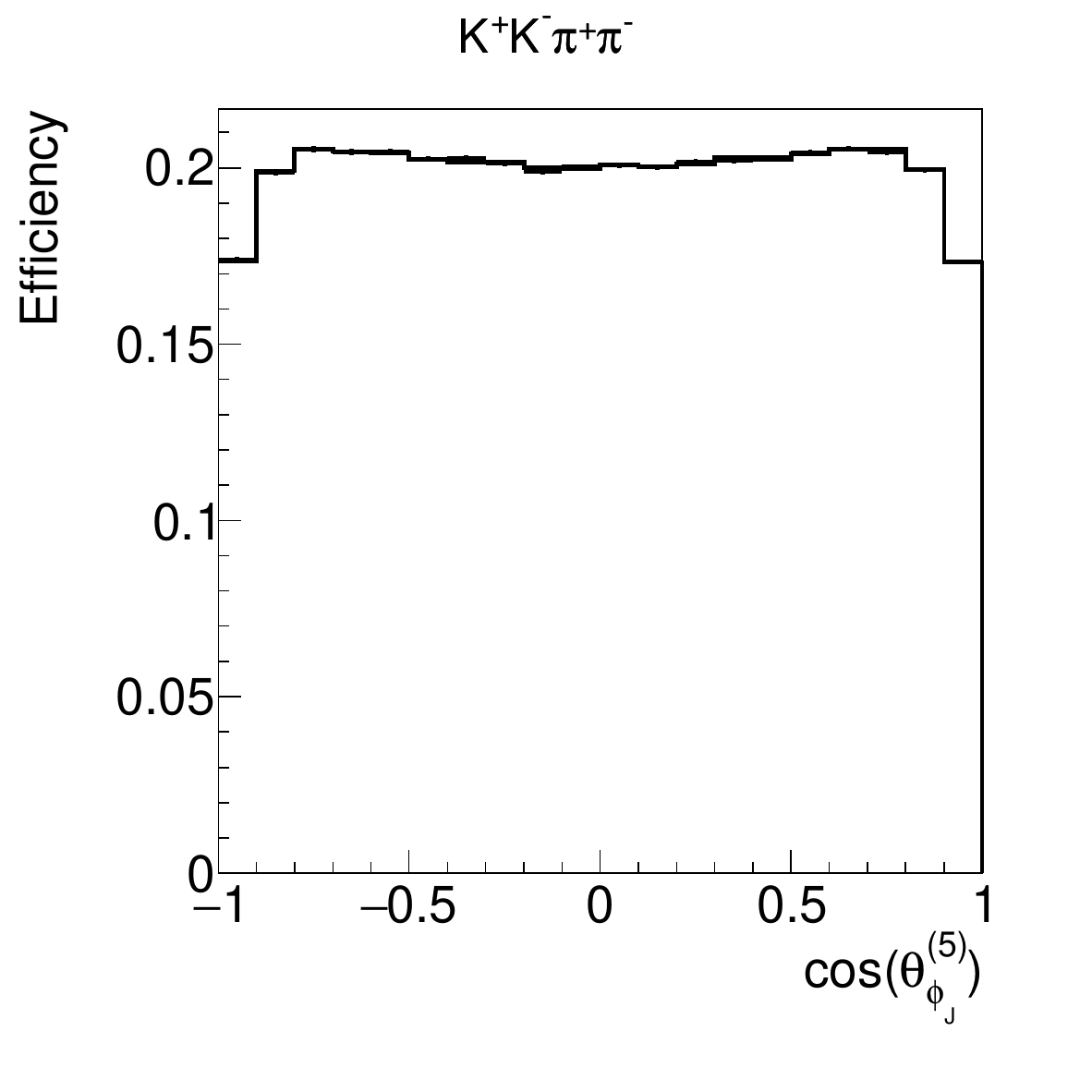}
\includegraphics[width=3.3cm]{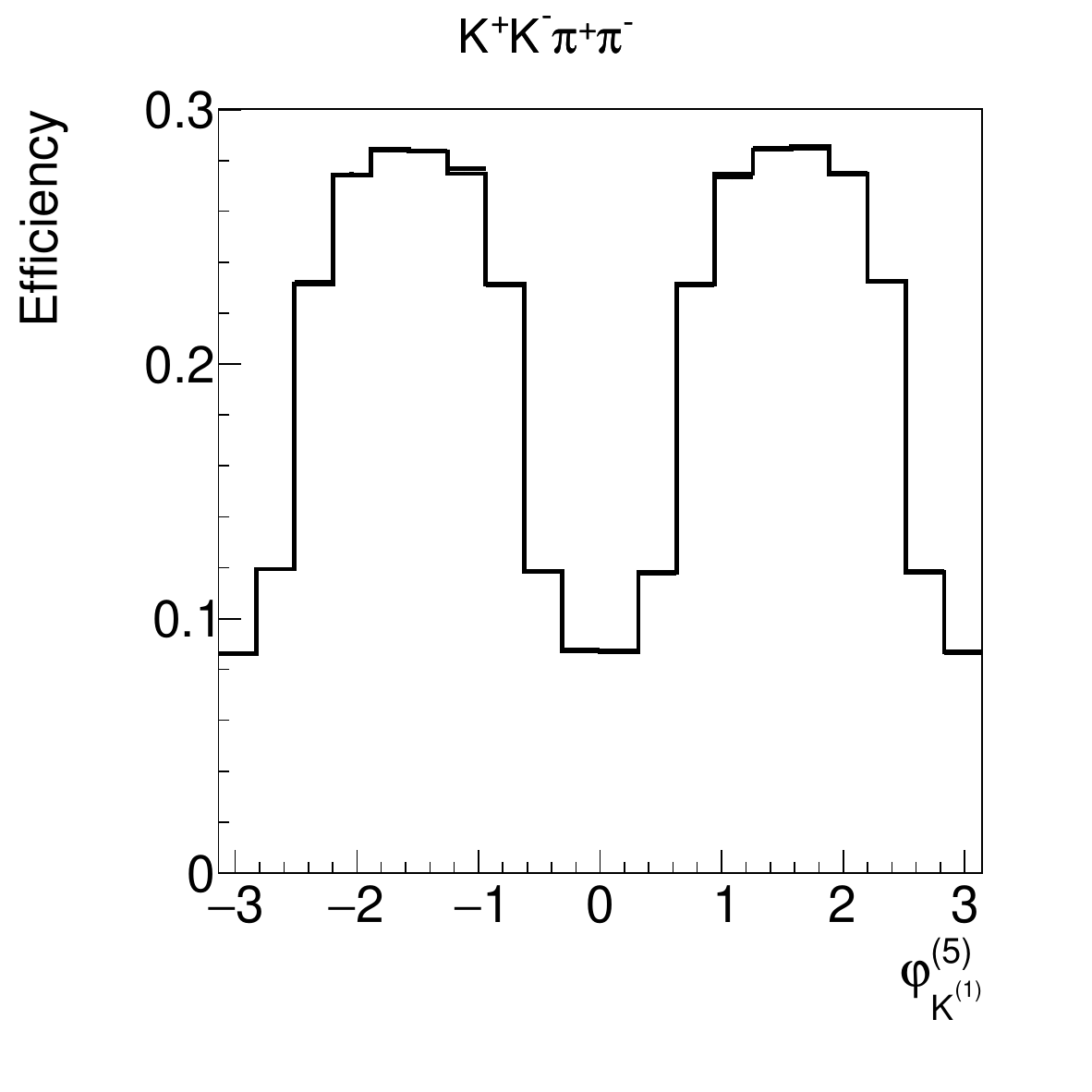} \\
\includegraphics[width=3.3cm]{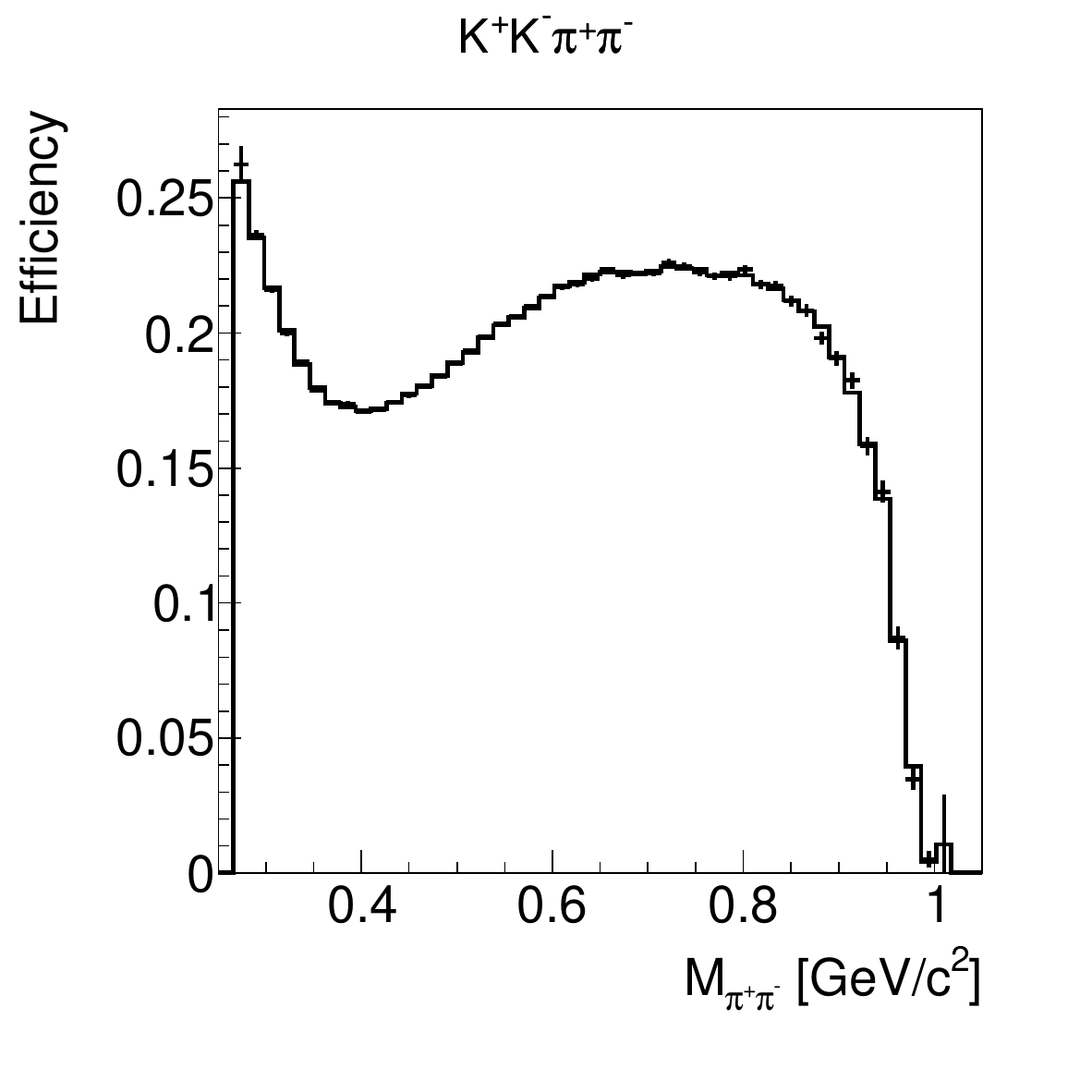}
\includegraphics[width=3.3cm]{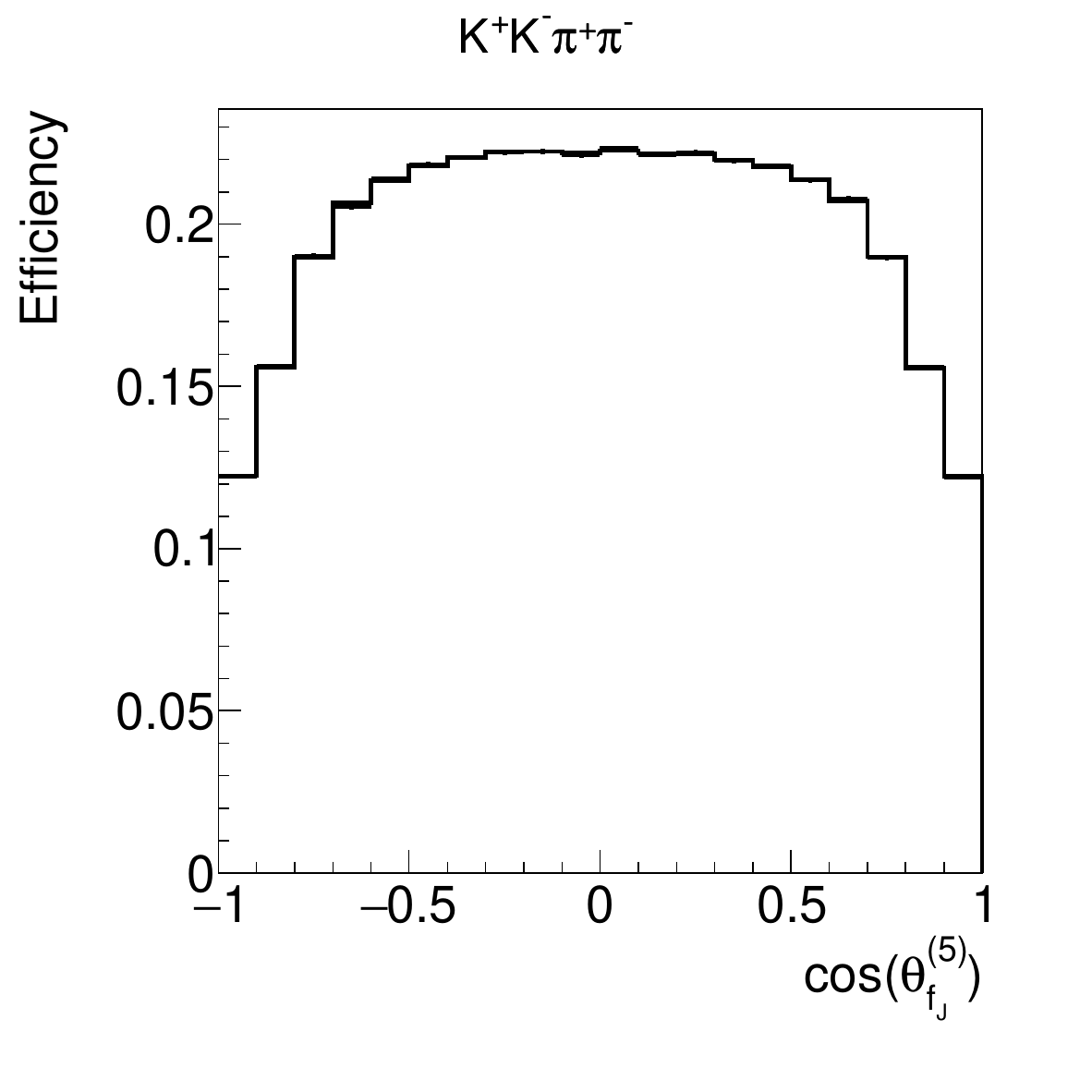}
\includegraphics[width=3.3cm]{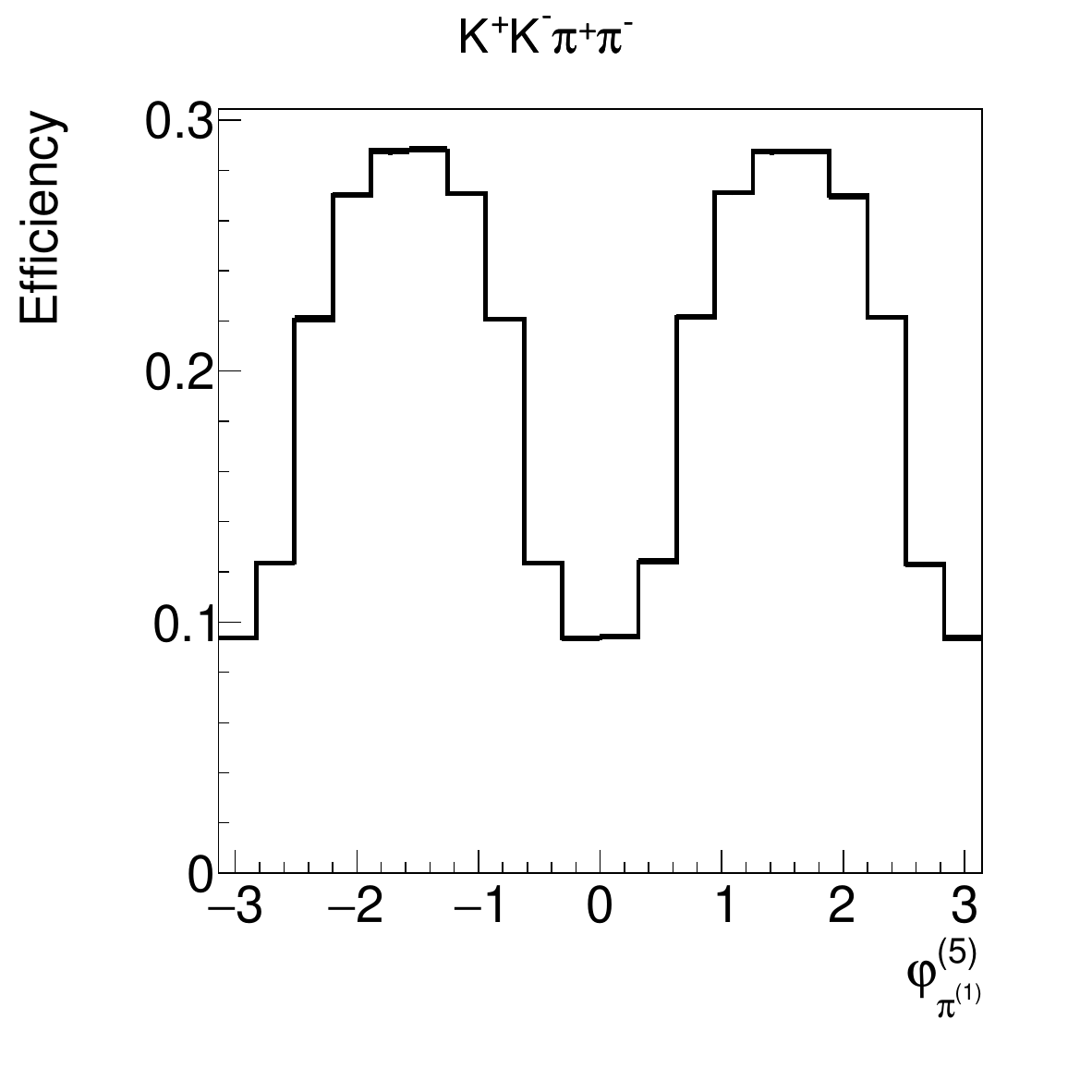} \\
\includegraphics[width=3.3cm]{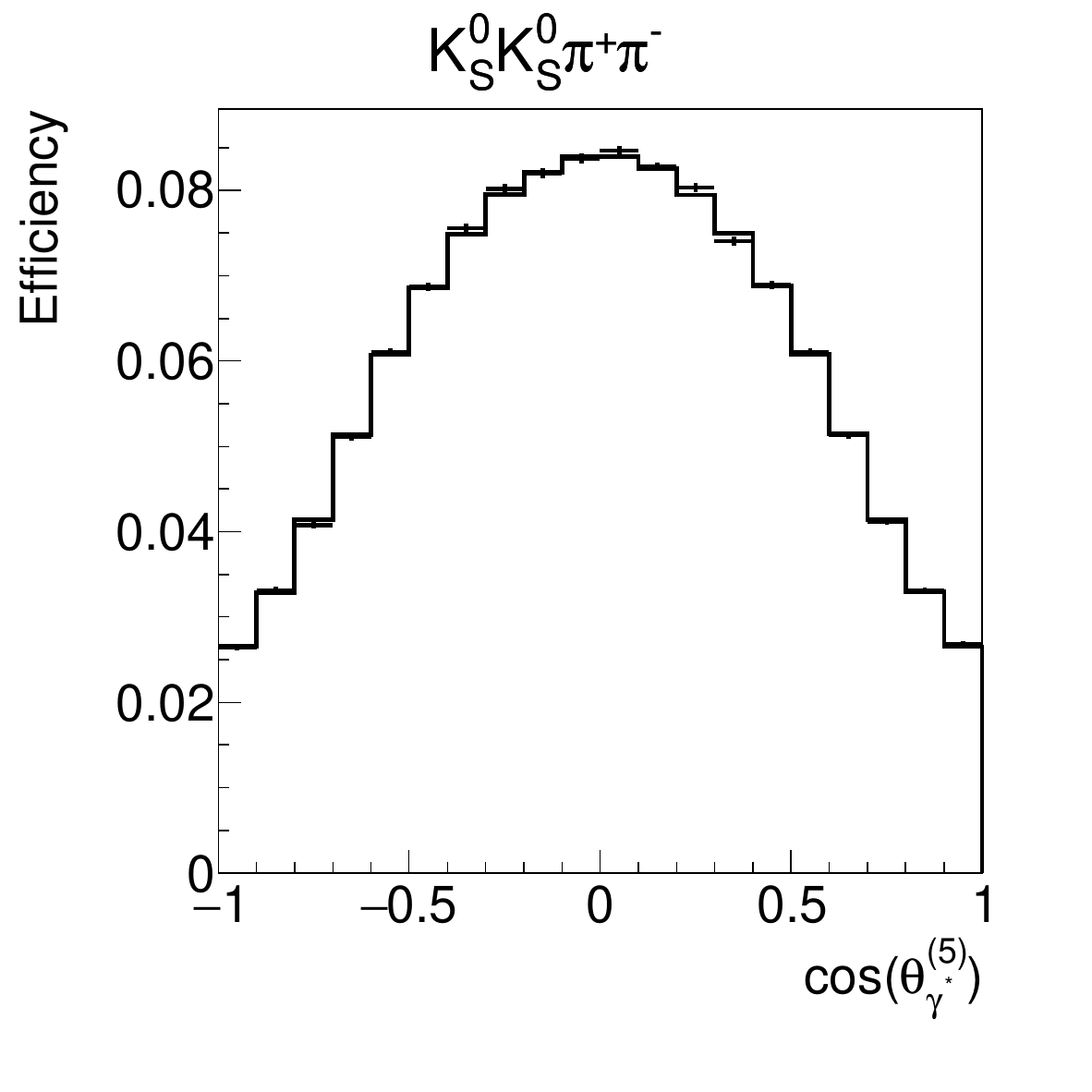}
\includegraphics[width=3.3cm]{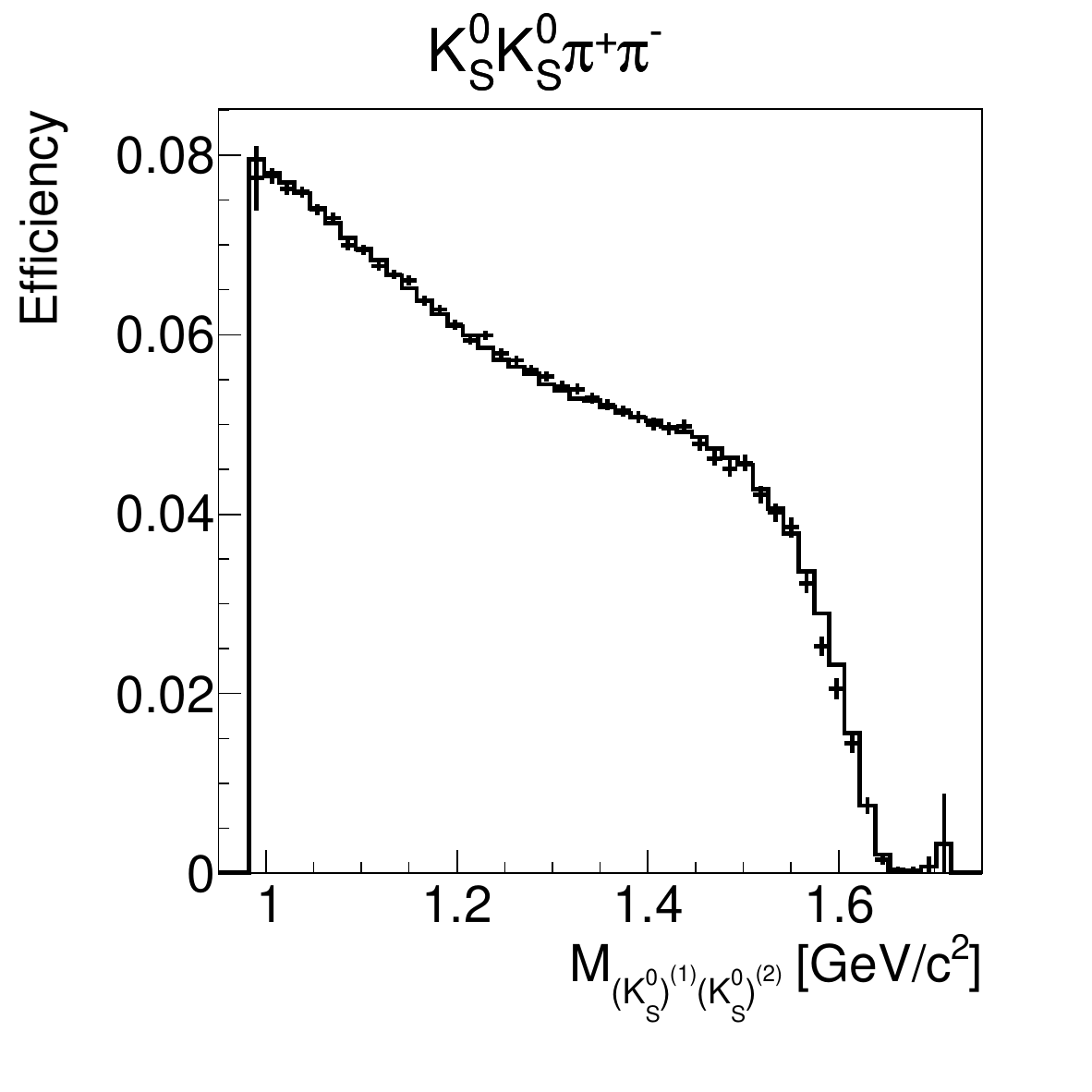}
\includegraphics[width=3.3cm]{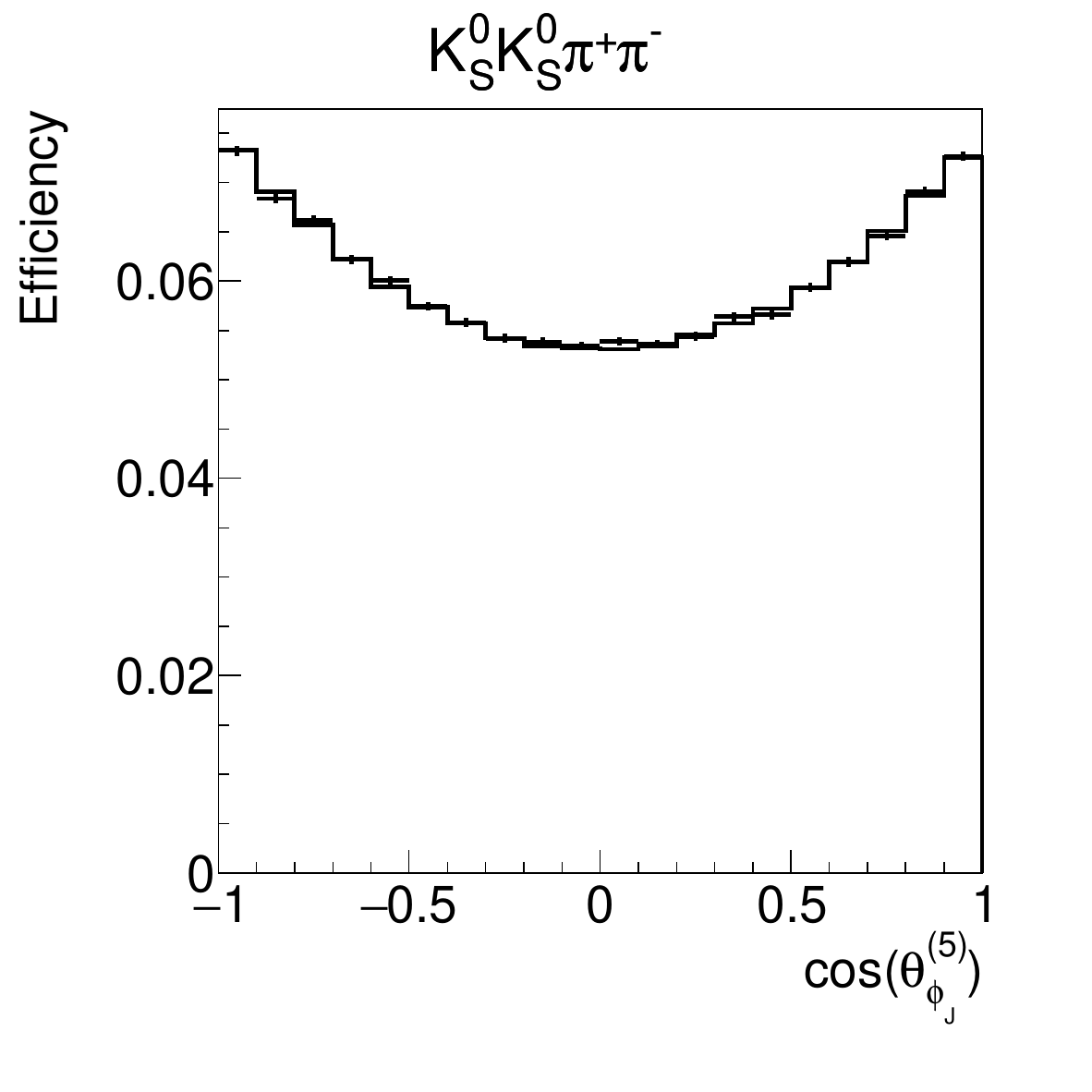}
\includegraphics[width=3.3cm]{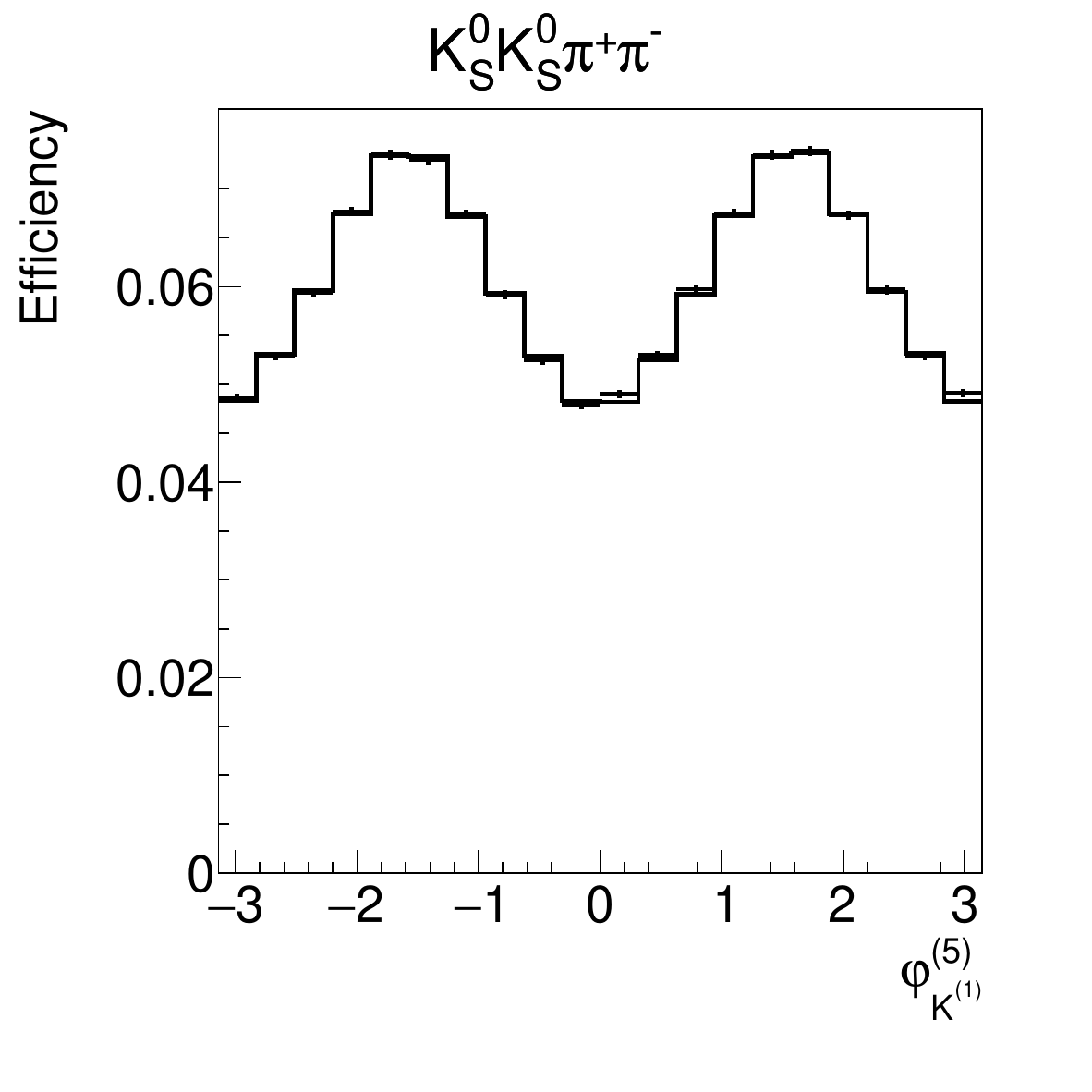} \\
\includegraphics[width=3.3cm]{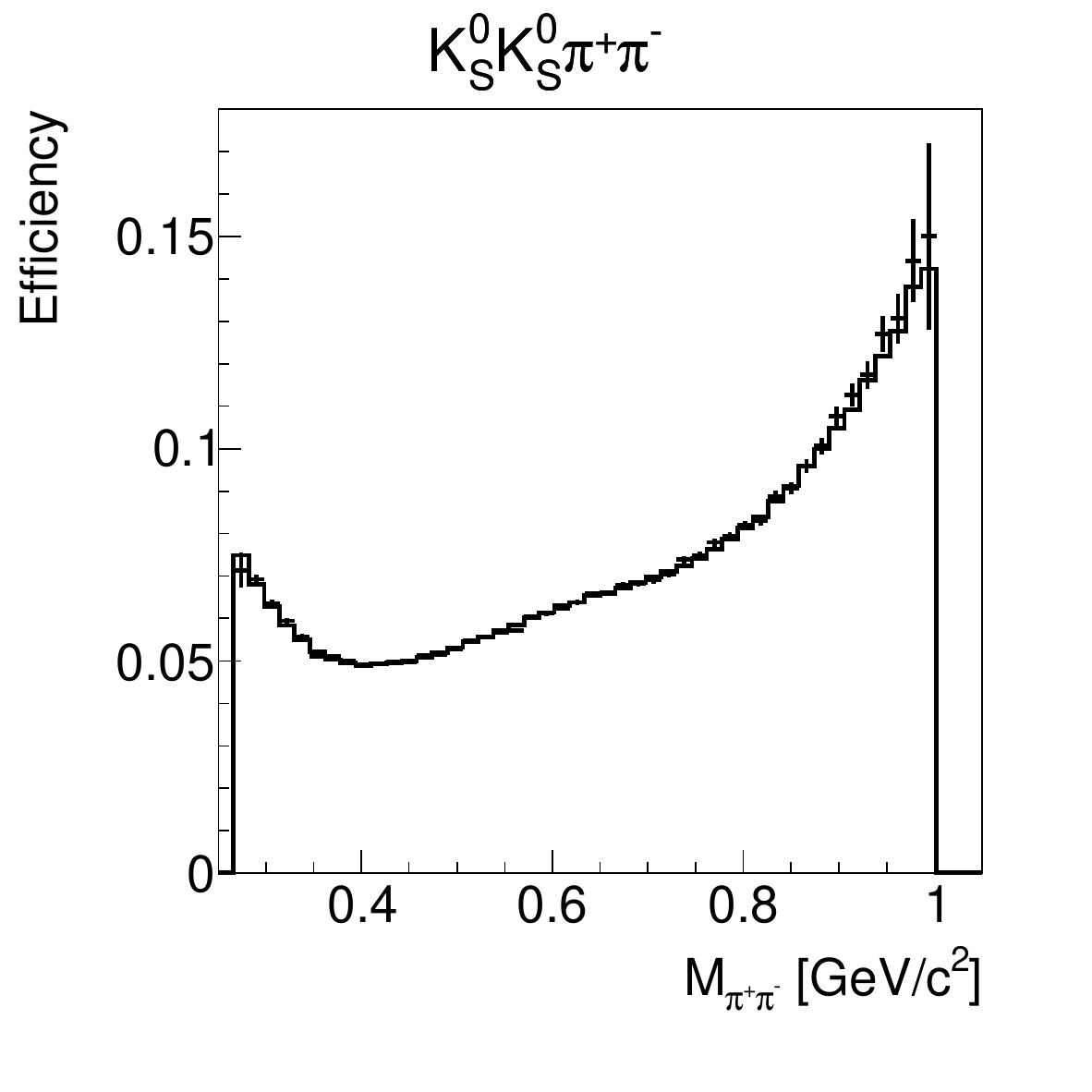}
\includegraphics[width=3.3cm]{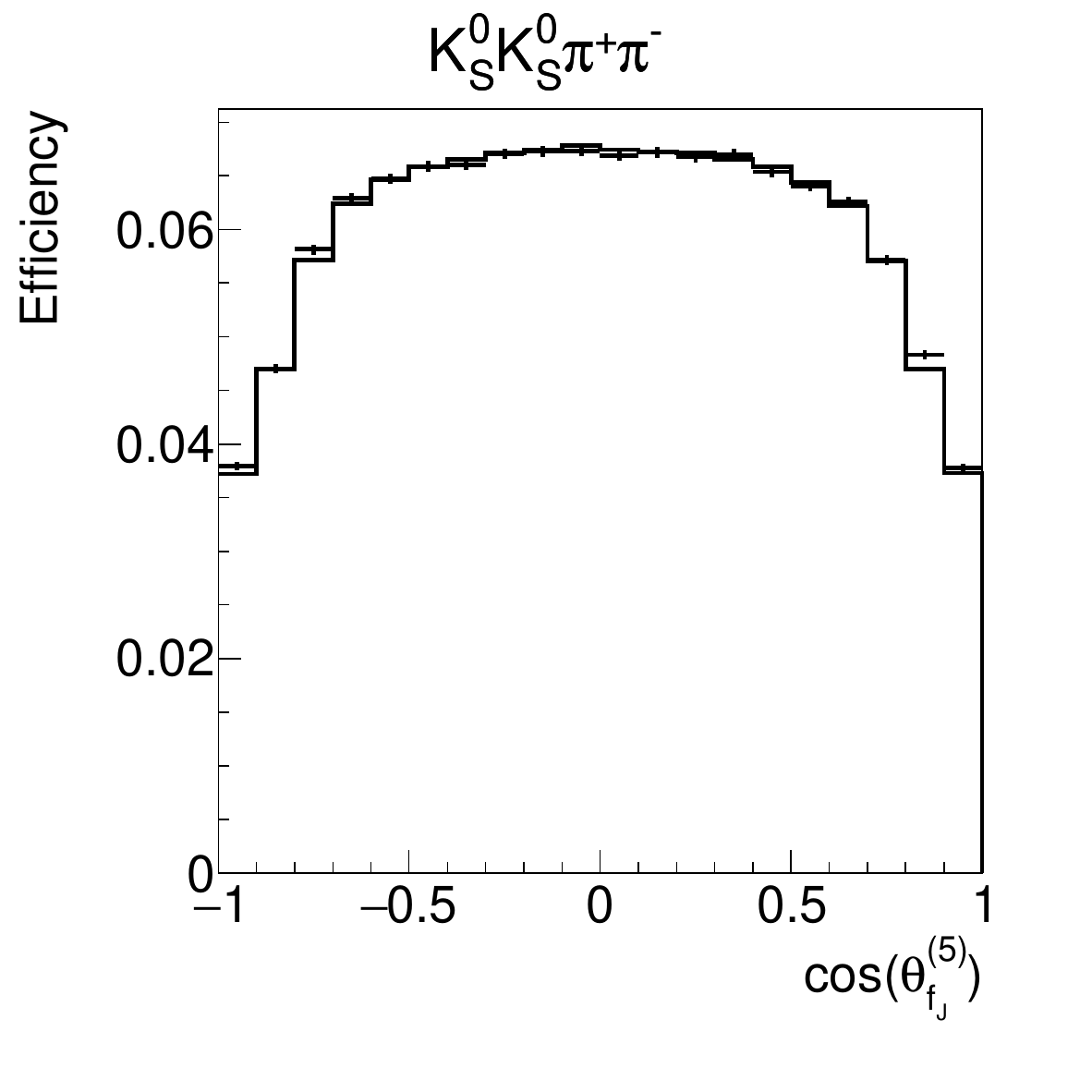}
\includegraphics[width=3.3cm]{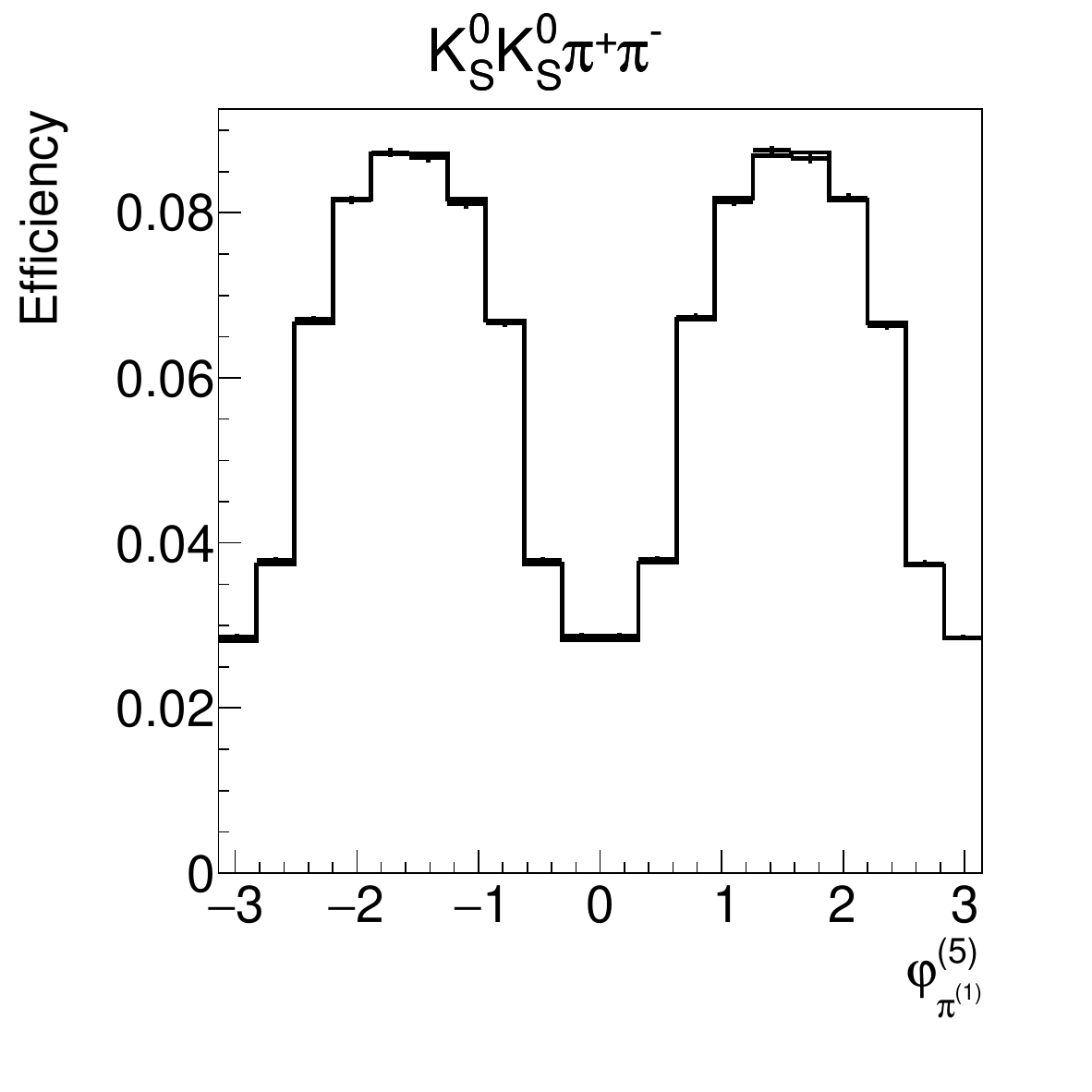} \\}
\caption{Projections of efficiency fit results onto
phase-space variables of the $\vpho \to \phij \fj$ decay chain
for the energy point
$\sqrt{s} = 2000\ \mev$. The points with error bars are efficiency (fraction
of reconstructed MC events) and the solid line is the fit result.}
\label{fig:efficiency}
\end{figure}

Projections of efficiency fit results onto
phase-space variables of the $\vpho \to \phij \fj$ decay chain
for the energy point $\sqrt{s} = 2000\ \mev$
are shown in Fig.~\ref{fig:efficiency}.

The efficiency fit provides an example of a fit with explicit gradient
calculation. Two different signal density functions with and without gradient
calculation need to be implemented, and minimization type should also be set:
\begin{lstlisting}[style=cppstyle]
        /* Set signal density. */
        res = vecampfit_fitter_set_signal_density(
                VECAMPFIT_FITTER_DENSITY_FUNCTION(
                eff_density_#model#_#process#_#sample#), 1);
        if (res != 0)
                return -1;
        /* Set signal density with gradient. */
        res = vecampfit_fitter_set_signal_density_gradient(
                VECAMPFIT_FITTER_DENSITY_GRADIENT_FUNCTION(
                eff_density_gradient_#model#_#process#_#sample#), 1);
        if (res != 0)
                return -1;
        /* Use explicit gradient calculation. */
        vecampfit_fitter_set_minimization_type(MINIMIZATION_GRADIENT);
\end{lstlisting}
The density function with gradient calculation has an additional
output argument for gradient:
\begin{lstlisting}[style=fortranstyle]
  MODULE SUBROUTINE EFF_DENSITY_GRADIENT_#MODEL#_#PROCESS#_#SAMPLE#( &
&     ED, GRADIENT, EV, PAR, PAR_BUF) BIND(C)
    USE, INTRINSIC :: ISO_C_BINDING
    USE VECAMPFIT_REAL_VECTOR_MODULE
    IMPLICIT NONE
    REAL(C_REAL_KIND), INTENT(OUT) :: ED(VECTOR_LENGTH)
    TYPE(VECAMPFIT_T_REAL_VECTOR), INTENT(OUT) :: &
&     GRADIENT(EPEM_KKPIPI_EFFICIENCY_MODEL_PARAMETERS)
    TYPE(EEB_#MODEL#_#PROCESS#_#SAMPLE#), INTENT(IN) :: EV
    REAL(C_REAL_KIND), INTENT(IN) :: &
&     PAR(EPEM_KKPIPI_EFFICIENCY_MODEL_PARAMETERS)
    TYPE(EPB_#MODEL#_#PROCESS#_#SAMPLE#), INTENT(IN) :: PAR_BUF
    ! The calculation is omitted.
  END SUBROUTINE
\end{lstlisting}
An array of real vectors defined similarly to the complex ones is used to
store the gradient because plain two-dimensional Fortran arrays
are not interoperable with C due to a different order or array elements.

At each minimization step, the \vecampfit{} fitter calls one of the density
functions depending on whether Minuit requests the gradient calculation
or not. The design with two different density functions was chosen
for performance reasons. Checking whether it is needed to calculate gradient
is not necessary and this allows for branchless implementation of the functions
with and without gradient calculation. On the other hand, this approach has
a severe disadvantage of code duplication: calculation of the density-function
value needs to be performed by both density functions. The implementation needs
validation which can be performed by comparison of the results of numerical
and direct calculation of the gradient. This check is implemented in Minuit
and it is performed when calling Minuit command ``SET GRADIENT''.
Additional validation functions are implemented in \vecampfit{} module
{\tth VECAMPFIT\_MINIMIZER}. The need for code duplication can be resolved
in the future by implementation of automatic gradient calculation.

Polynomials with arbitrary coefficients are necessary for parameterizations
of efficiency, resolution, and background. \vecampfit{} includes a tool
{\tth vecampfit-polynomial-parameters} for automatic generation of Fortran
code for polynomials. An example can be found in the file
{\tth efficiency\_correction\_polynomial.F90}.
Subroutine for calculation of the correction polynomial
$P_1(\Phi)$ from Eq.~\eqref{eq:efficiency} with optional second order
is generated by calling
\begin{lstlisting}[style=bashstyle]
vecampfit-polynomial-parameters 2 7 --minimum-order=1 --preprocessor-order \
  --subroutine
\end{lstlisting}
where the first two arguments are the polynomial order and
its number of dimensions, respectively.
Using other options, the polynomial-parameter tool can also generate
the version of the same subroutine with gradient calculation,
preprocessor definitions for polynomial parameters, parameter file with
their initial values, and C++ code for parameter releasing.

\subsection{Resolution parameterization and generation models}

The resolution is parameterized by a 12-dimensional Gaussian function:
\begin{equation}
R(\Phi^\text{(lab)}, {\Phi'}^\text{(lab)}) =
  \frac{1}{(2 \pi)^6 \sqrt{\det C}} \exp(-\frac{1}{2} x^T C^{-1} x),
\end{equation}
where $\Phi$ is the phase space for reconstructed kinematic parameters,
$\Phi'$ is the phase space for true (generator-level) kinematic parameters,
$C$ is covariance matrix and $x$ is the vector of differences between
true and reconstructed values:
\begin{equation}
x^T = (\Delta p_{\kI} / p_{\kI}, \Delta \cos \theta_{\kI}, \Delta \phi_{\kI},
  ...)
\end{equation}
Only diagonal covariance-matrix elements are used. Each element is a
first-order polynomial depending on the decay phase space $\Phi$.
The resolution density function is normalized by construction,
no numerical integration is necessary.

The resolution is parameterized using a reconstructed signal MC sample
with a uniform distribution of the generated events and both true and
reconstructed kinematic parameters recorded.
The reconstruction probability is simulated in the same way as
in Sec.~\ref{sec:kkpipi_efficiency}.

\begin{figure}
{\centering
\includegraphics[width=3.3cm]{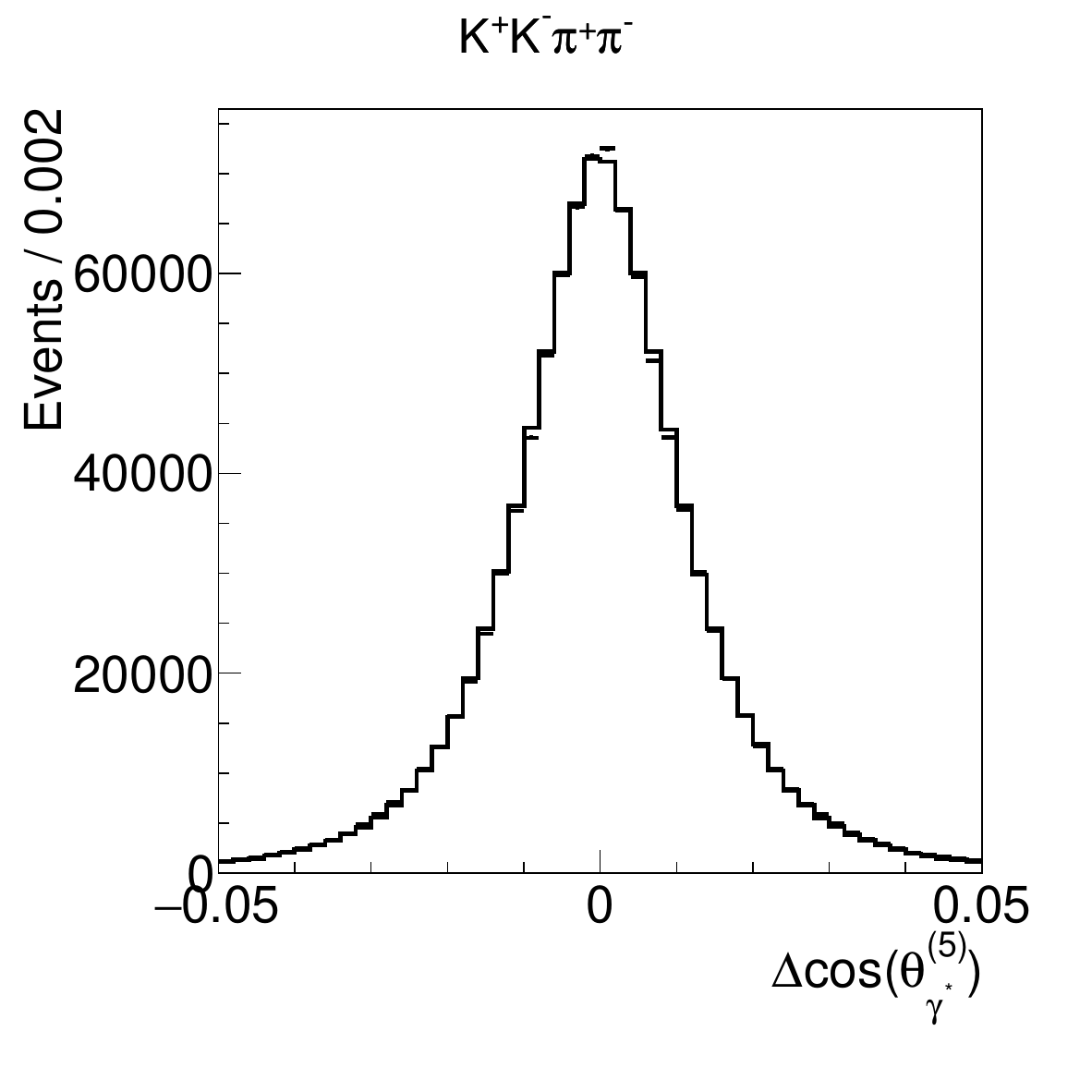}
\includegraphics[width=3.3cm]{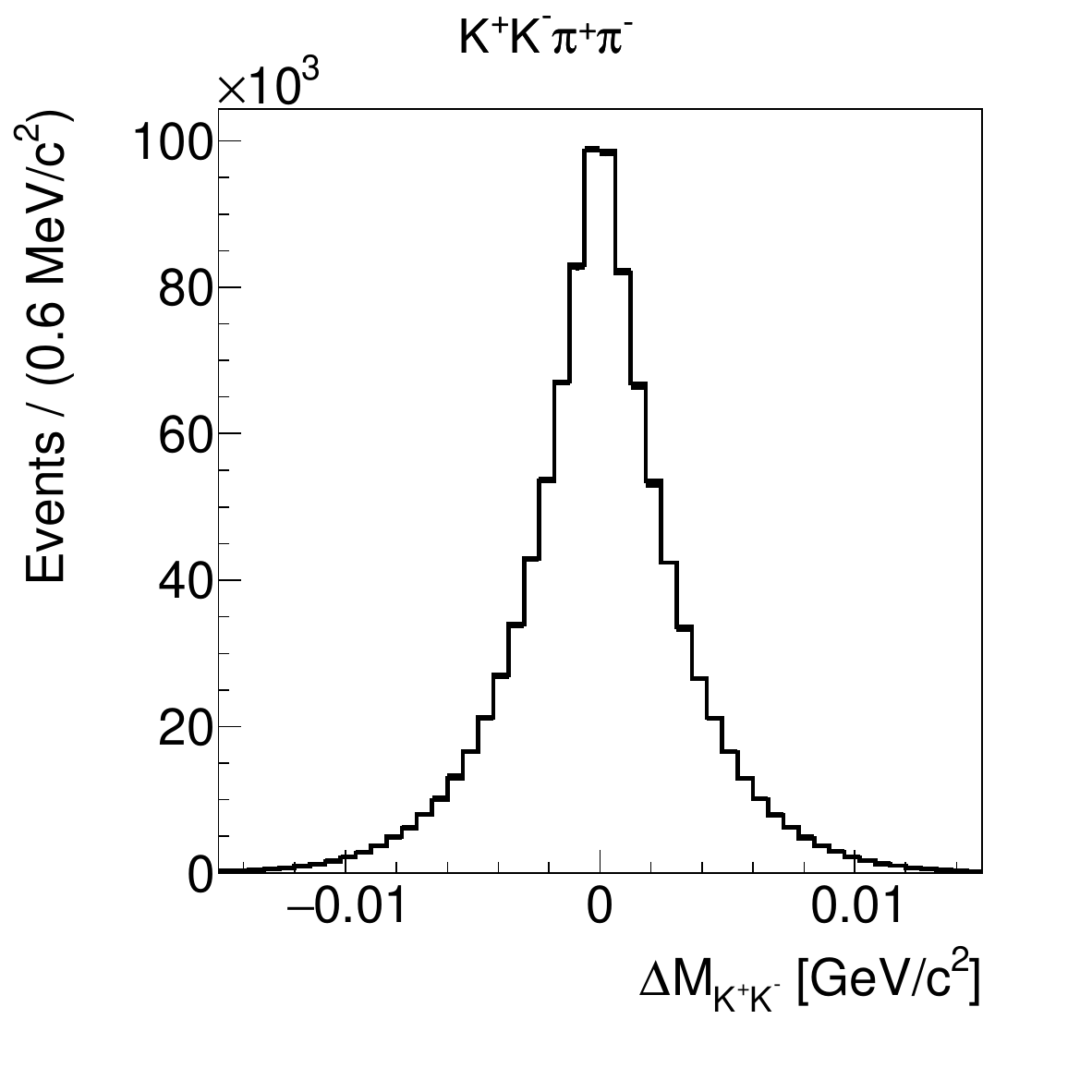}
\includegraphics[width=3.3cm]{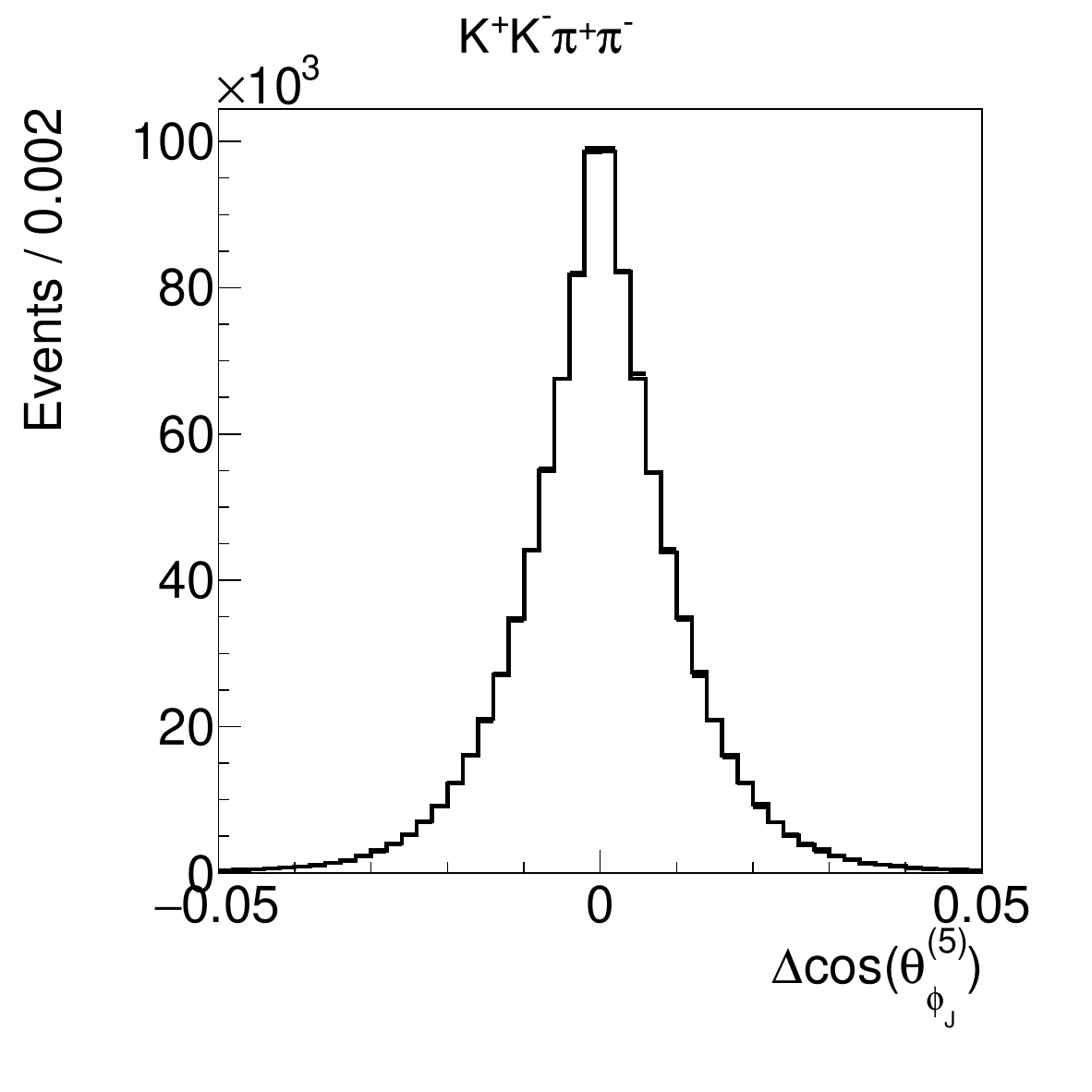}
\includegraphics[width=3.3cm]{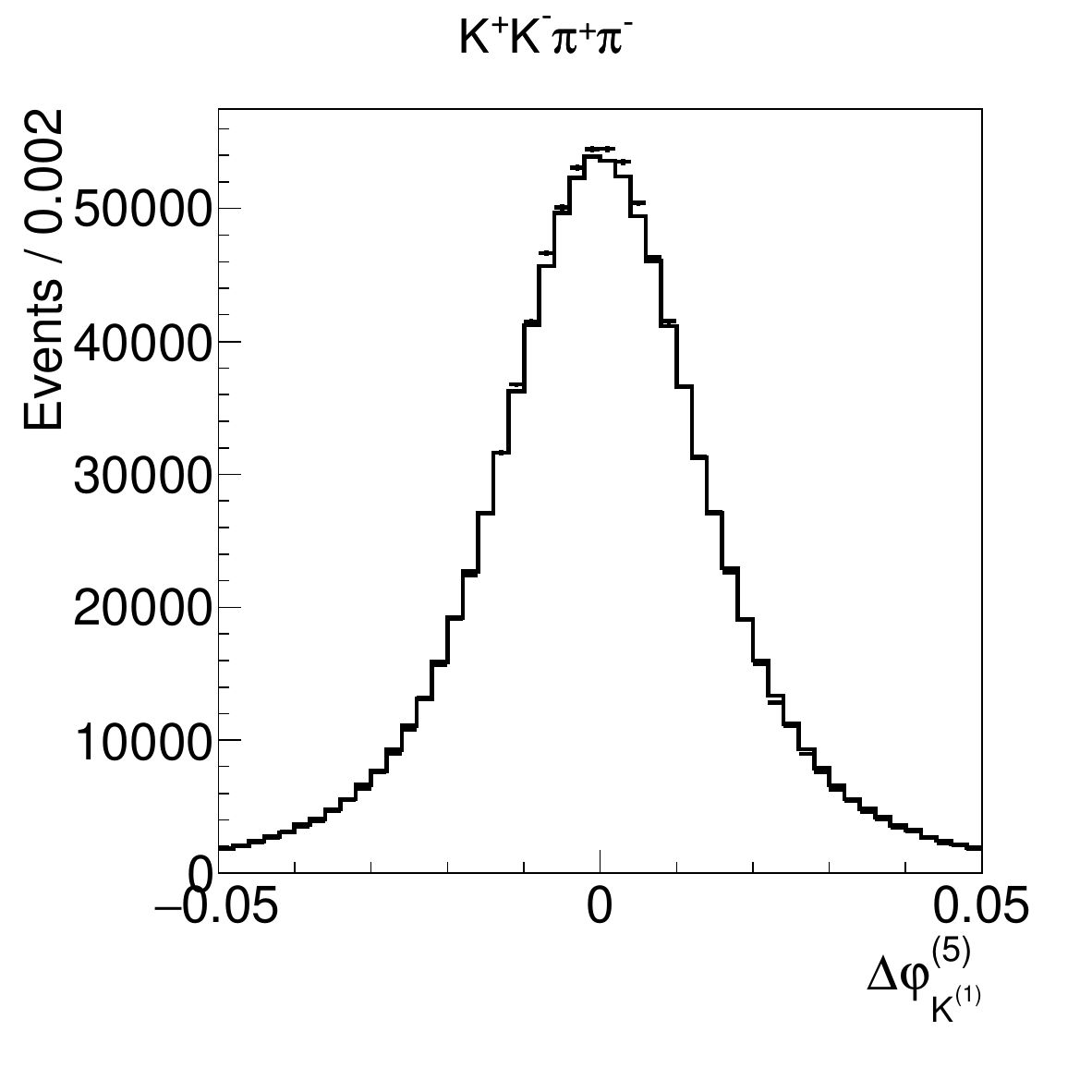} \\
\includegraphics[width=3.3cm]{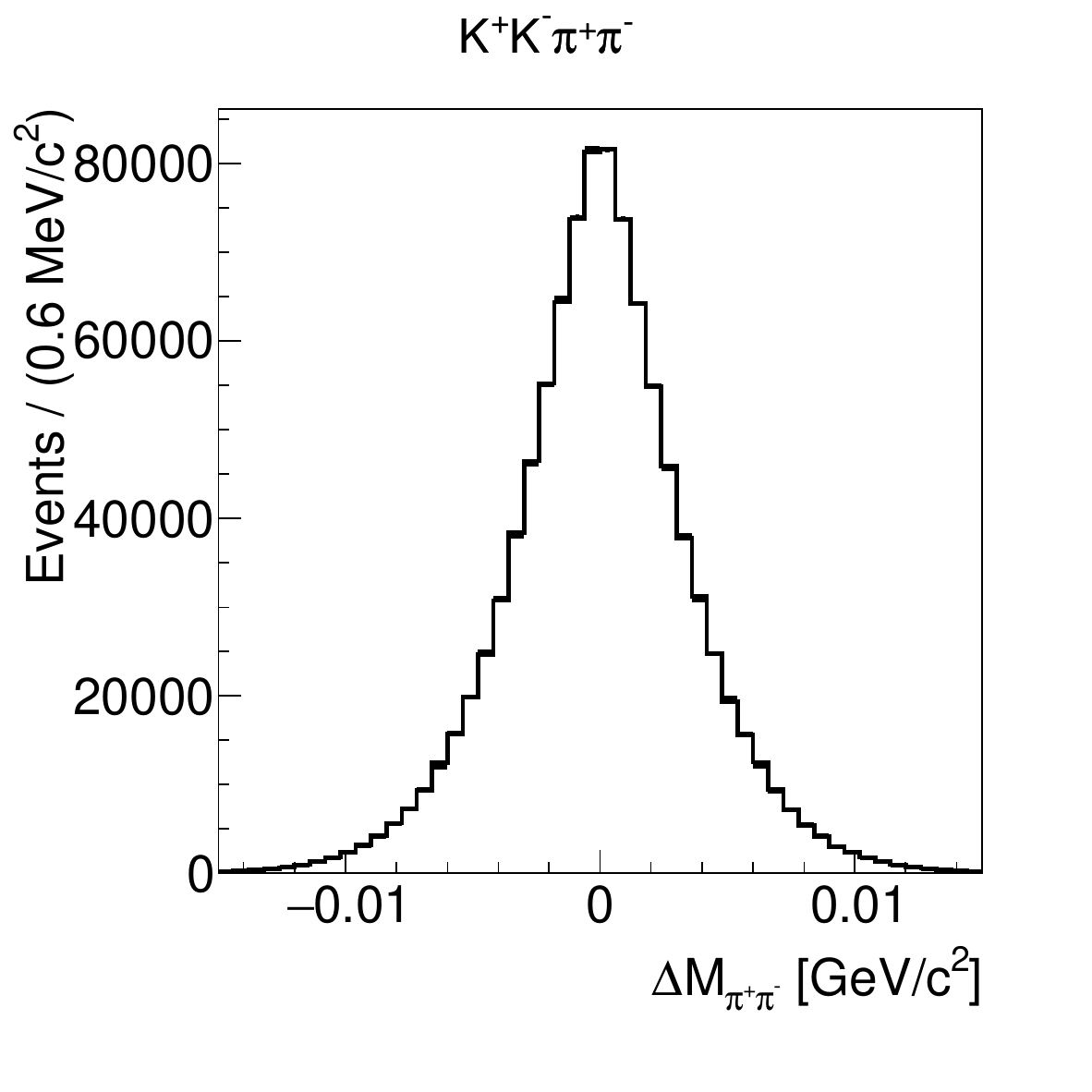}
\includegraphics[width=3.3cm]{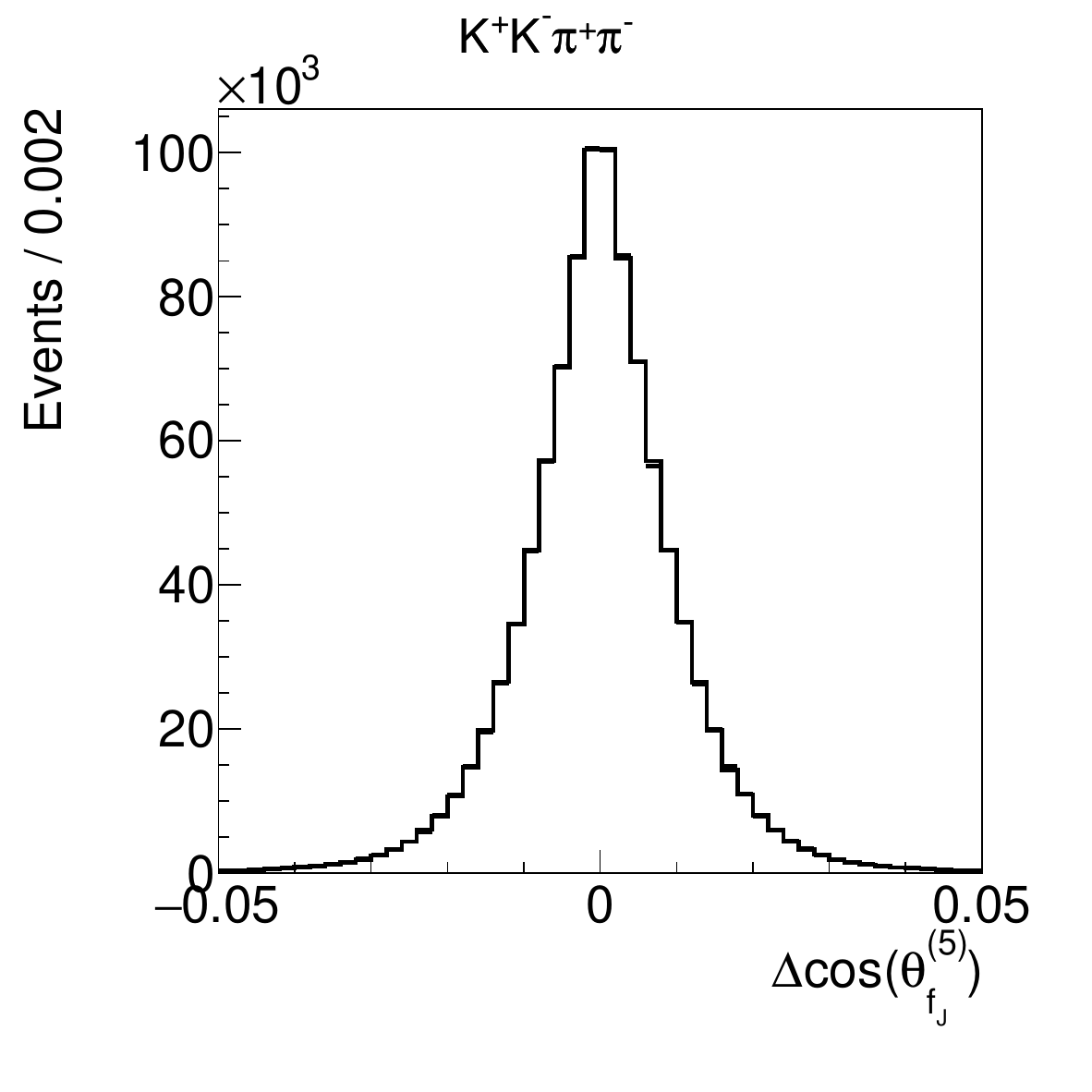}
\includegraphics[width=3.3cm]{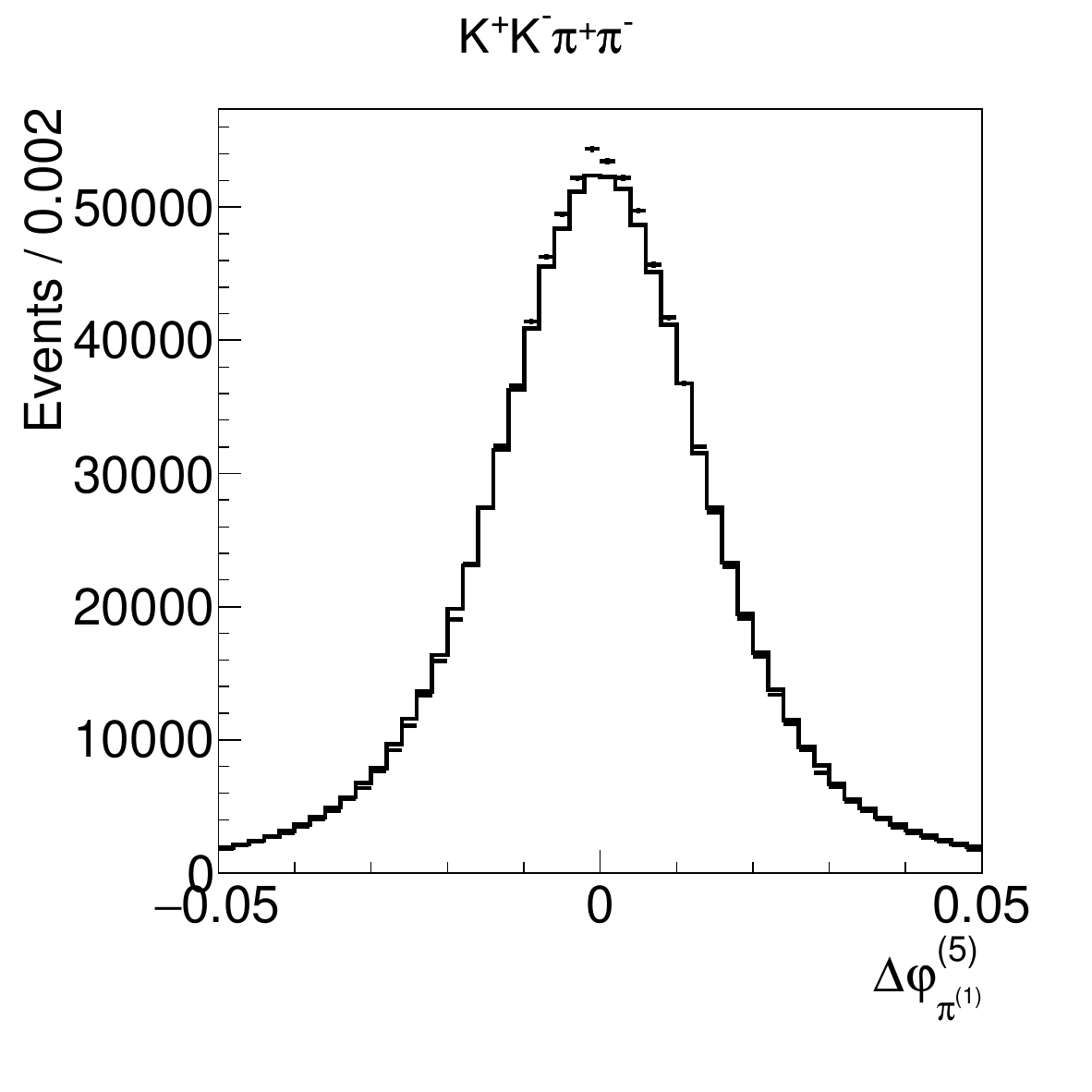} \\
\includegraphics[width=3.3cm]{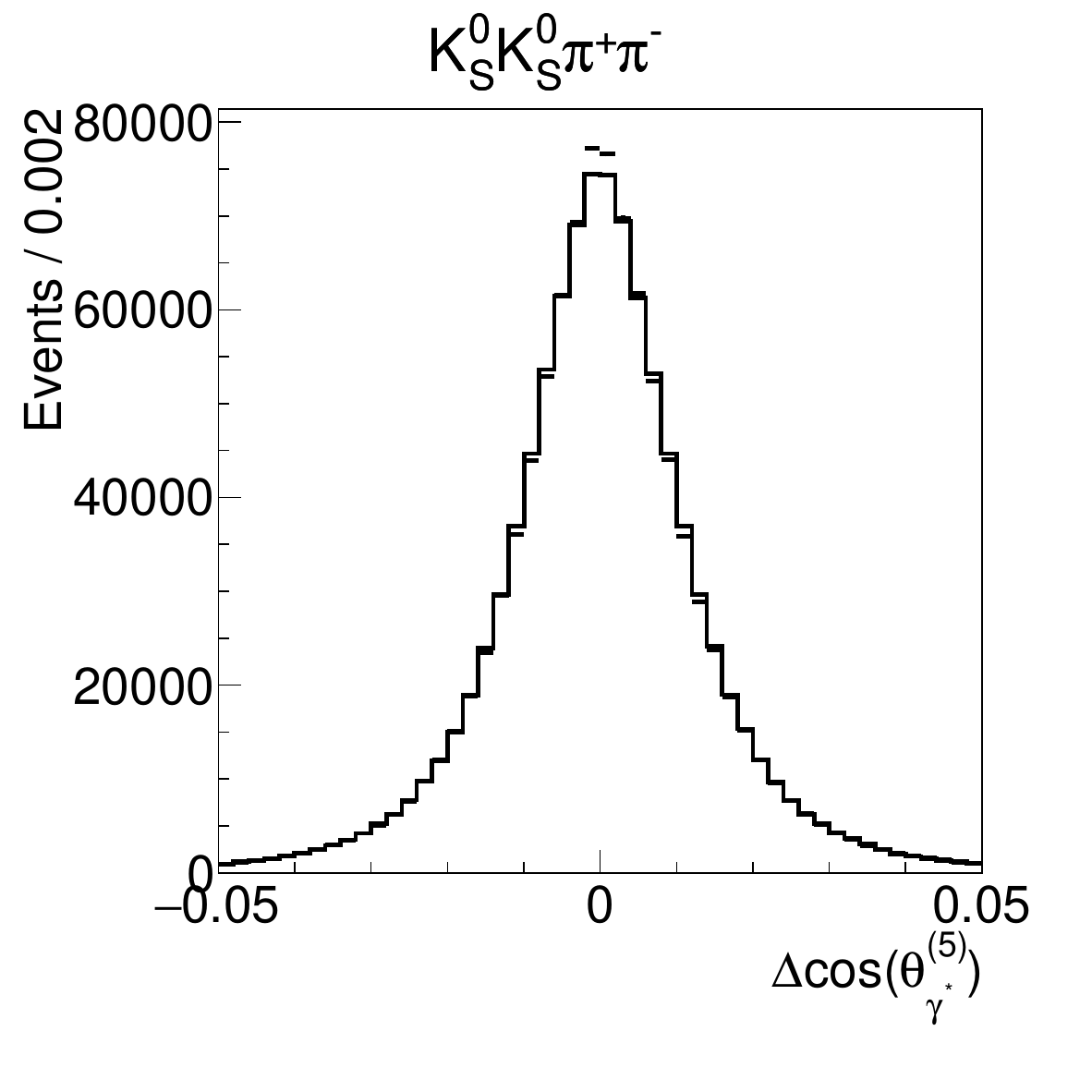}
\includegraphics[width=3.3cm]{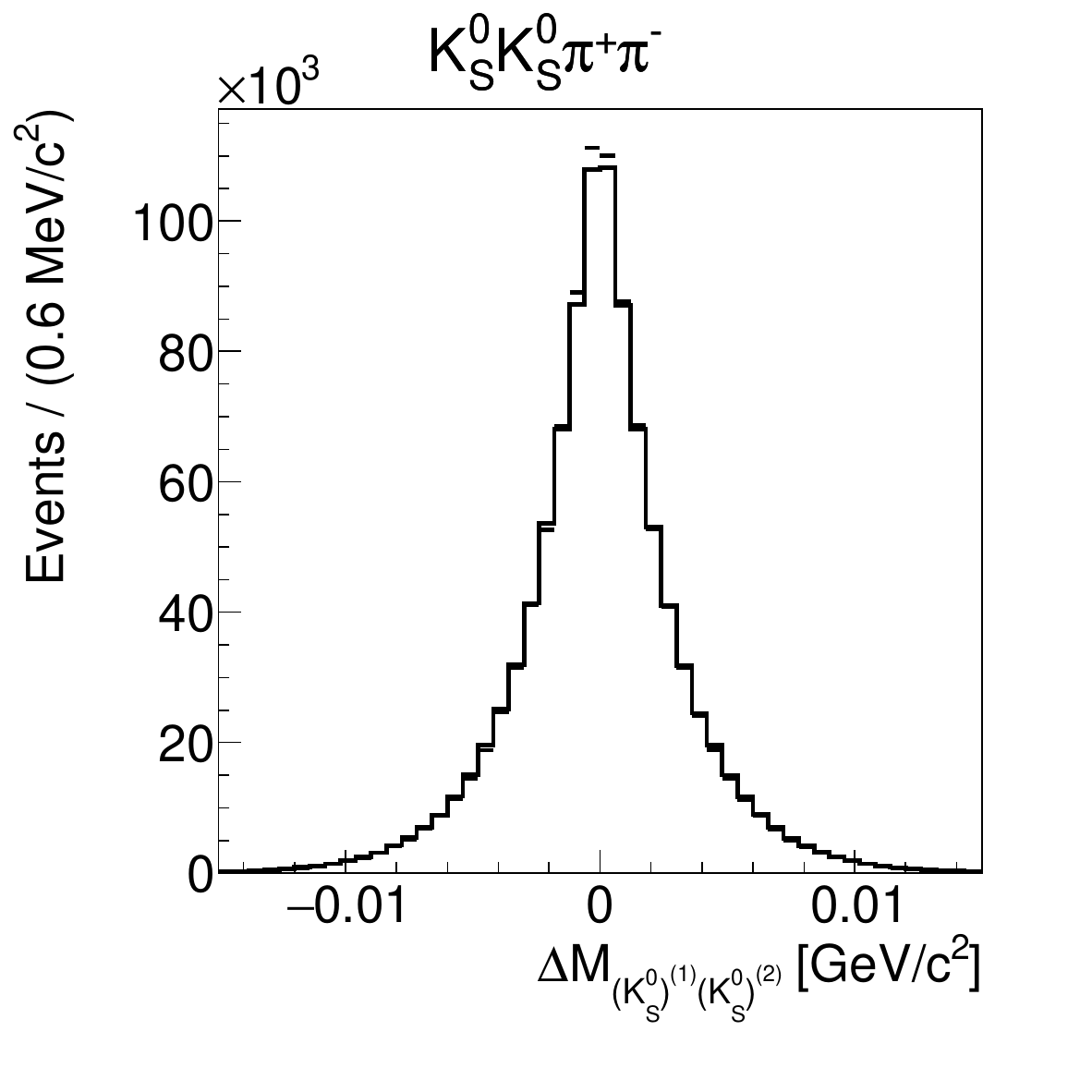}
\includegraphics[width=3.3cm]{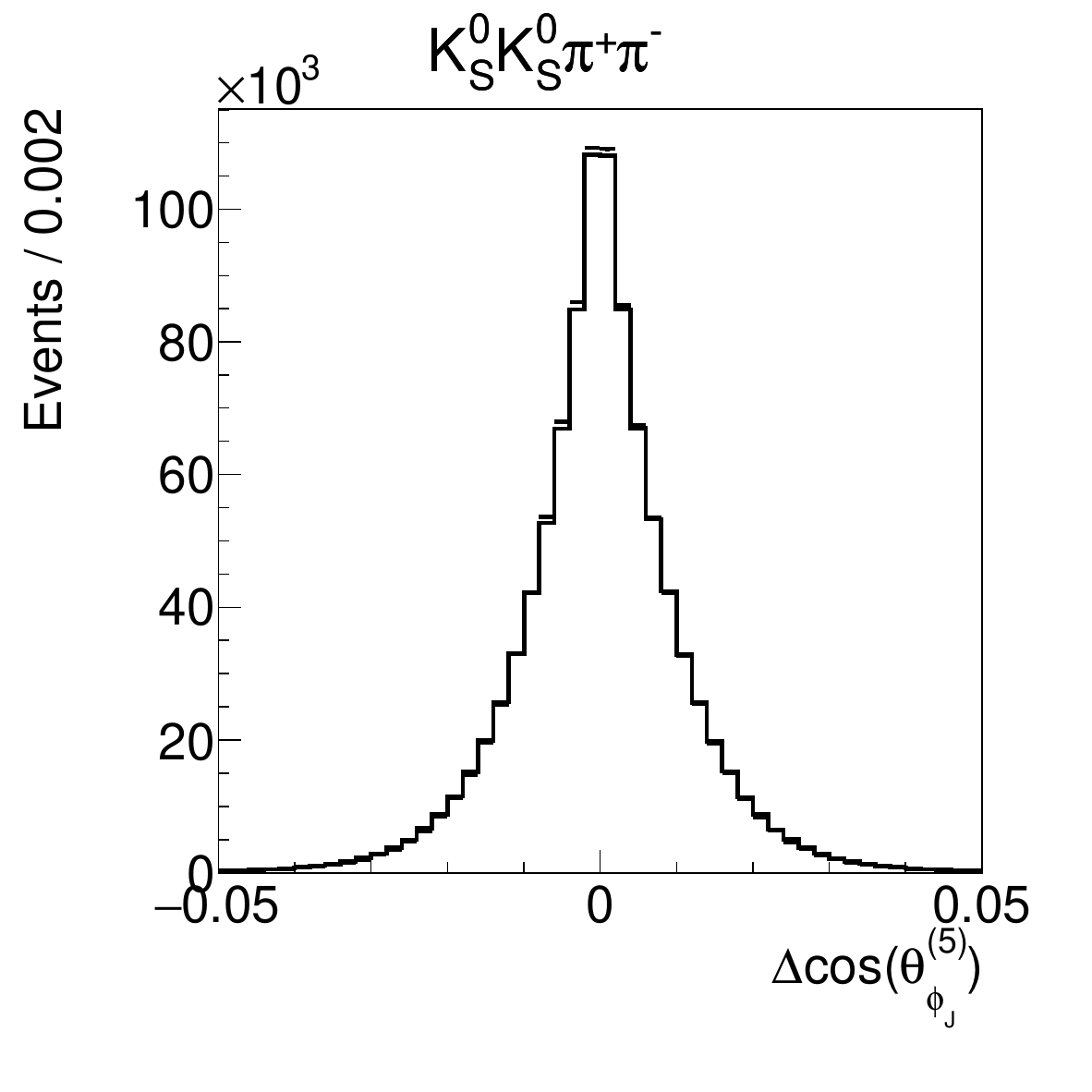}
\includegraphics[width=3.3cm]{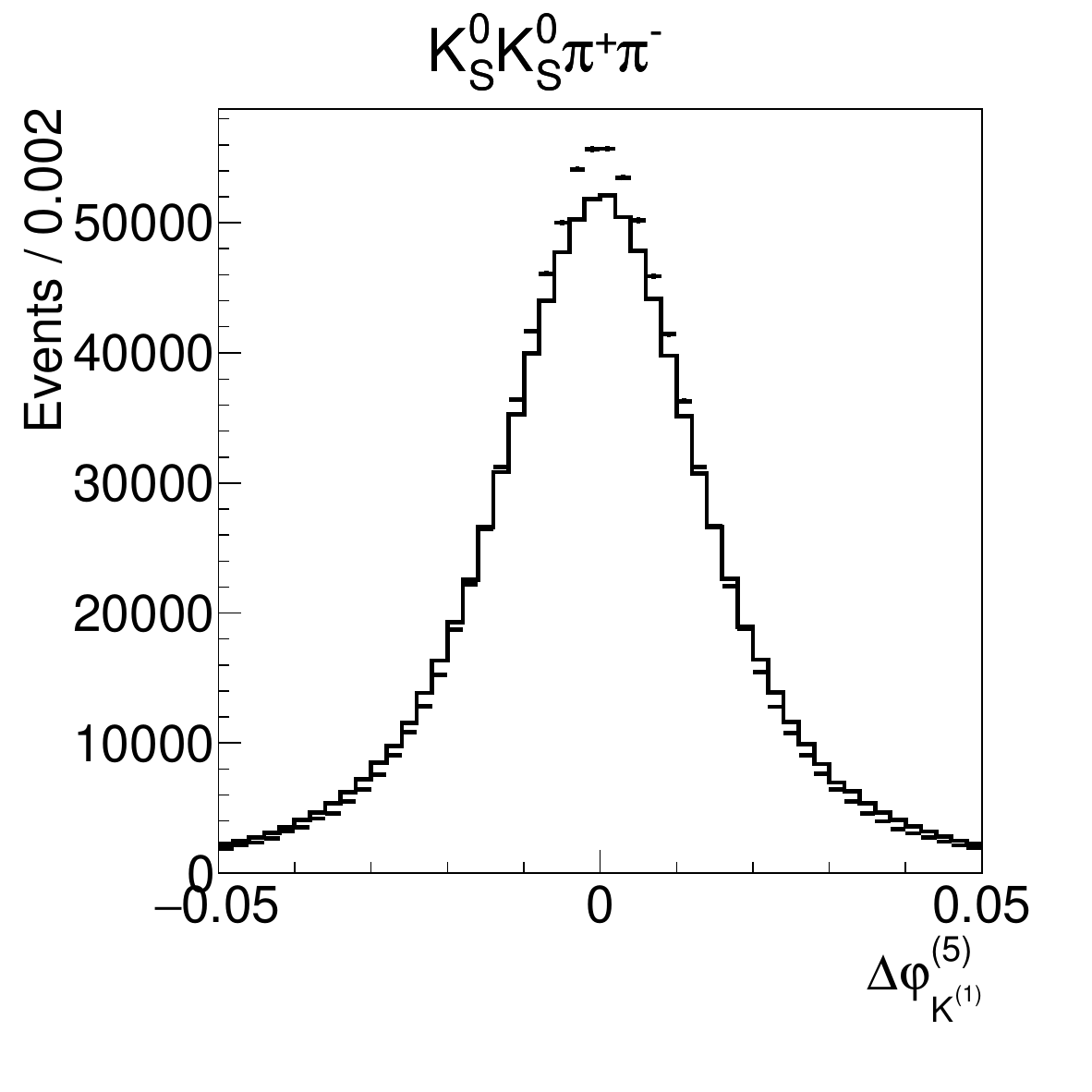} \\
\includegraphics[width=3.3cm]{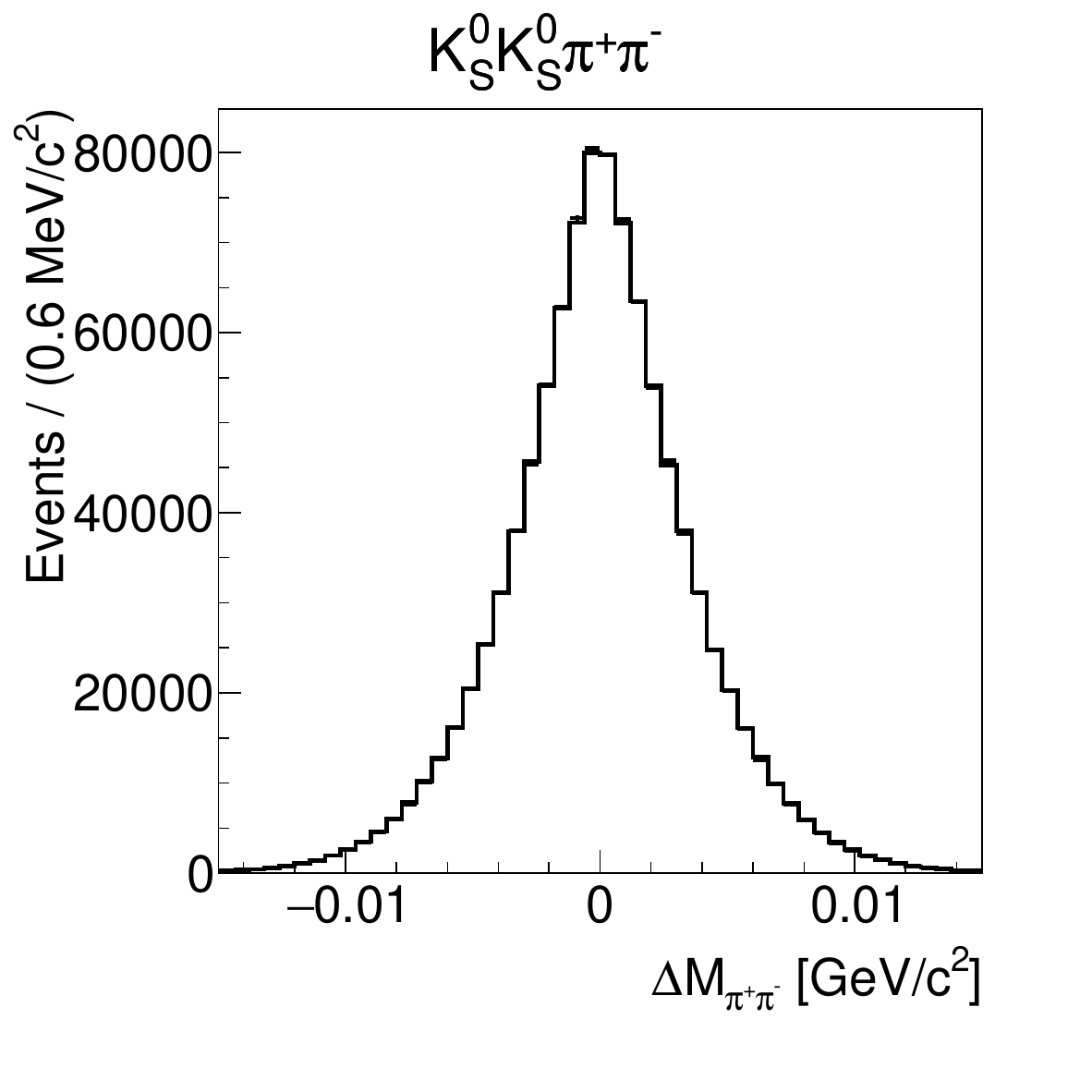}
\includegraphics[width=3.3cm]{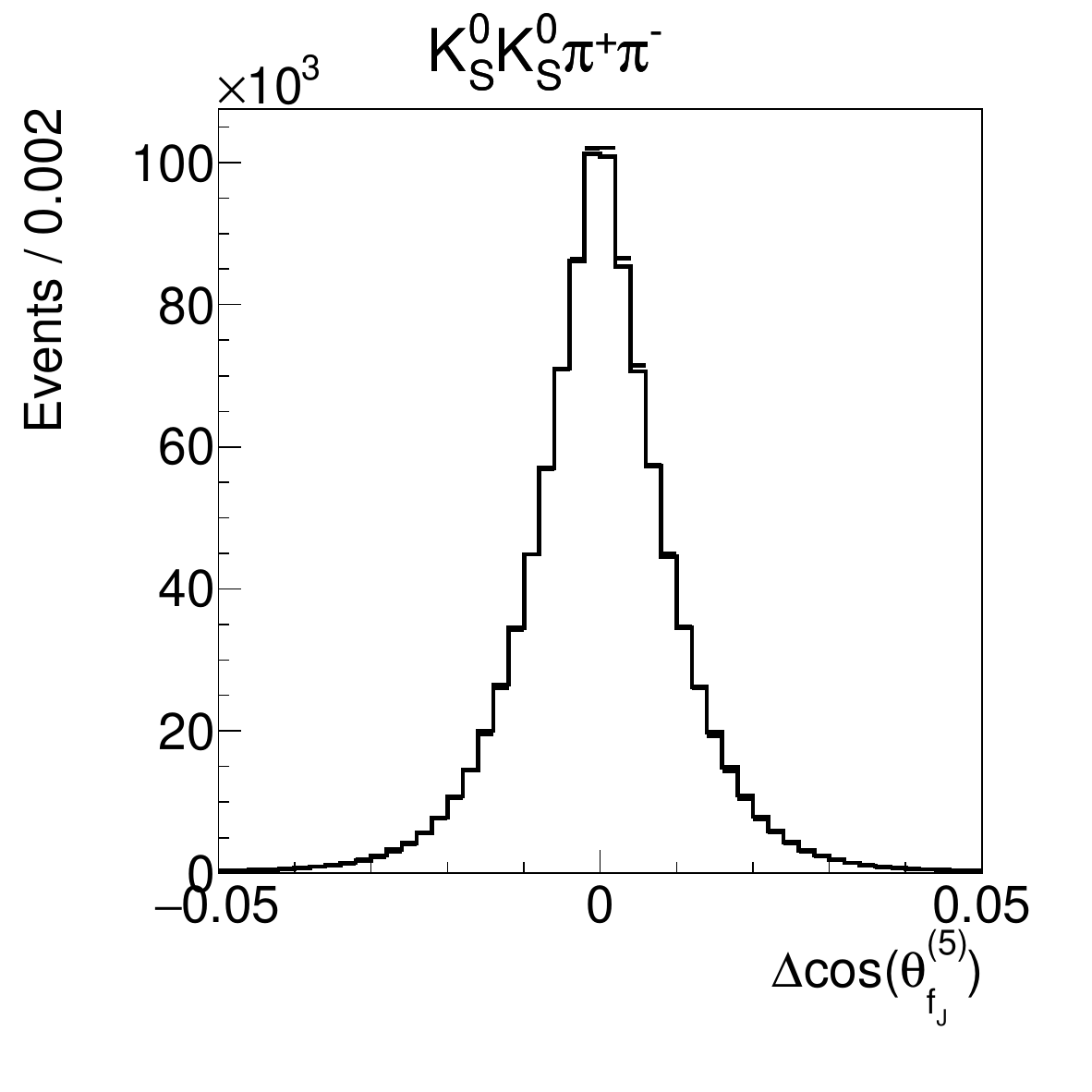}
\includegraphics[width=3.3cm]{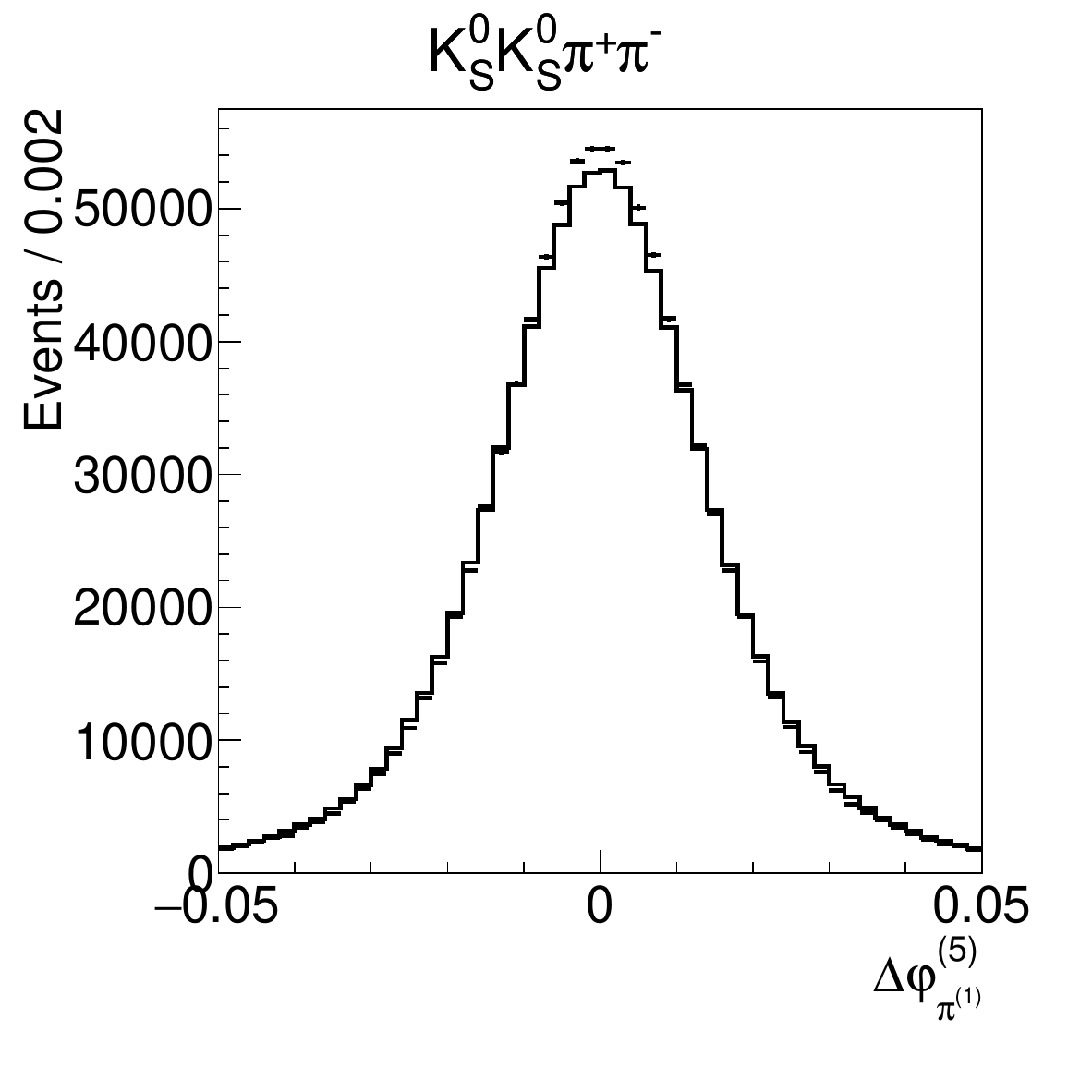} \\}
\caption{Resolution in
phase-space variables of the $\vpho \to \phij \fj$ decay chain
for the energy point $\sqrt{s} = 2000\ \mev$.
The points with error bars are resolution (difference between reconstructed
and generator-level value) and the solid line is the fit result.}
\label{fig:resolution}
\end{figure}

A different MC sample is necessary to represent the fit results.
The events are first generated with probability
density equal to the MC efficiency $E(\Phi^\text{(lab)},\Phi)$
determined at the previous step. Then, a multidimensional Gaussian generator
is used to generate the corresponding reconstructed values in accordance
with the resolution fit result.
Resolution in 
phase-space variables of the $\vpho \to \phij \fj$ decay chain
for the energy point $\sqrt{s} = 2000\ \mev$
is shown in in Fig.~\ref{fig:resolution}.

\vecampfit{} provides base classes for event generation:
{\tth generation\_model\_resolution} for generation of toy MC needed
to represent the results of resolution fits and {\tth generation\_model}
for generation of events distributed in accordance with the fitted model,
which can be used for event generation based on the fit results for any
other types of fits such as signal or background parameterization.
The user needs to implement filling of parameter buffer:
\begin{lstlisting}[style=cppstyle]
/**
 * Fill parameter buffer.
 */
virtual void fill_parameter_buffer() = 0;
\end{lstlisting}
which can be done by simply calling the filling function used by the fit.
For resolution fits, it is also necessary to implement generation of
reconstructed event from generator-level event:
\begin{lstlisting}[style=cppstyle]
/**
 * Generate reconstructed event from generator-level event
 * that needs to be already filled when calling this function.
 *
 * @param[in,out] event Event.
 */
virtual void generate_event(void *event) = 0;
\end{lstlisting}
The event class should contain both generator-level and reconstructed
kinematic information. For other types of fits, the user needs to implement
loading of maximum density which is used for normalization, and the density
function itself.
\begin{lstlisting}[style=cppstyle]
/**
 * Load maximum density.
 *
 * @param[in] fit_name Fit name.
 *
 * @return Zero on success, non-zero value on error.
 */
virtual int load_maximum_density(const char *fit_name) = 0;

/**
 * Calculate density function normalized so that its maximum value
 * is less than or equal to 1.
 *
 * @param[in] event Event.
 */
virtual double density(const void *event) = 0;
\end{lstlisting}
The density function can be implemented by calling
the density used by the fit with a choice of the necessary function if
the model has multiple ones selecting the necessary process, channel,
and data sample.

\subsection{Background and signal fits, simultaneous fitting}
\label{sec:epem_kkpipi_signal_fit}

\begin{figure}
\includegraphics[width=4.5cm]{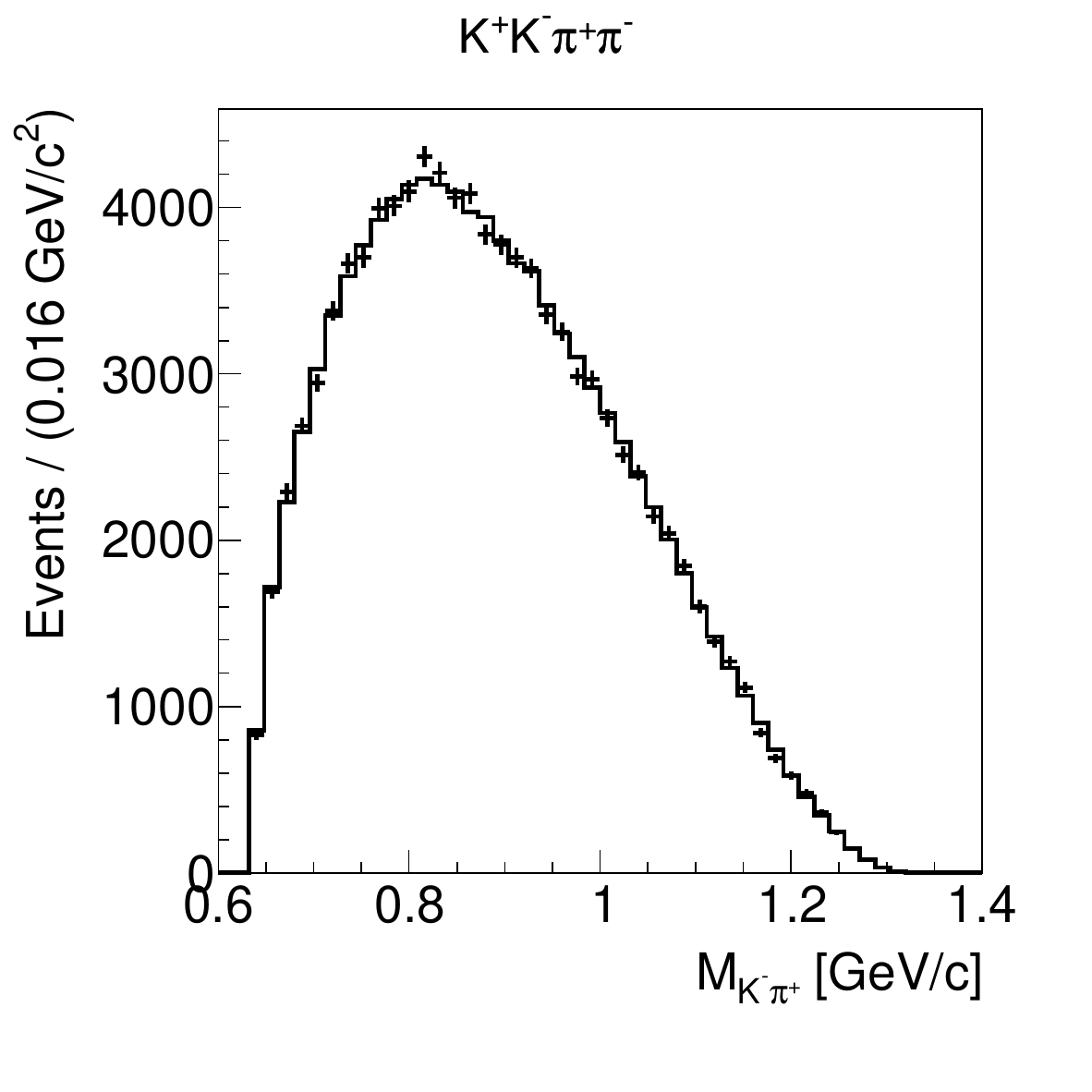}
\includegraphics[width=4.5cm]{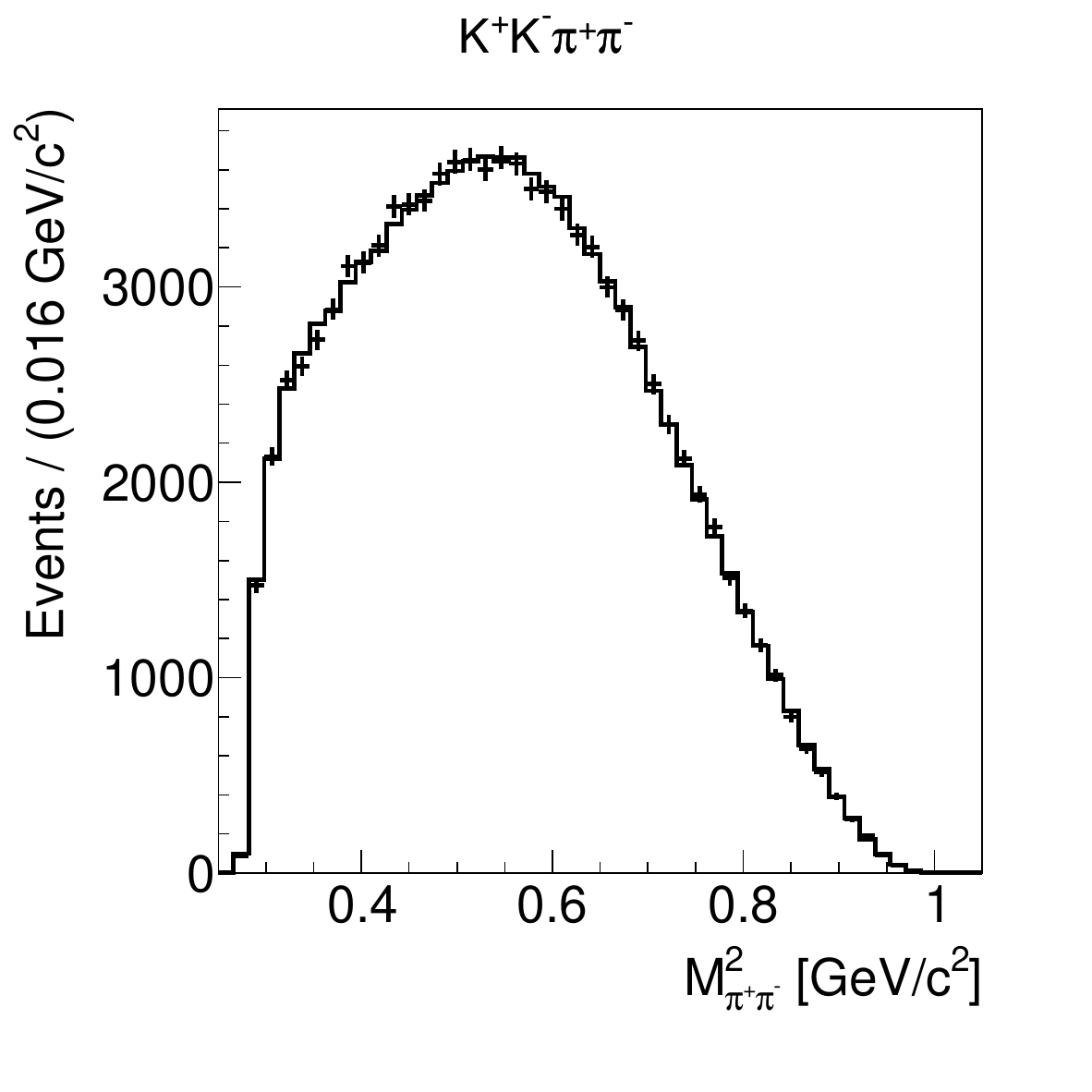}
\includegraphics[width=4.5cm]{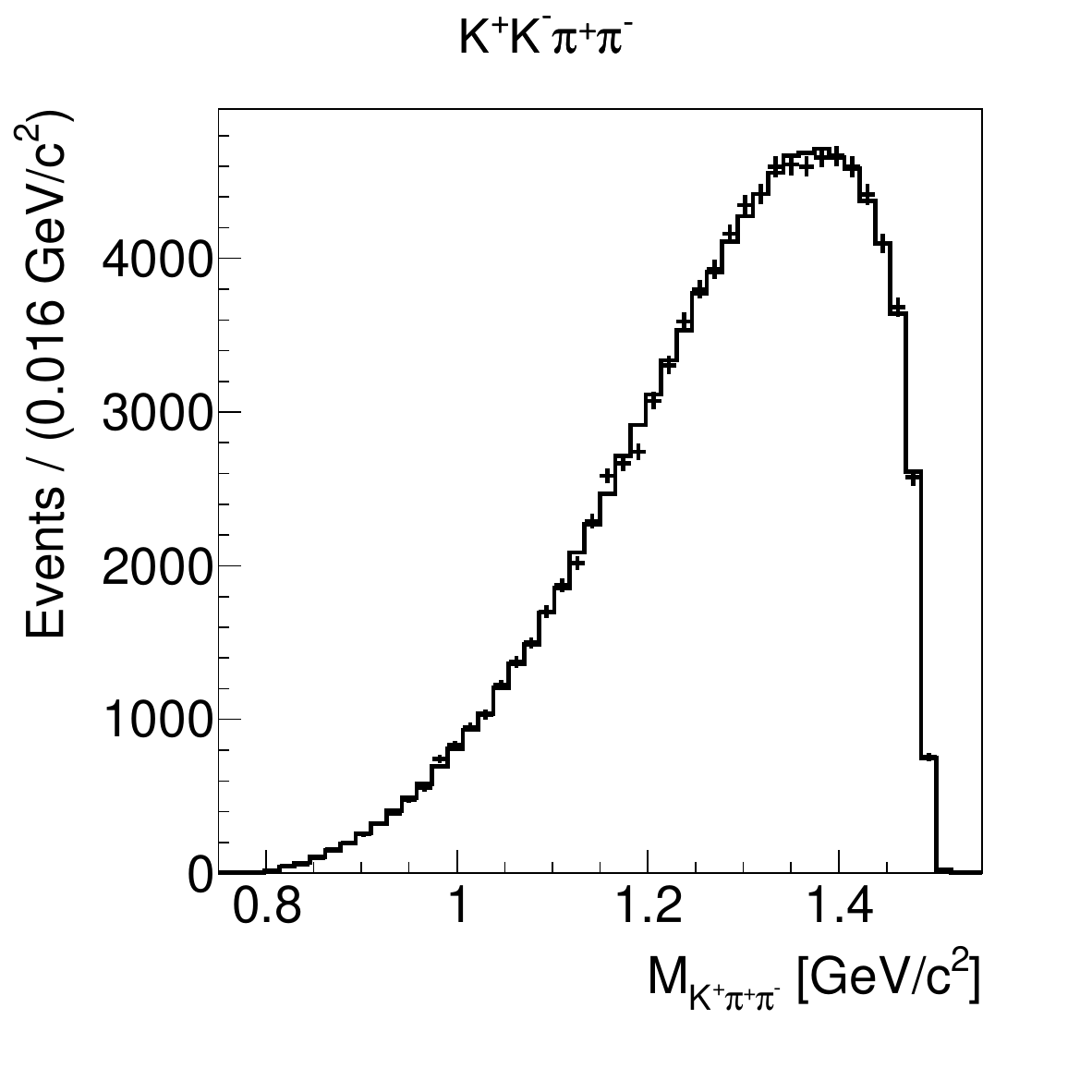} \\
\includegraphics[width=4.5cm]{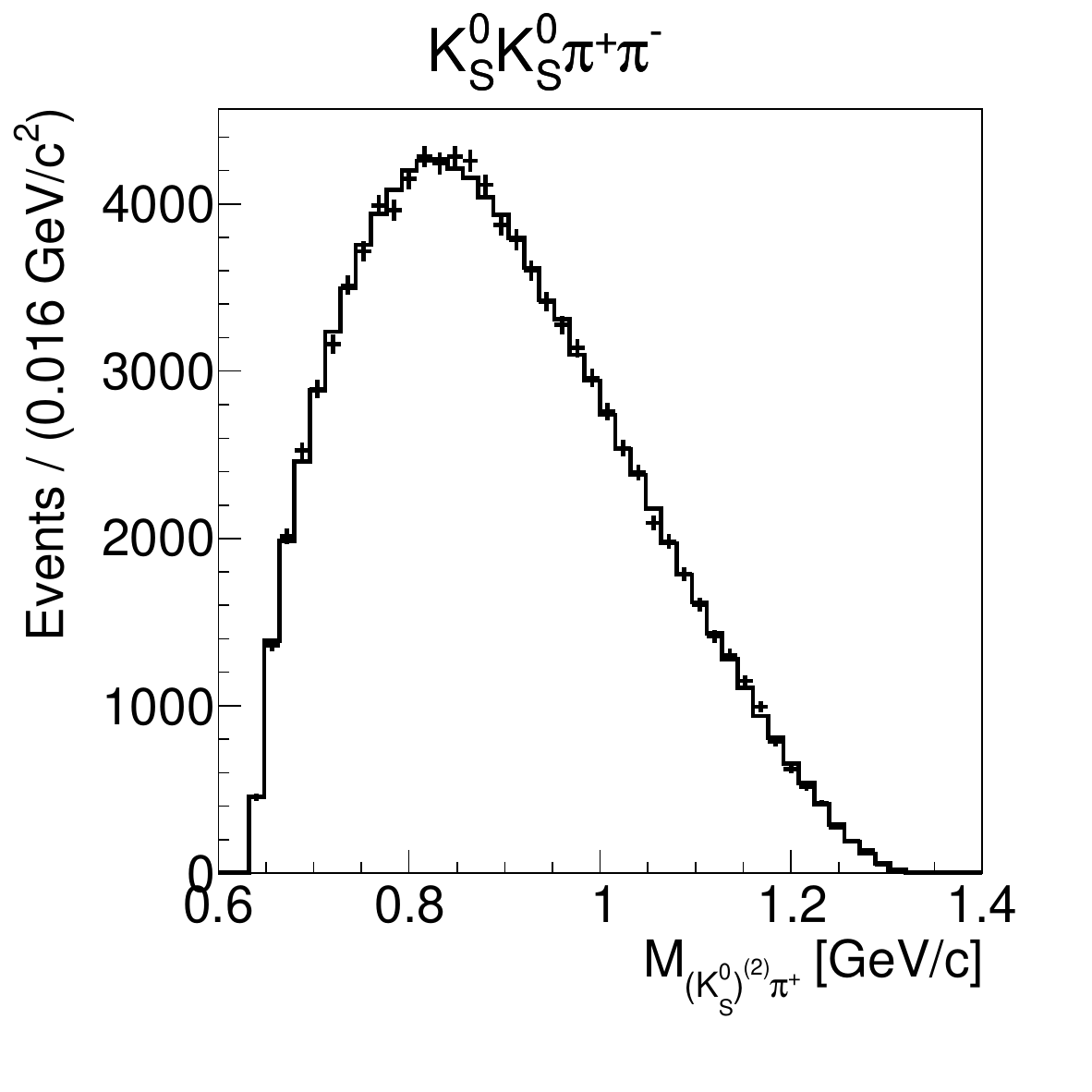}
\includegraphics[width=4.5cm]{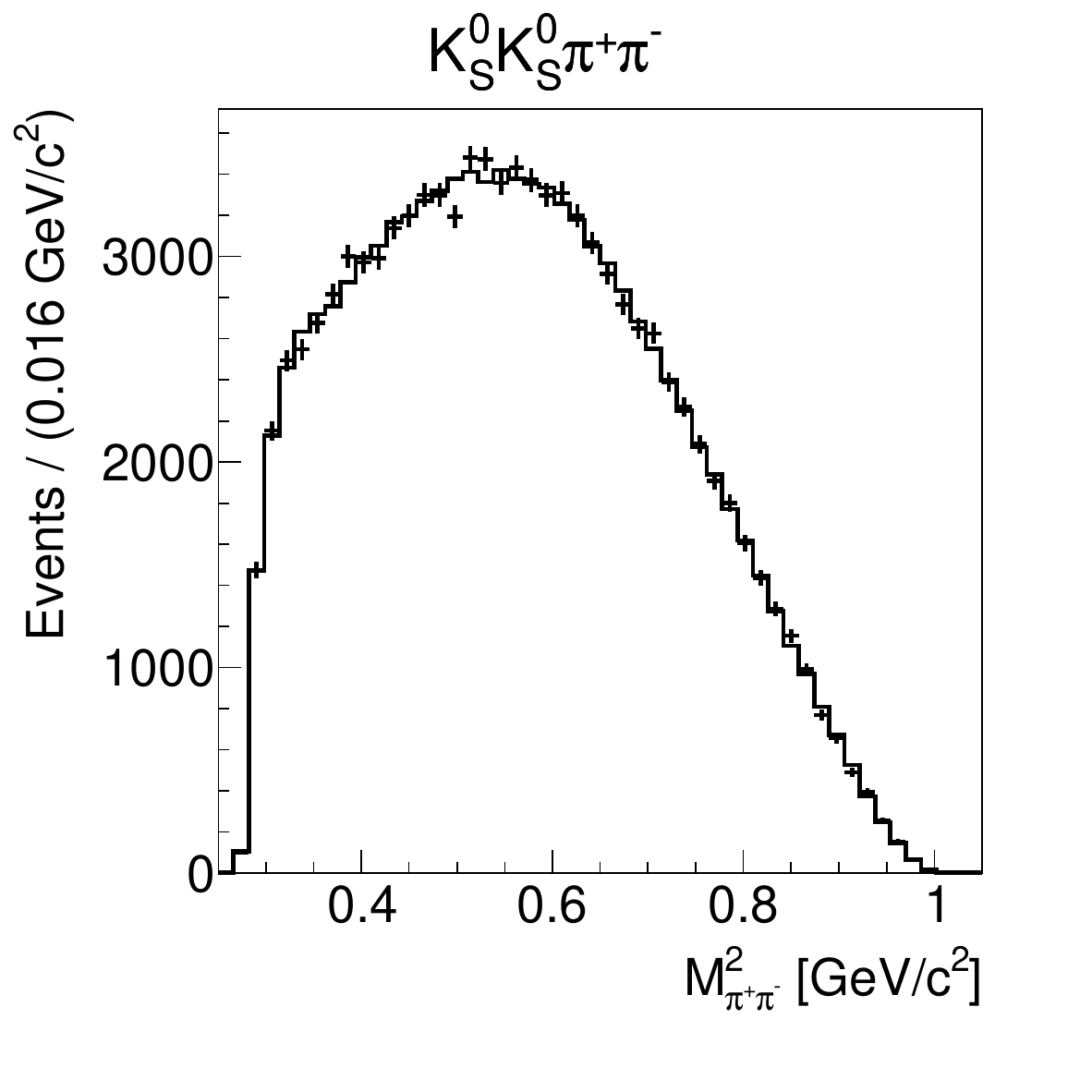}
\includegraphics[width=4.5cm]{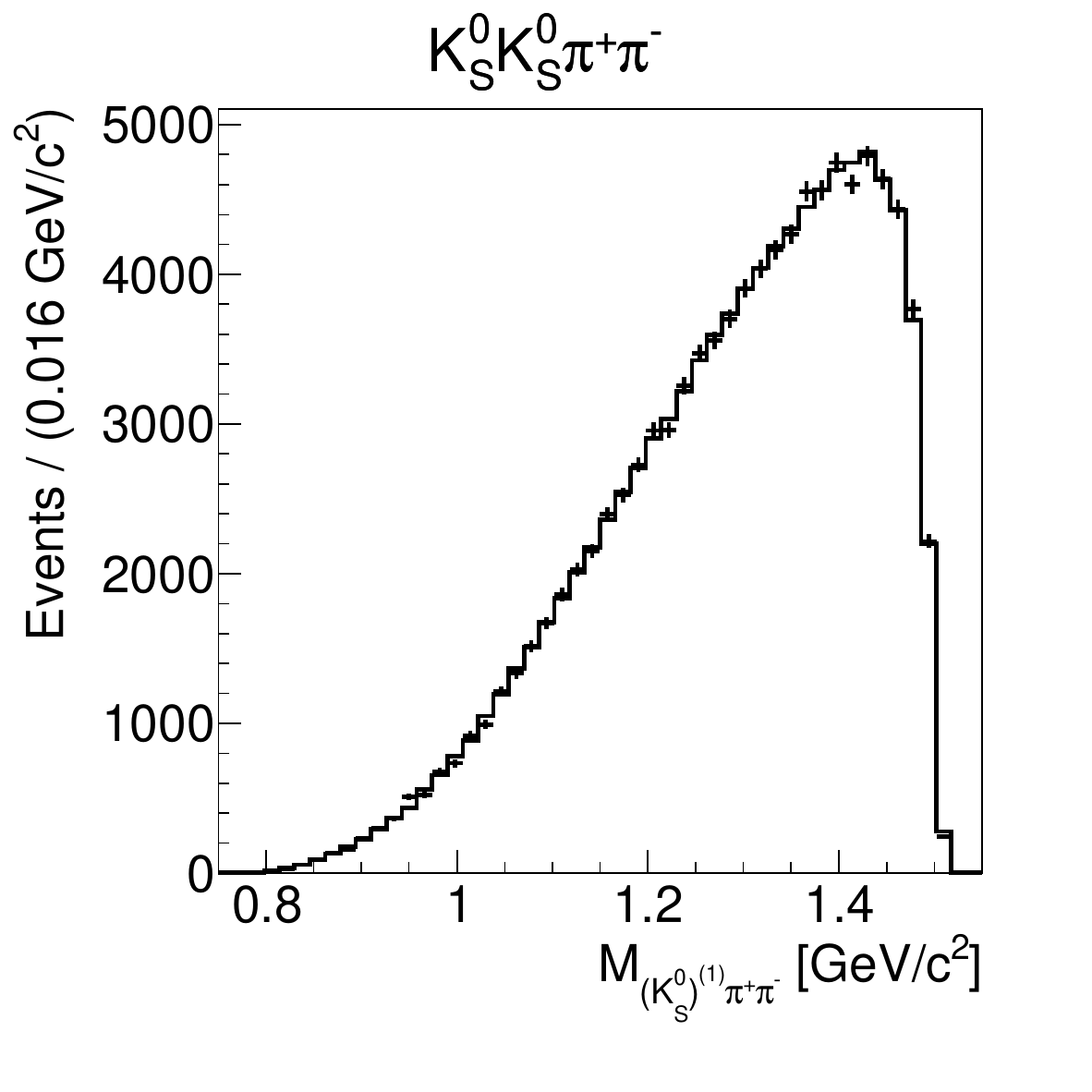} \\
\caption{Projections of background fit results onto $M_{\kI \piII}$,
$M_{\piI \piII}$, and $M_{\kI \piI \piII}$ for the energy point
$\sqrt{s} = 2000\ \mev$. The points with error bars are data
and the solid line is the fit result.}
\label{fig:background}
\end{figure}

The background is parameterized by a seven-dimensional second-order polynomial:
\begin{equation}
B(\Phi) = P_2(\Phi).
\end{equation}
Projections of background fit results onto $M_{\kI \piII}$, $M_{\piI \piII}$,
and $M_{\kI \piI \piII}$ for the energy point $\sqrt{s} = 2000\ \mev$
are shown in Fig.~\ref{fig:background}.

\vecampfit{} supports simultaneous fitting of multiple data sets.
Signal and background density function and event-buffer type, and background
parameter-buffer type are set separately for each simultaneous-fit point.
The fit parameters and signal parameter buffer are common for the entire fit.
Typical scenarios for simultaneous fitting include different experimental
periods with substantially different conditions such as Belle and Belle II;
different but related amplitudes such as $B$ decays with $\psi(2S) \to e^+ e^-$
or $\mu^+ \mu^-$ and $\psi(2S) \to J/\psi \pi^+ \pi^-$, or
with $D^{*0} \to D^0 \pi^0$ and $D^{*0} \to D^0 \gamma$; and
fit to $e^+ e^-$ annihilation processes at multiple different energies, which
is illustrated by the $\kkpipi$ example.

\begin{figure}
\includegraphics[width=4.5cm]{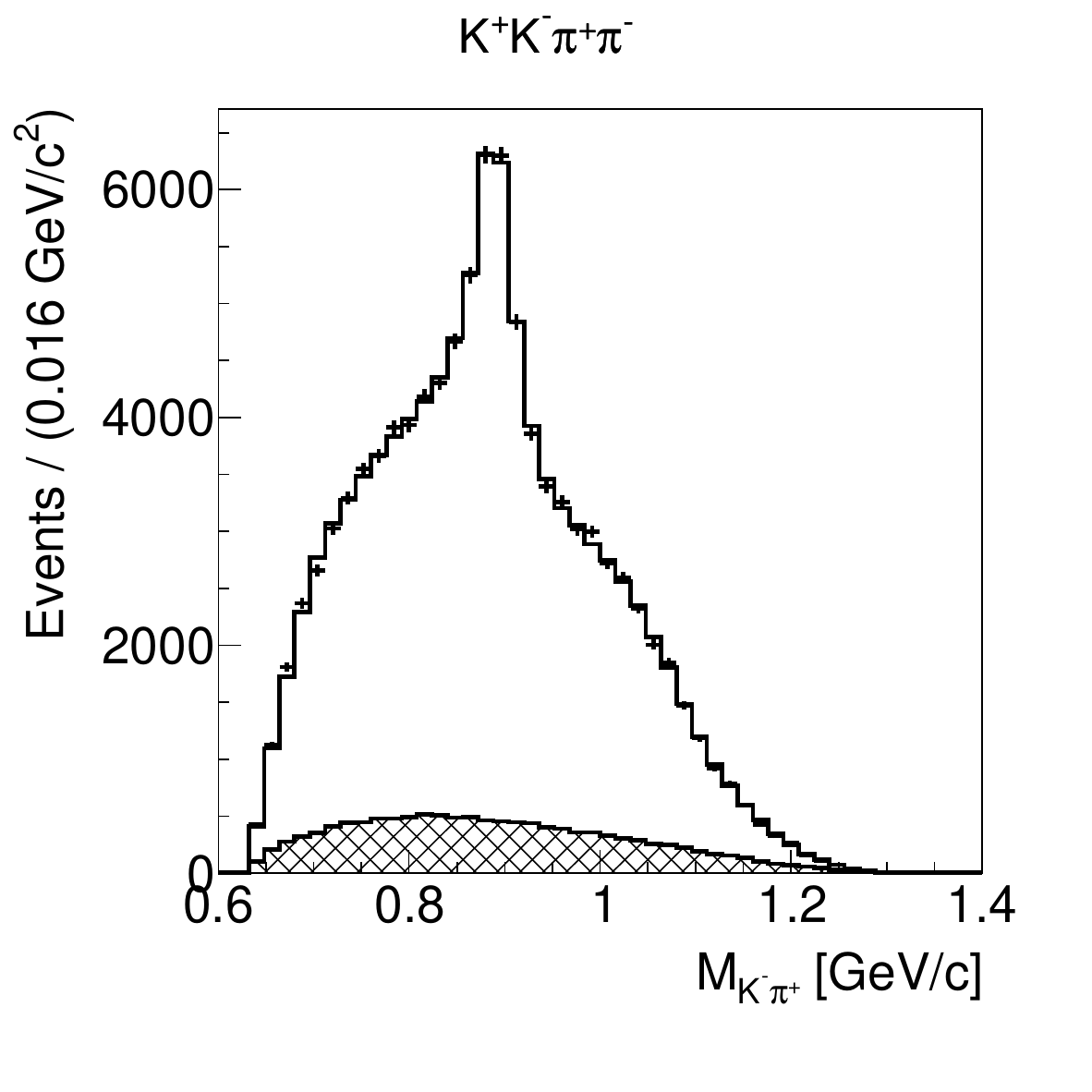}
\includegraphics[width=4.5cm]{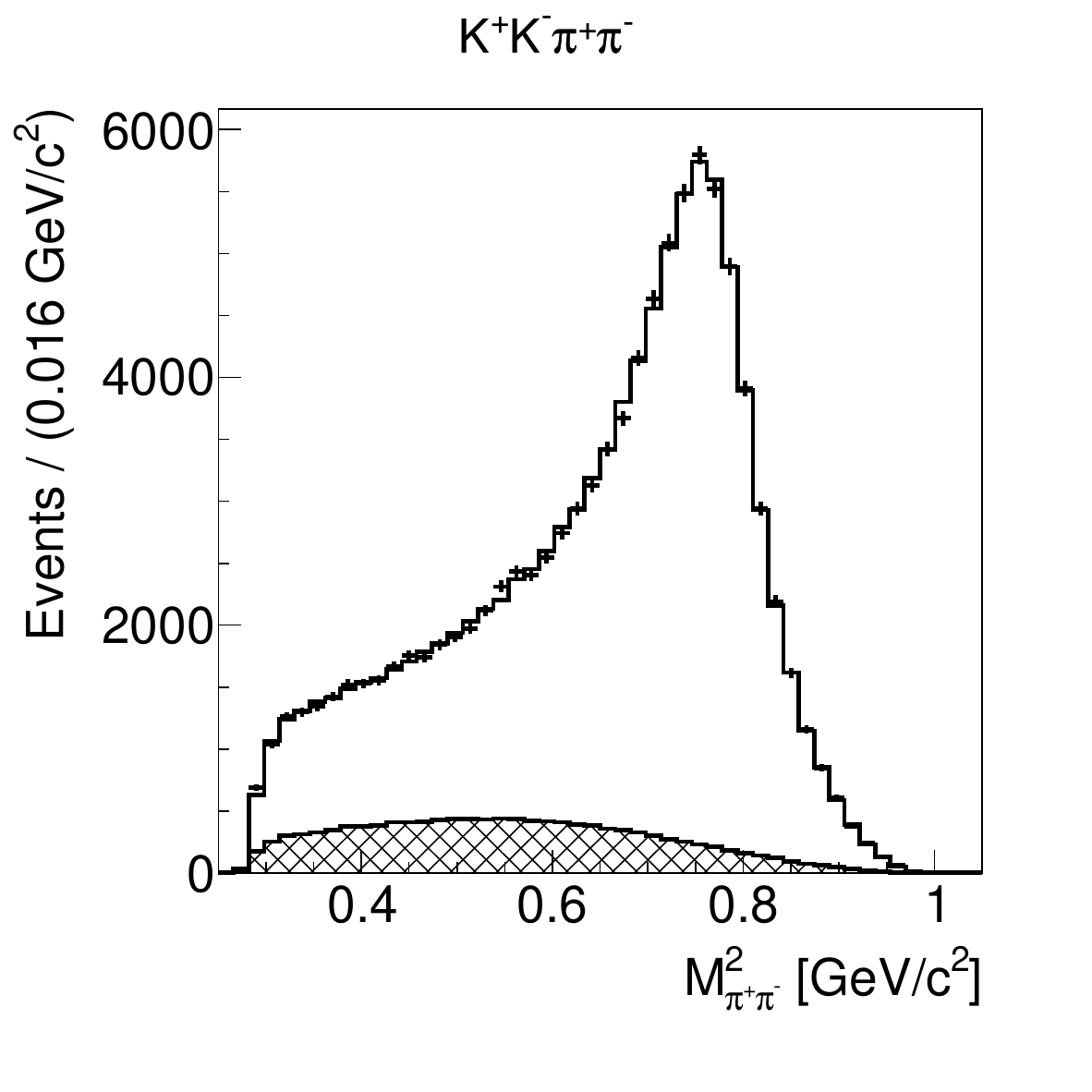}
\includegraphics[width=4.5cm]{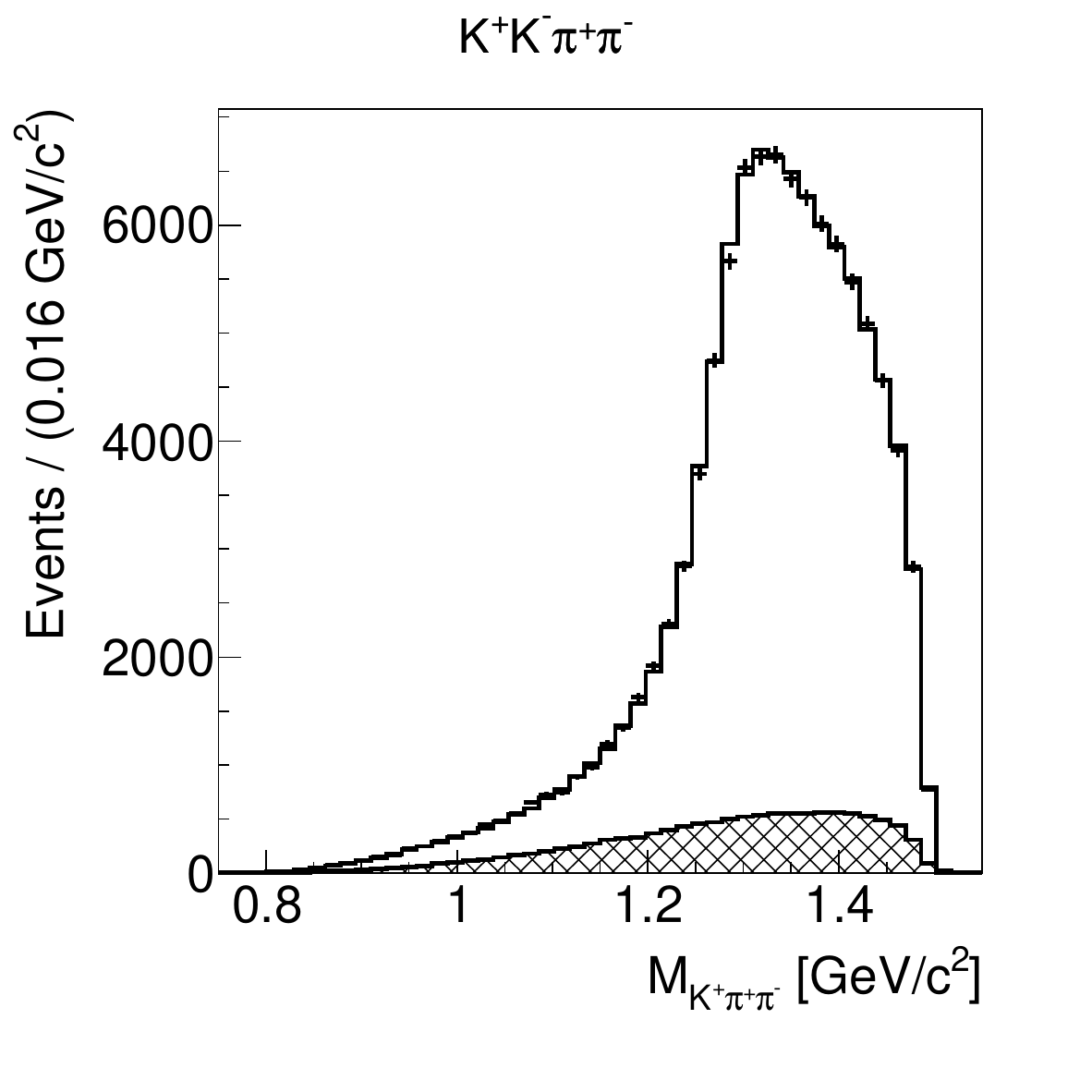} \\
\includegraphics[width=4.5cm]{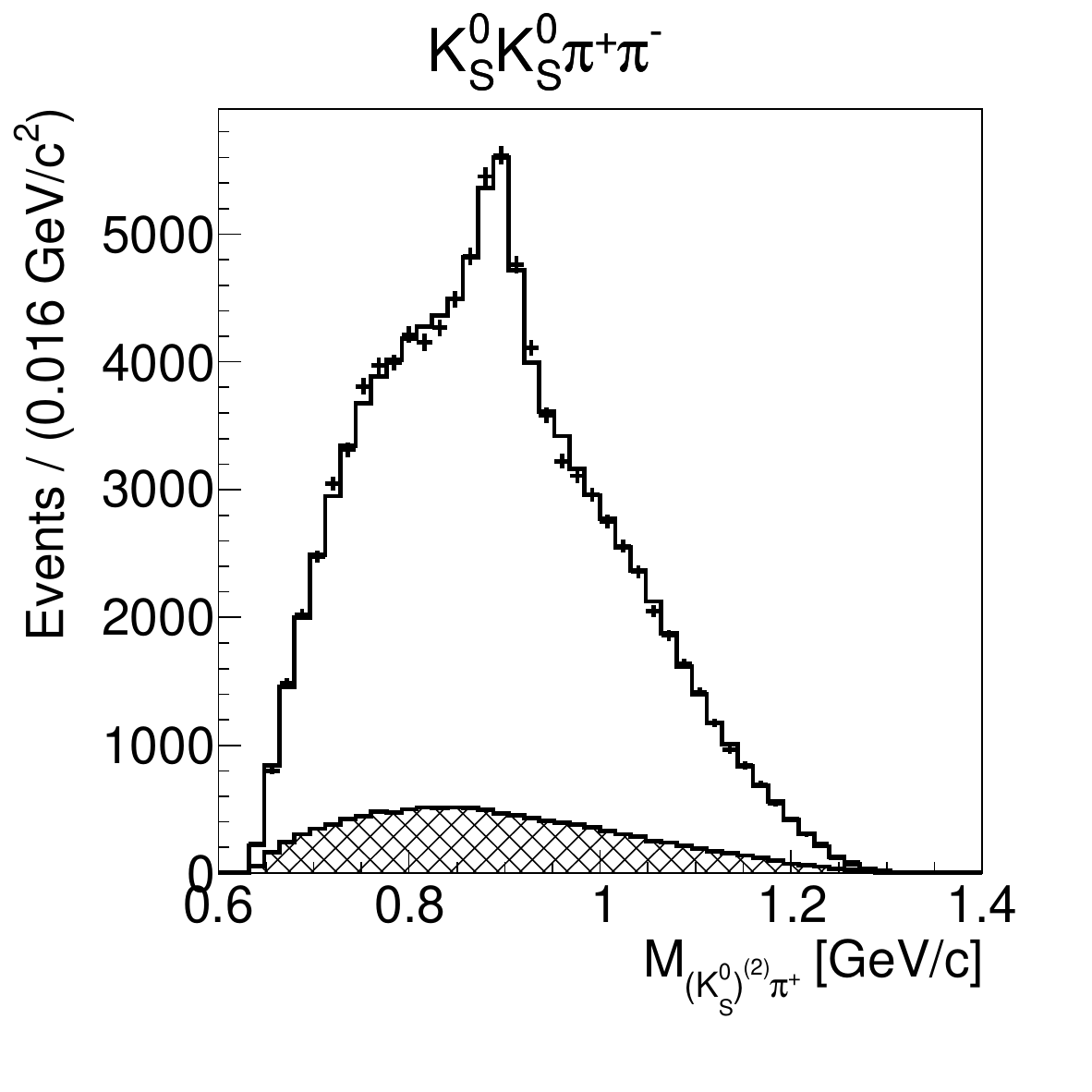}
\includegraphics[width=4.5cm]{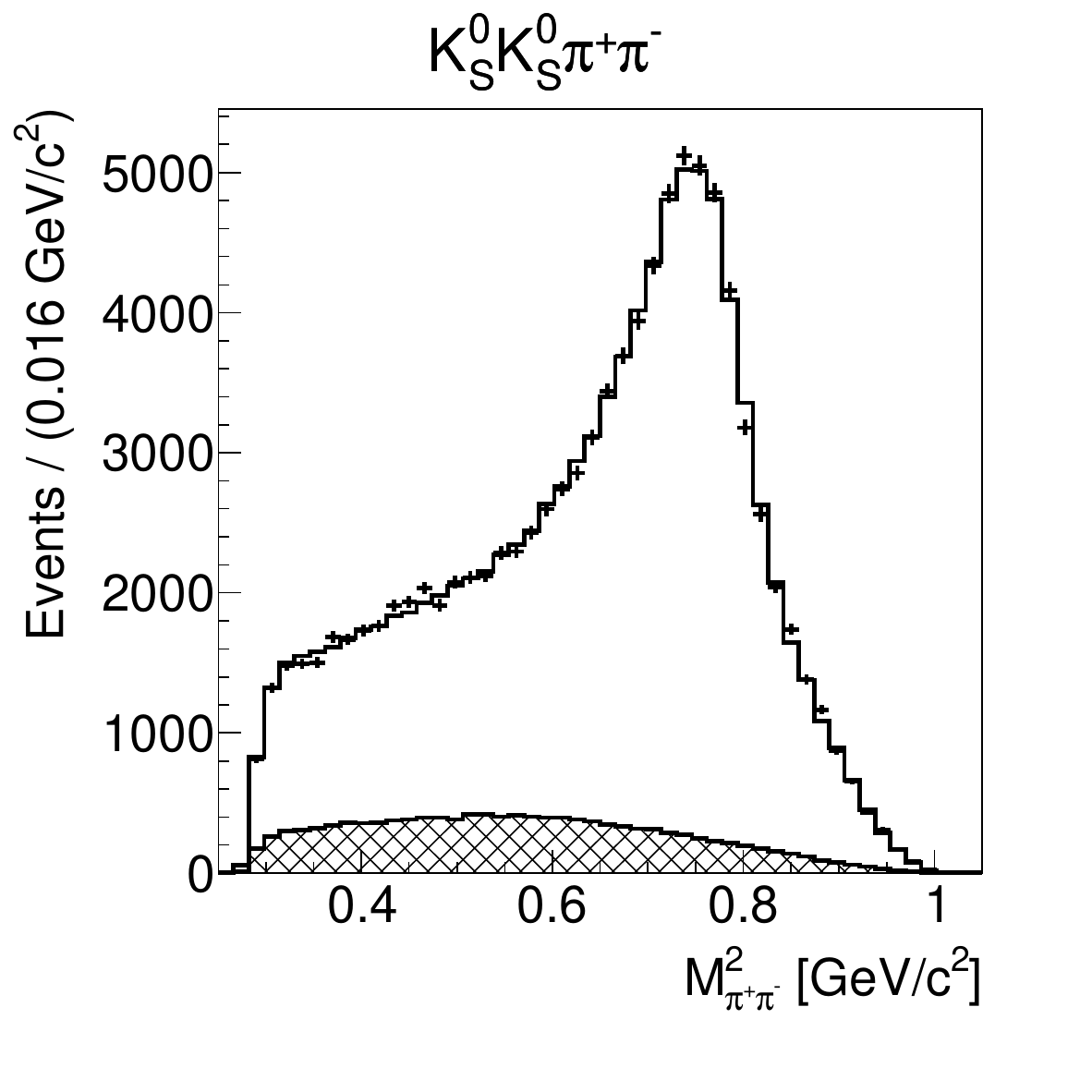}
\includegraphics[width=4.5cm]{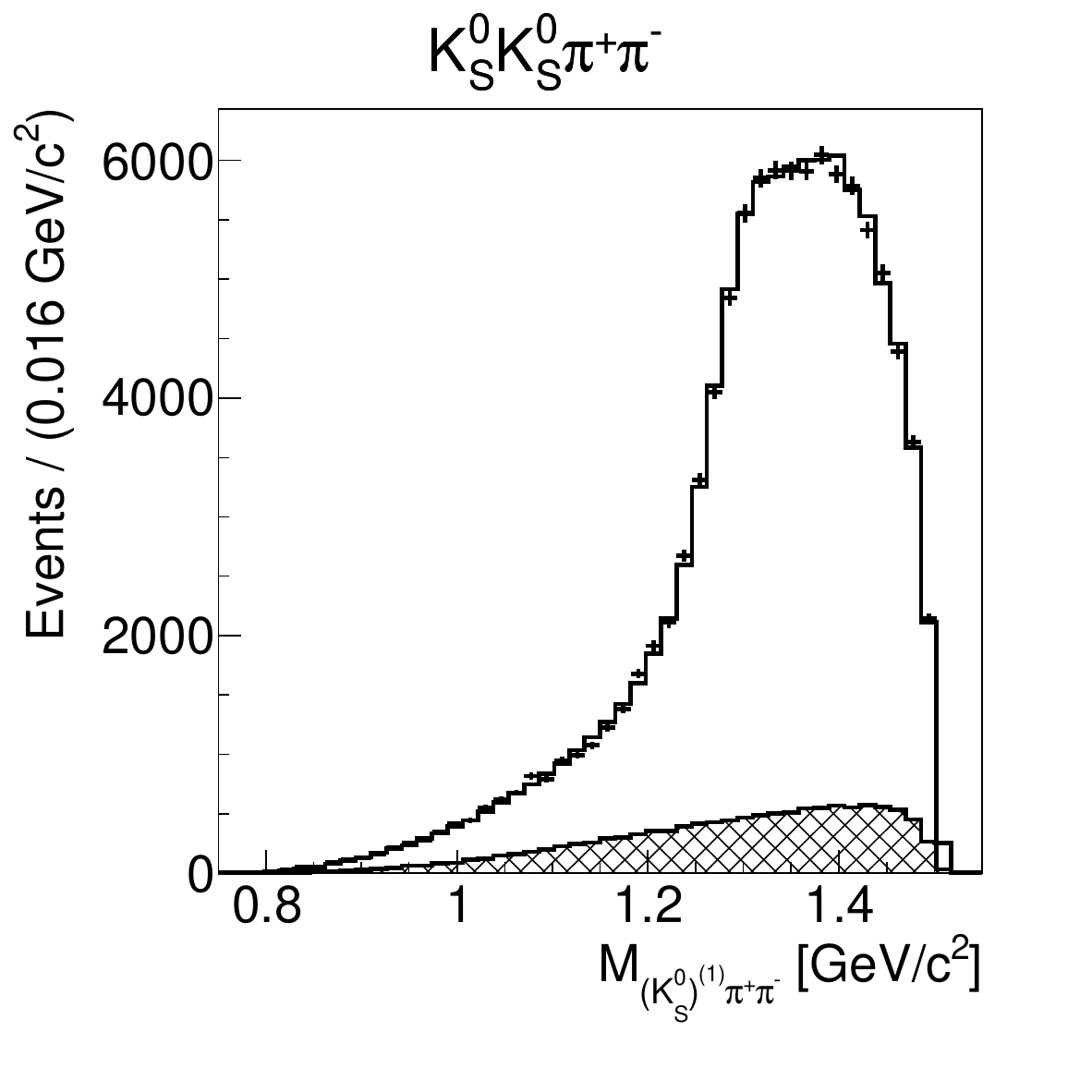} \\
\caption{Projections of signal fit results onto $M_{\kI \piII}$,
$M_{\piI \piII}$, and $M_{\kI \piI \piII}$ for the energy point
$\sqrt{s} = 2000\ \mev$. The points with error bars are pseudo-data,
the solid line is the fit result, and the dashed histogram
is the background contribution.}
\label{fig:signal}
\end{figure}

The signal fit is performed simultaneously
for $\kpkmpippim$ and $\kskspippim$ at five different energy points:
1800, 1850, 1900, 1950, and 2000 $\mev$; thus, there are 10 simultaneous-fit
points in total.
The production amplitudes $H^{(\vpho \to \kjI \kII)}_{\lambda_{\kjI}\,0}$
depend on the energy point, while the decay amplitudes such as
$H^{(\kjI \to \kstjIaII \kI)}_{\lambda_{\kstjIaII}\,0}$ are common for all
energy points.
Projections of signal fit results onto $M_{\kI \piII}$, $M_{\piI \piII}$,
and $M_{\kI \piI \piII}$ for the energy point $\sqrt{s} = 2000\ \mev$
are shown in Fig.~\ref{fig:signal}.

\vecampfit{} provides a structure {\tth vecampfit::simultaneous\_fit\_point}
containing functions and class definitions depending on the simultaneous-fit
point and fit-point process, experimental period, data sample, and channel.
A vector of simultaneous-fit point structures is declared in the signal
fitting program as
\begin{lstlisting}[style=cppstyle]
const std::vector<vecampfit::simultaneous_fit_point> c_simultaneous_fit_points {
#if USE_KPKMPIPPIM && USE_SAMPLE_1800_MEV
        vecampfit::simultaneous_fit_point{
        /* process, running_period, sample, channel */
        epem_kkpipi::process::c_kpkmpippim, 1, 1, 1,
        /* use_signal_process_background, use_cross_feed_background */
        false, {},
        /* signal */
        {0, fill_signal_event_buffer_#model#_kpkmpippim,
        sizeof(seb_#model#_kpkmpippim), VECAMPFIT_FITTER_DENSITY_FUNCTION(
        signal_density_#model#_kpkmpippim_1800_mev)},
        /* background */
        {BACKGROUND_FRACTION_KPKMPIPPIM_1800_MEV,
        fill_background_event_buffer_kpkmpippim_default_kpkmpippim_1800_mev,
        sizeof(beb_kpkmpippim_default_kpkmpippim_1800_mev),
        VECAMPFIT_FITTER_FILL_PARAMETER_BUFFER_FUNCTION(
        fill_background_pb_kpkmpippim_default_kpkmpippim_1800_mev),
        sizeof(bpb_kpkmpippim_default_kpkmpippim_1800_mev),
        VECAMPFIT_FITTER_DENSITY_FUNCTION(
        background_density_kpkmpippim_default_kpkmpippim_1800_mev)},
        /* signal_process_background, cross_feed_background */
        {}, {}},
#endif
	/* Other vector elements are omitted. */
}
\end{lstlisting}
The fitter initialization then requires simple loops over
the simultaneous-fit-point vector which do not require further checks
whether the point is included into the fit. For example, the functions
are initialized by
\begin{lstlisting}[style=cppstyle]
int set_simultaneous_fit_point_functions()
{
        int res;
        std::string file_name, channel_string;
        for (size_t i = 0; i < c_simultaneous_fit_points.size(); ++i) {
                const vecampfit::simultaneous_fit_point& point
                        = c_simultaneous_fit_points[i];
                int simultaneous_fit_point = i + 1;
                int background_source = 1;
                res = vecampfit_fitter_set_fill_signal_event_buffer(
                        point.signal.fill_event_buffer,
                        point.signal.sizeof_event_buffer,
                        simultaneous_fit_point);
                if (res != 0)
                        return -1;
		/* Further initialization is omitted. */
	}
}
\end{lstlisting}
Please note that one-based numbers are used for the simultaneous-fit points.

\section{Performance}
\label{sec:performance}

\subsection{Vector length and link-time optimization}

Performance tests are carried out at two machines. The first one has
an AMD Ryzen 3600 CPU with 6 cores and 12 threads operating
at the frequency of 3.6 GHz and 32 GB of RAM.
The second test machine has an Intel Core i3-10105 CPU
with 4 cores and 8 threads operating
at the frequency of 3.7 GHz and 8 GB of RAM.
Both test machines are running Debian 13.
All tests are performed in {\sc systemd} multi-user mode
without any GUI running. \vecampfit{} is compiled by {\sc gcc} version 15.2.0
with maximum possible optimization, including fast optimization settings
({\tth -Ofast} {\sc gcc} flag), aggressive inlining settings
({\tth --param inline-unit-growth=10000 --param max-inline-insns-auto=10000
--param large-function-growth=10000}),
link-time optimization (LTO) ({\tth -flto}),
and native code ({\tth -march=native}).

\begin{figure}
\centering
\includegraphics[width=6.5cm]{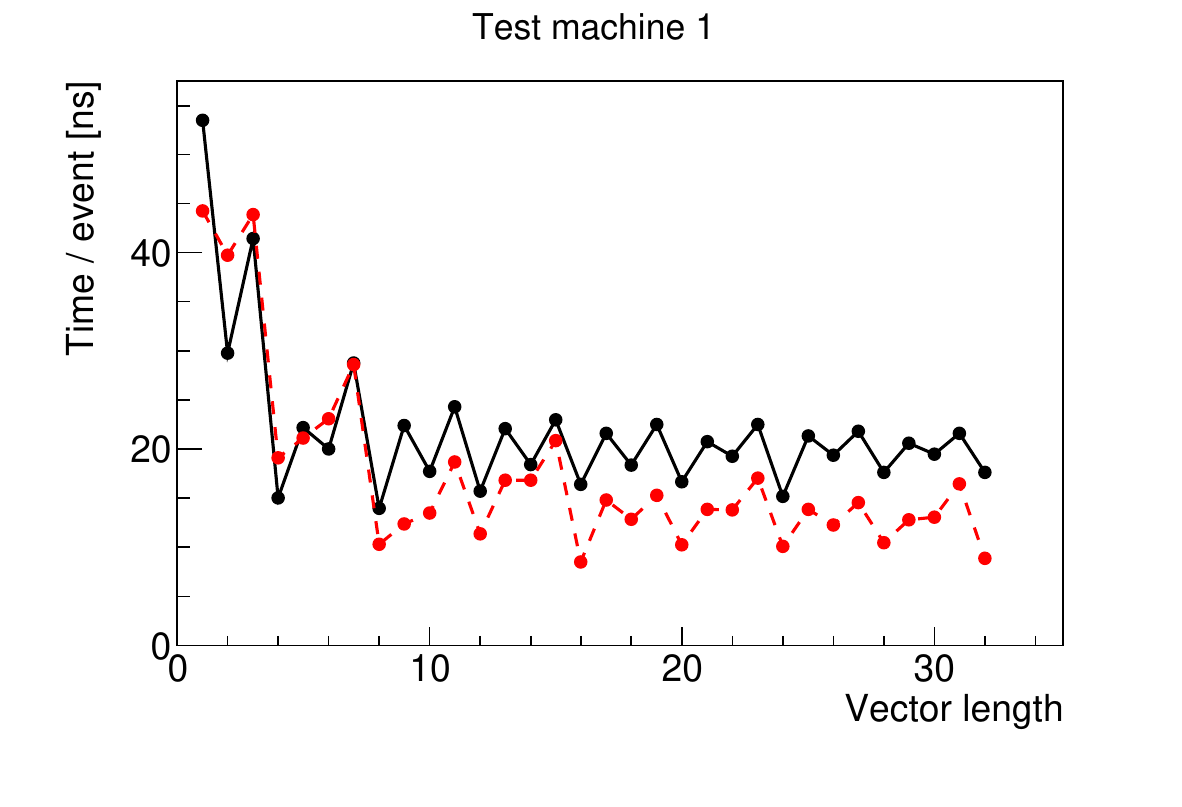}
\includegraphics[width=6.5cm]{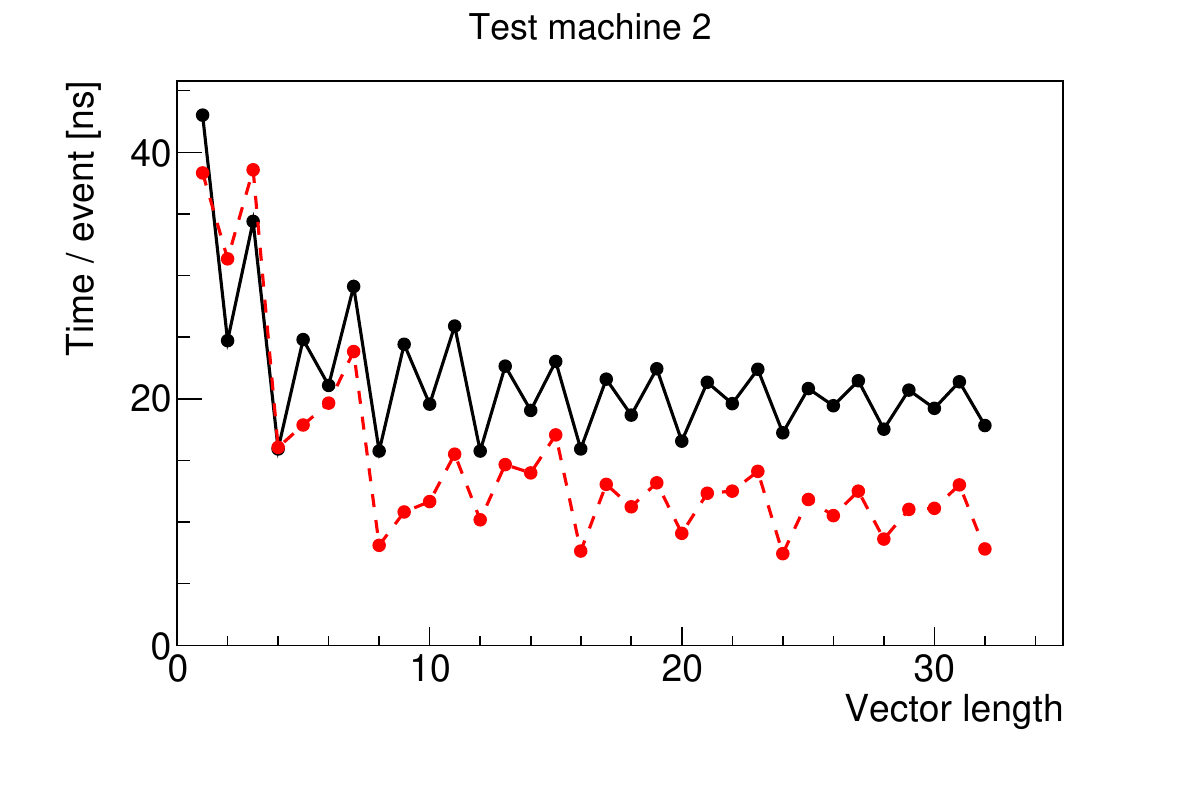}
\caption{Dependence of the performance of the $\kmpippiz$
example on the vector length for test machine 1 (left) and 2 (right).
The black solid line is the time per event per call for
double-precision calculations (Fortran type {\tth REAL(REAL64)},
C type {\tth double}) with LTO and the red dashed line
is the time per event for single-precision calculations (Fortran type
{\tth REAL(REAL32)}, C type {\tth float}) with LTO.}
\label{fig:performance_vector_length}
\end{figure}

Dependence of the performance on the vector length
is measured by execution of the $\kmpippiz$
example with both float and double floating-point types and all
vector lengths from 1 to 32. Since the dependence of the performance
on internal library settings is checked, the comparison is performed
for time per event per function call to eliminate the variation of execution
time caused by the varying number of the likelihood function calls.
The number of events used for fitting is 1000000 normalization Monte Carlo
events and 100000 signal pseudo-data events. One {\sc OpenMP} thread is used.
The results are shown in Fig.~\ref{fig:performance_vector_length}.
Local minima are observed for the vector lengths proportional to
the lengths of the largest vector register. For both test machines, there
are four 64-bit numbers or eight 32-bit numbers in the 256-bit AVX registers.

\begin{figure}
\centering
\includegraphics[width=6.5cm]{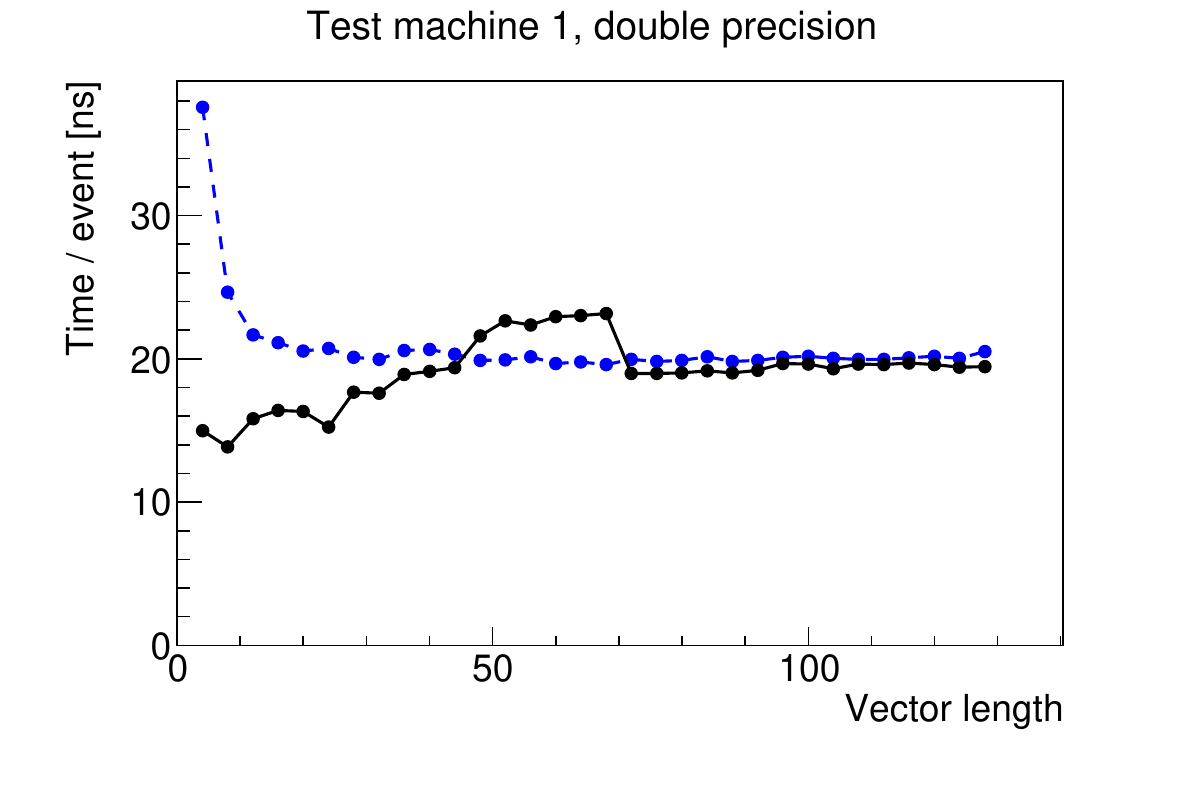}
\includegraphics[width=6.5cm]{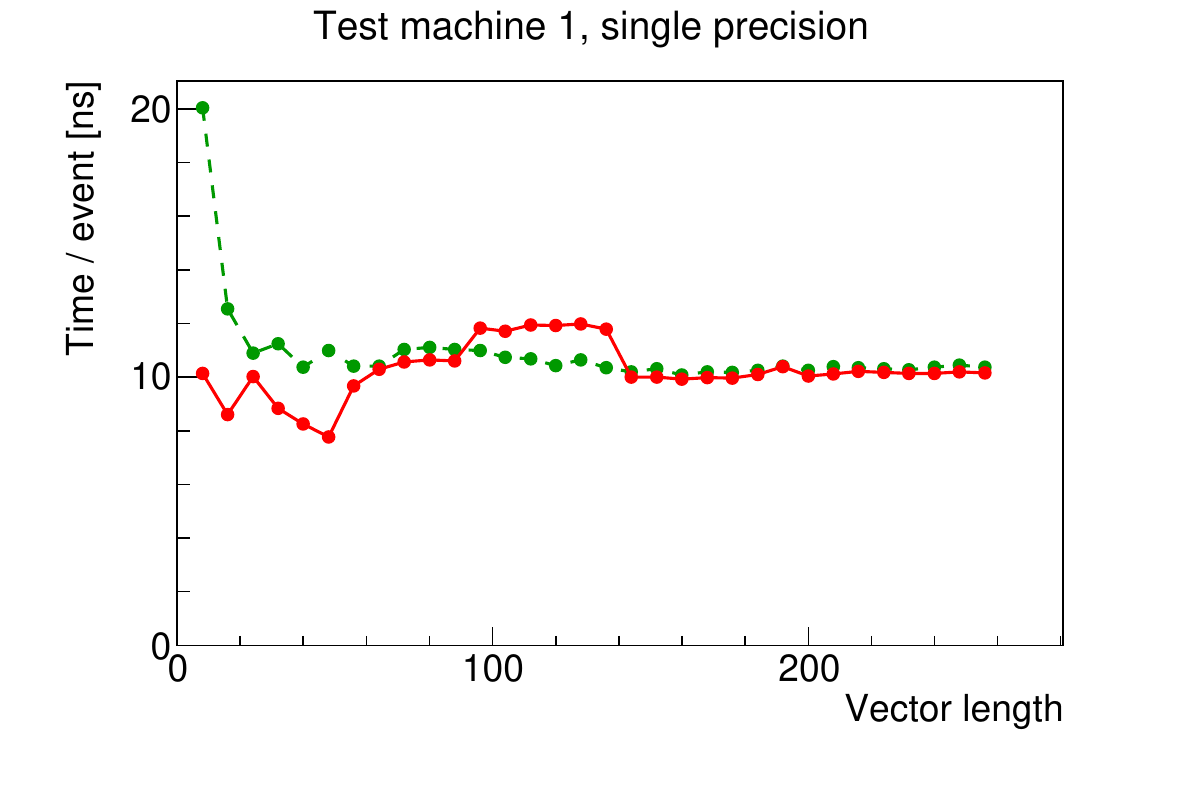} \\
\includegraphics[width=6.5cm]{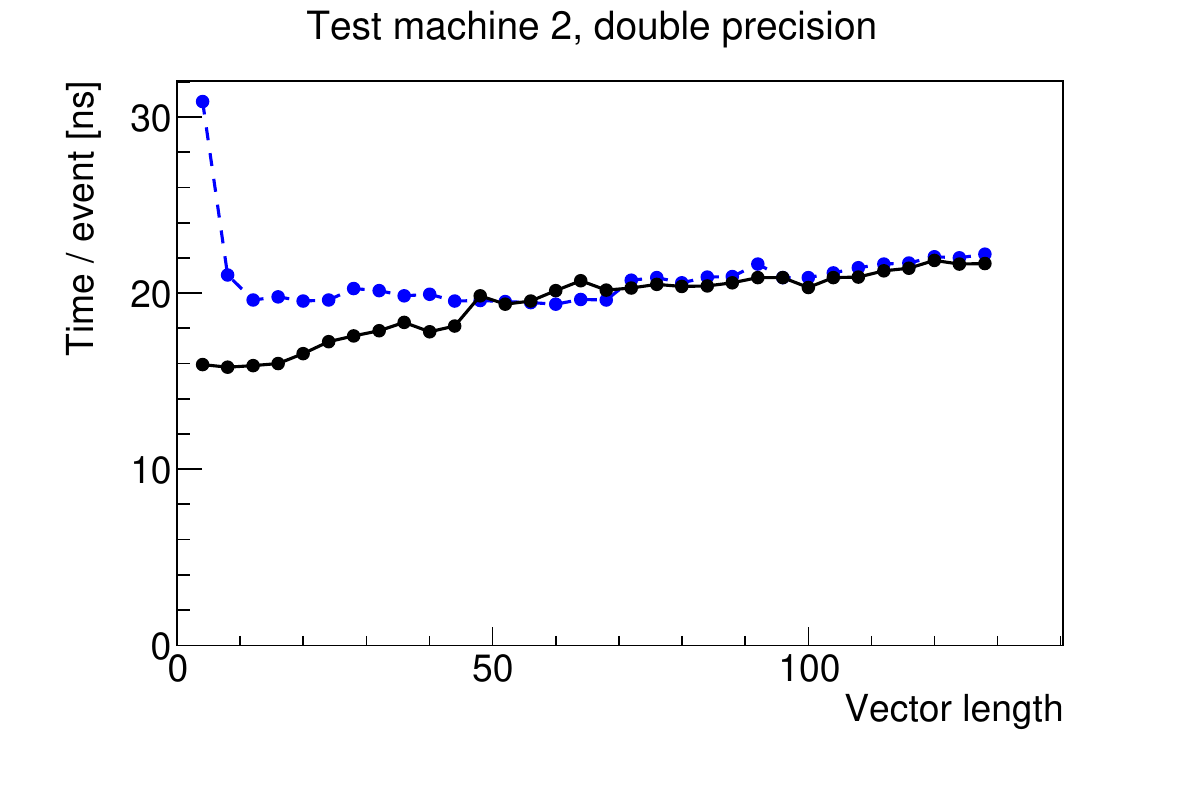}
\includegraphics[width=6.5cm]{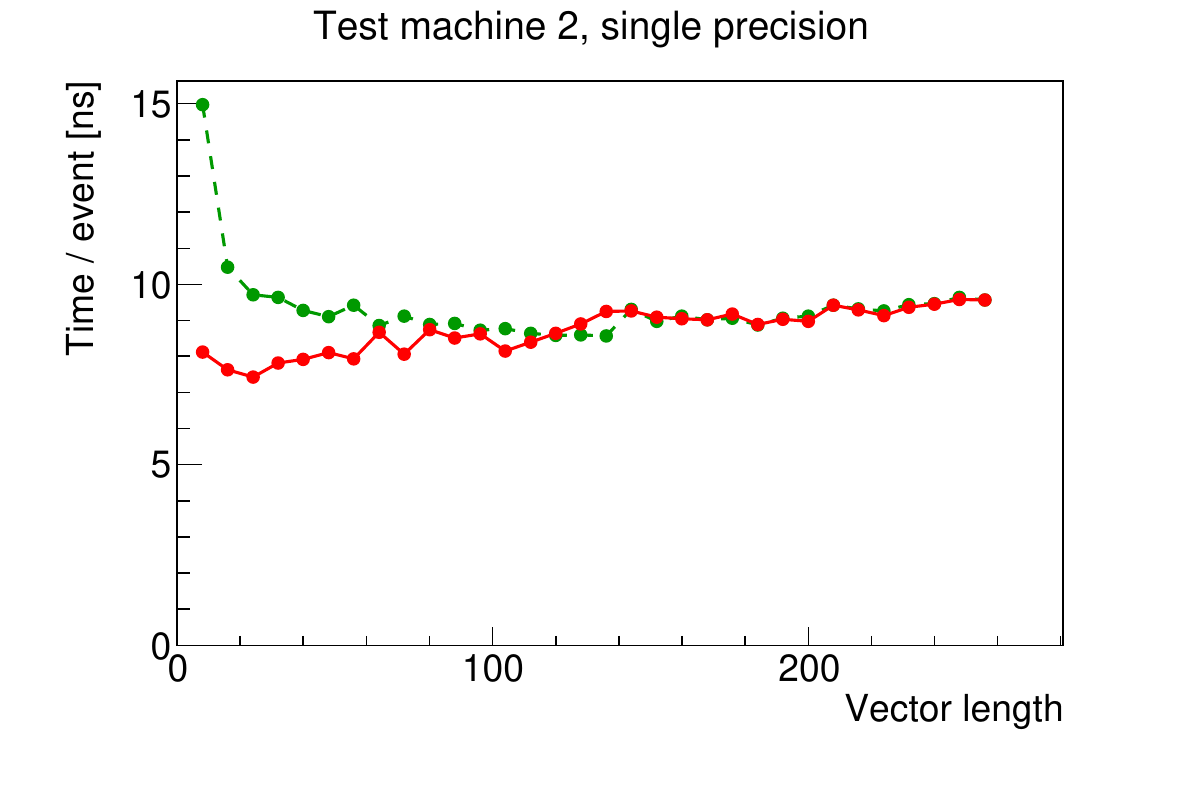} \\
\caption{Dependence of the performance of the $\kmpippiz$
example on the vector length for double-precision calculations (left)
and single-precision calculations (right) for test machine 1 (top) and 2
(bottom) for vector lengths proportional
to the length of AVX registers (4 for double-precision calculations and
8 for single-precision calculations).
The black solid line is the time per event per call for double-precision
calculations with LTO, the red solid line is the time per event per call for
single-precision calculations with LTO (the colors are the same as
in Fig.~\ref{fig:performance_vector_length}). The blue dashed line
is the time per event per call for double-precision calculations without LTO
and the green dashed line is the time per event per call for single-precision
calculations without LTO.}
\label{fig:performance_vector_length2}
\end{figure}

Additional tests are performed to determine the position of the global minimum.
The dependence of the execution time on the vector length is checked
again for vector lengths proportional to the length of the AVX registers:
8 for single-precision calculations and 4 for double-precision calculations.
To check the effect of link-time optimization, the same tests are also
performed for the $\kmpippiz$ example compiled
without link-time optimization.
The results are shown in Fig.~\ref{fig:performance_vector_length2}.
The preferred vector length is proportional to the length of
the largest vector register at the machine with a small
integer machine-dependent coefficient.
The preferred vector lengths and improvements over the scalar version
for both test machines are listed in Table~\ref{tab:vector_length}.
The improvement of the optimal-length vectorized calculation with respect
to the scalar version is somewhat less than the vector-register length,
e.g. the improvement for double precision is 2.7 and 3.9 for the test machines
1 and 2, respectively, while an AVX register contains four double-precision
numbers. Link-time optimization also results in an improvement:
the fitting programs compiled with LTO are found to be from
1.2 to 1.4 times faster than the programs compiled without it.
The preferred vector length is larger for fits without LTO.
When comparing the results for the same vector length,
the LTO effect reduces for larger vector lengths which are non-optimal.

\begin{table}
\caption{Preferred vector length, improvement over scalar version,
and effect of link-time optimization for the $\kmpippiz$ example.
The time is per event per call in nanoseconds.
The times correspond to the preferred vector length.}
\centering
\begin{tabular}{c|c|c|c|c}
Machine  & \multicolumn{2}{|c|}{Test machine 1} & \multicolumn{2}{|c}{Test machine 2} \\
\hline
Precision & Single & Double & Single & Double \\
\hline
Preferred vector length          &   48 &    8 &   24 &    8 \\
Preferred vector length (no LTO) &   40 &   68 &  120 &   60 \\
Scalar version time              & 44.3 & 53.6 & 38.3 & 43.0 \\
Vector version time              & 7.78 & 13.8 & 7.43 & 15.8 \\
Vector version time (no LTO)     & 10.4 & 19.6 & 8.59 & 21.0 \\
Ratio of scalar and vector times &  5.7 &  3.9 &  5.2 &  2.7 \\
Ratio of (no LTO / LTO) times    &  1.3 &  1.4 &  1.2 &  1.2 \\
\end{tabular}
\label{tab:vector_length}
\end{table}

\begin{figure}
\centering
\includegraphics[width=6.5cm]{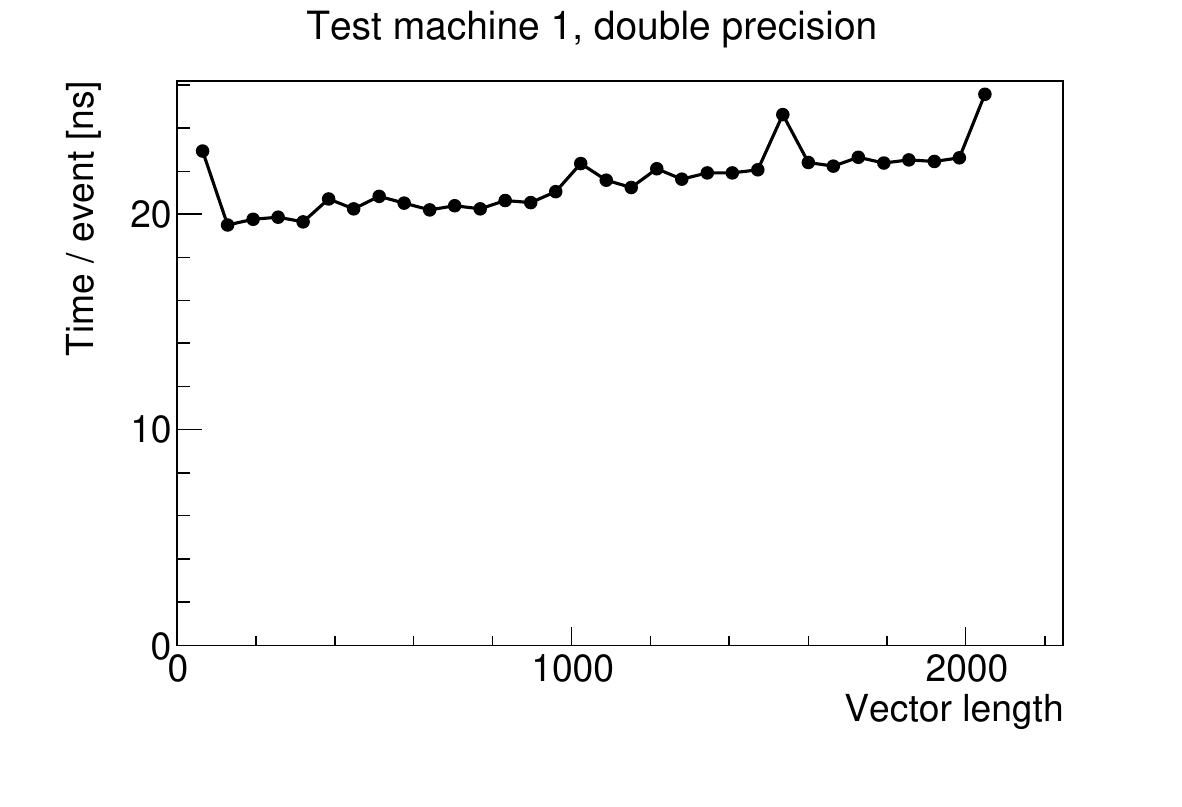}
\includegraphics[width=6.5cm]{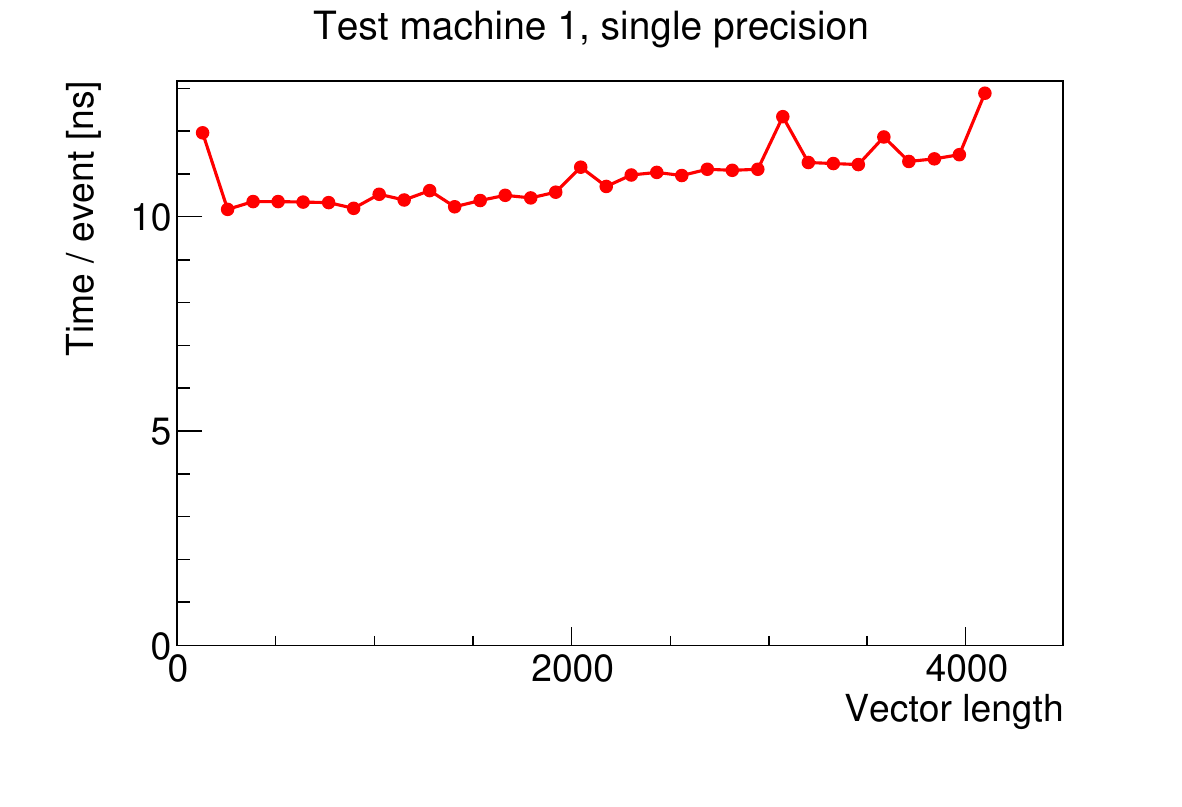} \\
\includegraphics[width=6.5cm]{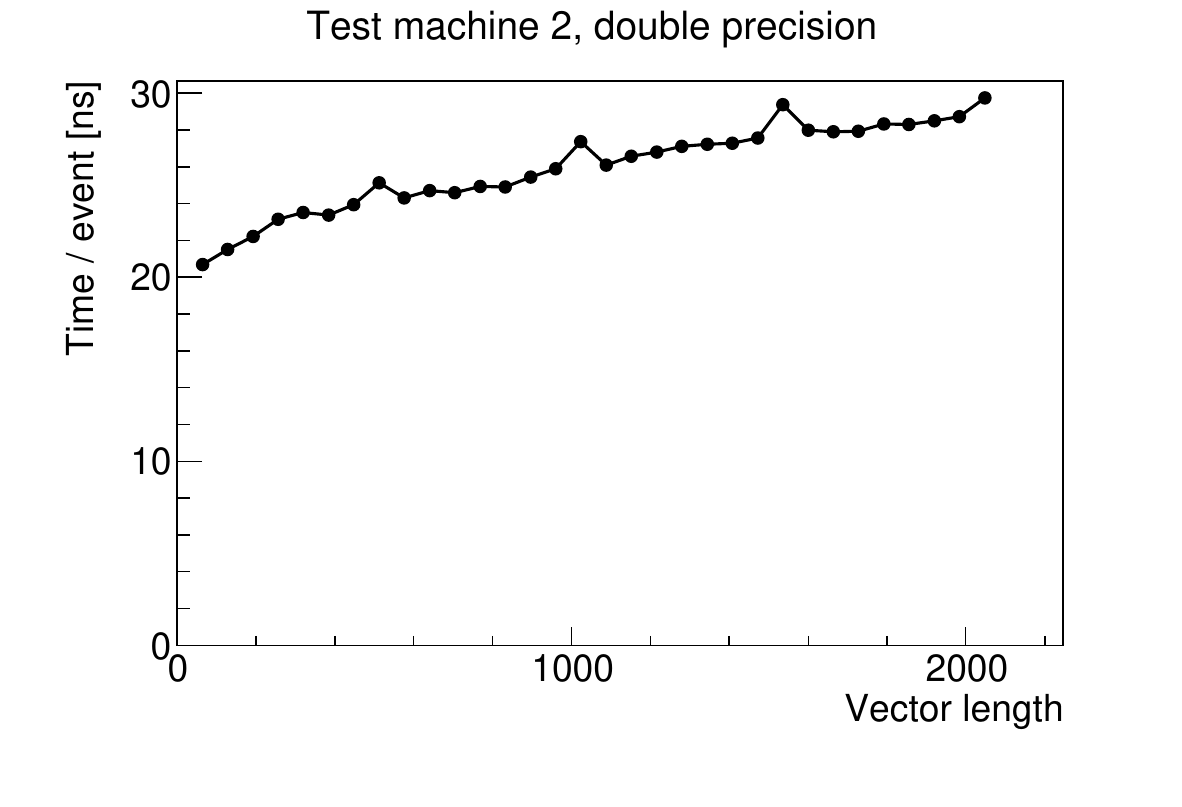}
\includegraphics[width=6.5cm]{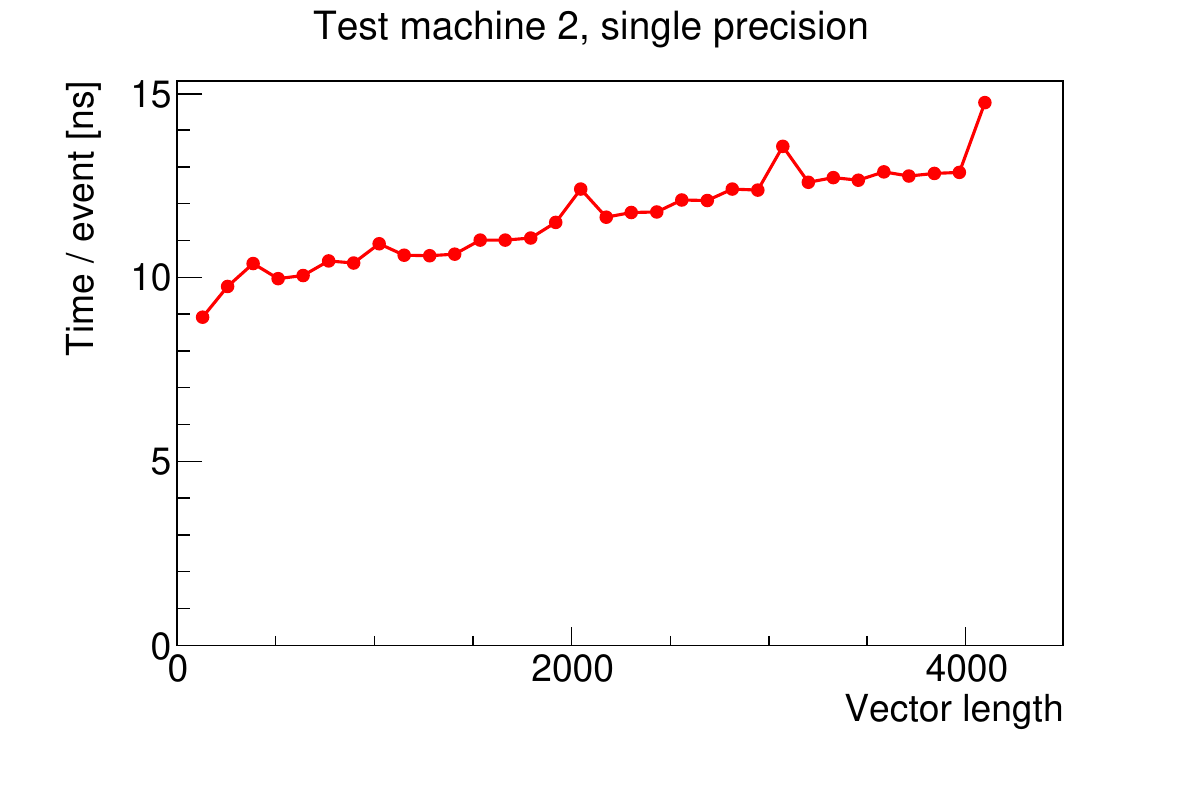} \\
\caption{Dependence of the performance of the $\kmpippiz$
example on the vector length for double-precision calculations (left)
and single-precision calculations (right) for test machine 1 (top) and 2
(bottom) for vector lengths proportional to 64 for double-precision
calculations and 128 for single-precision calculations.
The black solid line is the time per event per call for double-precision
calculations with LTO and the red solid line is the time per event per call for
single-precision calculations with LTO.}
\label{fig:performance_vector_length3}
\end{figure}

The dependence of the performance on the vector length for larger lengths
is checked by performing a similar test with vector lengths
proportional to 64 for double-precision calculations and 128
for single-precision calculations.
The results are shown in Fig.~\ref{fig:performance_vector_length3}.
A slow increase of calculation time is observed for both single and
double-precision calculations.

Based on the test results, the default set of vector lengths used when
compiling the library is chosen to be equal
to small powers of 2 and includes 1, 2, 4, 8, 16, 32, and 64.

\subsection{Comparison with \laura{}}
\label{sec:performance_laura}

\laura{} is a Dalitz-plot fitting library~\cite{Back:2017zqt}.
Version 3.8.0 is used for performance comparison.
A dedicated example {\tth laura\_genfit3k} is implemented in \vecampfit{}
to compare its performance with \laura{}.
This example performs a Dalitz analysis of a simplified model of the
decay $\kpkpkm$, reimplementing the \laura{} example {\tth GenFit3K}
as closely as possible.

The amplitude is calculated using covariant formalism.
Kinematic covariant factor for the decay chain $0 \to R + 3$, $R \to 1 + 2$
is given by
\begin{equation}
f(m_{12}) = \sqrt{1 + y^2} = \frac{m_0^2 + m_{12}^2 - m_3^2}{2 m_{12} m_0},
\label{eq:covariant_factor}
\end{equation}
where $m$ are invariant masses of particles or particle combinations specified
by the subscript~\cite{Filippini:1995yc,LHCb:2015klp}.
Covariant factors for specific spins of intermediate resonances are
\begin{equation}
\begin{aligned}
F_1^\text{(cov)}(m_{12}) &= \sqrt{1 + y^2} = f, \\
F_2^\text{(cov)}(m_{12}) &= \frac{3}{2} + y^2 = \frac{1}{2} + f^2, \\
F_3^\text{(cov)}(m_{12}) &= \sqrt{1 + y^2}(\frac{5}{2} + y^2) = f (\frac{3}{2} + f^2), \\
F_4^\text{(cov)}(m_{12}) &= \frac{35 + 40 y^2 + 8 y^4}{35} = \frac{3 + 24 f^2 + 8 f^4}{35}. \\
\end{aligned}
\label{eq:covariant_factor_j}
\end{equation}

The angular distribution depends on Legendre polynomials
\begin{equation}
\begin{aligned}
P_0^\text{(Leg)}(x) & = 1, \\
P_1^\text{(Leg)}(x) & = x, \\
P_n^\text{(Leg)}(x) & =
  \frac{1}{n}[(2 n - 1) x P_{n - 1}(x) - (n - 1) P_{n - 2}(x)]\text{ for $n \ge 2$}, \\
\end{aligned}
\end{equation}
which are normalized differently in \laura{}:
\begin{equation}
P_L^\text{(Leg \laura)}(x) = \frac{(-1)^{L} 2^{2 L}}{\binom{2 L}{L}}
  P_L^\text{(Leg)}(x).
\end{equation}
Full angular distribution for the decay chain $0 \to R + 3$, $R \to 1 + 2$
is given by
\begin{equation}
A_R^\text{(ang)}(\Phi) = (p^* q)^L F_L^\text{(cov)}(m_{12})
 P_L^\text{(Leg \laura)}(\cos \theta_R),
\end{equation}
where $p^*$ is the momentum of the particle $3$ in the rest frame of
the particle $0$, $q$ is the momentum of the particles $1$ and $2$ in
the rest frame of the intermediate resonance $R$, and $\theta_R$ is
the $R$ helicity angle.

The amplitude of the decay chain $\Bp \to \phi_J \kp$, $\phi_J \to \kp \km$ is
\begin{equation}
A_{\phi_J}(\Phi) = B_{\phi_J}^\text{(no $q$)}(m)
  A_{\phi_J}^\text{(ang)}(\Phi)
  F_{\Bp}^{(BW)}(q (F_J^\text{(cov)}(m_{12}))^\frac{1}{J}, r^{(B)}, J),
\end{equation}
where the Blatt-Weisskopf formfactor with scaled momentum is used for the
decay $\Bp \to \phi_J \kp$ and $r^{(B)}$ is $B$ effective radius.
The Breit-Wigner amplitude $B_{\phi_J}^\text{(no $q$)}$
is given by Eq.~\eqref{eq:breit_wigner_q_f_nophsp}.
The amplitude is
\begin{equation}
A(\Phi) = a_0 + \sum\limits_{\phi_J} a_{\phi_J}
\left[A_{\phi_J}^{(\Bp \to \phi_J \kp_{(1)})}(\Phi) +
A_R^{(\Bp \to \phi_J \kp_{(2)})}(\Phi)\right]
\end{equation}
where $a_0$ is constant nonresonant amplitude, $a_{\phi_J}$
are resonance complex amplitudes, and symmetrization over
two ways of selection of the decay chain
is performed. The signal density function is $S(\Phi) = |A(\Phi)|^2$.
Two intermediate $\phi_J$ resonances, $\phi(1020)$
and $f'_2(1525)$, are included into the model. The background is not included.

The \vecampfit{} and \laura{} implementations have the same 
physical model but differ by amplitude normalization method.
Normalization of each individual resonance is performed in \laura{}:
\begin{equation}
\int \left[A_{\phi_J}^{(\Bp \to \phi_J \kp_{(1)})}(\Phi) +
A_{\phi_J}^{(\Bp \to \phi_J \kp_{(2)})}(\Phi)\right]^2 d \Phi = 1,
\end{equation}
where integration is performed over the entire phase space (Dalitz plot).
\vecampfit{} does not provide any way to do the same and non-normalized
amplitudes are used. For both \laura{} and \vecampfit{},
the $\phi(1020)$ amplitude is fixed at $1.0 + 0.0 i$ and
the nonresonant and $f'_2(1525)$ amplitudes are free.
Thus, ratios of real and imaginary parts of the free amplitudes
should be exactly the same for both fit versions, providing a way for
validation of \vecampfit{} results.

\begin{figure}
{\centering
\includegraphics[width=6.7cm]{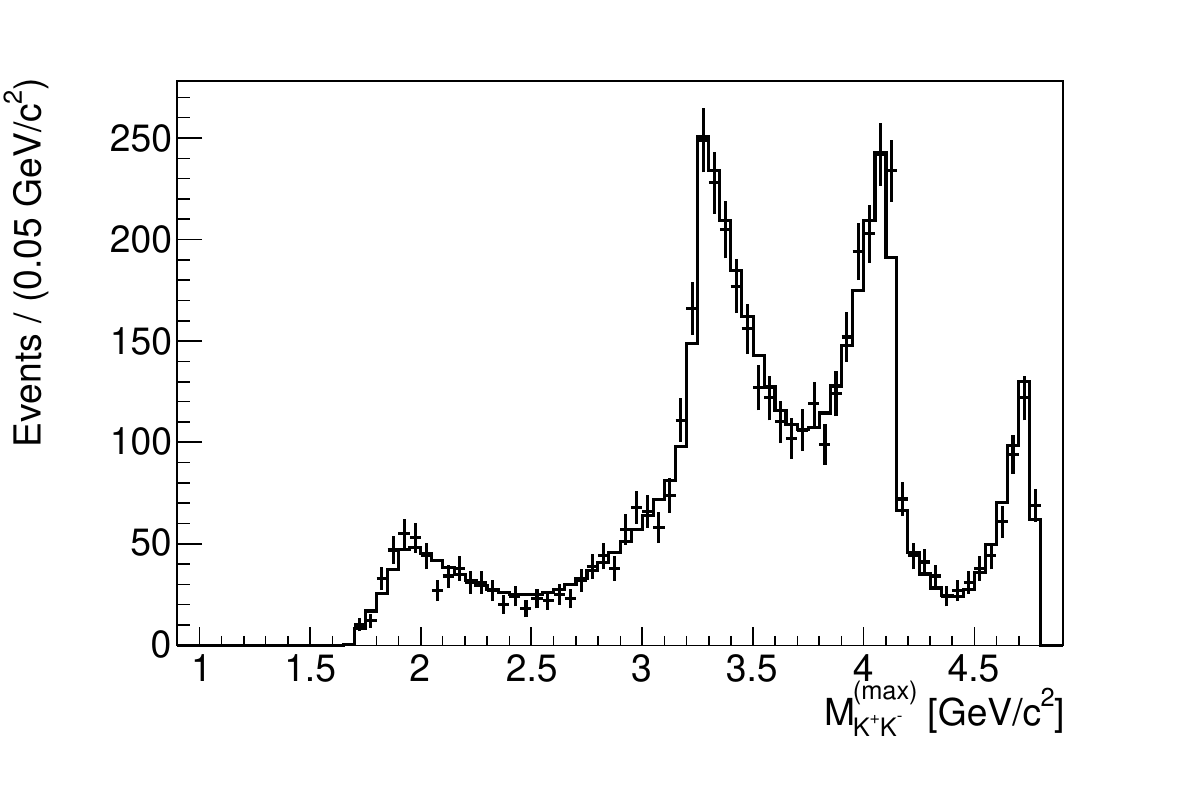}
\includegraphics[width=6.7cm]{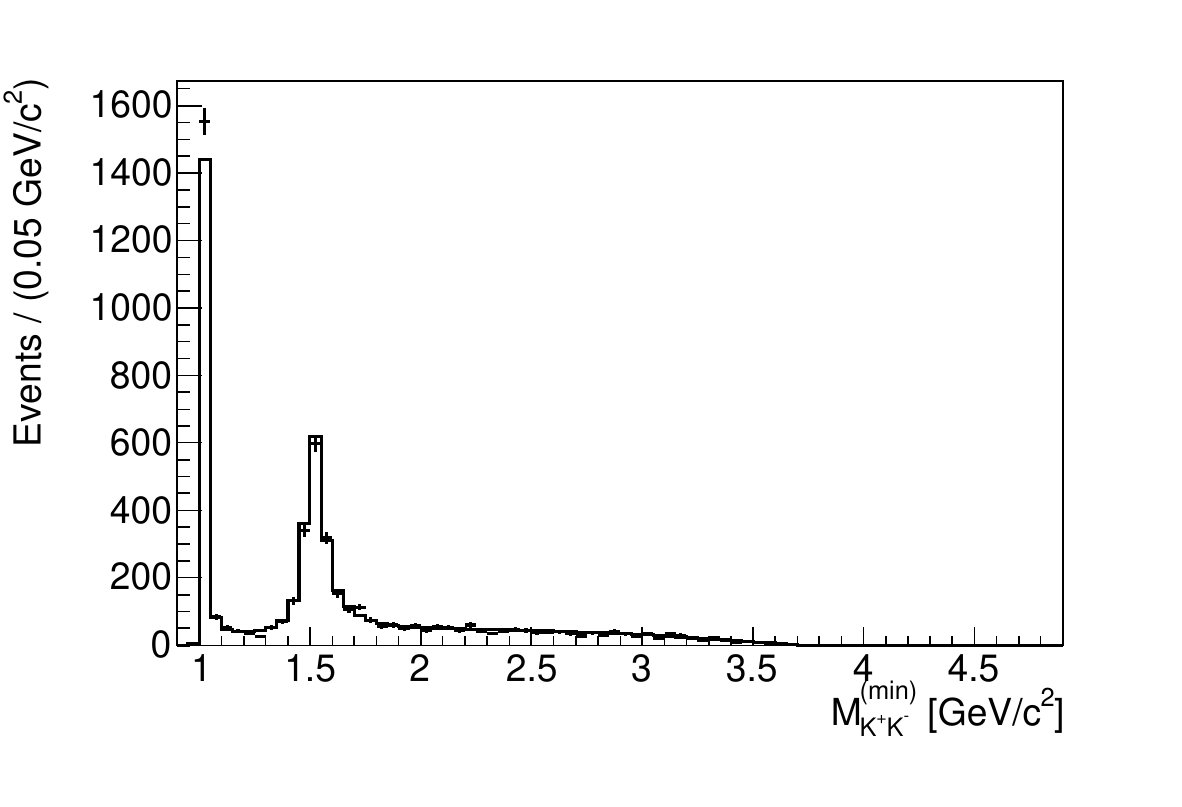} \\}
\caption{Results of fit to the $\kpkpkm$ signal distribution for
a pseudoexperiment generated by \laura{} example {\tth GenFit3K}.
The points with error bars are data, the solid line
is the fit result, and the hatched histograms are the background contribution.}
\label{fig:laura_genfit3k}
\end{figure}

The \vecampfit{} example is able to use pseudoexperiments generated
by \laura{} after conversion which stores each pseudoexperiment in
an individual file. The fit results for a pseudoexperiment with 5000
signal events generated by \laura{} are shown in Fig.~\ref{fig:laura_genfit3k}.

The \vecampfit{} example uses grid integration; the grid is manually set up
to be exactly the same as the grid created by \laura{} automatically.
Parallel calculations are implemented in \laura{} for independent data samples
only, while \vecampfit{} is able to perform parallel calculations using
different events from the same data sample. Thus, only one version
of the \laura{} example is used for comparison; the calculations are
performed in double precision by a single thread without vectorization.
Similarly, one thread and double precision are used for \vecampfit{};
several different vector lengths are tested.

Similar to the \laura{} testing procedure~\cite{Back:2017zqt},
50 pseudoexperiments are generated. \laura{} is compiled with maximum
optimization ({\tth -Ofast} {\sc gcc} flag), native-code generation
({\tth -march=native}), and link-time optimization ({\tth -flto}).
Unlike the previous performance tests, total execution time is used
for comparison with other libraries, since they have different internal
settings and available features. Initial values of the amplitudes are
randomized for both \laura{} and \vecampfit{}.
\vecampfit{} fit is performed in two versions with and without explicit
gradient calculation.

\begin{figure}
{\centering
\includegraphics[width=6.7cm]{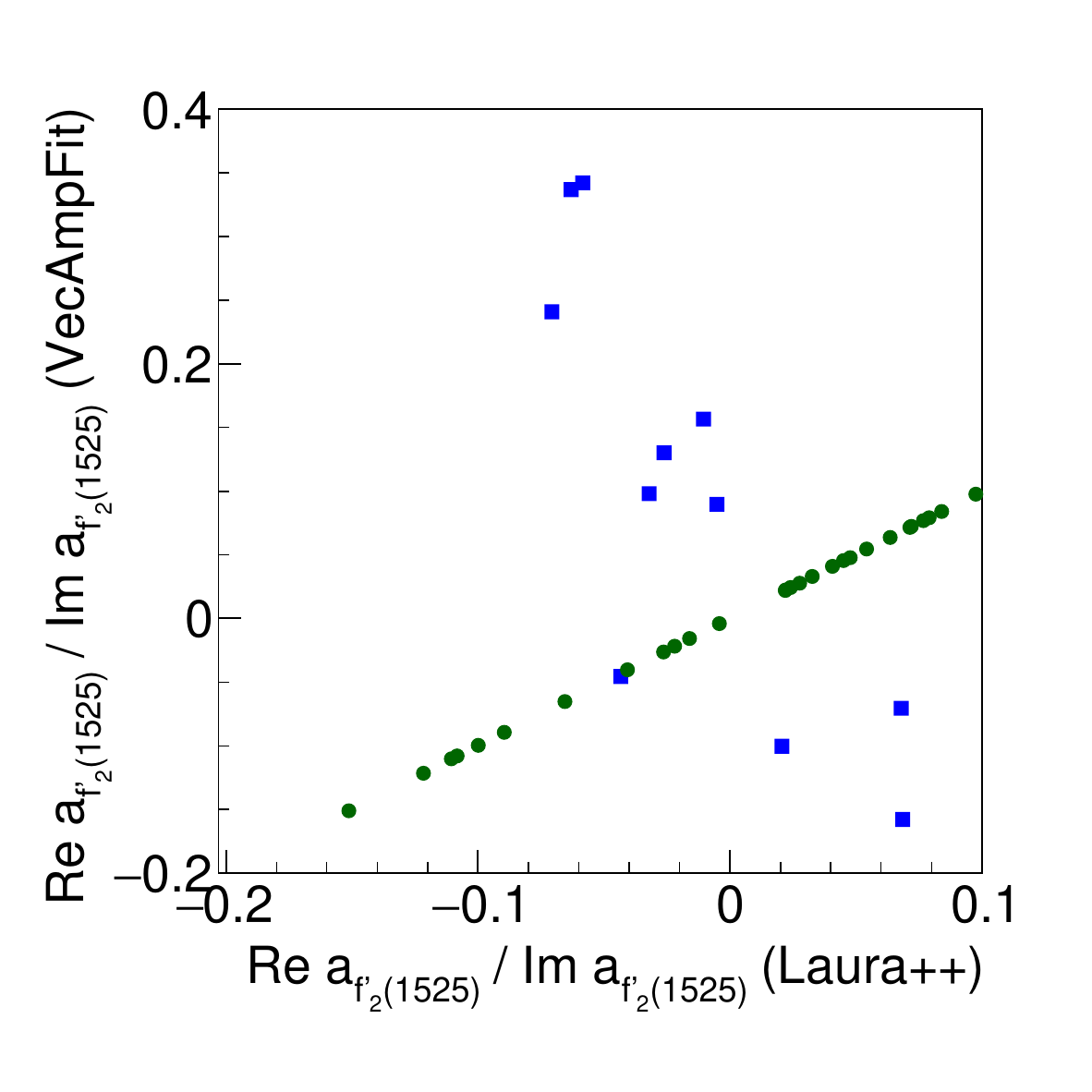}
\includegraphics[width=6.7cm]{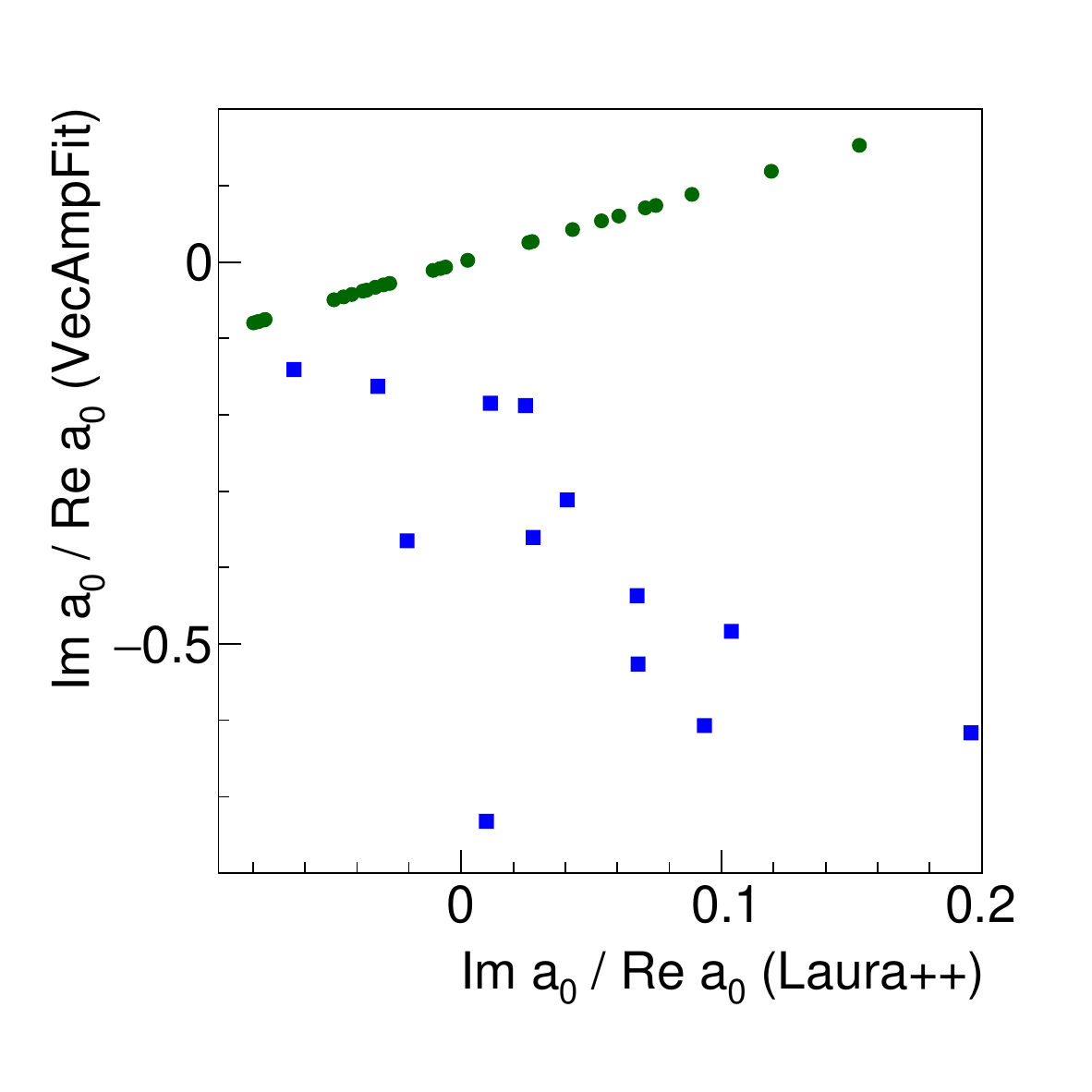} \\}
\caption{Comparison of ratios of real and imaginary parts
of the $f'_2(1525)$ amplitude and ratios of imaginary and real parts
of the nonresonant amplitude for \laura{} and \vecampfit{}.
Green circles are pseudoexperiments with both \laura{} and \vecampfit{}
converging to the true solution and blue squares are pseudoexperiments
with at least one of the fits converging to an alternative solution.}
\label{fig:laura_vecampfit_re_im}
\end{figure}

It is checked whether the fits converge to the same minimum by comparing
the resulting amplitudes. It is found that there is a second solution
with approximately inverse amplitude of the $f'_2(1525)$. Both \laura{}
and \vecampfit{} fits converge to the second solution for about 30\% of fits.
If they both converge to the true solution, then the results are the same.
Comparisons of ratios of real and imaginary parts of the $f'_2(1525)$ amplitude
and ratios of imaginary and real parts of the nonresonant amplitude
are shown in Fig.~\ref{fig:laura_vecampfit_re_im}. The \vecampfit{} fit
without gradient and with the vector length equal to four is used for
comparison.

\begin{figure}
{\centering
\includegraphics[width=6.7cm]{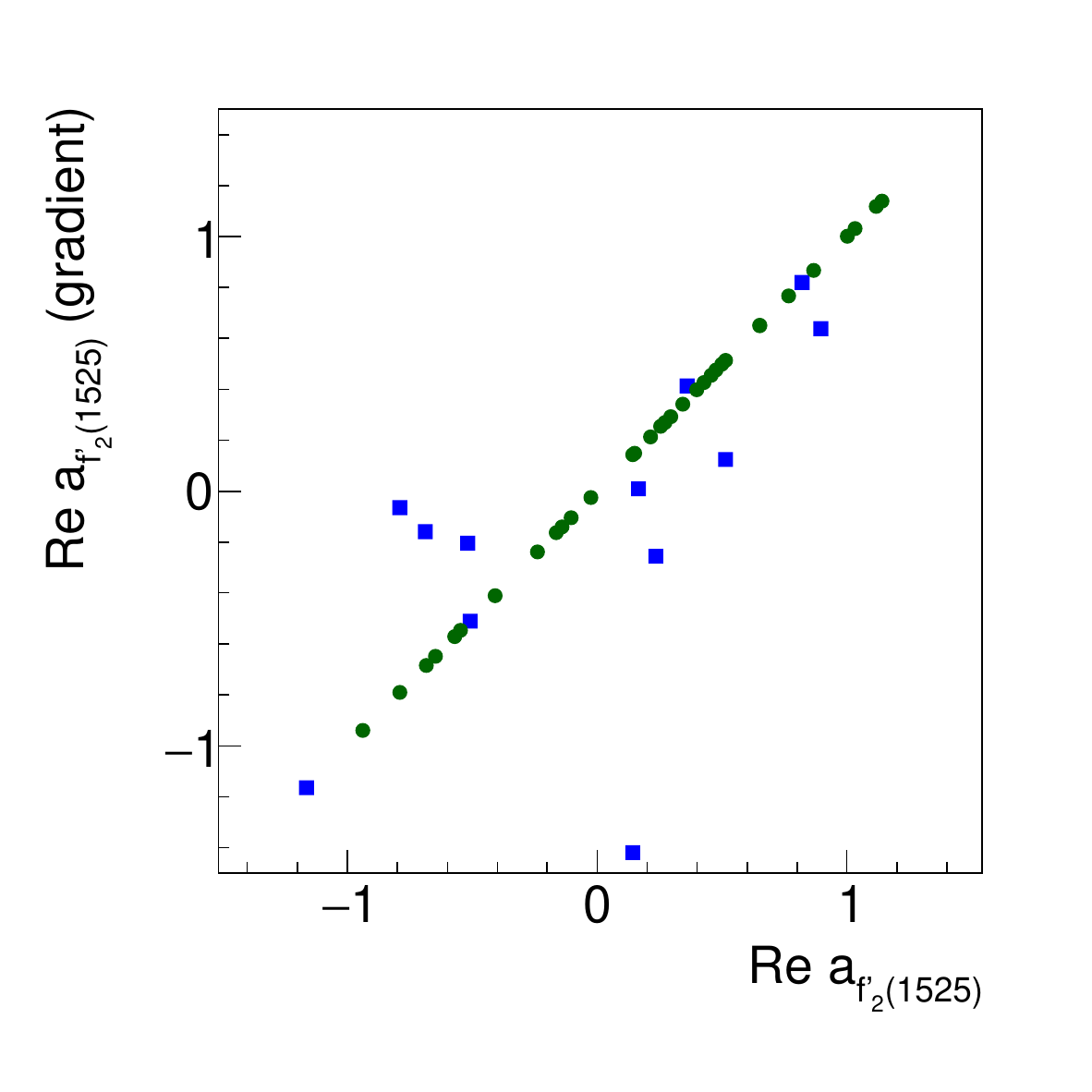}
\includegraphics[width=6.7cm]{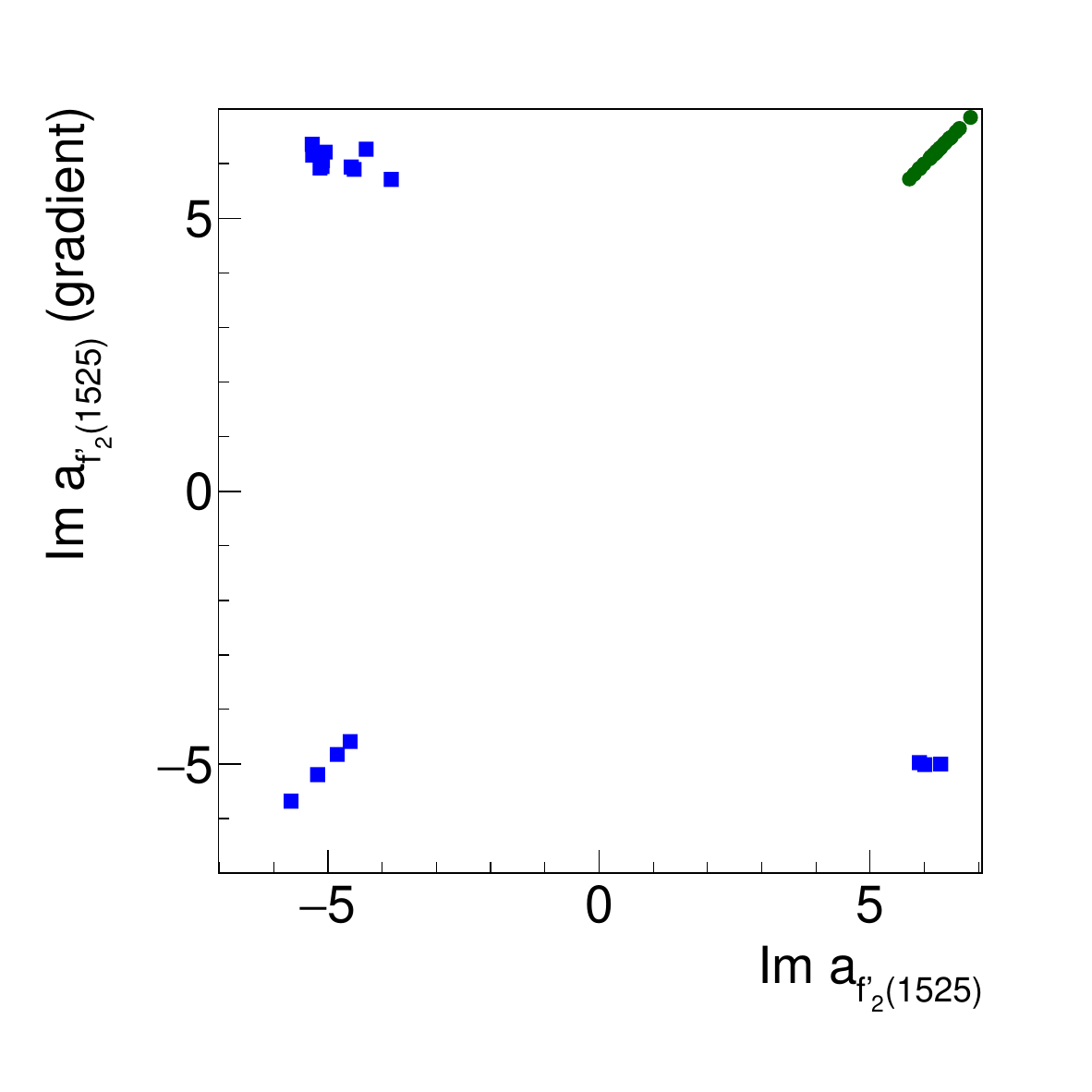} \\}
\caption{Comparison of real and imaginary parts
of the $f'_2(1525)$ amplitude for \vecampfit{} fits with and without gradient
and with the vector length equal to four.
Green circles are pseudoexperiments with both fits
converging to the true solution and blue squares are pseudoexperiments
with at least one of the fits converging to an alternative solution.}
\label{fig:laura_genfit3k_gradient_comparison}
\end{figure}

Comparison of \vecampfit{} fit results with and without gradient is
performed for validation of fitting with gradient calculation.
Fits with vector length equal to four are used.
Comparison of the real and imaginary parts of the $f'_2(1525)$ amplitude
is shown in Fig.~\ref{fig:laura_genfit3k_gradient_comparison}.
Similar to comparison with \laura{}, the results are the same
provided that both fits converge to the true minimum.
A similar comparison is performed for \vecampfit{} results obtained in
a single-thread and multiple-thread mode; it is found that the results
with a different number of threads are always the same.

The performance-test results are listed in Table~\ref{tab:laura}.
The \vecampfit{} fit is about 4 to 5 times faster than \laura{} fit
on both test machines, although note that a part
of this difference is caused by different normalization in \laura{} requiring
more calculations. Usage of gradient calculation reduces the fitting time
further, but the effect is not very large because this is a simple fit with
only 10 free parameters. Although the considered $\kpkpkm$ model
is very simple, runs fast, and does not require much optimization,
the performed comparison also estimates possible improvement from vectorization
and gradient calculation in more complex analyses. While \vecampfit{} fit
is faster, it is more difficult to implement due to the requirement
of manual implementation of vectorization and gradient calculation.

\begin{table}
\caption{Comparison of fitting time of the $\kpkpkm$
example for \vecampfit{} and \laura{} implementations. The time
is the total clock fitting time of 50 pseudoexperiments in seconds.
The ratios of \laura{} and \vecampfit{} time and \vecampfit{} times
with and without gradient is calculated for the optimal
\vecampfit{} vector length.}
\centering
\begin{tabular}{c|c|c}
Machine  & Machine 1 & Machine 2 \\
\hline
Generation (\laura{})                     &  179 &  130 \\
\laura                                    & 1435 & 1126 \\
\hline
\vecampfit{}, vector length 4             &  299 &  294 \\
\vecampfit{}, vector length 8             &  295 &  301 \\
\vecampfit{}, vector length 16            &  310 &  329 \\
\vecampfit{}, vector length 32            &  332 &  321 \\
\vecampfit{}, vector length 64            &  395 &  371 \\
\vecampfit{}, gradient, vector length 4   &  218 &  199 \\
\vecampfit{}, gradient, vector length 8   &  224 &  201 \\
\vecampfit{}, gradient, vector length 16  &  239 &  218 \\
\vecampfit{}, gradient, vector length 32  &  267 &  221 \\
\vecampfit{}, gradient, vector length 64  &  307 &  267 \\
\hline
\laura{} / \vecampfit{} without gradient  &  4.9  &  3.8 \\
\laura{} / \vecampfit{} with gradient     &  6.6  &  5.7 \\
\vecampfit{} (without / with gradient)    &  1.35 &  1.48 \\
\end{tabular}
\label{tab:laura}
\end{table}

\subsection{Comparison with {\sc TensorFlowAnalysis2}}

{\sc TensorFlowAnalysis2} (\tfa{}) is an amplitude-analysis framework based
on {\sc TensorFlow}~\cite{tfa2}. It does not have any tagged releases;
the development version with git commit
c4d03d8eb4 performed in December 2025
is used for performance comparison.
{\sc TensorFlow} version 2.17.0 and {\sc iminuit} 2.30.0 are used
for function evaluation and minimization, respectively.
Comparison between \vecampfit{} and \tfa{} is
performed using the $\kmpippiz$ example with 50 different
pseudoexperiments of 100000 events each and a common normalization sample
consisting of 1000000 events. The example is reimplemented using \tfa{},
and the pseudoexperiments are fitted with both \vecampfit{} and \tfa{}
fitting programs with random initial values of the amplitudes. The
amplitude caching
and parameter-releasing sequence are the same for \vecampfit{} and \tfa{} fits.
Note that the masses and widths of all resonances can be released, thus,
amplitudes of individual resonances cannot be added to the event buffer.

By default, \tensorflow{} sets the number of threads to the number of CPU
threads to achieve its maximum usage. This configuration is used for testing.
It is equivalent to {\sc OpenMP} behavior if the number of threads is not
set explicitly. Comparison of \vecampfit{} and \tfa{} performance
on CPU has been performed at the test machine 1.
By default, gradient is enabled for \tfa{} because it is computed automatically
and disabled for \vecampfit{} since it requires manual implementation,
but both versions with and without gradient calculation are tested for
both frameworks.
Versions with free amplitudes only and additional
releasing of masses or masses and widths are tested.
\tfa{} fits without JIT compilation are very slow on CPU
and they are not included into performance comparison.
Summary of performance tests is shown in Table~\ref{tab:vecampfit_tfa2}.
Explicit gradient calculation reduces the total CPU time, but for \tfa{}
it does not always improve the clock time.
The ratios of \tfa{} and \vecampfit{}
fitting times are shown in Table~\ref{tab:vecampfit_tfa2_ratio}.
For all fits, \vecampfit{} CPU time is about 5 times smaller,
while \vecampfit{} clock time is more than an order of magnitude better.
The larger difference in clock time is due to better utilization
of multiple threads by \vecampfit{}.

The \tfa{} fits require much more memory, most of which originates from
TensorFlow and Python interpreter heap allocations and shared-library code.
Particularly, comparison between the results with and without gradient
calculation shows that about 3 GB are necessary for \tensorflow{}-based
gradient calculation. \vecampfit{} only requires a buffer of
$N_\text{par} N_\text{threads} L$ floating-point values, where $N_\text{par}$
is the number of fit parameters, $N_\text{threads}$ is the maximum
number of OpenMP threads, and $L$ is the vector length. For the
$\kmpippiz$ example running with 12 threads, the gradient-buffer size
is only 28 kB. The difference of the memory usage contributes to the
large overall performance difference. In case of GPU offloading with gradient
calculation, \vecampfit{} also requires about 2 GB of additional memory
due to larger gradient-buffer size and increased GPU stack size.

\begin{table}
\caption{Comparison of fitting time of the $\kmpippiz$ example
for \vecampfit{} and \tfa{} implementations for CPU.
\vecampfit{} fit is performed in double precision with vector length 4.
\tfa{} performance without JIT compilation is significantly worse
and it is not included into comparison.
The clock and CPU times are in seconds and maximum memory usage is
in megabytes. The CPU time per event is the fit CPU time
divided by the numbers of pseudoexperiments and events.}
{ \centering
\begin{tabular}{c|c|c|c|c}
Test & Clock time & CPU time & CPU/event & Memory \\
\hline
\multicolumn{5}{c}{\vecampfit{}} \\
\hline
No gradient                     &   277 &  2470 & $4.5 \times 10^{-5}$ &  488 \\
No gradient, free $m$           &   434 &  4357 & $7.9 \times 10^{-5}$ &  488 \\
No gradient, free $m$, $\Gamma$ &   582 &  6089 & $11.1 \times 10^{-5}$ &  488 \\
Gradient                        &   195 &  1520 & $2.8 \times 10^{-5}$ &  488 \\
Gradient, free $m$              &   263 &  2291 & $4.2 \times 10^{-5}$ &  488 \\
Gradient, free $m$, $\Gamma$    &   289 &  2558 & $4.7 \times 10^{-5}$ &  488 \\
\hline
\multicolumn{5}{c}{\tfa{} with JIT compilation} \\
\hline
No gradient                     &  2898 &  8475 & $15.4 \times 10^{-5}$ & 1545 \\
No gradient, free $m$           &  4858 & 14584 & $26.5 \times 10^{-5}$ & 1750 \\
No gradient, free $m$, $\Gamma$ &  5947 & 18026 & $32.8 \times 10^{-5}$ & 1750 \\
Gradient                        &  3735 &  7934 & $14.4 \times 10^{-5}$ & 4650 \\
Gradient, free $m$              &  5549 & 12191 & $22.1 \times 10^{-5}$ & 4650 \\
Gradient, free $m$, $\Gamma$    &  5919 & 13299 & $24.2 \times 10^{-5}$ & 4650 \\
\end{tabular} \\}
\label{tab:vecampfit_tfa2}
\end{table}

\begin{figure}
{\centering
\includegraphics[width=6.7cm]{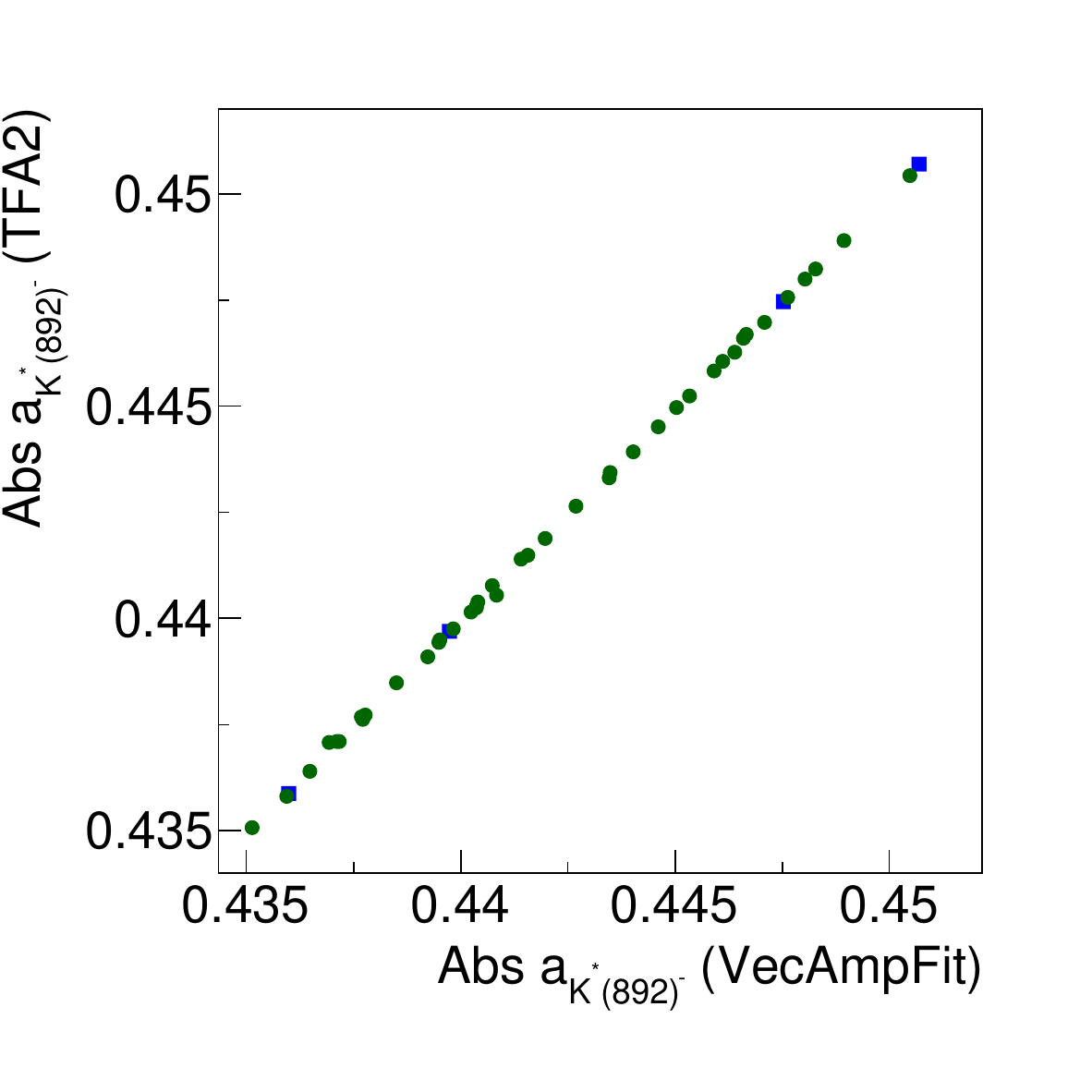}
\includegraphics[width=6.7cm]{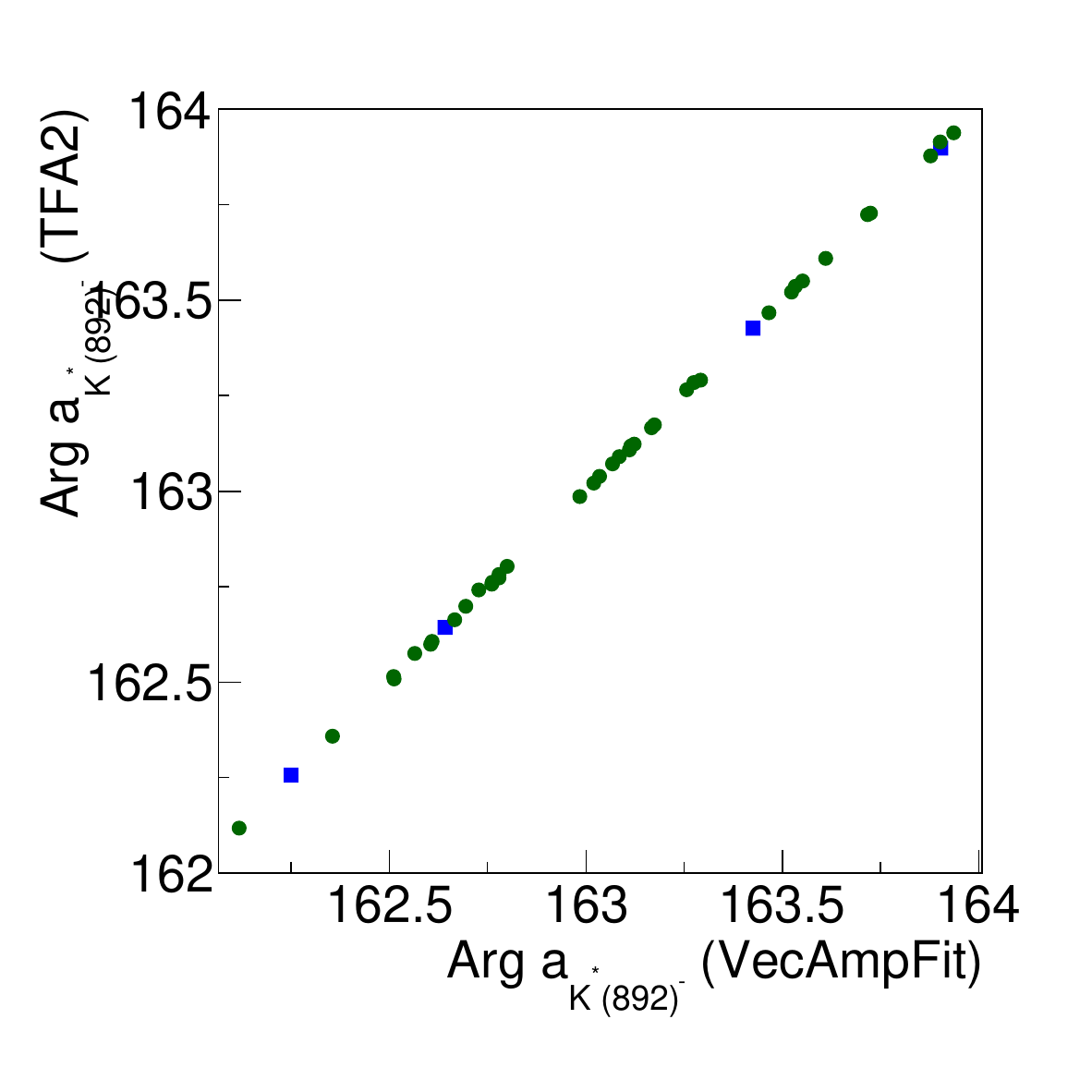} \\} 
\caption{Comparison of the absolute value and argument
of the $K^*(892)^-$ amplitude for \tfa{} and \vecampfit{} fits
with free resonance masses and widths.
Green circles are pseudoexperiments with both fits
converging to the true solution and blue squares are pseudoexperiments
with at least one of the fits converging to an alternative solution.}
\label{fig:tfa2_vecampfit_comparison}
\end{figure}

\begin{figure}
{\centering
\includegraphics[width=6.7cm]{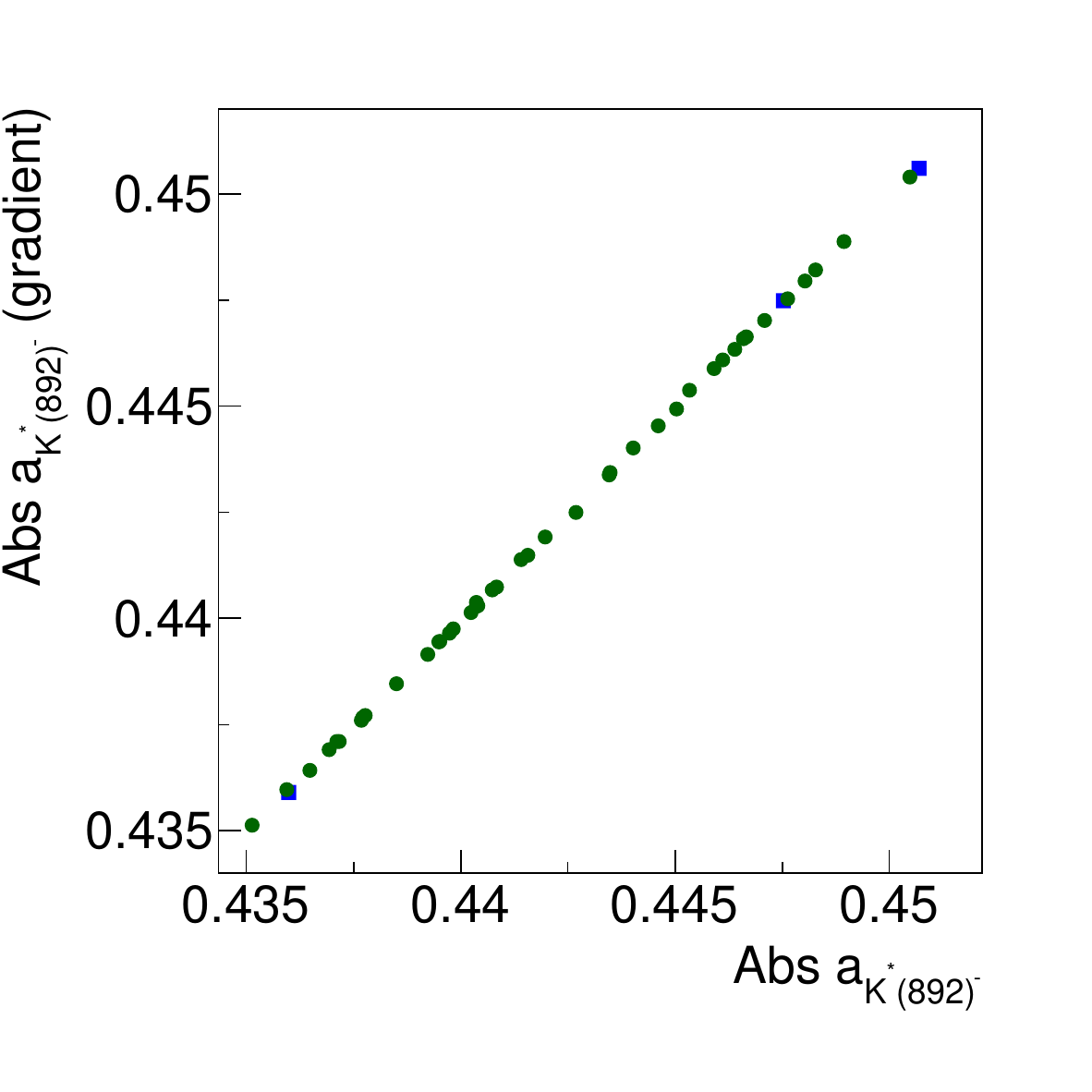}
\includegraphics[width=6.7cm]{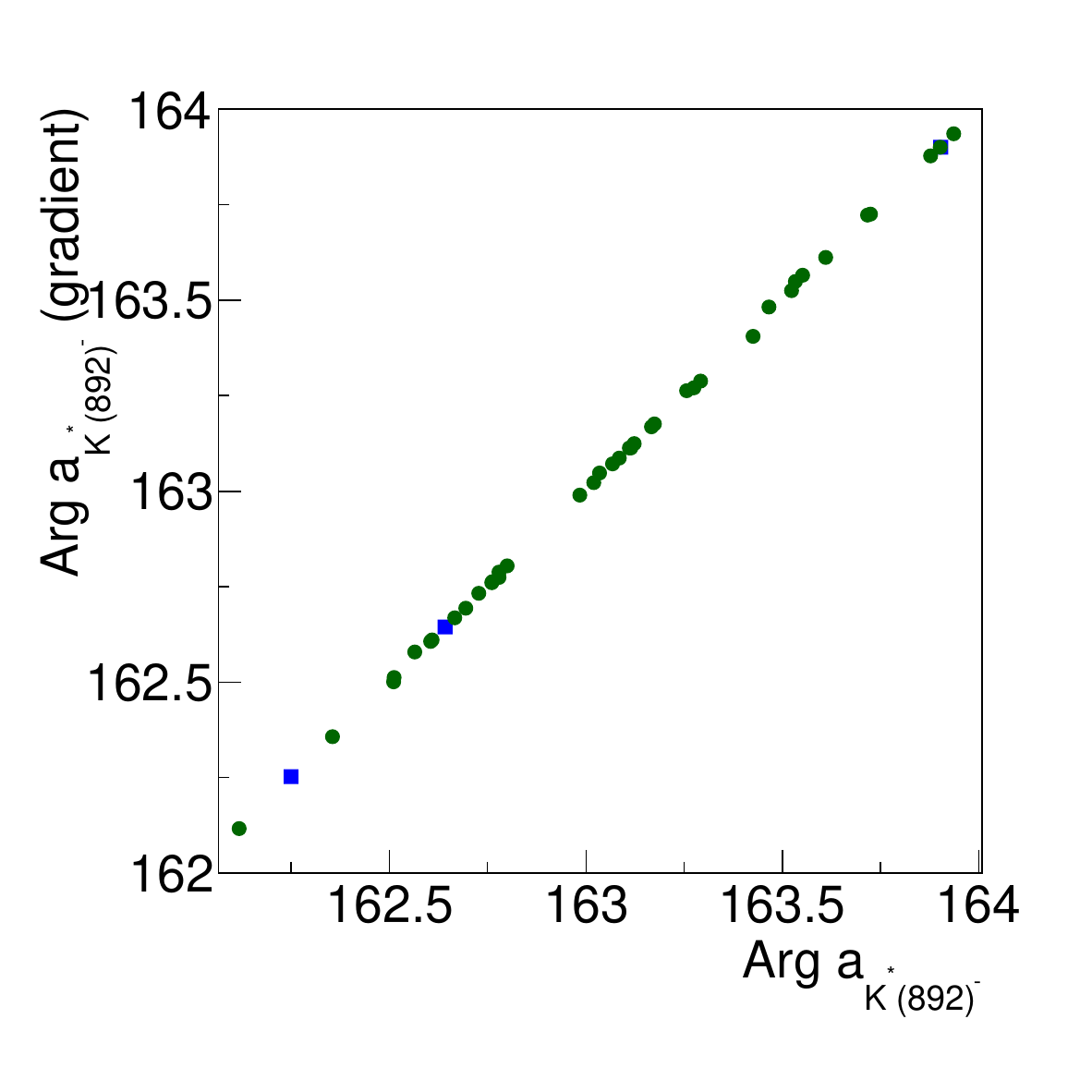} \\}
\caption{Comparison of the absolute value and argument
of the $K^*(892)^-$ amplitude for \vecampfit{} fits with and without gradient
with free resonance masses and widths.
Green circles are pseudoexperiments with both fits
converging to the true solution and blue squares are pseudoexperiments
with at least one of the fits converging to an alternative solution.}
\label{fig:dz_kmpippiz_gradient_comparison}
\end{figure}

Similar to \laura{} comparison, it is checked whether \vecampfit{} and
\tfa{} fits and \vecampfit{} fits with and without gradient calculation
converge to the same minimum.
Comparison of the absolute value and argument of the $K^*(892)^-$ amplitude
for \tfa{} and \vecampfit{} fits is shown
in Fig.~\ref{fig:tfa2_vecampfit_comparison} and comparison of the same
values for \vecampfit{} fits with and without gradient calculation
is shown in Fig.~\ref{fig:dz_kmpippiz_gradient_comparison}. Most fits converge
to the same minimum, but minor difference caused by the fits meeting
the convergence criteria and stopping at slightly different points near the
true minumum can be seen for some pseudoexperiments. The largest observed
difference between the likelihood values for the fits with and without
gradient calculation is 0.02.

\begin{table}
\caption{Ratios of the best \tfa{} and \vecampfit{} fitting times
for the $\kmpippiz$ example. The fits with free amplitudes,
free amplitudes and masses, and free amplitudes, masses, and widths have
15, 20, and 25 free parameters, respectively.
Clock times are compared for GPU fits.}
{ \centering
\begin{tabular}{c|c|c|c|c}
\multirow{2}{*}{Free parameters} & \multicolumn{2}{c|}{CPU} &
  \multirow{2}{*}{GT 710} & \multirow{2}{*}{A100} \\
\cline{2-3}
  & Clock time & CPU time & & \\
\hline
Amplitudes                & 14.9 & 5.2 & 1.1 & 3.3 \\
Amplitudes, $m$           & 18.5 & 5.3 & 0.9 & 2.8 \\
Amplitudes, $m$, $\Gamma$ & 20.5 & 5.2 & 0.7 & 2.5 \\
\end{tabular} \\}
\label{tab:vecampfit_tfa2_ratio}
\end{table}

Finally, \vecampfit{} and \tfa{} fitting times are compared for the case of
GPU offloading. The tests are performed at the first test
machine, which has an Nvidia GeForce GT 710 GPU with 2 GB of memory
(a very basic GPU that cannot be used for any real calculations,
but it is still possible to use it for testing), and a calculation server
equipped with an Nvidia A100. \tensorflow{} version 2.14.0, {\sc CUDA}
version 11.4.4 and {\sc CUDNN} version 8.7.0 are used by \tfa{} running
on the Nvidia GeForce GT 710. \tensorflow{} version 2.20.0, {\sc CUDA}
version 12.9.2 and {\sc CUDNN} version 9.24.1 are used by \tfa{} running
on the Nvidia A100.
Due to the limited performance of the Nvidia GeForce GT 710 GPU, the number
of pseudoexperiments used for comparison is reduced from 50 to 5,
the number of events in each pseudoexperiment is reduced from 100000 to 20000,
and the number of events in the normalization sample is reduced from 1000000
to 200000; overall, the amount of calculations is 50 times
smaller. The test using the Nvidia A100 performs the same calculations
as the CPU one. Comparison of \vecampfit{} CPU and GPU fit results
similar to the comparison of \vecampfit{} results with and without
gradient calculation shown in Fig.~\ref{fig:dz_kmpippiz_gradient_comparison}
is performed and it confirms that the CPU and GPU fit results are the same.
Optimal vector length is used for \vecampfit{}
GPU fits, which is found to be equal to four on the Nvidia GeForce GT 710
and eight on the Nvidia A100.

The test results are listed in Table~\ref{tab:vecampfit_tfa2_gpu_gt710}
for the Nvidia GeForce GT 710 and in Table~\ref{tab:vecampfit_tfa2_gpu_a100}
for the Nvidia A100. For both GPUs, the best \tfa{} fitting times are achieved
for the fits with JIT compilation and without gradient calculation.
\vecampfit{} fits without gradient calculation are faster than the
fits using the gradient. The ratios of the best \tfa{} and \vecampfit{}
fitting times are shown in Table~\ref{tab:vecampfit_tfa2_ratio}.
The performance is similar on the Nvidia GeForce GT 710, while on the
Nvidia A100 \vecampfit{} fits are 2.5 to 3.3 times faster.
The previous \vecampfit{} version (1.0.0) had much worse GPU performance
due to mixing of scalar OpenACC routine calls and vector loops in the same
main OpenACC parallel loop. The times of \vecampfit{} fits with gradient
calculation are similar to the fits without gradient on the Nvidia GeForce
GT 710; on the Nvidia A100, fitting with gradient is about 5 times slower.

\begin{table}
\caption{Comparison of fitting time of the $\kmpippiz$ example
for \vecampfit{} and \tfa{} implementations for the
Nvidia GeForce GT 710 GPU. The clock and CPU times
are in seconds and maximum memory usage is in megabytes.
The time per event is the fit clock time divided by the numbers
of pseudoexperiments and events.
The amount of calculations is 50 times less than for other tests
due to limited GPU performance.
\vecampfit{} memory usage is listed
for CPU (first value) and GPU (second value), while \tfa{} memory-usage
values are for CPU only: \tensorflow{} allocates almost entire available
GPU memory even if it is not actually needed, making a useful comparison
impossible.}
{ \centering
\begin{tabular}{c|c|c|c|c}
Test & Clock time & CPU time & Time/event & Memory \\
\hline
\multicolumn{5}{c}{\vecampfit{}} \\
\hline
No gradient                     &  355 &  353 & $1.61 \times 10^{-4}$ & 363 + 46 \\
No gradient, free $m$           &  696 &  694 & $3.16 \times 10^{-4}$ & 363 + 46 \\
No gradient, free $m$, $\Gamma$ &  983 &  981 & $4.47 \times 10^{-4}$ & 363 + 46 \\
Gradient                        &  466 &  463 & $2.12 \times 10^{-4}$ & 363 + 115 \\
Gradient, free $m$              &  837 &  835 & $3.80 \times 10^{-4}$ & 363 + 115 \\
Gradient, free $m$, $\Gamma$    &  857 &  855 & $3.89 \times 10^{-4}$ & 363 + 115 \\
\hline
\multicolumn{5}{c}{\tfa{} without JIT compilation} \\
\hline
No gradient                     & 1094 & 1109 & $4.97 \times 10^{-4}$ & 1504 \\
No gradient, free $m$           & 1924 & 1942 & $8.75 \times 10^{-4}$ & 1526 \\
No gradient, free $m$, $\Gamma$ & 2188 & 2206 & $9.95 \times 10^{-4}$ & 1524 \\
Gradient                        &  646 &  659 & $2.94 \times 10^{-4}$ & 1738 \\
Gradient, free $m$              & 1174 & 1189 & $5.33 \times 10^{-4}$ & 1850 \\
Gradient, free $m$, $\Gamma$    & 1219 & 1235 & $5.54 \times 10^{-4}$ & 1852 \\
\hline
\multicolumn{5}{c}{\tfa{} with JIT compilation} \\
\hline
No gradient                     &  394 &  409 & $1.79 \times 10^{-4}$ & 1951 \\
No gradient, free $m$           &  616 &  633 & $2.80 \times 10^{-4}$ & 1988 \\
No gradient, free $m$, $\Gamma$ &  707 &  724 & $3.22 \times 10^{-4}$ & 1987 \\
Gradient                        &  526 &  534 & $2.39 \times 10^{-4}$ & 2399 \\
Gradient, free $m$              &  \multicolumn{4}{c}{Not enough GPU memory} \\
Gradient, free $m$, $\Gamma$    &  \multicolumn{4}{c}{Not enough GPU memory} \\
\end{tabular} \\}
\label{tab:vecampfit_tfa2_gpu_gt710}
\end{table}

\begin{table}
\caption{Comparison of fitting time of the $\kmpippiz$ example
for \vecampfit{} and \tfa{} implementations for the Nvidia A100 GPU.
The clock and CPU times are in seconds and maximum memory usage is
in megabytes. The time per event is the fit clock time
divided by the numbers of pseudoexperiments and events.
\vecampfit{} memory usage is listed for CPU (first value) and GPU
(second value), while \tfa{} memory-usage values are for CPU only:
\tensorflow{} allocates almost entire available GPU memory even if it is not
actually needed, making a useful comparison impossible.}
{ \centering
\begin{tabular}{c|c|c|c|c}
Test & Clock time & CPU time & Time/event & Memory \\
\hline
\multicolumn{5}{c}{\vecampfit{}} \\
\hline
No gradient                     &   193 &   175 & $3.5 \times 10^{-6}$ & 564 + 2221 \\
No gradient, free $m$           &   273 &   261 & $5.0 \times 10^{-6}$ & 564 + 2221 \\
No gradient, free $m$, $\Gamma$ &   343 &   332 & $6.2 \times 10^{-6}$ & 564 + 2221 \\
Gradient                        &  1047 &  1030 & $19.0 \times 10^{-6}$ & 566 + 4161 \\
Gradient, free $m$              &  1549 &  1414 & $28.2 \times 10^{-6}$ & 566 + 4161 \\
Gradient, free $m$, $\Gamma$    &  1596 &  1544 & $29.0 \times 10^{-6}$ & 566 + 4161 \\
\hline
\multicolumn{5}{c}{\tfa{} without JIT compilation} \\
\hline
No gradient                     &   916 &  1339 & $16.7 \times 10^{-6}$ & 3750 \\
No gradient, free $m$           &  1353 &  1778 & $24.6 \times 10^{-6}$ & 3762 \\
No gradient, free $m$, $\Gamma$ &  1776 &  2005 & $32.3 \times 10^{-6}$ & 3771 \\
Gradient                        &   884 &  1307 & $16.0 \times 10^{-6}$ & 3972 \\
Gradient, free $m$              &  1264 &  1691 & $23.0 \times 10^{-6}$ & 4057 \\
Gradient, free $m$, $\Gamma$    &  1349 &  1778 & $24.5 \times 10^{-6}$ & 4064 \\
\hline
\multicolumn{5}{c}{\tfa{} with JIT compilation} \\
\hline
No gradient                     &   638 &  1088 & $11.6 \times 10^{-6}$ & 3816 \\
No gradient, free $m$           &   776 &  1324 & $14.1 \times 10^{-6}$ & 3853 \\
No gradient, free $m$, $\Gamma$ &   880 &  1436 & $16.0 \times 10^{-6}$ & 3853 \\
Gradient                        &  1408 &  1963 & $25.6 \times 10^{-6}$ & 4127 \\
Gradient, free $m$              &  1958 &  2671 & $35.6 \times 10^{-6}$ & 4301 \\
Gradient, free $m$, $\Gamma$    &  1979 &  2594 & $36.0 \times 10^{-6}$ & 4298 \\
\end{tabular} \\}
\label{tab:vecampfit_tfa2_gpu_a100}
\end{table}

\section{Conclusions}

A new library \vecampfit{} has been developed to perform amplitude analyses
of various processes in high-energy physics. It is designed to achieve high
performance of the analysis code by vectorization of data storage and
amplitude calculations and provides a number of vectorized subroutines
for physical quantities and mathematical functions.
The fitter supports explicit gradient calculation.
Simultaneous fitting of multiple data sets with related amplitudes is possible.
Specific tests have been carried out to compare \vecampfit{} performance
with \laura{} and {\sc TensorFlowAnalysis2}
and show a major improvement compared to both on CPU.
The performance of \vecampfit{} fits with GPU offloading
is similar or better than {\sc TensorFlowAnalysis2} depending on GPU.
However, offloading with gradient calculation may be much slower than
the offloading without it and requires further improvement;
\vecampfit{} offloading support is planned to be improved in future releases.

\appendix

\section{\vecampfit{} subprogram overview}
\label{sec:subprogram_overview}

\subsection{Auxiliary library parts}

\vecampfit{} Fortran processing tool {\tth vecampfit-fortran} performs parsing
and syntax analysis of a subset of the Fortran 2018 standard used by
the library. It is used during building to generate interface parts
of Fortran subprograms implemented in submodules
for their inclusion into modules, C headers containing declarations of
C-interoperable subprograms, and documentation.
The Fortran subprograms are documented in the code using {\sc Doxygen}-like
syntax.
It is possible to specify the scalar version of vectorized subprograms
and C interface if the subprogram has them. The documentation is generated in
LaTeX format and compiled into a PDF file during building.

The library {\tth vecampfit-real} contains
operations on real-number representation implemented in the modules
{\tth VECAMPFIT\_REAL\_NUMBER\_VECTOR\_\#LENGTH\#} and
{\tth VECAMPFIT\_REAL\_NUMBER\_BASE}. They include determination of
the real-number type (finite, infinity, or NaN),
extraction and setting of the mantissa, exponent, or sign bit,
scaling by a power of two by changing the exponent.
The library also includes
vectorized subprograms for conversion between integer and real numbers
from the module {\tth VECAMPFIT\_TYPE\_CONVERSION\_VECTOR\_\#LENGTH\#}
and transfer subprograms for getting of low and high parts of the types
{\tth INT(INT64)} and {\tth REAL(REAL128)} implemented in the module
{\tth VECAMPFIT\_TRANSFER}.

Arbitrary-precision real-number calculations
implemented in {\tth vecampfit-arbitrary-precision} are mostly written in C,
the interface being provided by the module {\tth VECAMPFIT\_APR}.
The arithmetic operations: addition, subtraction, multiplication, and division
are implemented with guaranteed precision and rounding. Rounding modes
to the nearest number, to zero, away from zero, to larger values, and
to smaller values are supported.
Only precisions proportional to 64 bits are currently supported.
The code is validated by comparing the calculation results for random
input values with the GNU {\sc MPFR} library~\cite{mpfr_2007}
(the {\sc MPFR} library itself could not be used because \vecampfit{}
is licensed under permissive Apache License version 2.0 preventing linking
with the {\sc MPFR} code licensed under the GNU GPL).
No difference is found. Performance of addition and subtraction is similar,
but performance of multiplication and division
is much worse since short multiplication and division algorithms used in
{\sc MPFR}~\cite{shortdiv} are not implemented in \vecampfit{}.
Error, exponential, logarithm, and power functions are implemented
using the arbitrary-precision arithmetic without guaranteed precision.
Due to its limitations, \vecampfit{} arbitrary-precision subsystem
is currently not recommended to be used as a general-purpose
arbitrary-precision library, however, it is sufficient for the needs
of \vecampfit{} itself.
The arbitrary-precision arithmetic operations and functions are used
for calculation of mathematical constants described
in Sec.~\ref{sec:mathematics}
and series coefficients for vectorized calculation of the logarithm
and exponential functions. Additionally, the arbitrary-precision calculations
are used for calculation of integration points and their weights
for Gauss-Legendre quadrature using the Newton's method with
the starting values from Ref.~\cite{petras_1999}.

\subsection{Complex-vector type}

The usual {\tth COMPLEX} type is not well suited for vectorization,
because the real and imaginary parts of complex-array elements are stored
in adjacent memory regions for each array element.
In order to get a vector of real or imaginary parts,
it is necessary to perform permutations of the data stored in memory.
Thus, a new type for complex vectors with real and imaginary parts stored
in separate arrays is introduced in modules
{\tth VECAMPFIT\_COMPLEX\_VECTOR\_\#LENGTH\#}:
\begin{lstlisting}[style=fortranstyle]
TYPE TYPE_WITH_KIND_LENGTH(COMPLEX_VECTOR)
  !> Real parts.
  REAL(__KIND) RE(__LENGTH)
  !> Imaginary parts.
  REAL(__KIND) IM(__LENGTH)
END TYPE
\end{lstlisting}
This type is used for calculations involving complex amplitudes, for example,
to store the results of the vectorized calculation of the Breit-Wigner
amplitude. The complex-vector modules provide vectorized subprograms
for calculations with complex vectors including addition, subtraction,
and multiplication of a complex vector and another complex vector,
real array, real variable, or complex variable;
assignment of a real or complex array or constant,
negation, inversion ($1 / x$), conjugation, absolute value,
and physical-sheet or unphysical-sheet square root.
Non-vectorized subprograms include getting and setting of individual elements
and sum of vector elements.

\subsection{Mathematical subprograms}
\label{sec:mathematics}

The module {\tth VECAMPFIT\_MATHEMATICAL\_CONSTANTS} provides several
mathematical constants in three precisions: {\tth REAL32}, {\tth REAL64},
and {\tth REAL128}. The constants are calculated by the library itself during
building using its arbitrary-precision operations. The list of constants
currently includes $\sqrt{2}$, $\ln 2$, $e$, $\pi$, $\sqrt{\pi / 2}$,
$\sqrt{\pi}$, $\sqrt{2 \pi}$, and coefficients for conversion between degrees
and radians.

The modules {\tth VECAMPFIT\_MATHEMATICAL\_FUNCTIONS\_VECTOR\_\#LENGTH\#}
contain implementations of general mathematical functions and
the module {\tth VECAMPFIT\_MATHEMATICAL\_FUNCTIONS} implements their
scalar versions. Similarly, other vectorized modules described below have
the corresponding scalar versions.
The mathematical functions include vectorized versions of exponential, floor,
logarithm, and power functions, calculation of
the Gaussian function, asymmetric Gaussian function,
asymmetric Gaussian functions with various non-Gaussian tails
(exponential, exponential and power-law, power-law, and power-law transiting
to exponential ones), Chebyshev polynomials~\cite{chebyshev},
K\"all\'en lambda function~\cite{Kallen:1964lxa}, and logistic function.

The modules {\tth VECAMPFIT\_GEOMETRY\_VECTOR\_\#LENGTH\#}
contain a function for determination of the difference between
two azimuthal angles in the range $[-\pi,\pi)$.
Linear interpolation is implemented in the modules
{\tth VECAMPFIT\_INTERPOLATION\_VECTOR\_\#LENGTH\#}.
The modules {\tth VECAMPFIT\_LINEAR\_ALGEBRA\_VECTOR\_\#LENGTH\#} contain
$LU$ decomposition and calculation of determinant of a vector of $2 \times 2$
complex matrices. The corresponding scalar module
includes conversion between compact and full forms of a symmetric matrix
and calculation of a symmetric-matrix eigenvectors and eigenvalues
performed using LAPACK~\cite{lapack}.
The module {\tth VECAMPFIT\_NUMERIC\_INTEGRATION} implements one-dimensional
Simpson integration, adaptive multidimensional Genz-Malik
integration~\cite{genz_malik_1980}, and a subroutine getting Gauss-Legendre
quadrature data: integration points and their weights.

Statistical functions are included into two modules: generalization
of Feldman-Cousins confidence intervals~\cite{Feldman:1997qc}
to asymmetric-Gaussian likelihood is implemented in the module
{\tth VECAMPFIT\_STATISTICS}, and the module {\tth VECAMPFIT\_MIXED\_SAMPLE}
contains subroutines for the mixed-sample
method~\cite{mixed_sample_schilling,mixed_sample_henze}.

\subsection{Physical subprograms}

Kinematic subprograms are implemented in the modules
{\tth VECAMPFIT\_KINEMATICS\_VECTOR\_\#LENGTH\#}
and include the covariant factor
given by Eq.~\eqref{eq:covariant_factor} (subroutine {\tth COVARIANT\_FACTOR}),
momentum of a daughter particle in the mother rest frame
for the decay $0 \to 1 + 2$ (subroutine {\tth DECAY\_MOMENTUM}) given by
\begin{equation}
q = \frac{\sqrt{(m_0^2 - (m_1 + m_2)^2)(m_0^2 - (m_1 - m_2)^2)}}{2 m_0},
\label{eq:decay_momentum}
\end{equation}
and closely-related two-body phase-space factor
({\tth PHASE\_SPACE\_2BODY\_COMPLEX}) given by
\begin{equation}
\rho = \frac{1}{16 \pi}\frac{\sqrt{(s - (m_1 + m_2)^2)(s - (m_1 - m_2)^2)}}{s},
\label{eq:phase_space_factor}
\end{equation}
where the Mandelstam variable $s$ is complex as this function is intended
to be used for determination of $T$-matrix poles.
The scalar module {\tth VECAMPFIT\_KINEMATICS} also has functions for conversion
of helicity-formalism variables into four-momenta and
Dalitz-plot subprograms: determination of cosine of the helicity angle,
range of the squared invariant mass $m_{23}^2$
depending on the squared invariant mass $m_{12}^2$,
and checking whether the squared invariant masses $m_{12}^2$ and $m_{23}^2$
are into the allowed kinematic region of the Dalitz plot.
Three and four-dimensional vectors are implemented in the modules
{\tth VECAMPFIT\_VECTOR3} and {\tth VECAMPFIT\_VECTOR4}.

The modules {\tth VECAMPFIT\_HIGH\_ENERGY\_PHYSICS\_VECTOR\_\#LENGTH\#}
contain high-energy-physics subprograms including the Blatt-Weisskopf
formfactors, various forms of the Breit-Wigner amplitude,
covariant factors given by Eq.~\eqref{eq:covariant_factor_j}
(subroutines {\tth COVARIANT\_FACTOR*}),
and $K$-matrix amplitude parameterization~\cite{Chung:1995dx}.
The Blatt-Weisskopf formfactors~\cite{blattweisskopf} are given by
Eq.~\eqref{eq:blatt_weisskopf}; squared formfactors and their ratios
(subroutines {\tth BLATT\_WEISSKOPF*}) are included into the library.

The Breit-Wigner amplitudes include the amplitude with known daughter momenta
and formfactors with a phase-space factor given by
Eq.~\eqref{eq:breit_wigner_q_f} (subroutines {\tth BREIT\_WIGNER\_Q\_F*})
and without the phase-space factor given by
Eq.~\eqref{eq:breit_wigner_q_f_nophsp}
(subroutines {\tth BREIT\_WIGNER\_Q\_F\_NOPHSP*}).
The Breit-Wigner amplitude with a constant width and invariant mass
in the imaginary part (subroutine {\tth BREIT\_WIGNER\_CONSTANT\_WIDTH\_M})
is given by
\begin{equation}
B(m) = \frac{1}{m_0^2 - m^2 - i m \Gamma_0}
\end{equation}
and similar amplitude with resonance mass in the imaginary part
({\tth BREIT\_WIGNER\_CONSTANT\_WIDTH\_M0})
\begin{equation}
B(m) = \frac{1}{m_0^2 - m^2 - i m_0 \Gamma_0}.
\end{equation}
The general form of the Breit-Wigner amplitude (subroutine
{\tth BREIT\_WIGNER\_GENERAL}) is given by
\begin{equation}
\begin{aligned}
B(m) & = \frac{\mathcal{N}(m)}
{m_0^2 - m^2 - i \mathcal{G}(m)}, \\
\end{aligned}
\end{equation}
where $m_0$ is the resonance mass, $m = \sqrt{s}$ is the invariant mass,
$\mathcal{N}(m) = \alpha g_a n_a(m)$ is the product of coupling and formfactor
coupling for the current decay channel,
and $\mathcal{G}(m) = m_0 \Gamma(m) = \sum_b g_b^2 \rho_b(m) n_b^2(m)$
is the sum of products of squared formfactor, phase-space factor,
and squared coupling for all channels~\cite{ParticleDataGroup:2024cfk}.
The Breit-Wigner matrix element (subroutine {\tth BREIT\_WIGNER\_M2})
is given by
\begin{equation}
B(m) = \frac{\left(\frac{q}{q_0}\right)^{2L + 1}
\frac{F_L^2(q)}{F_L^2(q_0)}}
{(m_0^2 - m^2)^2 + m_0^2 \Gamma^2(m)}.
\label{eq:breit_wigner_m2}
\end{equation}

Another group of high-energy-physics subprograms is related to $K$-matrix
amplitude~\cite{Chung:1995dx}. The $K$-matrix is given by
\begin{equation}
K_{\alpha \beta} = b_{\alpha \beta}
                   + \sum\limits_R \frac{g_\alpha g_\beta}{m_R^2 - s},
\end{equation}
where $b$ is the nonresonant part, $g$ are resonance couplings, and
$m_R$ are their bare masses.
The $P$-vector describing the production amplitude with the appropriate
pole structure is given by
\begin{equation}
P_{\alpha} = a_{\alpha} + \sum\limits_R A_R \frac{g_\alpha}{m_R^2 - s},
\end{equation}
where $a$ is the nonresonant production amplitude and $A_R$ is
resonance production amplitude. Subroutines for addition of poles
to both $K$-matrix and $P$-vector are available for one- and two-channel
$K$-matrices.

The $K$-matrix amplitude is given by
\begin{equation}
A = n [I - i K \rho n^2]^{-1} P,
\end{equation}
where $I$ is identity matrix, $\rho$ is diagonal matrix
with elements equal to phase-space factor given by
Eq.~\eqref{eq:phase_space_factor}, and $n$ is a diagonal kinematic-factor
matrix containing the centrifugal barrier near threshold and formfactor,
which is calculated as
\begin{equation}
n_k = \left(\frac{q}{q_0}\right)^L \frac{F(q)}{F(q_0)},
\end{equation}
where $k$ is channel number, $q$ is decay momentum, $q_0$ is momentum scale,
$L$ is the angular momentum, and $F$ is a formfactor.
Subroutines for calculation of the matrix $I - i K \rho n^2$ and their version
for $S$ wave without $n^2$ are available for one- and two-channel
$K$-matrices.

Momentum analytic continuation
(subroutine {\tth MOMENTUM\_ANALYTIC\_CONTINUATION}) is given by
\begin{equation}
q_\text{cont} =
\begin{dcases*}
q & for $m_0 \ge m_1 + m_2$, \\
i \sqrt{-|q|^2} & for $m_0 < m_1 + m_2$, \\
\end{dcases*}
\end{equation}
where $|q|^2$ is calculated as in Eq.~\eqref{eq:decay_momentum} with both
its sides squared. The same analytic continuation is used
in the Flatt\'e distribution~\cite{Flatte:1976xu}.

Quantum-mechanics subprograms are implemented in the module
{\tth VECAMPFIT\_QUANTUM\_MECHANICS}. Wigner $d$ functions~\cite{wigner_1931}
(function {\tth D\_FUNCTION}) are given by
\begin{equation}
\begin{aligned}
d^{j}_{m_1\,m_2}(\theta) = & \sqrt{(j + m_1)!(j - m_1)!(j + m_2)!(j - m_2)!} \\
& \times
\sum\limits_{k = \max(0,m - m_2)}^{\min(j + m_1, j - m_2)}
\frac{(-1)^k (\cos \frac{\theta}{2})^{2 j + m_2 - m_1 - 2 k}
  (-\sin \frac{\theta}{2})^{m_1 - m_2 + 2 k}}{
  (j - m_1 - k)!(j + m_2 - k)!(k + m_1 - m_2)!k!}. \\
\end{aligned}
\end{equation}
The module also contains several subroutines for calculation
of specific $d$ functions from $\cos \theta$ and $\sin \theta$.
The subroutine {\tth ANGULAR\_MOMENTUM\_RANGE} returns minimum and
maximum angular momentum allowed in a two-body decay taking into account
spin and parity of all particles and whether the parity is conserved.

\subsection{Other data-analysis functions}

The module {\tth VECAMPFIT\_GENERATION} implements subroutines
for generation of Monte Carlo for two-body and multi-body decays
with uniform phase-space distribution. The algorithm for generation of
multi-body decays is described in~\cite{James:1968gu}.

It is necessary to know the masses of the initial and final-state particles
for calculation of the amplitudes. The masses of weakly-decaying and narrow
states are provided as constants in \vecampfit, while the parameters of
the intermediate resonances are supposed to be represented by
the fit parameters. Parameter variables {\tth PDG\_2020},
{\tth PDG\_2022}, and {\tth PDG\_2024} are defined in the module
{\tth VECAMPFIT\_PARTICLE\_DATA},
containing the 2020~\cite{ParticleDataGroup:2020ssz},
2022~\cite{ParticleDataGroup:2022pth}, and 2024~\cite{ParticleDataGroup:2024cfk}
Particle Data Group data, respectively.

The fitter is written in Fortran but the data are stored using ROOT
C++ libraries. \vecampfit{} has several ROOT interface modules
providing Fortran interfaces for the classes {\tth TMinuit}, {\tth TFile},
{\tth TRandom1}, and {\tth TTree}. A small part of \vecampfit{} is written
for usage with ROOT, including calculation of helicity and azimuthal angles
using ROOT Lorentz vectors and data object for storage of value with errors.

\section{Derivation of $\kkpipi$ amplitude}
\label{sec:kkpipi_amplitude}

\begin{figure}
{\centering
\includegraphics[width=6cm]{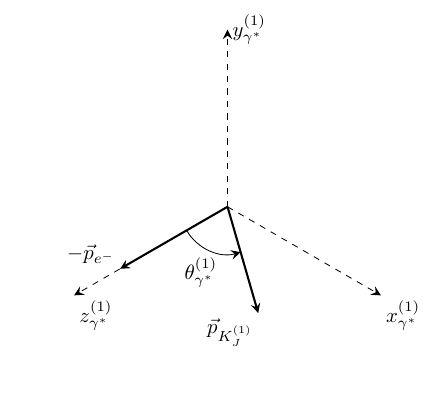} \\}
\caption{Definition of the $\vpho$ helicity angle (in the $\vpho$ rest frame).}
\label{fig:thetavpho1}
\end{figure}

Helicity amplitude is calculated for the decay chain $\vpho \to \kjI \kII$,
$\kjI \to \kI \piI \piII$ with
$\kjI \to \kstjIaI \piII$, $\kstjIaI \to \kI \piI$;
$\kjI \to \kstjIaII \piI$, $\kstjIaII \to \kI \piII$; or
$\kjI \to \rhoj \kI$, $\rhoj \to \piI \piII$.
The initial coordinate system is chosen in the virtual-photon rest frame
in such a way that the $z$ axis has the opposite direction to the electron beam.
The definition of the virtual-photon helicity angle $\theta_{\vpho}^{(1)}$
is shown in Fig.~\ref{fig:thetavpho1}. The superscript in parentheses
denotes the number of the decay chain.
This angle is given by
\begin{equation}
\cos \theta_{\vpho}^{(1)} = \frac{-({\vec{p}_{\elm}}\cdot{\vec{p}_{\kjI}})}
{|\vec{p}_{\elm}||\vec{p}_{\kjI}|},
\end{equation}
where $\vec{p}_{\elp}$ and $\vec{p}_{\kjI}$ are the momenta of $\elm$
and $\kjI$ in the $\vpho$ rest frame, respectively.
All other helicity angles defined below
can be calculated in a similar way.

\begin{figure}
{\centering
\includegraphics[width=6cm]{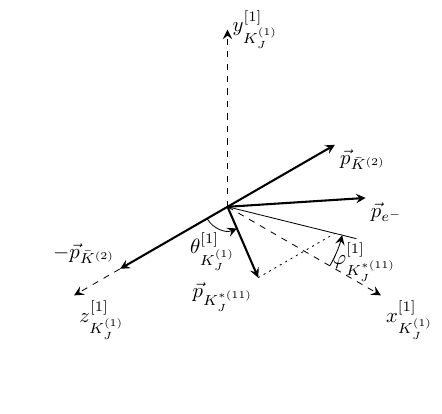} \\}
\caption{Definitions of the $\kjI$ helicity angle and
the $\kstjIaI$ azimuthal angle (in the $\kjI$ rest frame).}
\label{fig:thetakj1}
\end{figure}

The definitions of the $\kjI$ helicity angle $\theta_{\kjI}^{[1]}$ and
the $\kstjIaI$ azimuthal angle $\varphi_{\kstjIaI}^{[1]}$ are shown in
Fig.~\ref{fig:thetakj1}. The superscript in square
brackets denotes the number of the $\kjI$ decay chain.

\begin{figure}
{\centering
\includegraphics[width=6cm]{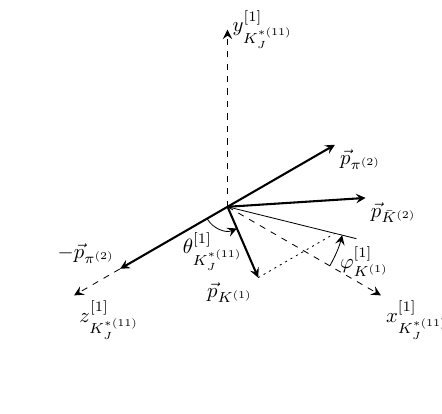} \\}
\caption{Definitions of the $\kstjIaI$ helicity angle and
the $\kI$ azimuthal angle (in the $\kstjIaI$ rest frame).}
\label{fig:thetakstj11}
\end{figure}

The azimuthal angle $\varphi_{\kstjIaI}^{[1]}$ is given by
\begin{equation}
\cos \varphi_{\kstjIaI}^{[1]} = \frac{({\vec{a}_{\elm}}\cdot{\vec{a}_{\kstjIaI}})}{|\vec{a}_{\elm}||\vec{a}_{\kstjIaI}|}, \\
\ \sin \varphi_{\kstjIaI}^{[1]} = \frac{-([\vec{p}_{\kII}\times\vec{a}_{\elm}]\cdot{\vec{a}_{\kstjIaI}})}{|\vec{p}_{\kII}||\vec{a}_{\elm}||\vec{a}_{\kstjIaI}|}, \\
\end{equation}
where
\begin{equation}
\vec{a}_{\elm} = \vec{p}_{\elm} - \frac{({\vec{p}_{\elm}}\cdot{\vec{p}_{\kII}})}{|\vec{p}_{\kII}|^2} \vec{p}_{\kII}, \\
\ \vec{a}_{\kstjIaI} = \vec{p}_{\kstjIaI} - \frac{({\vec{p}_{\kstjIaI}}\cdot{\vec{p}_{\kII}})}{|\vec{p}_{\kII}|^2} \vec{p}_{\kII}, \\
\end{equation}
where $\vec{p}_{\elm}$, $\vec{p}_{\kII}$, and $\vec{p}_{\kstjIaI}$
are the momenta of $\elm$, $\kII$, and $\kstjIaI$ in the $\kjI$ rest frame,
respectively. All other azimuthal angles defined below can be calculated
similarly.
The definitions of the $\kstjIaI$ helicity angle $\theta_{\kstjIaI}^{[1]}$ and
the $\kI$ azimuthal angle $\varphi_{\kI}^{[1]}$ are shown in
Fig.~\ref{fig:thetakstj11}.

The amplitude of the decay chain $\kjI \to \kstjIaI \piII$,
$\kstjIaI \to \kI \piI$ is
\begin{equation}
\begin{aligned}
A^{(\kjI \to \kstjIaI \piII)}_{\lambda_{\kjI}\,\lambda_{\kstjIaI}}(\Phi) = &
  H^{(\kjI \to \kstjIaI \piII)}_{\lambda_{\kstjIaI}\,0}
  \Dfun{J(\kjI)\,*}{\lambda_{\kjI}}{\lambda_{\kstjIaI}}{\varphi_{\kstjIaI}^{[1]}}{\theta_{\kjI}^{[1]}}{0} \\
& \times
  H^{(\kstjIaI \to \kI \piI)}_{0\,0}
  \Dfun{J(\kstjIaI)\,*}{\lambda_{\kstjIaI}}{0}{\varphi_{\kI}^{[1]}}{\theta_{\kstjIaI}^{[1]}}{0} \\
= &
  H^{(\kjI \to \kstjIaI \piII)}_{\lambda_{\kstjIaI}\,0}
  H^{(\kstjIaI \to \kI \piI)}_{0\,0}
  e^{i \lambda_{\kjI} \varphi_{\kstjIaI}^{[1]}} \\
& \times
  \dfun{J(\kjI)}{\lambda_{\kjI}}{\lambda_{\kstjIaI}}{\theta_{\kjI}^{[1]}}
  e^{i \lambda_{\kstjIaI} \varphi_{\kI}^{[1]}}
  \dfun{J(\kstjIaI)}{\lambda_{\kstjIaI}}{0}{\theta_{\kstjIaI}^{[1]}}. \\
\end{aligned}
\end{equation}
The selection rule is
\begin{equation}
|\lambda_{\kstjIaI}| \leq J(\kjI).
\end{equation}
The helicity amplitudes are related due to parity conservation:
\begin{equation}
H^{(\kjI \to \kstjIaI \piII)}_{-\lambda_{\kstjIaI}\,0} =
- P(\kjI) P(\kstjIaI) (-1)^{J(\kjI) - J(\kstjIaI)}
H^{(\kjI \to \kstjIaI \piII)}_{\lambda_{\kstjIaI}\,0}.
\end{equation}
The amplitude $H^{(\kstjIaI \to \kI \piI)}_{0\,0}$ can be omitted if one
needs to perform a fit to a single $\kkpipi$ process, but it is not possible
if it is necessary to fit multiple isospin-related processes simultaneously.

The amplitude including mass dependence summed over intermediate $\kstjIaI$
states is
\begin{equation}
\begin{aligned}
& A^{(\kjI \to \kstjIaI \piII)}_{\lambda_{\kjI}}(\Phi) \\
& \qquad =
  \sum\limits_{\kstjIaI}
  B_{\kstjIaI}(M_{\kI \piI}) \\
& \qquad\quad \times
\sum\limits_{\lambda_{\kstjIaI} = -J(\kjI)}^{J(\kjI)}
  H^{(\kjI \to \kstjIaI \piII)}_{\lambda_{\kstjIaI}\,0}
  H^{(\kstjIaI \to \kI \piI)}_{0\,0}
  e^{i \lambda_{\kjI} \varphi_{\kstjIaI}^{[1]}} \\
& \qquad\qquad\qquad\qquad\qquad \times
  \dfun{J(\kjI)}{\lambda_{\kjI}}{\lambda_{\kstjIaI}}{\theta_{\kjI}^{[1]}}
  e^{i \lambda_{\kstjIaI} \varphi_{\kI}^{[1]}}
  \dfun{J(\kstjIaI)}{\lambda_{\kstjIaI}}{0}{\theta_{\kstjIaI}^{[1]}} \\
& \qquad =
  e^{i \lambda_{\kjI} \varphi_{\kstjIaI}^{[1]}}
\sum\limits_{\lambda_{\kstjIaI} = -J(\kjI)}^{J(\kjI)}
  \dfun{J(\kjI)}{\lambda_{\kjI}}{\lambda_{\kstjIaI}}{\theta_{\kjI}^{[1]}} \\
& \qquad\quad \times
\sum\limits_{J(\kstjIaI)}
  e^{i \lambda_{\kstjIaI} \varphi_{\kI}^{[1]}}
  \dfun{J(\kstjIaI)}{\lambda_{\kstjIaI}}{0}{\theta_{\kstjIaI}^{[1]}} \\
& \qquad\qquad\qquad\quad \times
\sum\limits_{\kstjIaI}
  H^{(\kjI \to \kstjIaI \piII)}_{\lambda_{\kstjIaI}\,0}
  H^{(\kstjIaI \to \kI \piI)}_{0\,0}
  B_{\kstjIaI}(M_{\kI \piI}).
\end{aligned}
\end{equation}

\begin{figure}
{\centering
\includegraphics[width=6cm]{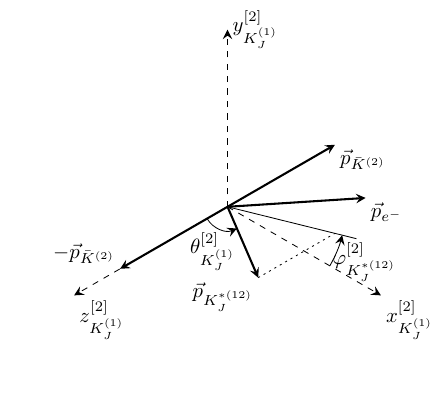} \\}
\caption{Definitions of the $\kjI$ helicity angle and
the $\kstjIaII$ azimuthal angle (in the $\kjI$ rest frame).}
\label{fig:thetakj2}
\end{figure}

\begin{figure}
{\centering
\includegraphics[width=6cm]{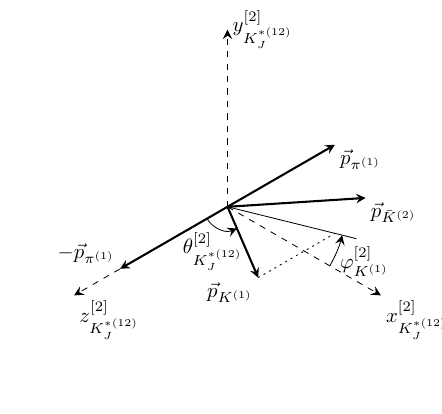} \\}
\caption{Definitions of the $\kstjIaII$ helicity angle and
the $\kI$ azimuthal angle (in the $\kstjIaII$ rest frame).}
\label{fig:thetakstj12}
\end{figure}

The definitions of the $\kjI$ helicity angle $\theta_{\kjI}^{[2]}$ and
the $\kstjIaII$ azimuthal angle $\varphi_{\kstjIaII}^{[2]}$ are shown in
Fig.~\ref{fig:thetakj2}.
The definitions of the $\kstjIaII$ helicity angle $\theta_{\kstjIaII}^{[2]}$ and
the $\kI$ azimuthal angle $\varphi_{\kI}^{[2]}$ are shown in
Fig.~\ref{fig:thetakstj12}.

The amplitude of the decay chain $\kjI \to \kstjIaII \piI$,
$\kstjIaII \to \kI \piII$ is
\begin{equation}
\begin{aligned}
A^{(\kjI \to \kstjIaII \piI)}_{\lambda_{\kjI}\,\lambda_{\kstjIaII}}(\Phi) = &
  H^{(\kjI \to \kstjIaII \piI)}_{\lambda_{\kstjIaII}\,0}
  \Dfun{J(\kjI)\,*}{\lambda_{\kjI}}{\lambda_{\kstjIaII}}{\varphi_{\kstjIaII}^{[2]}}{\theta_{\kjI}^{[2]}}{0} \\
& \times
  H^{(\kstjIaII \to \kI \piII)}_{0\,0}
  \Dfun{J(\kstjIaII)\,*}{\lambda_{\kstjIaII}}{0}{\varphi_{\kI}^{[2]}}{\theta_{\kstjIaII}^{[2]}}{0} \\
= &
  H^{(\kjI \to \kstjIaII \piI)}_{\lambda_{\kstjIaII}\,0}
  H^{(\kstjIaII \to \kI \piII)}_{0\,0}
  e^{i \lambda_{\kjI} \varphi_{\kstjIaII}^{[2]}} \\
& \times
  \dfun{J(\kjI)}{\lambda_{\kjI}}{\lambda_{\kstjIaII}}{\theta_{\kjI}^{[2]}}
  e^{i \lambda_{\kstjIaII} \varphi_{\kI}^{[2]}}
  \dfun{J(\kstjIaII)}{\lambda_{\kstjIaII}}{0}{\theta_{\kstjIaII}^{[2]}}. \\
\end{aligned}
\end{equation}
The selection rule is
\begin{equation}
|\lambda_{\kstjIaII}| \leq J(\kjI).
\end{equation}
The helicity amplitudes are related due to parity conservation:
\begin{equation}
H^{(\kjI \to \kstjIaII \piI)}_{-\lambda_{\kstjIaII}\,0} =
- P(\kjI) P(\kstjIaII) (-1)^{J(\kjI) - J(\kstjIaII)}
H^{(\kjI \to \kstjIaII \piI)}_{\lambda_{\kstjIaII}\,0}.
\end{equation}
The amplitude $H^{(\kstjIaII \to \kI \piII)}_{0\,0}$ can be omitted if one
needs to perform a fit to a single $\kkpipi$ process, but it is not possible
if it is necessary to fit multiple isospin-related processes simultaneously.

The amplitude including mass dependence summed over intermediate $\kstjIaII$
states is
\begin{equation}
\begin{aligned}
& A^{(\kjI \to \kstjIaII \piI)}_{\lambda_{\kjI}}(\Phi) \\
& \qquad =
\sum\limits_{\kstjIaII}
  B_{\kstjIaII}(M_{\kI \piII}) \\
& \qquad\quad \times
\sum\limits_{\lambda_{\kstjIaII} = -J(\kjI)}^{J(\kjI)}
  H^{(\kjI \to \kstjIaII \piI)}_{\lambda_{\kstjIaII}\,0}
  H^{(\kstjIaII \to \kI \piII)}_{0\,0}
  e^{i \lambda_{\kjI} \varphi_{\kstjIaII}^{[2]}} \\
& \qquad\qquad\qquad\qquad\qquad \times
  \dfun{J(\kjI)}{\lambda_{\kjI}}{\lambda_{\kstjIaII}}{\theta_{\kjI}^{[2]}}
  e^{i \lambda_{\kstjIaII} \varphi_{\kI}^{[2]}}
  \dfun{J(\kstjIaII)}{\lambda_{\kstjIaII}}{0}{\theta_{\kstjIaII}^{[2]}} \\
& \qquad =
  e^{i \lambda_{\kjI} \varphi_{\kstjIaII}^{[2]}}
\sum\limits_{\lambda_{\kstjIaII} = -J(\kjI)}^{J(\kjI)}
  \dfun{J(\kjI)}{\lambda_{\kjI}}{\lambda_{\kstjIaII}}{\theta_{\kjI}^{[2]}} \\
& \qquad\quad \times
\sum\limits_{J(\kstjIaII)}
  e^{i \lambda_{\kstjIaII} \varphi_{\kI}^{[2]}}
  \dfun{J(\kstjIaII)}{\lambda_{\kstjIaII}}{0}{\theta_{\kstjIaII}^{[2]}} \\
& \qquad\qquad\qquad\quad \times
\sum\limits_{\kstjIaII}
  H^{(\kjI \to \kstjIaII \piI)}_{\lambda_{\kstjIaII}\,0}
  H^{(\kstjIaII \to \kI \piII)}_{0\,0}
  B_{\kstjIaII}(M_{\kI \piII}).
\end{aligned}
\end{equation}

\begin{figure}
{\centering
\includegraphics[width=6cm]{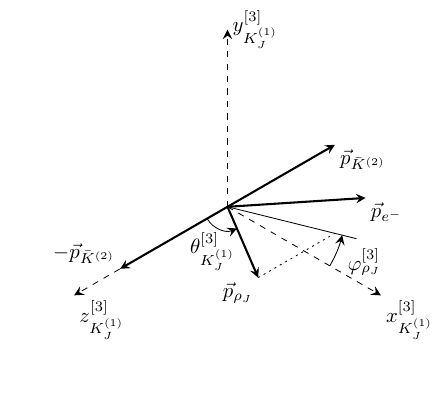} \\}
\caption{Definitions of the $\kjI$ helicity angle and
the $\rhoj$ azimuthal angle (in the $\kjI$ rest frame).}
\label{fig:thetakj3}
\end{figure}

\begin{figure}
{\centering
\includegraphics[width=6cm]{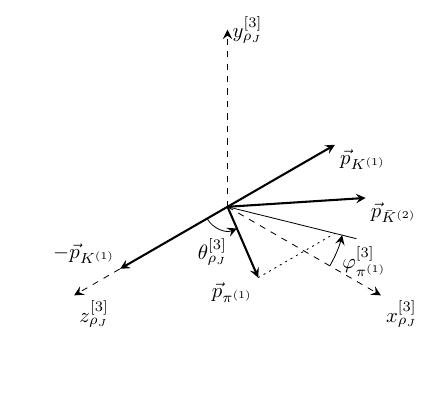} \\}
\caption{Definitions of the $\rhoj$ helicity angle and
the $\piI$ azimuthal angle (in the $\rhoj$ rest frame).}
\label{fig:thetarhoj}
\end{figure}

The definitions of the $\kjI$ helicity angle $\theta_{\kjI}^{[3]}$ and
the $\rhoj$ azimuthal angle $\varphi_{\rhoj}^{[3]}$ are shown in
Fig.~\ref{fig:thetakj3}.
The definitions of the $\rhoj$ helicity angle $\theta_{\rhoj}^{[3]}$ and
the $\piI$ azimuthal angle $\varphi_{\piI}^{[3]}$ are shown in
Fig.~\ref{fig:thetarhoj}.

The amplitude of the decay chain $\kjI \to \rhoj \kI$, $\rhoj \to \piI \piII$
is
\begin{equation}
\begin{aligned}
A^{(\kjI \to \rhoj \kI)}_{\lambda_{\kjI}\,\lambda_{\rhoj}}(\Phi) = &
  H^{(\kjI \to \rhoj \kI)}_{\lambda_{\rhoj}\,0}
  \Dfun{J(\kjI)\,*}{\lambda_{\kjI}}{\lambda_{\rhoj}}{\varphi_{\rhoj}^{[3]}}{\theta_{\kjI}^{[3]}}{0} \\
& \times
  H^{(\rhoj \to \piI \piII)}_{0\,0}
  \Dfun{J(\rhoj)\,*}{\lambda_{\rhoj}}{0}{\varphi_{\piI}^{[3]}}{\theta_{\rhoj}^{[3]}}{0} \\
= &
  H^{(\kjI \to \rhoj \kI)}_{\lambda_{\rhoj}\,0}
  H^{(\rhoj \to \piI \piII)}_{0\,0}
  e^{i \lambda_{\kjI} \varphi_{\rhoj}^{[3]}} \\
& \times
  \dfun{J(\kjI)}{\lambda_{\kjI}}{\lambda_{\rhoj}}{\theta_{\kjI}^{[3]}}
  e^{i \lambda_{\rhoj} \varphi_{\piI}^{[3]}}
  \dfun{J(\rhoj)}{\lambda_{\rhoj}}{0}{\theta_{\rhoj}^{[3]}}, \\
\end{aligned}
\end{equation}
where $H$ are helicity amplitudes of two-body decays
and $\lambda$ is the helicity of the particle specified by the index.
The selection rule is
\begin{equation}
|\lambda_{\rhoj}| \leq J(\kjI).
\end{equation}
The helicity amplitudes are related due to parity conservation:
\begin{equation}
H^{(\kjI \to \rhoj \kI)}_{-\lambda_{\rhoj}\,0} =
- P(\kjI) P(\rhoj) (-1)^{J(\kjI) - J(\rhoj)}
H^{(\kjI \to \rhoj \kI)}_{\lambda_{\rhoj}\,0}.
\end{equation}
The amplitude $H^{(\rhoj \to \piI \piII)}_{0\,0}$ can be omitted if one
needs to perform a fit to a single $\kkpipi$ process, but it is not possible
if it is necessary to fit multiple isospin-related processes simultaneously.

The amplitude including mass dependence summed over intermediate $\rhoj$
states is
\begin{equation}
\begin{aligned}
& A^{(\kjI \to \rhoj \kI)}_{\lambda_{\kjI}}(\Phi) \\
& \qquad = 
\sum\limits_{\rhoj}
  B_{\rhoj}(M_{\piI \piII})
\sum\limits_{\lambda_{\rhoj} = -J(\kjI)}^{J(\kjI)}
  H^{(\kjI \to \rhoj \kI)}_{\lambda_{\rhoj}\,0}
  H^{(\rhoj \to \piI \piII)}_{0\,0}
  e^{i \lambda_{\kjI} \varphi_{\rhoj}^{[3]}} \\
& \qquad\qquad\qquad\qquad\qquad\qquad\qquad\qquad \times
  \dfun{J(\kjI)}{\lambda_{\kjI}}{\lambda_{\rhoj}}{\theta_{\kjI}^{[3]}}
  e^{i \lambda_{\rhoj} \varphi_{\piI}^{[3]}}
  \dfun{J(\rhoj)}{\lambda_{\rhoj}}{0}{\theta_{\rhoj}^{[3]}} \\
& \qquad =
  e^{i \lambda_{\kjI} \varphi_{\rhoj}^{[3]}}
\sum\limits_{\lambda_{\rhoj} = -J(\kjI)}^{J(\kjI)}
  \dfun{J(\kjI)}{\lambda_{\kjI}}{\lambda_{\rhoj}}{\theta_{\kjI}^{[3]}} \\
& \qquad\quad \times
\sum\limits_{J(\rhoj)}
  e^{i \lambda_{\rhoj} \varphi_{\piI}^{[3]}}
  \dfun{J(\rhoj)}{\lambda_{\rhoj}}{0}{\theta_{\rhoj}^{[3]}}
\sum\limits_{\rhoj}
  H^{(\kjI \to \rhoj \kI)}_{\lambda_{\rhoj}\,0}
  H^{(\rhoj \to \piI \piII)}_{0\,0}
  B_{\rhoj}(M_{\piI \piII}). \\
\end{aligned}
\end{equation}

The amplitude of the decay $\kjI \to \kI \piI \piII$ is
\begin{equation}
\begin{aligned}
& A^{(\kjI \to \kI \piI \piII)}_{\lambda_{\kjI}}(\Phi) \\
& \qquad = B_{\kjI}(M_{\kI \piI \piII}) \\
& \qquad\quad \times \left(
  A^{(\kjI \to \kstjIaI \piII)}_{\lambda_{\kjI}}(\Phi)
+ A^{(\kjI \to \kstjIaII \piI)}_{\lambda_{\kjI}}(\Phi)
+ A^{(\kjI \to \rhoj \kI)}_{\lambda_{\kjI}}(\Phi)\right). \\
\end{aligned}
\end{equation}

The amplitude of the decay chain $\vpho \to \kjI \kII$,
$\kjI \to \kI \piI \piII$ is
\begin{equation}
\begin{aligned}
A^{\vpho \to \kjI \kII}_{\lambda_{\vpho}\,\lambda_{\kjI}}(\Phi) = &
  H^{(\vpho \to \kjI \kII)}_{\lambda_{\kjI}\,0}
  \Dfun{1\,*}{\lambda_{\vpho}}{\lambda_{\kjI}}{0}{\theta_{\vpho}^{(1)}}{0}
  A^{(\kjI \to \kI \piI \piII)}_{\lambda_{\kjI}}(\Phi) \\
= &
  H^{(\vpho \to \kjI \kII)}_{\lambda_{\kjI}\,0}
  \dfun{1}{\lambda_{\vpho}}{\lambda_{\kjI}}{\theta_{\vpho}^{(1)}}
  A^{(\kjI \to \kI \piI \piII)}_{\lambda_{\kjI}}(\Phi),
\end{aligned}
\end{equation}
and its sum over $\lambda_{\kjI}$ and intermediate $\kjI$ states is
\begin{equation}
\begin{aligned}
& A^{\vpho \to \kjI \kII}_{\lambda_{\vpho}}(\Phi) \\
& \qquad = 
\sum\limits_{\lambda_{\kjI}=-1,0,1}
  \dfun{1}{\lambda_{\vpho}}{\lambda_{\kjI}}{\theta_{\vpho}^{(1)}}
\sum\limits_{\kjI}
  H^{(\vpho \to \kjI \kII)}_{\lambda_{\kjI}\,0}
  B_{\kjI}(M_{\kI \piI \piII}) \\
& \qquad\quad \times \left(
  A^{(\kjI \to \kstjIaI \piII)}_{\lambda_{\kjI}}(\Phi)
+ A^{(\kjI \to \kstjIaII \piI)}_{\lambda_{\kjI}}(\Phi)
+ A^{(\kjI \to \rhoj \kI)}_{\lambda_{\kjI}}(\Phi)\right). \\
\end{aligned}
\end{equation}

\bibliographystyle{elsarticle-num}
\bibliography{vecampfit-bibliography.bib}

\end{document}